\documentclass[12pt]{ruthesis}
\usepackage{cite}
\usepackage{amsmath}
\usepackage{amssymb}
\usepackage{latexsym}
\usepackage{theorem}
\usepackage{graphicx}
\usepackage{textcomp}
\usepackage{multirow}
\usepackage[hidelinks]{hyperref}
\usepackage[utf8]{inputenc}
\usepackage{longtable}
\usepackage{pdflscape}
\usepackage{calligra}
\usepackage{emptypage}

\newcommand {\pt}{\ensuremath{p_\mathrm{T}}}
\newcommand {\PKzS}{\ensuremath{\mathrm{K^0_S}}}
\newcommand {\PgL}{\ensuremath{\Lambda}}
\newcommand {\PagL}{\ensuremath{\overline{\Lambda}}}
\newcommand {\noff}{\ensuremath{N_\mathrm{trk}^\mathrm{offline}}}
\newcommand {\nonline}{\ensuremath{N_\mathrm{trk}^\mathrm{online}}}
\newcommand {\ncorr}{\ensuremath{N_\mathrm{trk}^\mathrm{corrected}}}
\newcommand{\ngen}{\ensuremath{N_\text{trk}^\text{Gen}}}

\newcommand {\roots}{\ensuremath{\sqrt{s}}}
\newcommand {\rootsNN}  {\ensuremath{\sqrt{s_{_{NN}}}}}
\newcommand {\deta}     {\ensuremath{\Delta\eta}}
\newcommand {\dphi}     {\ensuremath{\Delta\phi}}
\newcommand {\pttrg}       {\ensuremath{p_\mathrm{T}^{\mathrm{trig}}}}
\newcommand {\ptass}       {\ensuremath{p_\mathrm{T}^{\mathrm{assoc}}}}
\newcommand {\ptref}       {\ensuremath{p_\mathrm{T}^{\mathrm{ref}}}}
\newcommand {\vnsig}    {\ensuremath{v_n^\mathrm{signal}}}
\newcommand {\vnbkg}    {\ensuremath{v_n^\mathrm{bkg}}}
\newcommand {\vnobs}    {\ensuremath{v_n^\mathrm{obs}}}
\newcommand{\ket}       {\ensuremath{KE_{\mathrm{T}}}}
\newcommand {\vsecsig}    {\ensuremath{v_2^\mathrm{signal}}}

\newcommand {\vtrdsig}    {\ensuremath{v_3^\mathrm{signal}}}

\theoremheaderfont{\itshape} {\theoremstyle{break}
} \theoremstyle{break}
 \theoremstyle{break}
 {\theoremstyle{plain}
  \theorembodyfont{\rmfamily}  }
{\theoremstyle{plain}
  \theorembodyfont{\rmfamily}  }

\title{Collective long-range particle correlations in proton-proton and proton-nucleus collisions at the LHC with the CMS detector}
\ctitle{Collective long-range particle correlations in proton-proton and proton-nucleus collisions at the LHC with the CMS detector}
\author{Zhenyu Chen}
\department{Physics and Astronomy}
\school{Rice University}
\degree{Doctor of Philosophy}

\committee {
        Wei Li, Chair \\
        Assistant Professor of Physics and Astronomy \and
        Frank Geurts \\
        Associate Professor of Physics and Astronomy \and
        Stephen Semmes\\
        Noah Harding Professor of Mathematics
}

\address{Houston, Texas}
\donemonth{June} \doneyear{2017} \makeindex
\begin{document}

  \begin{frontmatter}
   \pagenumbering{roman}
   \maketitle
   \null\newpage
	\thispagestyle{empty}
   \thispagestyle{empty}
\begin{abstract}

The observation of long-range two-particle angular correlations (known as the ``ridge'') in high final-state particle multiplicity (high-multiplicity) proton-proton
(pp) and proton-lead (pPb) collisions at the LHC has opened up new opportunities for studying novel dynamics of particle
production in small, high-density quantum chromodynamic (QCD) systems.
Such a correlation structure was first observed in relativistic nucleus-nucleus (AA) collisions at RHIC and the LHC. 
While extensive studies in AA collisions have suggested that the hydrodynamic collective flow
of a strongly interacting and expanding medium is responsible for these long-range correlation phenomenon, 
the nature of the ``ridge'' in pp and pPb collisions still remains poorly understood. 
A better understanding of the underlying particle correlation mechanisms requires detailed study of the properties of two-particle angular correlations in pp and pPb collisions. 
In particular, their dependence on particle species, and other aspects related to their possible collective nature, are the key to scrutinize various theoretical interpretations. 

Measurements of two--particle angular correlations of inclusive charged particles as well as identified 
strange hadron ($\PKzS$ or $\PgL$/$\PagL$) in pp and pPb collisions are presented over a wide
range in pseudorapidity and full azimuth. The data were collected using the CMS detector at the LHC, 
with nucleon-nucleon center-of-mass energy of 5.02 TeV for pPb collisions and 5, 7, 13 TeV for pp collisions.
The results
are compared to semi-peripheral PbPb collision data at center-of-mass energy of 2.76 TeV,
covering similar charged-particle multiplicity ranges of the events.
The observed azimuthal
correlations at large relative pseudorapidity are used to extract the second-order ($v_2$) and third-order ($v_3$) anisotropy Fourier
harmonics as functions of the charged-particle multiplicity in the event and
the transverse momentum (\pt) of the particles.

For high-multiplicity pp and pPb events, a clear particle species
dependence of $v_2$ is observed. For $\pt \lesssim 2$ GeV/c, the $v_2$ values of $\PKzS$ particles (lighter in mass)
are larger than those of $\PgL$/$\PagL$ particles at the same \pt. 
Such behavior is consistent with expectations in hydrodynamic models where a common velocity field is developed among all particles in the collision. 
When divided by the number of constituent quarks and compared at the same transverse
kinetic energy per quark, $v_2$ for $\PKzS$ particles are observed
to be consistent with those for $\PgL$/$\PagL$ particles in pp and pPb collisions over a broad range of particle transverse kinetic energy.
In AA collisions, this scaling behavior is conjectured to be related to quark recombination, which postulates that collective flow is developed among constituent quarks before they combine into final-state hadrons.

For high-multiplicity pp collisions at 13 TeV, the $v_2$ values obtained for inclusive charged particles with two-, 
four- and six-particle correlations are found to be comparable within uncertainties. 
This behavior is similar to what was observed in pPb and PbPb collisions. 
Together with the particle species dependence of $v_2$, 
these measurements provide strong evidence for the collective nature of the long-range correlations observed in pp collisions.

\end{abstract}

   \thispagestyle{empty}
   \vspace*{7cm}
   {\fontfamily{calligra}\selectfont \centerline{To my beloved wife, Yuan Zhao, and my parents, Ming Chen and Guihong Di.} }
   \clearpage
   \null\newpage
   \thispagestyle{empty}
   \tableofcontents
   \null\newpage
   \null\newpage
   \thispagestyle{empty}
  \end{frontmatter}
\pagenumbering{arabic}

\linespacing{1.7}

\chapter{Introduction}
\label{ch:Intro}

\section{Quarks, gluons and hadrons}
\label{sec:quark}
Quarks and gluons, together called partons, are the fundamental constituents of nuclear matter. 
By mediating the strong force between quarks through the color field, gluons hold quarks together to make composite particles known as hadrons, 
in a similar way as molecules are held together by the electromagnetic force mediated by photons. 
Quantum chromodynamics (QCD) is the theory of the strong force interactions, which is a non-abelian gauge theory 
with two peculiar features: 
\begin{itemize}
\item Confinement. As a quark-antiquark pair becomes separated, a narrow string of color field is formed between them. This is different from the behavior of the electric field between opposite charge pairs which extends and diminishes at large distance. Because of such behavior of the color field, as the separation increases, the strong force between the pair of quarks is almost constant regardless of the distance. The gluon binding potential between quark and antiquark is therefore proportional to the separation distance. At certain point of quark pair separation, it is more energetically favorable to create a new quark-antiquark pair instead of extending the string further. When such a new quark pair is created, the color field is separated into two regions that each region forms a hadron itself. This process prevents the creation of isolated, free quarks. Confinement refers to the nature that quarks in a group cannot be separated from their parent hadron. Based on the number of quarks, most of the hadrons are categorized into two families: baryons made of three quarks and mesons made of one quark and one antiquark. Recently, experimental evidences have been observed for tetraquark~\cite{Ablikim:2013mio,Wang:2014hta,Aaij:2014jqa,D0:2016mwd,Aaij:2016nsc,Aaij:2016iza} (composed of two quarks and two antiquarks) and pentaquark~\cite{Aaij:2015tga} (composed of four quarks and one antiquark).
\item Asymptotic freedom. As a result of the non-abelian gauge theory of QCD, the binding energy between quarks becomes weaker as energy exchanged in an interaction increases or distance between quarks decreases. This fundamental property of QCD predicts that quarks and gluons can exist in a deconfined state at high temperature or density named Quark Gluon Plasma (QGP)~\cite{Shuryak:1980tp}, which will be discussed in the following sections.
\end{itemize}

\section{Quark Gluon Plasma}
\label{sec:qgp}

The deconfined state of quarks and gluons is expected to be created with high nuclear densities. 
As the nuclear matter density increases, the hadrons are compressed together. 
Once the distance between hadrons is smaller than the radius of a single quark-antiquark pair, the quarks are not able to identify the original antiquark partner. 
This is similar to the Debye screening effect in electric plasma. 
Each quark is surrounded by numerous other quarks and gluons in a dense medium of quarks and gluons. 
The effective color field potential between quark-antiquark pair is screened such that the quark sees smaller effective color charge carried by the antiquark, resulting in less binding energy between the pair. 
As the nuclear matter density increases, the binding energy eventually drops to zero and the quarks are free to move over extended volume compared to the original volume of the hadron. 
Such a phenomena is referred as deconfinement, and the medium created is called the Quark Gluon Plasma. 
Similarly, the QGP can also be created with high temperature, as the increase in energy density of nuclear matter would result in creation of numerous quark-antiquark pairs from the vacuum, which also leads to the screening effect. 

\begin{figure}[htb]
\centering
\includegraphics[width=\linewidth]{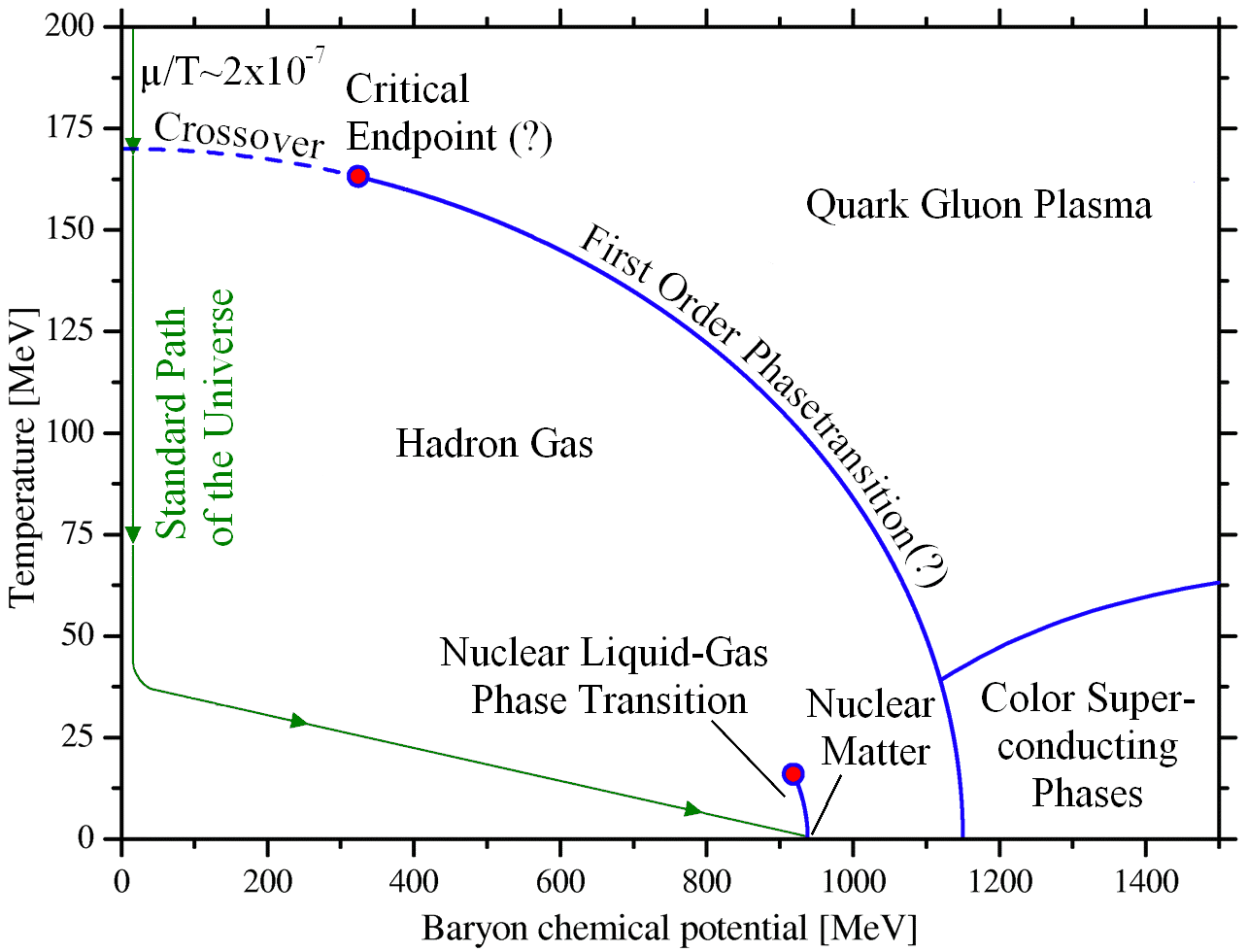}
  \caption{ \label{fig:qcdphase}
     Sketch of a possible QCD phase diagram~\cite{Boeckel:2011yj}. The green curve shows the commonly accepted standard evolution path of the universe as calculated e.g. in ~\cite{Fromerth:2002wb}.
     }
\end{figure}

Besides the formation of QGP, the QCD theory also provides understanding on the phase diagram for nuclear matter.
Figure~\ref{fig:qcdphase} summarizes the state-of-the-art QCD phase diagram including conjectures which are not fully established. 
Note here the QCD phase diagram is using chemical potential, proportional to the net baryon density, instead of nuclear matter density. 
At present, relatively firm statements can be made only in limited cases at finite T with small chemical potential and at asymptotically high chemical potential ($\gg \sim 200$ MeV). 
At low chemical potential region (around 0), the transition from hadrons to the QGP is predicted to be a cross-over by Lattice QCD calculations~\cite{Aoki:2006br,Bazavov:2011nk,Borsanyi:2011bn}, which occurs at critical temperature around 157 MeV.
At asymptotically high chemical potential region, the transition is predicted to be first order, while it is believed that there is a critical point connecting the two regions of phase transition.  
Apart from hadrons and QGP, a third form QCD phase is also predicted at high chemical potential and low temperature. 
It is referred as color superconductor which is believed to be the state of matter inside neutron stars~\cite{Alford:1997zt}. 

\section{Heavy ion collisions}
\label{sec:HIcol}

Ultra-relativistic heavy ion collisions were proposed to be one of the means to create QGP in the laboratory~\cite{Bjorken:1982qr}. 
Two nuclei are accelerated close to the speed of light and collide with each other. 
Tremendous amount of energy is deposited into the collision region through multiple inelastic nucleon-nucleon interactions. 
If the energy density reaches the value of phase transition, a QGP is expected to form. 

Nowadays, experiments at the Relativistic Heavy Ion Collider (RHIC) and the Large Hadron Collider (LHC) are the main facilities to study the formation and properties of QGP.
The QGP that is potentially created in trillion electron volts (TeV) energy level collisions at the LHC has low baryon chemical potential. 
This is because at large collision energies, baryons inside the colliding nucleus or ions will recede away from the center of mass without being completely stopped, leaving behind a QGP with very little net-baryon content. 
As the collision energy increases, baryon chemical potential gets smaller and initial temperature gets larger, which brings us close to what universe is believed to be shortly after the big-bang, shown as green curve in Fig.~\ref{fig:qcdphase}, 
despite that the QGP in early universe has much larger temperature compared to what can be reached at accelerators. 
On the other hand, QGP created in higher chemical potential region can be studied by analysing data from heavy ion collisions with lower energies at RHIC.

\begin{figure}[htb]
\centering
\includegraphics[width=0.8\linewidth]{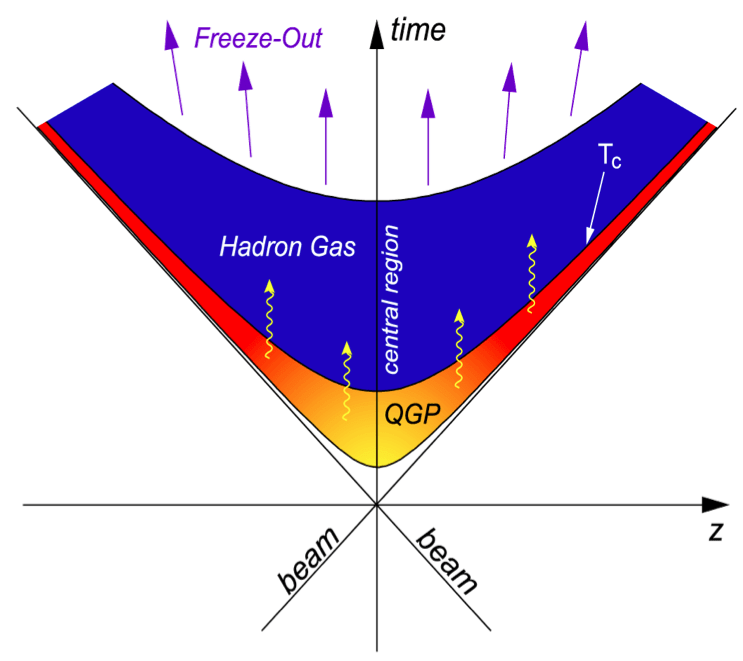}
  \caption{ \label{fig:HIevolve}
     Sketch of the space-time evolution (with only the longitudinal dimension z corresponding to the beam direction) of an ultra-relativistic heavy-ion collision. Taken from Ref.~\cite{Rosnet:2015rva}.
     }
\end{figure}

The relativistic heavy ion collision system evolves in several space-time stages as demonstrated in Fig.~\ref{fig:HIevolve}. 
The collision happens through multiple parton-parton scatterings.
Although the dynamics of the system is not well understood right after the collision, the QGP is expected to form within $\tau \approx$ 1fm/c after the collision~\cite{Bjorken:1982qr}. 
Further partonic scatterings inside the QGP quickly bring it to thermal equilibrium in a very short time~\cite{Harris:1996zx,Adams:2005dq}. 
As the scattering continues, the system expands in three dimensions while the temperature and chemical potential decrease. 
Once the temperature and chemical potential reach the phase transition critical values, the system starts to turn into a hadron gas, which happens at $\tau \approx$ 10 fm/c~\cite{Bass:2000ib}. 
After hadronization, the hadrons continue to interact with each other inelastically. 
When the inelastic hadronic interactions cease, particle species is frozen. Such a stage is called the chemical freeze-out. 
Elastic scatterings between particles continue until the stage of the kinetic freeze-out when all interactions between particles stop. 
The final state particles then free stream and reach the detector, carrying the information about the QGP and its evolution through various stages.

\subsection{Evidence of QGP in AA collisions}

As of today, QGP is believed to be created in nucleus-nucleus (AA) collisions at SPS, RHIC and the LHC. 
First measurements at RHIC from gold-gold (AuAu) collisions indicated that this new form of matter behaves almost as a perfect fluid with minimum viscosity~\cite{Adams:2005dq,Adcox:2004mh}. 
Such behavior was later confirmed by studies of lead-lead (PbPb) collisions at the LHC. 
The following paragraphs summarize the key experimental evidences of the existence of a fluid-like QGP matter in AA collisions. 

\paragraph{Quarkonium suppression.} Quarkonium is the name given to particles composed of a heavy quark and it's antiquark. 
Among the quarkonium family, $J/\Psi$ ($c\bar{c}$) and $\Upsilon$ ($b\bar{b}$) are the most studied particles in heavy-ion collisions. 
Due to the small binding energy and the color screening of the quark-antiquark potential in the hot and dense QGP medium, 
they are expected to dissociate. 
Therefore, if QGP is present in AA collisions, the production of $J/\Psi$ and $\Upsilon$ should be suppressed comparing to the production in pp collisions. 
Such a suppression has been demonstrated at RHIC~\cite{Tang:2010zza,Adamczyk:2012ey,Adamczyk:2013tvk,Adamczyk:2013poh,Adamczyk:2016dzv,Adamczyk:2016srz,Adare:2008sh,Adare:2011yf,Adare:2012wf,Aidala:2014bqx,Adare:2014hje,Adare:2015hva} and also at the LHC~\cite{Aad:2010aa,Abelev:2012rv,Abelev:2013ila,Abelev:2014nua,Adam:2015rba,Adam:2015isa,Adam:2016rdg,Chatrchyan:2011pe,Chatrchyan:2012np,Chatrchyan:2012lxa,Khachatryan:2014bva,Khachatryan:2016ypw,Khachatryan:2016xxp,Sirunyan:2016znt}.

\paragraph{Parton-medium interaction.} In high energy particle collisions, a parton in the projectile interacts with a parton in the target. 
A hard scattering is a process when the momentum transferred in the interaction is relatively large. 
In a hard scattering, the final partons gain large transverse energy and thus fragments into a shower of partons. 
These partons eventually hadronize into a cluster of hadrons which is called a jet. 
If a QGP medium is created, the hard scattering partons would exchange energy with the medium, and thus the energy of those partons and their fragmentation functions are modified compared to the case in vacuum. 
Those modifications have been observed at RHIC~\cite{Alver:2005nb,Adcox:2002pe,Adler:2003au,Adler:2005ee,Adare:2009vd,Afanasiev:2009aa,Adare:2010mq,Adare:2012qi,Adler:2002tq,Adams:2006yt,Abelev:2009ab,Adamczyk:2013jei,Adamczyk:2017yhe} and the LHC~\cite{Aad:2010bu,Aad:2012vca,Aad:2013sla,Aad:2014wha,Aad:2014bxa,Aad:2015wga,Aamodt:2010jd,Abelev:2012hxa,Abelev:2013kqa,Abelev:2014laa,Adam:2015ewa,Adam:2015doa,Adam:2015kca,Chatrchyan:2011sx,Chatrchyan:2012nia,Chatrchyan:2012gt,Chatrchyan:2012gw,Chatrchyan:2013kwa,Chatrchyan:2013exa,Chatrchyan:2014ava,Khachatryan:2015lha,Khachatryan:2016erx,Khachatryan:2016tfj,Khachatryan:2016odn,Sirunyan:2017jic}, through the study of jet energy modification (known as jet quenching), jet fragmentation function modification and high-\pt\ particle suppression. 

\paragraph{Collective flow.} Collective flow refers to the fact that particles move in a way which can be described by collective motion. It is considered to be strong evidence for a perfect-fluid-like medium created in heavy ion collisions.  
Analyses and results presented in this thesis are related to collective flow. Therefore it is discussed in more detail in Sec.~\ref{sec:flow}.

\subsection{Centrality classification}
\label{subsec:centrality}

\begin{figure}[htb]
\centering
\includegraphics[width=0.8\linewidth]{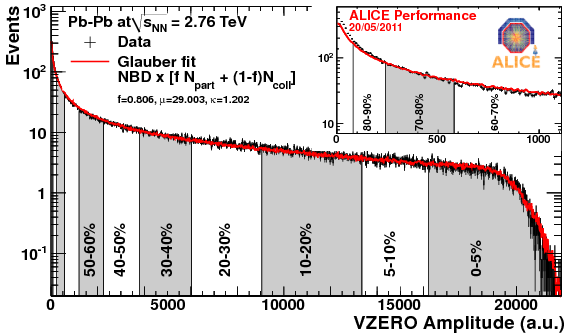}
  \caption{ \label{fig:ALICEcent}
     Distribution of the sum of amplitudes in the VZERO scintillators. The centrality classes are indicated in the figure. The inset shows a zoom of the most peripheral region. Taken from Ref.~\cite{Collaboration:2011rta}.
     }
\end{figure}

The size and evolution of the QGP medium created in a heavy ion collision depends on collision energy and geometry. 
Two nuclei do not always collide with each other head-on. 
The collision can happen with only a fraction of nuclei overlapping each other.
The impact parameter is used to quantify the collision geometry, defined as the distance between the centers of two colliding nuclei. 
Events with small impact parameters are called central events, while those with large impact parameters are called peripheral events. 
However, the impact parameters cannot be measured directly in heavy ion collision. 
Instead, experiments characterize AA collisions based on the total energy or particle multiplicity measured in the detector (often in the forward region)~\cite{Tuo:2014oca,Abelev:2013qoq}. 
Fig.~\ref{fig:ALICEcent} shows an example of centrality classification in PbPb collisions by ALICE collaboration with their forward VZERO detector~\cite{Collaboration:2011rta}. 
The VZERO amplitude distribution is used to divide the data sample into bins corresponding to the centrality fraction, where 0\% corresponds to most central collisions and 100\% corresponds to most peripheral collisions. 
With more energy deposited into the collision region, QGP is more likely to form in central events than in peripheral events. 

\section{Collective flow}
\label{sec:flow}

Collectivity in the context of heavy ion collisions means that a group of emitted particles exhibit a common velocity field or moves in a common direction. 
The common features of all emitted particles in a heavy ion collision is referred as collective flow, which can be indicators for the underlying nuclear matter phase space distribution. 
Collective flow can be categorized into several types: the longitudinal flow, the symmetric radial flow, and the azimuthal anisotropic flow. 
The collective motion of the particles in the direction defined by the beam is described by the longitudinal flow, which is not discussed in this thesis. 
The symmetric radial flow and the azimuthal anisotropic flow will be discussed in the following subsections.

\subsection{Radial flow} 
\label{subsec:radialflow}

\begin{figure}[htb]
\centering
\includegraphics[width=0.8\linewidth]{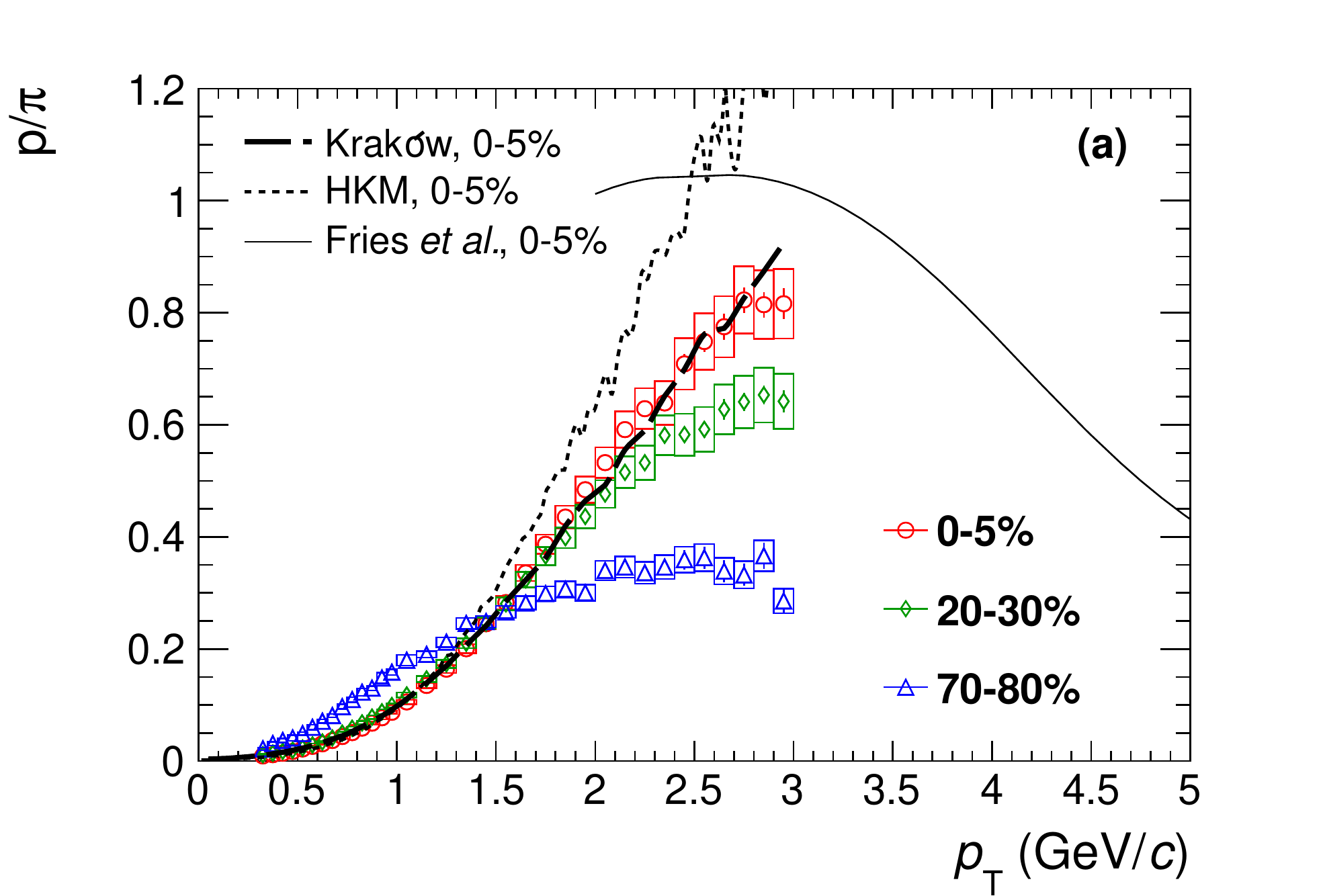}
  \caption{ \label{fig:radialflow}
     p/$\pi$ ratio as a function of transverse momentum for different centrality bins in 2.76 TeV PbPb collision measured by ALICE collaboration. Solid and dash lines are theory predictions. Taken from Ref.~\cite{Abelev:2013vea}.
     }
\end{figure}

Radial flow characterizes particles that are emitted from a source with a common velocity field and spherical symmetry. 
In heavy ion collision where a QGP is formed, a non-zero radial flow exists due to the radial expansion of the hot and dense medium driven by radial pressure gradient. 
Particles emitted from the collision experience a common velocity boost in the radial direction. 
The boost enhances particle momentum proportional to their mass. 
This effect is more prominent in central than in peripheral collisions because the higher energy density in the central collision results in a stronger boost. 
Therefore, it is expected that particle production ratio between a heavier particle and a lighter one to increase as a function of centrality at intermediate momentum  with a corresponding depletion at low momentum. 
Observation of such pattern has been made in AuAu~\cite{Abelev:2008ab} and PbPb~\cite{Abelev:2013vea} collisions. 
Fig.~\ref{fig:radialflow} shows an example measurement in PbPb collisions by ALICE~\cite{Abelev:2013vea}. 

\subsection{Azimuthal anisotropic flow} 
\label{subsec:vnflow}

\begin{figure}[htb]
\centering
\includegraphics[width=0.8\linewidth]{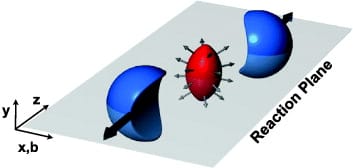}
  \caption{ \label{fig:v2shape}
     Almond-shaped interaction volume after a non-central collision of two nuclei, taken from Ref.~\cite{Snellings:2011sz}.
     }
\end{figure}

In a non-central heavy ion collision, the geometry of the overlap collision region in transverse plane has a almond shape in spatial coordinates, as illustrated in Fig.~\ref{fig:v2shape}. 
The collision region has a short axis parallel to the vector connecting the center of two nuclei. 
Together with the beam direction, the short axis vector defines a plane in 3D space called reaction plane, its azimuthal angle is denoted as $\Psi_{RP}$. 
Due to this initial geometry, the pressure gradient is asymmetric in azimuthal angle. 
The particles which are along the reaction plane are subject to a larger pressure gradient than the particles perpendicular to it. 
Through the expansion of QGP medium, azimuthal anisotropy is developed in final state momentum space, in a way that particles are boosted stronger in the reaction plane direction. 
The response of the final momentum anisotropy to the initial geometry depends on the interaction strength among the constituents. 
The stronger they interact, the larger momentum anisotropy develops. 
On the other hand, if the constituents are not interacting, i.e. they are not aware of the initial spatial geometry of the system, the momentum space would be uniform in azimuthal angle. 

Azimuthal anisotropic flow refers to the measurements of the momentum anisotropy of QGP medium. 
It is conveniently characterized by a Fourier expansion of the particle distributions, 
\begin{equation}
\frac{dN}{d\phi} = \frac{N}{2\pi} \left(1+2\sum_{n=1}^{\infty} v_{n}\cos[n(\phi-\Psi_{n})]\right),
\end{equation}
\noindent where $E$ is the energy of the particle, $p$ is the momentum, \pt\ is the transverse momentum, $\phi$ is the azimuthal angle, $y$ is the rapidity, and $\Psi_{RP}$ is the reaction plane angle. 
The sine terms in the Fourier expansion vanish due to reflection symmetry with respect to the reaction plane. 
The Fourier coefficients are given by 
\begin{equation}
v_{n} = <\cos[n(\phi-\Psi_{RP})]>,
\end{equation}
\noindent where the angular brackets denote an average over the particles summed over all events.
The $v_1$ and $v_2$ coefficients are known as the directed flow and elliptic flow. 

\paragraph{Directed flow.} Directed flow ($v_1$) describes collective sideward motion of produced particles and nuclear fragments.
It is believed to be mainly formed at early stages of the collisions and hence carries information on the early pressure gradients in the evolving medium~\cite{Herrmann:1999wu,Sorge:1996pc}. 
The $v_1$ coefficient has been studied as function of $y$ in heavy ion collisions at AGS and SPS~\cite{Barrette:1996rs,Alt:2003ab,Reisdorf:1997fx}, as well as at RHIC and the LHC~\cite{Back:2005pc,Adams:2005ca,Abelev:2008jga,Selyuzhenkov:2011zj,Adamczyk:2014ipa}.	
At low collision energies ($\approx$ \rootsNN\ 10 GeV), the results are consistent with predictions from a baryon stopping picture~\cite{Snellings:1999bt}, where a small negative slope of $v_1$ results as a function of rapidity for pions and an opposite slope for protons are observed. 
For high-energy collisions, both pions and protons have negative slope of $v_1$ near mid-rapidity, which is inconsistent with baryon stopping picture but consistent with predictions based on hydrodynamic expansion of a highly compressed, disk-shaped QGP medium, with the plane of disk initially tilted with respect to the beam direction~\cite{Bozek:2010bi}. 
Therefore, the $v_1$ measurements are considered as signature of QGP formation in high-energy heavy ion collision.

\begin{figure}[htb]
\centering
\includegraphics[width=0.8\linewidth]{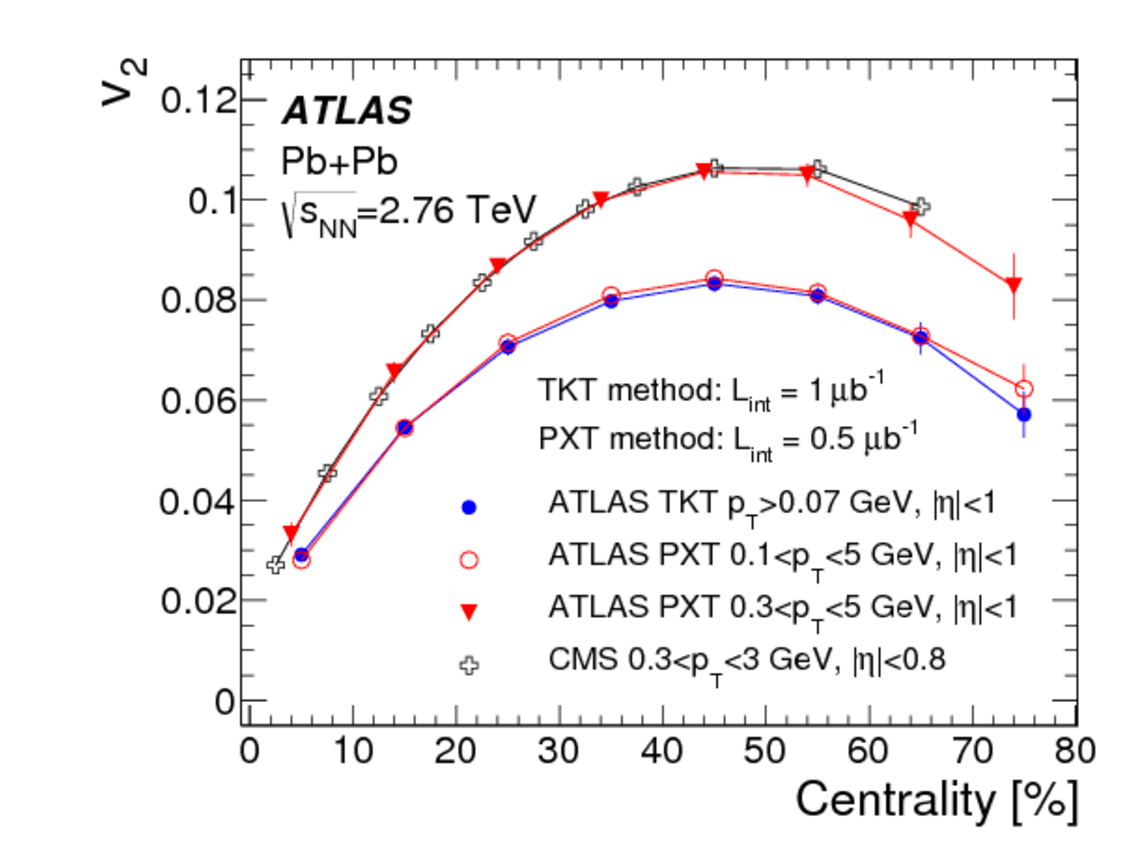}
  \caption{ \label{fig:v2vscent}
     Centrality dependence of elliptic flow, $v_2$, measured for $|\eta|<1$ and integrated over transverse momenta, \pt, for different charged-particle reconstruction methods as described in Ref.~\cite{Aad:2014eoa}. Also shown are $v_2$ measurements by CMS integrated over $0.3<\pt<3$ GeV/c and $|\eta|<0.8$ \cite{Chatrchyan:2012ta} (open crosses).
     }
\end{figure}

\begin{figure}[htb]
\centering
\includegraphics[width=0.95\linewidth]{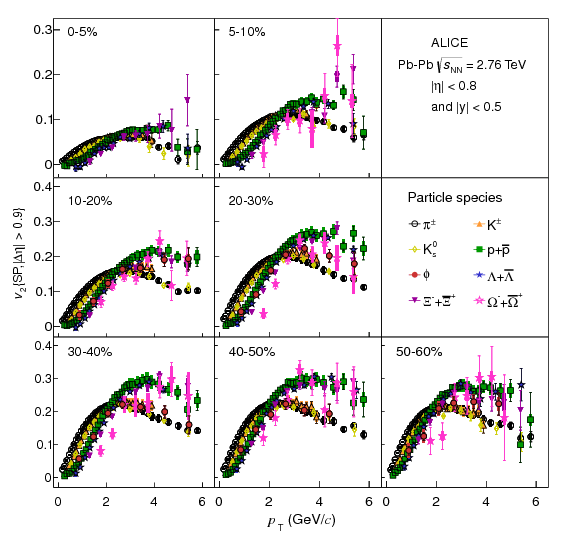}
  \caption{ \label{fig:v2massorder}
     The \pt\ differential $v_2$ for different particle species grouped by centrality class of PbPb collisions at 2.76 TeV, taken from Ref.~\cite{Abelev:2014pua}.
     }
\end{figure}

\begin{figure}[htb]
\centering
\includegraphics[width=0.95\linewidth]{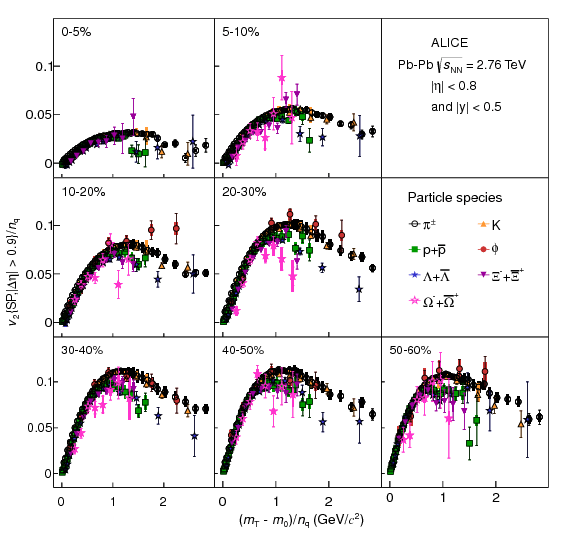}
  \caption{ \label{fig:ALICEncq}
     The $(m_T-m_0)/n_q$ dependence of $v_2/n_q$ for different particle species for Pb--Pb collisions at 2.76 TeV, taken from Ref.~\cite{Abelev:2014pua}.
     }
\end{figure}

\paragraph{Elliptic flow.} Elliptic flow ($v_2$) is a fundamental observable which directly reflects the initial spatial anisotropy of the nuclear overlapping region in the transverse plane defined as the plane perpendicular to the reaction plane and the beam direction. 
The large elliptic flow observed in AA collisions at top RHIC and LHC energies provides compelling evidence for strongly interacting matter which appears to behave like a perfect fluid when compared to hydrodynamics models~\cite{Gyulassy:2004zy}.
At those energies, elliptic flow tends to enhance momentum of emitted particles along the direction of the reaction plane. 
The strength of momentum enhancement, i.e. magnitude of measured $v_2$, is proportional to the initial eccentricity of the collision region, defined as
\begin{equation}
\epsilon_2 = \frac{<y^2-x^2>}{<y^2+x^2>},
\end{equation}
\noindent where ($x,y$) is the transverse plane spatial position of a participant nucleon inside the colliding nuclei, 
and the angular brackets are the average over all participant nucleons with unity weight. 
This proportionality results in a decrease of $v_2$ values from peripheral to central events. 
However,  there is a competing effect related to particle density of the collision systems. 
Comparing to central collisions, systems created in peripheral collisions tend to be more dilute. 
The initial eccentricity is less reflected in the final state particle momentum anisotropy due to the lack of interaction between particles in dilute systems. 
Combining the effects from initial eccentricity and particle density, $v_2$ in AA collision is expected to be small in most central collisions where initial eccentricity is small, and increase towards peripheral collisions, but decrease again in the very peripheral region due to low particle density. 
Fig.~\ref{fig:v2vscent} shows the $v_2$ results as function of centrality in 2.76 TeV PbPb collision measured by ATLAS collaboration~\cite{Aad:2014eoa}, which is consistent with the expectation. 
The particle species dependence of $v_2$ is of special interest. 
As discussed in Sec.~\ref{subsec:radialflow}, particles with different mass are momentum-boosted by the QGP medium with different strengths. 
The particle-species-dependent boost results in a stronger depletion of low \pt\ particles for heavier particles, which leads to a stronger decrease in the particle density. 
Therefore, $v_2$ of heavier particles is expected to be smaller than that of lighter particles at same \pt\ value. 
Such a observation has been made in AA collisions at RHIC and the LHC~\cite{Abelev:2014pua}, an example is shown in Fig.~\ref{fig:v2massorder}.
Furthermore, a universal scaling is discovered if $v_2$ per constituent quark ($n_q$) is plotted against transverse kinetic energy per constituent quark ($(m_T-m_0)/n_q$, where $m_T = \sqrt{m_0^2+\pt^2}$ and $m_0$ is the particle rest mass). It is denoted as number of constituent quark scaling (NCQ scaling). 
As shown in Fig.~\ref{fig:ALICEncq}, such a scaling indicates that all quarks share the same $v_2$, which is a strong support for deconfinement and that collectivity is developed in the partonic stage.

\begin{figure}[htb]
\centering
\includegraphics[width=0.75\linewidth]{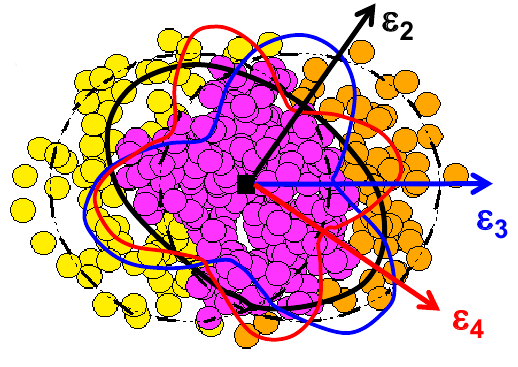}
  \caption{ \label{fig:lumpyIS}
     A demonstration of the non-uniform initial geometry of AA collision.
     }
\end{figure}

\paragraph{Participant fluctuations and higher-order flow.} Nowadays it is well-known that the event-by-event fluctuations in the initial geometry in heavy ion collisions lead to a lumpy initial state~\cite{Alver:2010gr,Teaney:2010vd}, as shown in Fig.~\ref{fig:lumpyIS}. 
Quantum fluctuations of the nucleon position inside a nuclei result in a non-uniform collision region instead of a smooth ellipse as in Fig.~\ref{fig:v2shape}. 
This non-uniform region can be decomposed into different shapes with different order of azimuthal asymmetry $\epsilon_n$. 
The odd-order $v_n$s are of particular interest, since they are purely created by the fluctuations in the initial state instead of the almond shape introduced by the nuclei. 
In AA collisions, higher order $v_n$ has been studied in detail at RHIC and the LHC~\cite{Abelev:2014pua}.
Similar behaviors as for $v_2$ have been observed such as mass ordering and NCQ scaling.

\clearpage

\section{Hydrodynamics in heavy ion collisions}
\label{sec:hydro}

The dynamics of the QGP expansion and collective flow can be described using QCD with Lagrangian density
\begin{equation}
L = \bar{\Psi}_i\left(i\gamma_{\mu}D^{\mu}_{ij}-m\delta_{ij}\right)\Psi_{j}-\frac{1}{4}F_{\mu\nu\alpha}F^{\mu\nu\alpha},
\end{equation}
\noindent where $\Psi_{i}$ is a quark field ($i =1,2,3$ is the color index for quarks), $D^{\mu}$ is a covariant derivative, $m$ is a quark mass, $F^{\mu\nu\alpha}$ is a field strength tensor of gluons, and $\alpha=1,2,...,8$ is the color index for gluons. 
Although this Lagrangian looks very simple, prediction in the heavy ion collision system is difficult. 
The complexity arises from non-linearity of gluon interactions, dynamical many body system and color confinement. 
All together, they make it almost impossible to do any precise QCD calculation in heavy ion collision. 
Therefore, to connect the first principle with phenomena, hydrodynamics (hydro) is introduced as a phenomenological approach to describe the heavy ion collision data. 

In hydrodynamical description, the space time evolution of QCD matter is determined  by conservation laws. 
The basic equations are energy-momentum conservation 
\begin{equation}
\partial_\mu T^{\mu\nu} = 0, 
\end{equation}
\noindent where $T^{\mu\nu}$ is the energy-momentum tensor and the current conservation 
\begin{equation}
\partial_\mu N_{i}^{\mu} = 0, 
\end{equation}
\noindent where $N_{i}^{\mu}$ is the conserved current in heavy ion collision such as baryon number, strangeness, and electric charge. 
In the relativistic ideal fluid approximation with zero viscosity, the equations can be solved analytically, with the assumption of boost invariant expansion and a homogeneous medium in the transverse plane~\cite{Bjorken:1982qr}. 
Once viscosity of the relativistic fluid is taken into consideration, the decomposition of energy-momentum tensor gets rather lengthy~\cite{Baier:2007ix,Israel:1976tn,Israel:1979wp}. 
Numerical hydrodynamic frameworks are needed to treat the dynamics of the initial matter properly, and to incorporate event-by-event differences in the initial collision geometry. 
Hydrodynamic frameworks which keep the assumption of boost invariant expansion and solve the medium evolution only in transverse plane and time are called 2+1D~\cite{Song:2007ux,Song:2013gia}. 
Because the boost-invariant assumption starts to fail at large rapidity in heavy ion collisions~\cite{Back:2001bq,Chatrchyan:2012mb}, they describe experimental data well at mid-rapidity but starts to deviate when comparing to measurements with large rapidity. 
Therefore, the state-of-art hydrodynamic frameworks are 3+1D including the longitudinal dynamics as well~\cite{Karpenko:2013wva,Schenke:2011bn}. 

\begin{figure}[htb]
\centering
\includegraphics[width=0.3\linewidth]{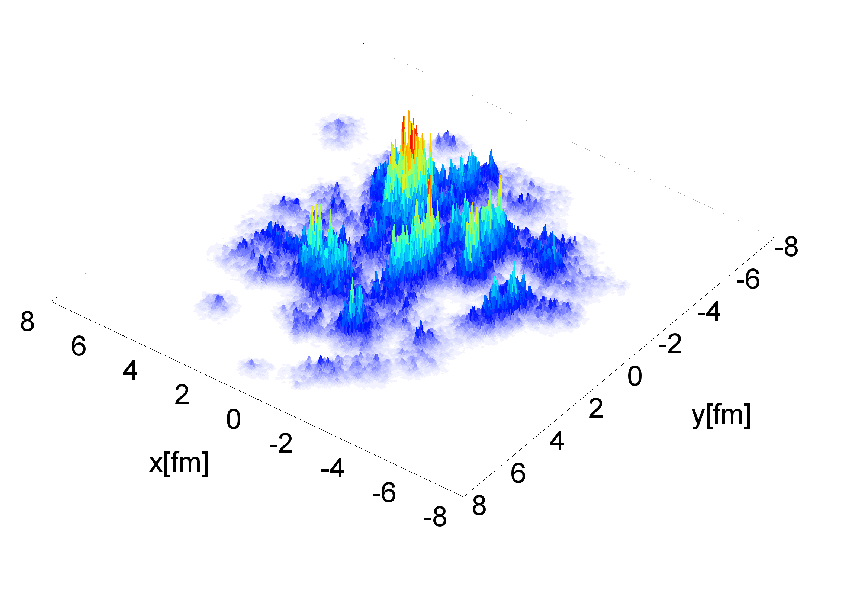}
\includegraphics[width=0.3\linewidth]{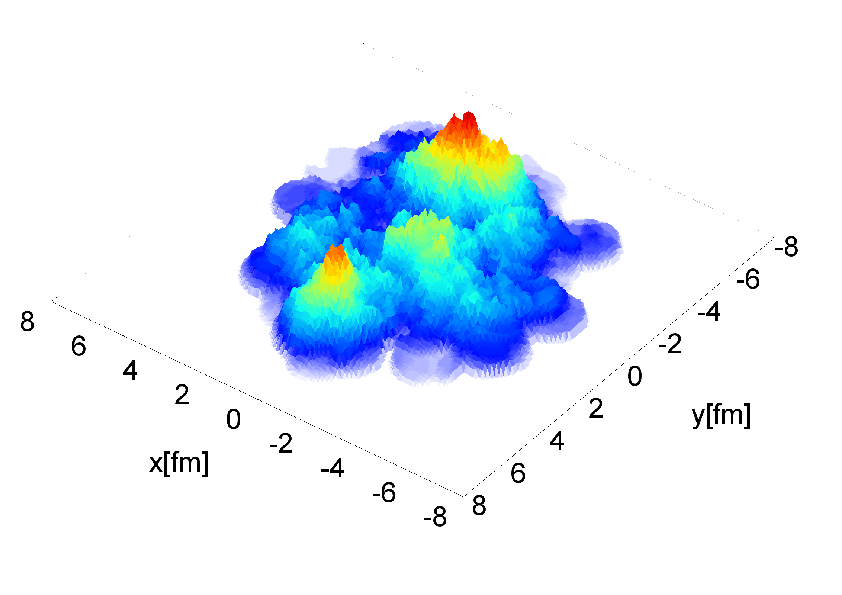}
\includegraphics[width=0.3\linewidth]{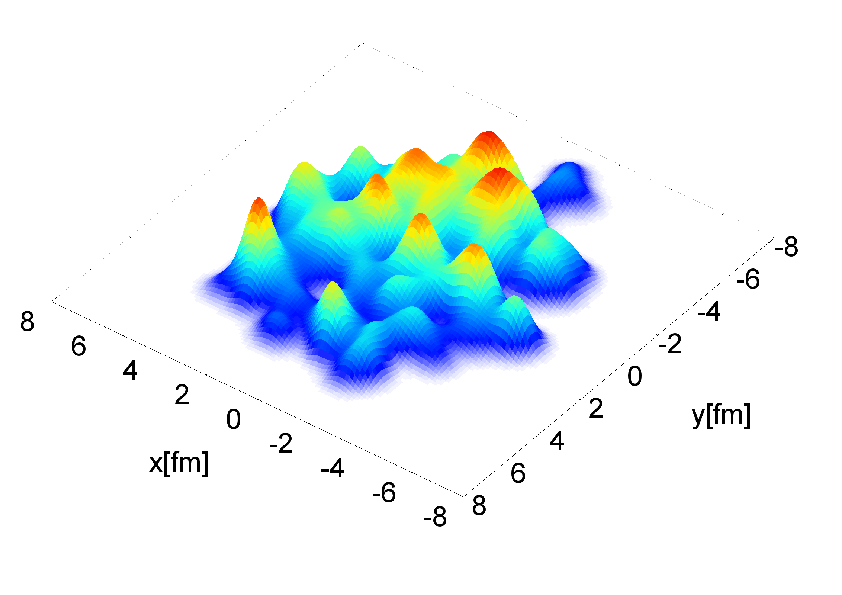}
\includegraphics[width=0.4\linewidth]{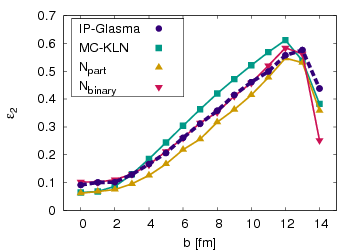}
\includegraphics[width=0.4\linewidth]{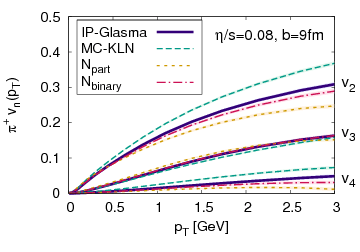}
  \caption{ \label{fig:lumpychange}
     Top: Initial energy density (arbitrary units) in the transverse plane in three different heavy-ion collision events: from left to right, IP-Glasma, MC-KLN and MC-Glauber models.
     Bottom: $\epsilon_2$ (left) and $v_2$ (right) from different initial states. 
     Taken from Ref.~\cite{Schenke:2012wb}.
     }
\end{figure}

Hydrodynamics requires a system size ($L$) much larger than the mean free path ($\lambda$) among the interacting particles, $L \gg \lambda$. 
The initial stage in a heavy ion collision where the requirement is not fulfilled lies outside the domain of applicability of the hydrodynamic description.
Therefore, the initial conditions of the medium evolution are commonly modelled by dedicated models in two different approaches. 
One of them is to use the energy density obtained from numerical relativity solutions to AdS/CFT~\cite{vanderSchee:2013pia,Casalderrey-Solana:2013aba,Chesler:2013urd} before the equilibrium, and the other approach is to use Color Glass Condensate (CGC)~\cite{Gelis:2010nm} and evolving it with Glasma gluon filed solutions~\cite{Gale:2012rq,Schenke:2012wb}. 
Different initial states are shown to have large effects on experimental observables, such as the $v_n$ values~\cite{Schenke:2012wb}. 
The difference in lumpiness of initial geometry results in large variation in the measured $v_2$ values as shown in Fig.~\ref{fig:lumpychange}. 
As of today, how well the initial condition models describe the true pre-equilibrium phase of the collision is still an open question. 

Hydrodynamic description is applicable 
during the expansion of the medium, until the point that the nuclear matter density becomes too dilute that $L \gg \lambda$ can be no longer fulfilled.
Relying on the fact that the entropy density, energy density, particle density and temperature profiles are directly related, hydrodynamic frameworks assume the medium decouple on a surface of constant temperature and convert the fluid cells to hadrons. 
This results in a sudden freeze-out where the mean free path drops from infinite to zero, which is purely artificial.
The better approach is carried out in hybrid models~\cite{Song:2010aq,Lokhtin:2008xi,Petersen:2008dd}, after the hadrons are converted, they are handed to microscopic models which continues to model interaction between hadrons until a kinetic freeze-out is reached.
 
 \begin{figure}[htb]
\centering
\includegraphics[width=0.32\linewidth]{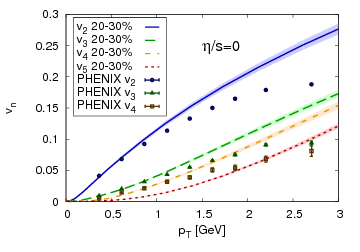}
\includegraphics[width=0.32\linewidth]{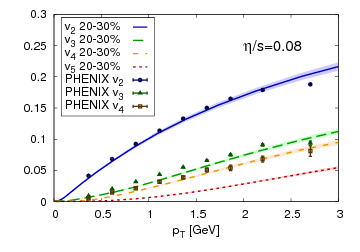}
\includegraphics[width=0.32\linewidth]{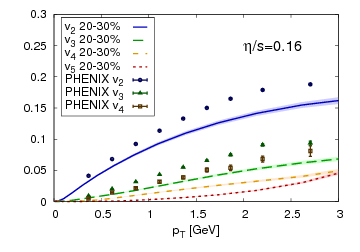}
  \caption{ \label{fig:viscous}
     \pt-differential $v_2$ to $v_5$ from ideal hydrodynamics (left), viscous hydrodynamics with $\eta/s$=0.08 (middle), and $\eta/s$=0.16 (right). Results are averaged over 200 events each. Experimental data from PHENIX~\cite{Adare:2011tg}.
     Taken from Ref.~\cite{Schenke:2011bn}.
     }
\end{figure}
 
Once hydrodynamics turns out to describe the experimental measurements well, observables which are not directly measurable can be extracted from its output. 
The shear viscosity over entropy density ratio, $\eta/s$, is given as an input to hydrodynamics calculations. 
Comparing calculations with different $\eta/s$ values to experimental results allow the determination of the properties of the medium. 
Fig.~\ref{fig:viscous} shows a comparison between hydrodynamics calculation and experimental data for $v_n$. 
The $\eta/s$ values extracted at RHIC and LHC energies is $\approx 0.08-0.12$ and $\approx 0.16-0.20$, respectively~\cite{Gale:2013da,Heinz:2015tua}. 
The surprisingly low $\eta/s$ value, close to the $1/4\pi$ minimum viscosity bound from first principle calculations~\cite{Policastro:2001yc}, is a strong evidence that the created QGP medium behaves like a perfect fluid. 
Furthermore, the local temperature or energy density of the medium can also be extracted from hydro calculations. 
In the current picture of jet-medium interaction, the energy density is a key input for simulations of energy loss of a parton~\cite{Gyulassy:2003mc,Kovner:2003zj}. 
In the context of quarkonium suppression, if one quarkonium is expected to melt above certain temperature, the local temperature extracted from hydro is extremely useful to tell whether it melts at a fixed position in the medium. 
Therefore, hydro in heavy ion collision does not only describes expansion and collective flow of the medium but also provides important information for other phenomena. 

\clearpage

\section{QGP in small systems}
\label{sec:qgpsmall}

Besides AA collision, smaller collision systems such as proton-lead (pPb) and proton-proton (pp) collisions are also studied at the accelerators.
Recently, results on many of the experimental observables in these small collision systems are found to be strikingly similar to the results from AA collisions. 

\begin{figure}[htb]
\centering
\includegraphics[width=0.3\linewidth]{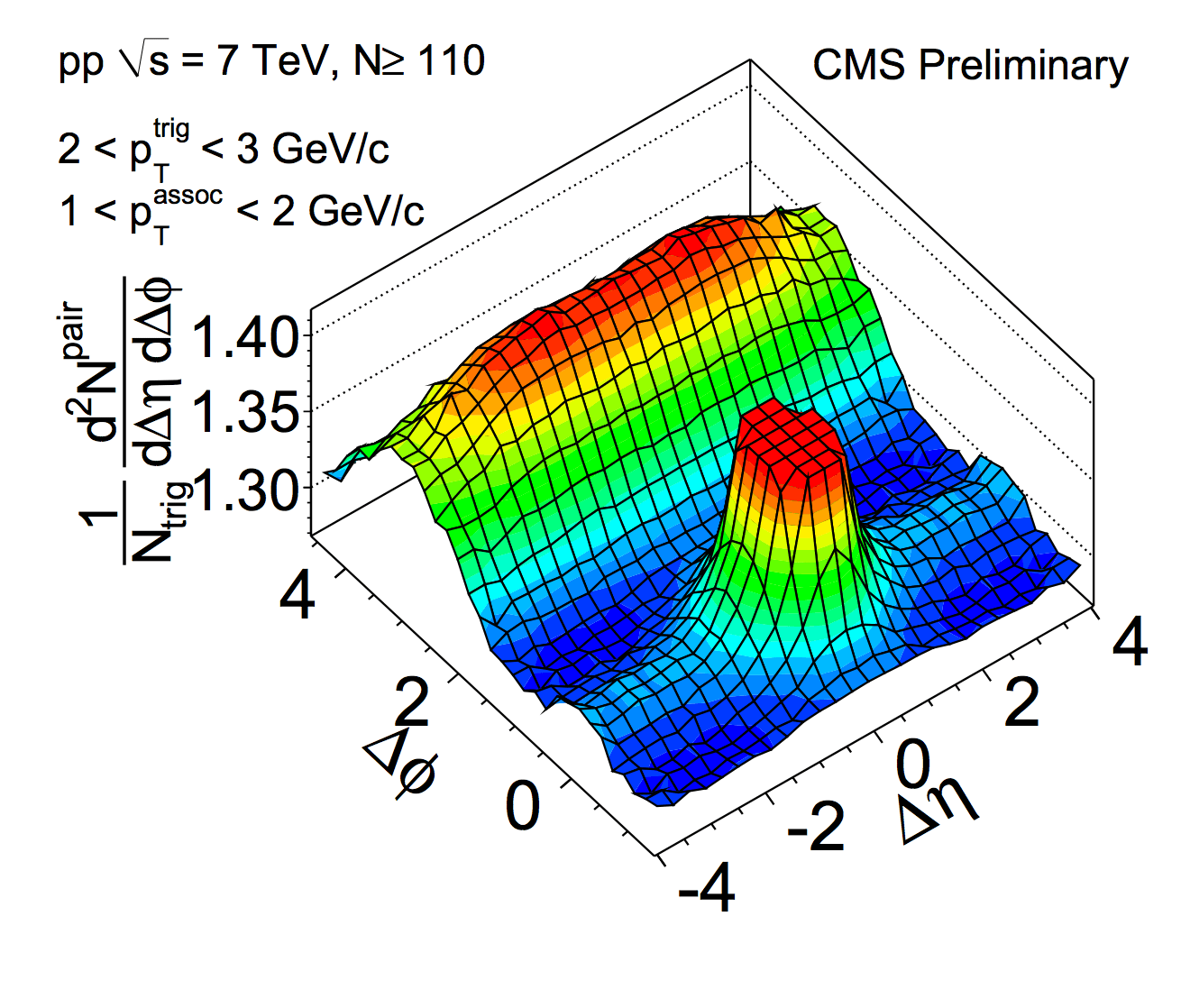}
\includegraphics[width=0.3\linewidth]{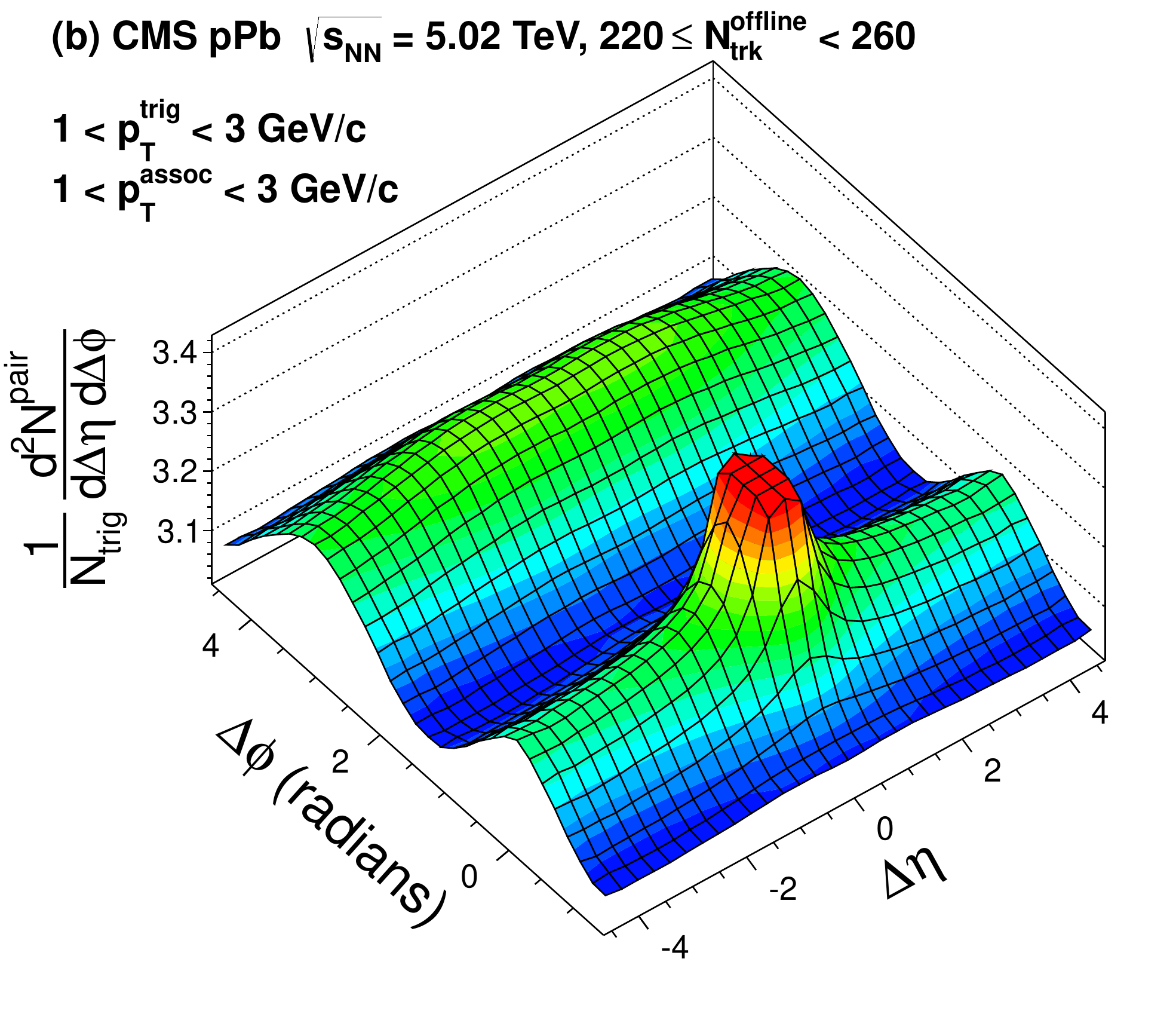}
\includegraphics[width=0.3\linewidth]{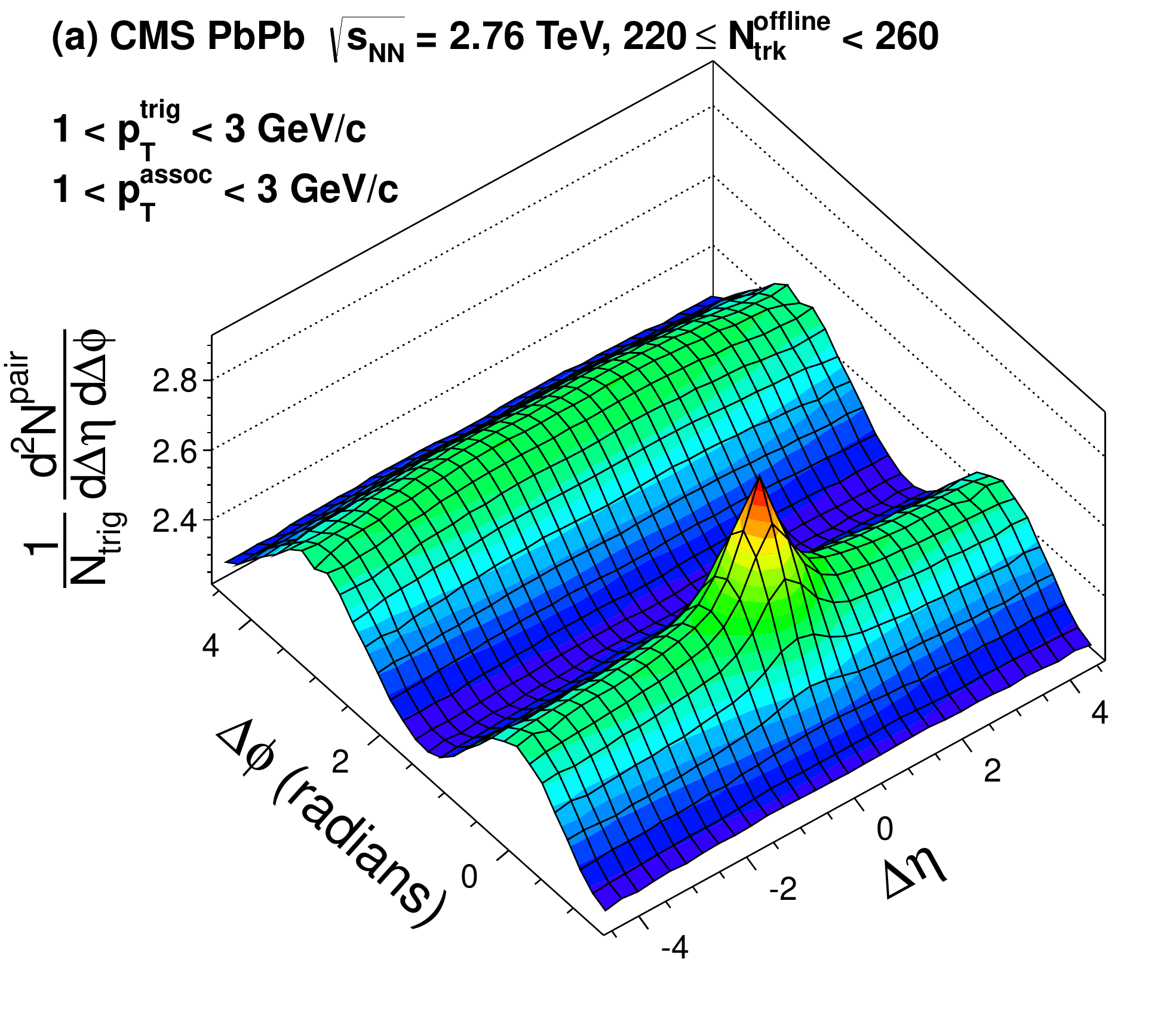}
  \caption{ \label{fig:ppridge}
	The two-particle $\Delta\phi$-$\Delta\eta$ correlation functions
  for 7 TeV pp (left), 5.02 TeV pPb (middle) and 2.76 TeV PbPb (right) collisions for pairs of charged particles.
     Taken from Ref.~\cite{Khachatryan:2010gv,Chatrchyan:2013nka}.
     }
\end{figure}

In 2010, the observation of long-range two-particle azimuthal correlations at large relative pseudorapidity in high final-state particle multiplicity (high-multiplicity) pp collisions at the LHC~\cite{Khachatryan:2010gv} opened up new opportunities for studying novel dynamics of particle production in small, high-density QCD systems. 
The key feature, known as ``ridge", is an enhanced structure on the near-side (relative azimuthal angle $|\Delta\phi|\approx0$) of two-particle $\Delta\phi$-$\Delta\eta$ correlation functions that extends over a wide range in relative pseudorapidity as shown in Fig.~\ref{fig:ppridge} (left).
This phenomenon resembles similar effects observed in AA collisions (Fig.~\ref{fig:ppridge}, right), which results from the expansion of the QGP medium. 
Later in 2012, the same ridge is also seen in high multiplicity pPb collisions~\cite{CMS:2012qk,Abelev:2012ola,Aad:2012gla,Aaij:2015qcq} (Fig.~\ref{fig:ppridge}, middle).
These measurements question the heavy ion community about the existence of QGP in small collision systems. 

\begin{figure}[htb]
\centering
\includegraphics[width=0.45\linewidth]{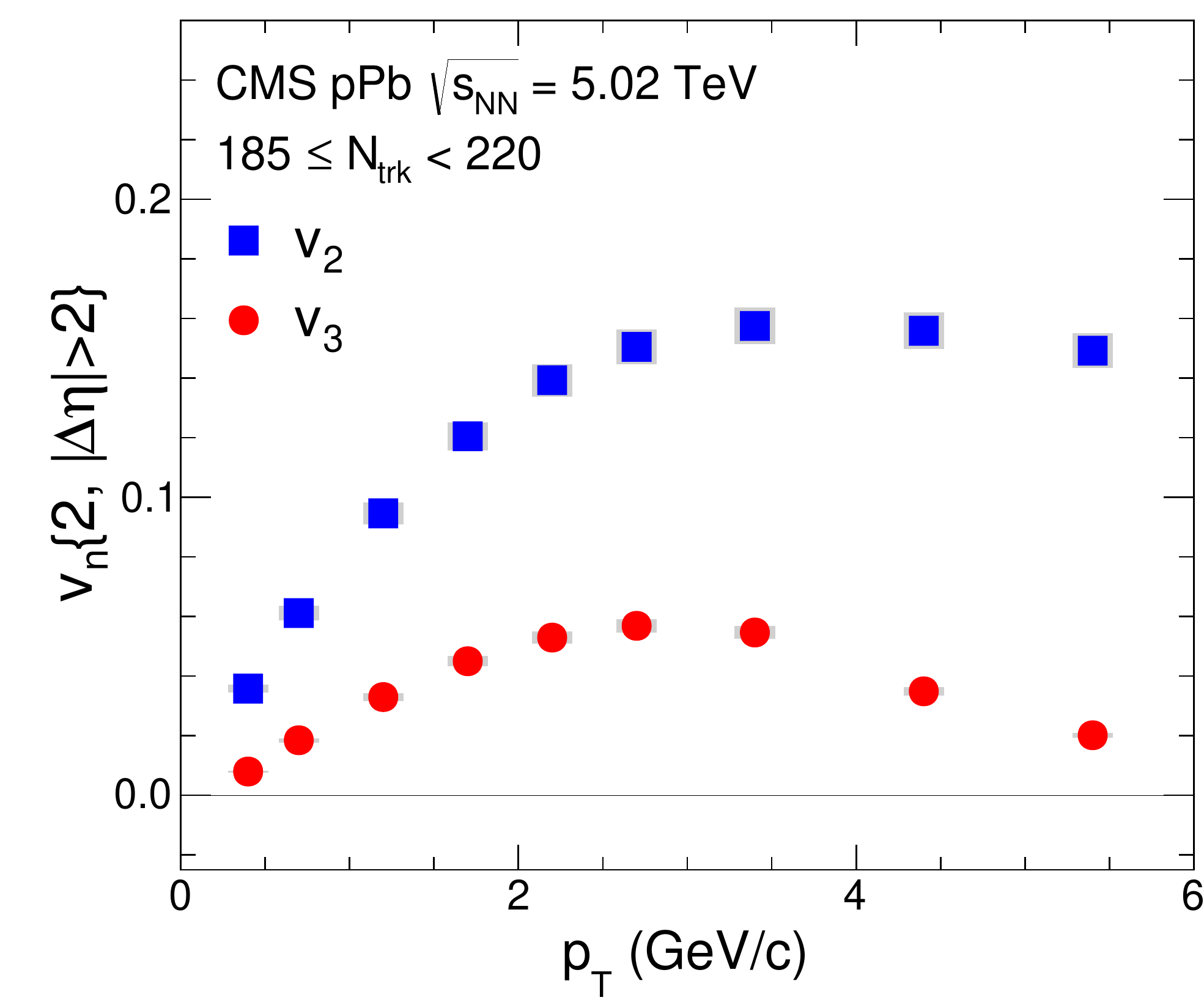}
\includegraphics[width=0.45\linewidth]{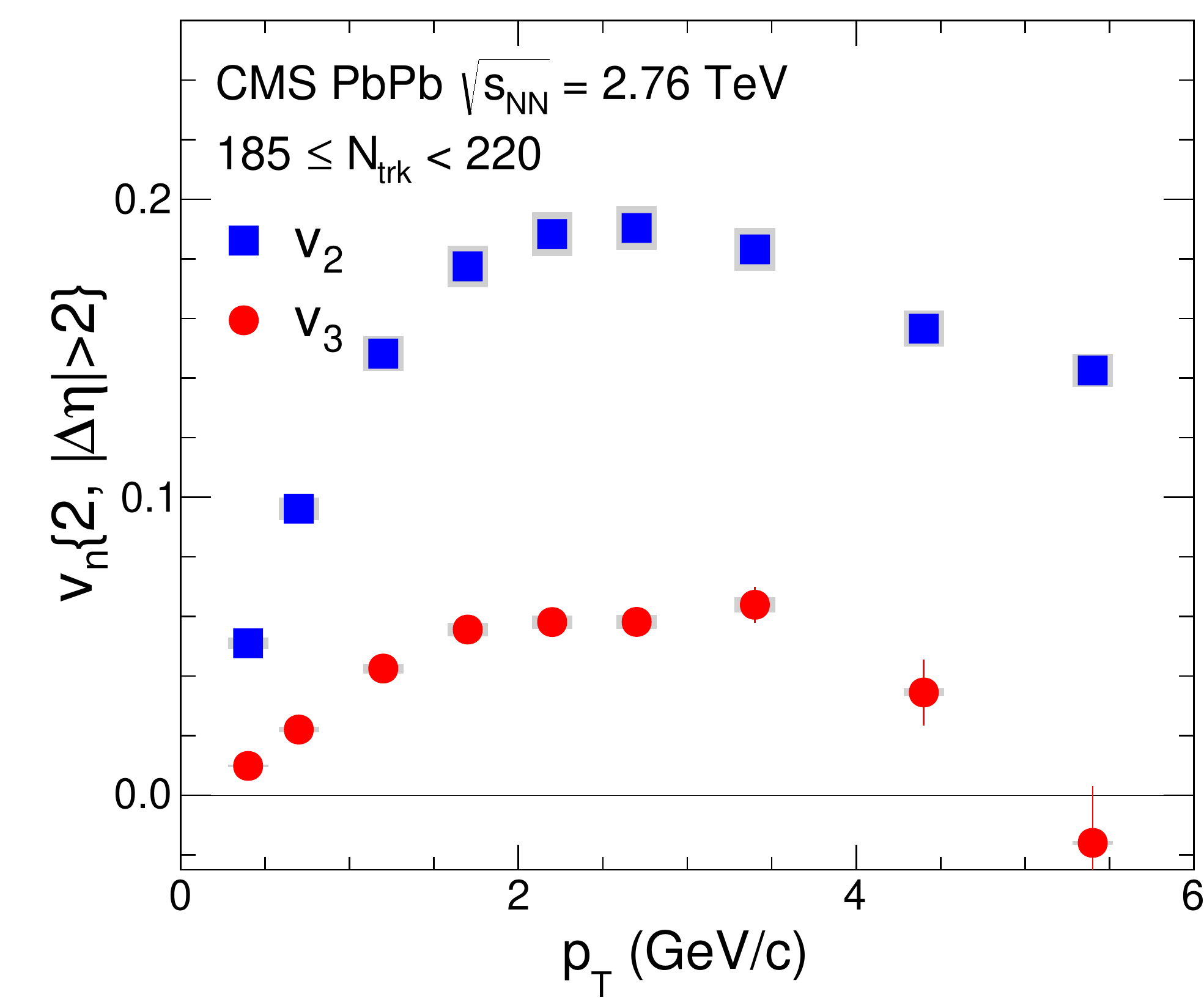}
  \caption{ \label{fig:v2v3pPb}
     $v_2$ and $v_3$ results as function of \pt\ for 5.02 TeV pPb collisions (left) and 2.76 TeV PbPb collisions (right) for $185 \leq \noff < 220$. Taken from Ref.~\cite{Chatrchyan:2013nka}.
     }
\end{figure}

\begin{figure}[htb]
\centering
\includegraphics[width=0.8\linewidth]{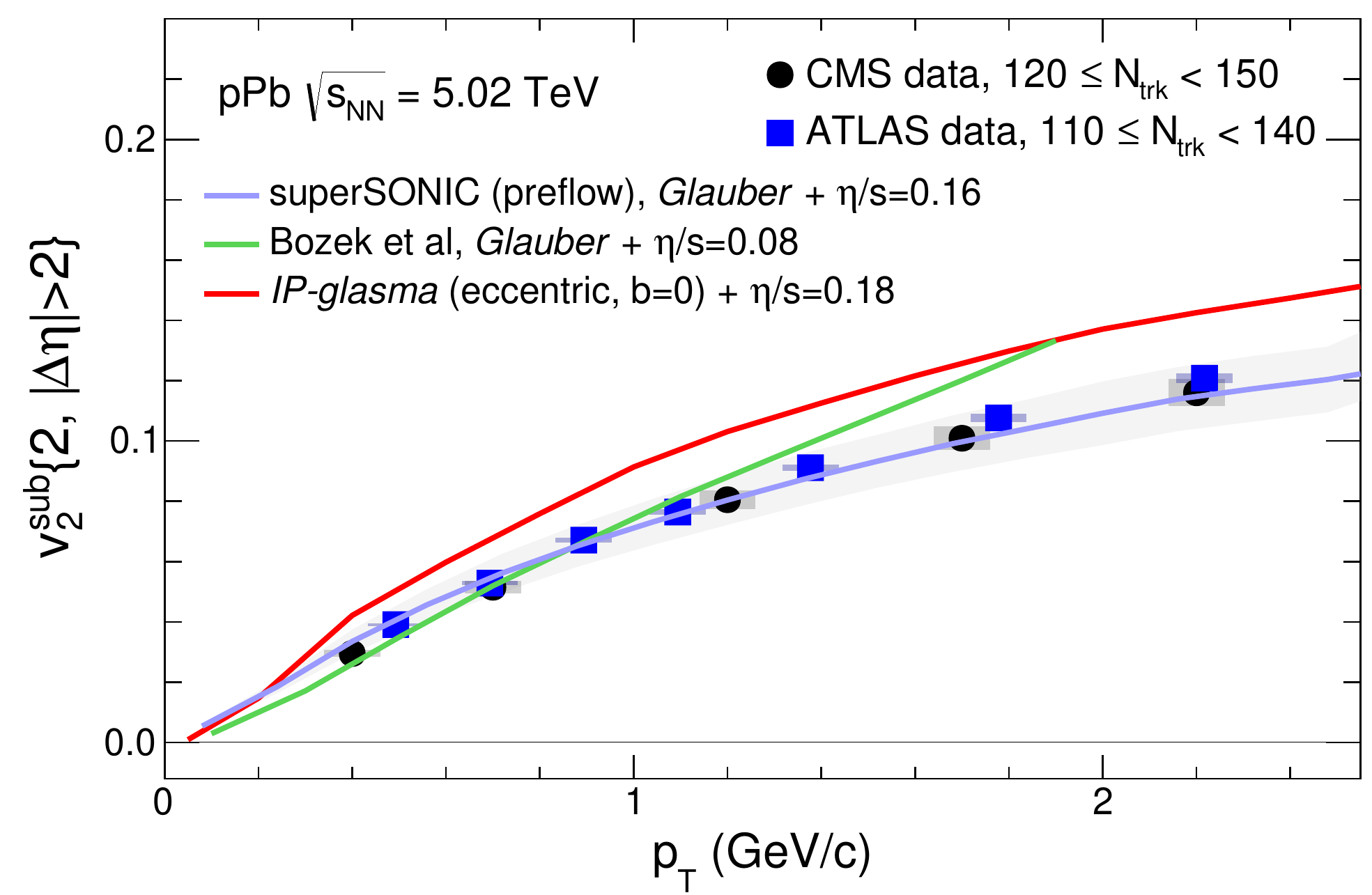}
  \caption{ \label{fig:v2v3datahydro}
     Experimental results for $v_2$ in 5.02 TeV pPb collision from ATLAS and CMS compared to three different hydrodynamic calculations: A prediction with MC Glauber initial condition~\cite{Bozek:2014wpa}, results from superSONIC with pre-flow~\cite{Romatschke:2015gxa}, and IP-Glasma+MUSIC calculation~\cite{Dusling:2015gta}.
     }
\end{figure}

The magnitude of ridge in pPb collisions is much larger than the pp ridge at same multiplicity and becomes comparable to that seen in PbPb collisions. 
Motivated by the study of flow harmonics in AA collisions, the ridge in pPb has been analysed using the same Fourier decomposition. 
The $v_2$ and $v_3$ are extracted from the correlations as a function of \pt\ in high multiplicity pPb collisions at 5.02 TeV, shown in Fig.~\ref{fig:v2v3pPb} together with results from PbPb collision at 2.76 TeV at same multiplicity. 
The $v_2$, $v_3$ values first rise with \pt\ up to around 3 GeV/c and then fall off toward higher \pt, a behavior very similar to PbPb collisions. 
This similarity might indicate a common origin of the ridge phenomenon in the two collision systems. 
Hydrodynamic calculations aiming at the prediction and description of experimental data has become available~\cite{Werner:2013ipa,Bozek:2013ska,Schenke:2014zha,Werner:2010ny,Werner:2010ss,Bozek:2011if,Bozek:2012gr,Werner:2013tya}, in particular in pPb collisions. 
Qualitative agreement between calculation and experimental data has been shown in \pt-differential $v_2$, as shown in Fig.~\ref{fig:v2v3datahydro}.
\noindent However, due to the system size being significantly smaller, the hydrodynamics interpretation from AA collisions may be questionable in small systems. 
The applicability of hydrodynamics has to be investigated with more detailed measurement. 
Meanwhile, alternative models based on gluon interactions in the initial stage can also qualitatively describe the general trend of the data~\cite{Schenke:2015aqa}. 

The analyses presented in this thesis provide study of detailed properties of collective flow in pPb collisions (Chapter~\ref{ch:resultpPb}) in order to shed light on the possible QGP formation, and furthermore extend the study to proton-proton (pp) collisions (Chapter~\ref{ch:resultpp}) to reveal evidence of the existence of a collective medium. 

\section{Overview of this thesis}
\label{sec:overview}

This thesis presents results on inclusive charged particle and identified strange hadron ($\PKzS$ or $\PgL$/$\PagL$) two-particle angular correlations in pPb collisions at 5.02 TeV and pp collisions at 5, 7, and 13 TeV over a wide range in pseudorapidity and full azimuth. 
The observed azimuthal correlation at large relative pseudorapidity are used to extract the second-order ($v_2$) and third order ($v_3$) anisotropy harmonics. 
These quantities are studied as a function of the charged-particle multiplicity in collision events and the transverse momentum of the particles. 

The experimental setup of CMS detector, as well as the LHC accelerator, are described in Ch.~\ref{ch:Detector}. 
The trigger and data acquisition system of CMS is introduced in Ch.~\ref{ch:trigger}, as well as the triggers used for the analyses in this thesis, particularly the high multiplicity triggers that enable the precise $v_n$ measurements.
The data used in this work collected by the CMS detector is described in Ch.~\ref{ch:datamc}. 
The reconstruction of $\PKzS$, $\PgL$/$\PagL$ and inclusive charged particles are discussed in Ch.~\ref{ch:Reco}.
Ch.~\ref{ch:evtsel} focus on the offline event selection procedure, including the pileup rejection algorithm. 
The two-particle correlation technique is described in Ch.~\ref{ch:technique} in detail, together with the procedure of $v_n$ extraction for identified particles. 
Final results are presented in Ch.~\ref{ch:resultpPb} for pPb collisions and in Ch.~\ref{ch:resultpp} for pp collisions as well as their connection to the theoretical interpretations. 
Ch.~\ref{ch:resultpp} also includes the discussion of jet contribution correction to $v_n$ results and provide a comparison between correction methods used by CMS and ATLAS.
Ch.~\ref{ch:conclusion} provides a summary of the work presented in this thesis.

\cleardoublepage
\chapter{The CMS experiment at the LHC}
\label{ch:Detector}
The production of elementary particles can be studied under controlled conditions through particle accelerators and colliders. 
Electrons, protons, or heavy nuclei are accelerated and brought to collision either one on another or on a fixed target. 
The elementary particles produced in the collisions are registered and memorized by the particle detectors.

The analysis presented in this thesis is based on the data collected by the Compact Muon Solenoid (CMS) experiment at the Large Hadron Collider (LHC). 

\section{The LHC}
\begin{figure}
\centering
\mbox{\includegraphics[width=0.8\linewidth]{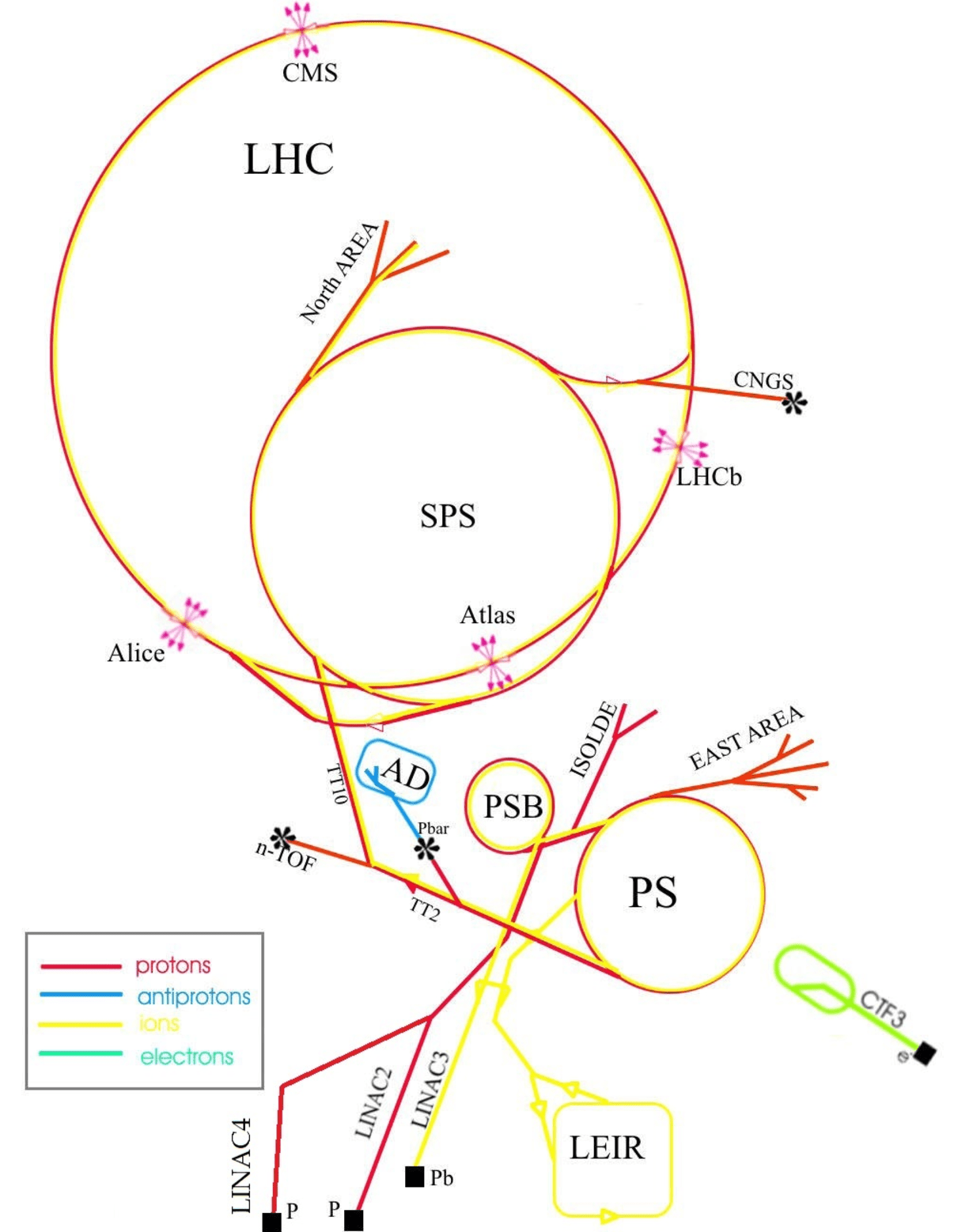}}
  \caption{ \label{fig:LHCoverview} 
     The CERN accelerator complex~\cite{LHCplot}.
   }
\end{figure}

The LHC~\cite{1748-0221-3-08-S08001} is the world's largest and most powerful particle collider ever built. 
It is a two-ring superconducting hadron accelerator and collider which is a part of CERN's (European Organization for Nuclear Research) accelerator complex. 
It is designed to collide proton beams with a nominal energy of 7 TeV per beam (i.e. center-of-mass energy of \roots = 14 TeV), 
and heavy ion beams with a nominal energy of 2.76 TeV per nucleon for lead (Pb) nuclei. 
Instead of directly accelerating the particles from low to the maximum energy at the LHC, the process is optimized through a chain of pre-accelerators. 
A schematic overview of CERN accelerator complex is shown in Fig.~\ref{fig:LHCoverview}, where the particles are accelerated as following:
\begin{itemize}
\item Proton: The protons from the $H_2$ source enter the LINAC2 linear accelerator and exit with an energy of 50 MeV. They are accelerated more in the Proton Synchrotron Booster (PSB) to 1.4 GeV. The Proton Synchrotron (PS) follows the PSB and accelerates the protons to 25 GeV and injects them to the Super Proton Synchrotron (SPS). The SPS raises the proton energy again to 450 GeV and deliver them to the LHC where the maximum energy is achieved.
\item Heavy ion: Currently, the LHC is capable to accelerate only the Pb nuclei. Starting from a source of vaporized lead, the Pb ions enter LINAC3 and get accelerated to an energy of 4.2 MeV. They are then collected and accelerated in the Low Energy Ion Ring (LEIR) to 72 MeV. After being injected to the PS from LEIR, the same route to maximum energy is taken as the protons.
\end{itemize} 

\clearpage

\section{The CMS experiment}
\begin{figure}
\centering
\mbox{\includegraphics[width=\linewidth]{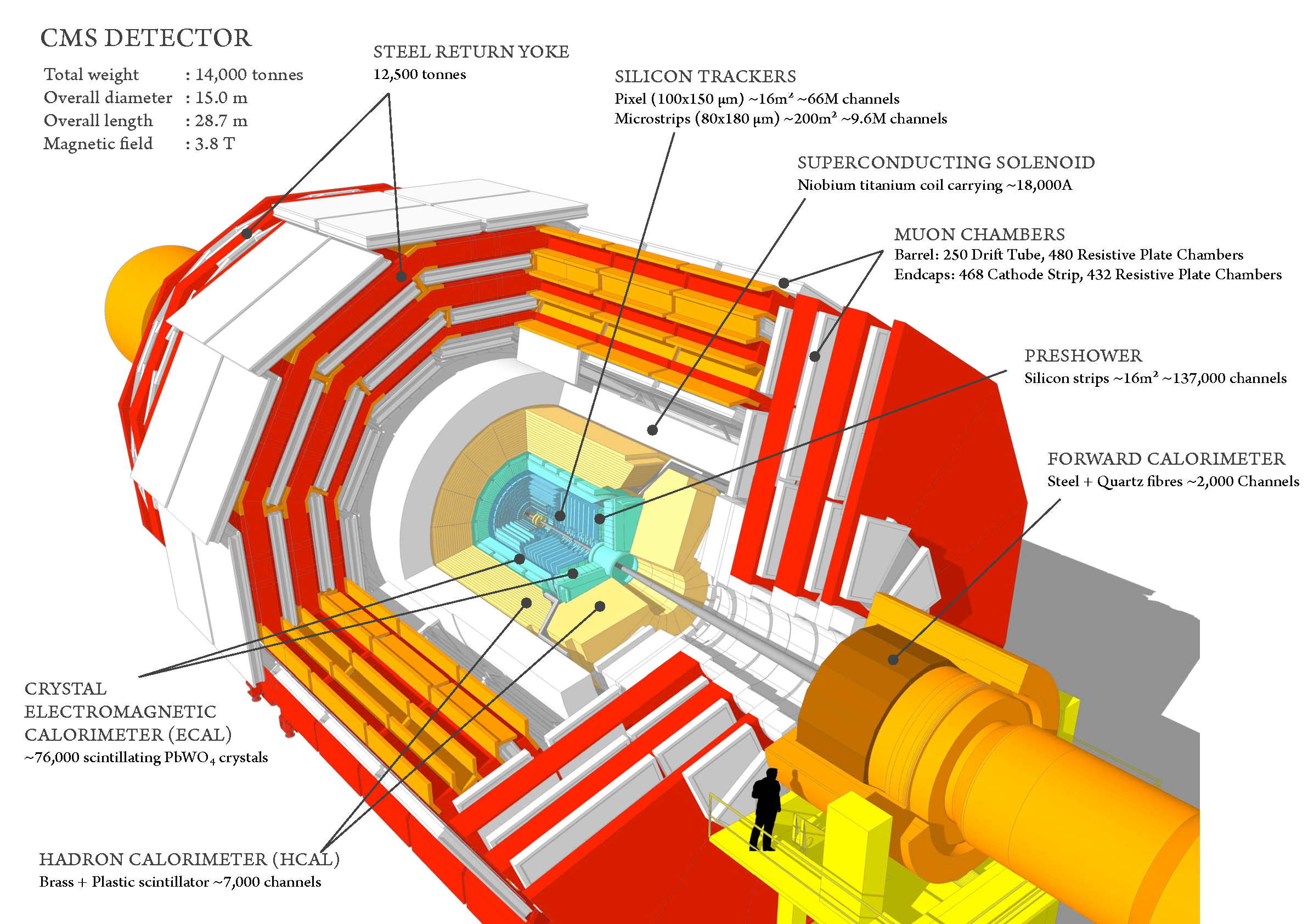}}
  \caption{ \label{fig:CMSoverview} 
     A cutaway view of the CMS detector~\cite{Sakuma:2013jqa}.
   }
\end{figure}

The CMS detector is one of the four experiments placed on the ring of the LHC. 
It is a general purpose detector whose main goal is to explore physics at the TeV scale. 
As stated in the name, the detector consists of layers of solenoid structure, which are sub-detector parts of different functionality. 
Figure~\ref{fig:CMSoverview} shows a schematic view of CMS detector, the structure from inner to outer is formed including the following detector parts: 
\begin{itemize}
\item The inner silicon tracking system insures good particle momentum and spatial resolution.
\item The electromagnetic calorimeter (ECAL) allows accurate measurement of the energy of leptons and photons.
\item The hadronic calorimeter (HCAL) allows precise measurement of the energy of hadrons.
\item The solenoid magnet with a strong magnetic field of 3.8 T makes the determination of high momentum particle possible.
\item The muon system provides excellent muon identification.
\end{itemize}
\noindent  More detailed description on the sub-detector used in the analysis presented in this thesis will be given in the following subsection.

A common coordinate system definition is important for analysing data derived from each sub-detector consistently. 
The coordinate system adopted by CMS has a center at the nominal collision point inside the detector. 
The x-axis is defined to point towards the center of the LHC ring, the y-axis is defined to point straight upward and the z-axis is defined to point along counter clockwise direction of the LHC ring. 
For the spherical coordinates, the azimuthal angle $\phi$ and the radial coordinate $r$ is measured in the x-y plane from the x-axis. 
The polar angle $\theta$ is measured from the z-axis. 

In experimental particle physics, it is more convenient to use pseudorapidity, $\eta$, instead of the polar angle $\theta$. 
It is defined as 
\begin{equation}
\eta = -ln \tan(\frac{\theta}{2}).
\end{equation}
The other convenient variable which is often used in data analysis is the transverse momentum (\pt) of the objects.

\clearpage

\subsection{Silicon tracking system}
The silicon tracking system is used in the finding of position of collision vertex, in the reconstruction of charged particles (described in Section~\ref{sec:tracking}) and in the reconstruction of $V^{0}$ particles (described in Section~\ref{sec:V0}). 
Therefore, it has central importance for the analysis presented in this thesis.

\begin{figure}
\centering
\mbox{\includegraphics[width=\linewidth]{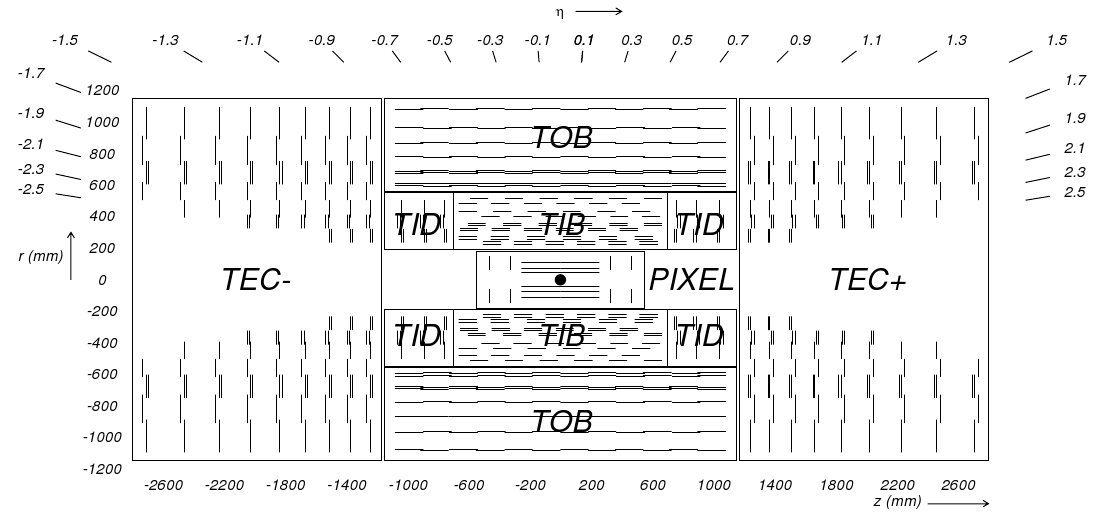}}
  \caption{ \label{fig:tracker} 
     View of the CMS tracker in the $rz$-plane~\cite{Sprenger:2010ss}. Each line in the strip tracker represents a silicon strip detector, whereas lines in the pixel tracker represent ladders and petals on which the detectors are mounted in the barrel and endcaps, respectively.
   }
\end{figure}

The tracking system is composed of an inner silicon pixel detector and an outer silicon strip detector. 
Both of the two detectors cover a pseudorapidity range of $|\eta| < 2.5$. The layout of the tracking system is shown in Fig.~\ref{fig:tracker}.

\begin{figure}
\centering
\mbox{\includegraphics[width=\linewidth]{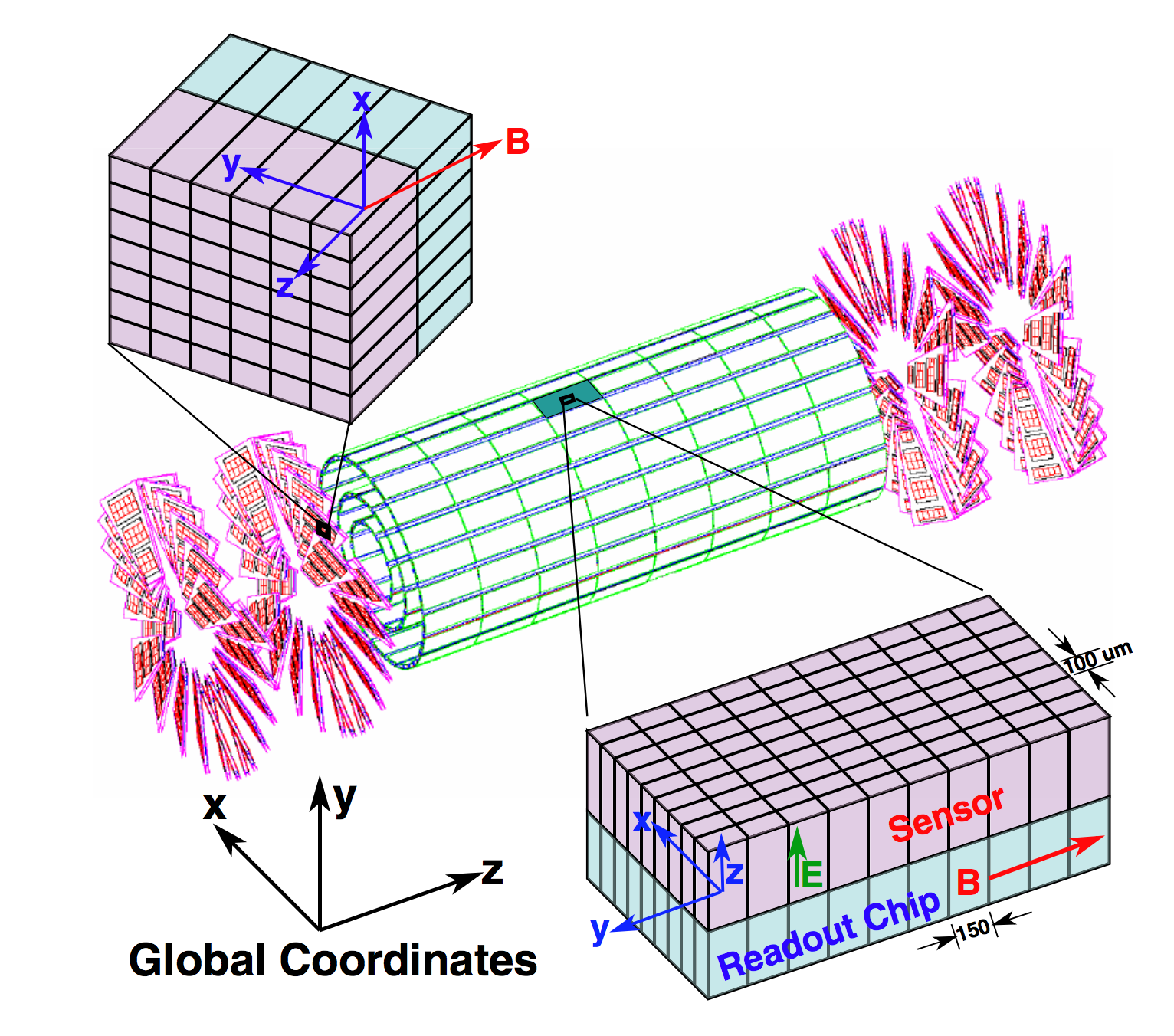}}
  \caption{ \label{fig:pixel} 
     View of the CMS silicon pixel tracker.
   }
\end{figure}

\paragraph{Silicon pixel detector.}
The silicon pixel detector is the inner most detector of CMS, consisting of 3 concentric cylindrical barrel layers and two layers of fan-blade disks at either end (shown in Fig.~\ref{fig:pixel})~\cite{Giurgiu:2008ir}. 
It is designed to provide high precision 3D determinations of track trajectory points. 
The three barrel layers are located at radii of 4.3 cm, 7.3 cm and 10.2 cm to the interaction point, and have an active length of 53 cm. 
The two layers of disks cover the region between radii 4.8 cm and 14.4 cm, at longitudinal distance of 35.5 cm and 48.5 cm from the interaction point. 
This geometry layout ensures particle passage through 3 layers of detector in the region $|\eta| < 2.2$ and 2 layers of detector in the region $|\eta| < 2.5$. 
The entire pixel detector is composed of 1440 pixel modules with 65 million pixels. Each pixel, with an area of 100 $\mu$m $\times$ 150 $\mu$m, oriented in the azimuthal direction in the barrel and the radial direction in the forward disks. 
The electrons created by ionization during the passage of charged particles (track hits) in the barrel region are significantly Lorentz drifted in the 3.8 T magnetic field of CMS. This drift results in charge sharing on different readout modules. 
The weighted center of the charge distribution can be calculated from the analogue readout which provide much better spatial resolution than a binary readout. 
To ensure the use of Lorentz drift at the forward disks, the blades are rotated by 20 degrees about their radial axes to produce a vertical component of magnetic filed with respect to the electric field in the pixels.
The entire pixel detector is operating at a temperature of -15\textdegree{}C to limit the impact of radiation damage and to minimize leakage current.

\paragraph{Silicon strip detector.}
As shown in Fig.~\ref{fig:tracker}, the silicon strip detector is composed of tracker inner barrel (TIB), tracker inner disk (TID), tracker outer barrel (TOB) and tracker outer endcap (TEC). 
A total of 15148 silicon strip modules with 10 million strips are arranged in 10 barrel layers extending outward to radii 1.1 m and 12 disks on each side of the barrel to cover the region $|\eta| < 2.5$. 
The active detector area is about 200 m$^2$ which makes it the largest silicon tracker ever built. 
Instead of providing 2D information of track hits in $\phi$ and $z$ direction as the pixel detector, the silicon detector provides only 1D information. 
However, if two layers of strip detectors are placed on either side of a module with an angle, the double-sided module can obtain 2D information. 
Both single-sided (single line in Fig.~\ref{fig:tracker}) and double-sided modules (double line in Fig.~\ref{fig:tracker}) are used in the silicon detector at various physical locations, to maximize the performance with a limited material budget. 
Due to the complex layout of the silicon tracker, particle with different kinematics leave trajectories coincide with different number of layers. 
Particles passing through more layers have higher probability to be reconstructed then those passing through less layers, which results in a non-uniform track reconstruction efficiency as function of pseudorapidity which will be shown in Sec.~\ref{sec:trackeff}.

\subsection{Calorimeter system}
\label{subsec:calosyst}

The CMS calorimeter system aims to find the energies of emerging particles in order to build up a picture of collision events. 
The system provides precise measure of photon, electron and jet energies and with the hermetic design allows the measurements of missing transverse energy for neutrinos. 
From inner to outer, it is composed of ECAL and HCAL. 

\paragraph{Electromagnetic Calorimeter} Among the particles emitted in a collision, electrons and photons are of particular interest because of their use in finding the Higgs boson and other new particles. These particles are measured within the ECAL, which is made up of a barrel section and two endcap disks. In order to handle the 3.8 T magnetic field of CMS and the high radiation level induced by collisions, lead tungstate crystal is chosen. Such a crystal is made of metal primarily, but with a touch of oxygen in its crystalline form, it is highly transparent and produces light in fast, short and well-defined photon bursts in proportional to the energy of particle passing through. The cylindrical barrel contains 61200 crystals formed into 36 modules with a depth of 25.8 radiation lengths (the crystal has radiation length of 0.89 cm). The flat endcap disks seal off the barrel at either end and are made up of around 15000 crystals with a depth of 24.7 radiation length. The barrel section covers $|\eta|<1.479$ while the endcap disks extend the range to $|\eta|<3$.

The ECAL also contains Preshower detectors in front of the endcap disks to provide extra spatial resolution at those regions. The Preshower detectors are placed starting at 298.5 cm from the center of CMS and ending at 316.5 cm. They consists of two lead radiators, about 2 and 1 radiation lengths thick respectively, each followed by a layer of silicon microstrip detectors. The two layers have their strips orthogonal to each other to provide 3D spatial resolution of the particle shower initiated by photons or electrons hitting the lead radiators. 

\paragraph{Hadron Calorimeter} The HCAL measures the energy of hadrons and provides indirect measurement of the presence of non-interacting uncharged particles such as neutrinos through the missing transverse energy. It is a sampling calorimeter made of repeating layers of dense absorber and tiles of plastic scintillator. An interaction occurs producing numerous secondary particles when a hadronic particle hits a plate of absorber. As these secondary particles flow through layers of absorbers they produce more particles which results in a cascade. The particles pass through the alternating layers of active scintillators causing them to emit light which are collected up and amplified for a measurement of the initial particle's energy. Similar to ECAL, the HCAL consists of a barrel section and two endcap disks. The barrel reaches $|\eta|$ of 1.3 while the endcap disks extend to $|\eta|$ of 3. 

The HCAL has two hadronic forward calorimeters (HF) positioned at either end of CMS to cover the $|\eta|$ range of 3 to 5. The HF receives large fraction of particle energy contained in the collision hence must be made very resistant to radiation.  Therefore, it is built with steel absorbers and quartz fibers where detection of signal is done with Cherenkov light produced in the fibers. The HF is very important for heavy ion collisions as it is used to select collision events (described in Sec.~\ref{sec:MBtrigger}) and to determine centrality (described in Sec.~\ref{subsec:centrality}).

\cleardoublepage
\chapter{Trigger and data acquisition}
\label{ch:trigger}

The CMS trigger and data acquisition (DAQ) system for the selection of good collision events and events with specific physics interests is described in this chapter. 
Sec.~\ref{sec:trigger} provides description of the CMS trigger and DAQ system. 
The trigger for good collision events and high multiplicity events is discussed in Sec.~\ref{sec:MBtrigger}-\ref{sec:theHMtrigger}. 
The upgrade of high multiplicity trigger for 2016 and 2017 data taking is discussed in Sec.~\ref{subsec:HMtriggerUp}

\section{The CMS trigger and data acquisition system}
\label{sec:trigger}

\begin{figure}[htb]
	\centering
	\includegraphics[width=0.5\textwidth]{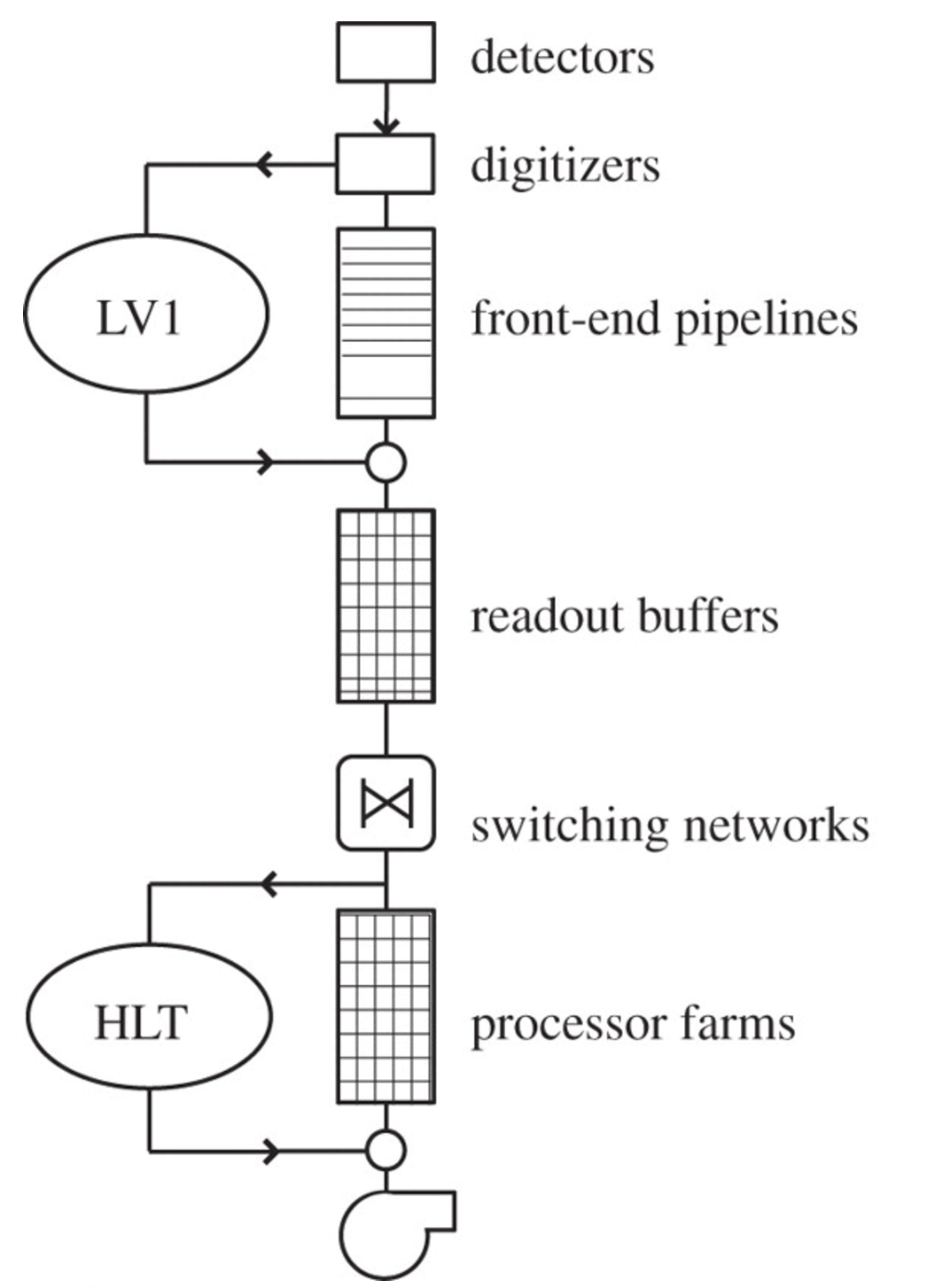}
	\caption{Schematic of the functionality and data flow through the DAQ system. Take from Ref.~\cite{Cittolin:2012zz}.}
        \label{fig:CMSDAQ}
\end{figure}

For nominal data taking, the LHC is delivering particle collision events at a rate on the order of MHz. 
This results in an enormous amount of data from all the collision events, and make it impossible to store all the information. 
The trigger and data acquisition (DAQ) system~\cite{Cittolin:578006} is designed to filter out only the events which contains interesting physics processes. 
Figure~\ref{fig:CMSDAQ} shows a schematic of the function of the full trigger and DAQ system.
The DAQ has the task to transport the data from about 650 front ends at the detector side, through the trigger system for processing and filtering of events, to the storage units.  
The CMS trigger system utilizes two levels of selections, the level-1 (L1) trigger and the high-level trigger (HLT).
Based on the decision of the trigger system, an event is stored or skipped. 
The stored events are written to a temporary disk buffer before being transferred to the computing center (Tier 0) at CERN for offline processing.

\paragraph{Level-1 trigger.}
The level-1 trigger is composed of custom hardware processors~\cite{Dasu:2000ge}. 
Its input comes from sub-detectors such as ECAL, HCAL, muon detectors and beam monitoring detectors. 
In order to handle the large event rate the LHC delivers to the detector, the system is built to select the most interesting events in a fixed time interval of less than 4 $\mu$s. 
Because of this limitation of data processing time, L1 triggers with user defined algorithms use information only from the calorimeters and muon detectors to select events containing candidate physics objects, e.g. total transverse energy ($E_{TT}$), or ionization deposits consistent with a muon, or energy deposit consistent with a jet, or energy clusters consistent with an electron, photon, $\tau$ lepton. 
The L1 output rate is limited to 100 kHz for pp collisions and 5 kHz for PbPb collisions by the upper limit imposed by the CMS readout electronics. 
In order to fit in this limit, the thresholds of the L1 triggers can be adjusted during the data taking in response to the instantaneous collision rate delivered by the LHC. Alternatively, the output rate can be adjusted by prescaling the number of events that pass the selection criteria of specific algorithms, which is done by randomly skip events in an N event interval where N is the prescale factor.

\paragraph{High level trigger.}
Events passing the L1 triggers are then passed to the HLT system composed of numerous triggers.
The triggers, implemented in software, are algorithms exploiting the full event information to make choice based on primer analysis of fully reconstructed physics objects. 
They read the event information from the front-end electronics memory, analyse them and forward the accepted events to the storage. 
The HLT output rate is mainly limited by the data transfer bandwidth from the detector to Tier0 and the data processing time needed by the trigger algorithms. 
The triggers are running with a computer farm of more than 16000 CPU cores, imposing a processing time limit of about 160 ms assuming the L1 input rate is 100 kHz. 
The disk buffer used to store data before they are transferred to Tier 0 has a bandwidth limit of around 8 GB/s. 
During stable operation, i.e. amount of data transferred into disk buffer is almost equal to the amount of data transferred out to Tier 0,
this imposes a limit of HLT output of 4 GB/s. 
Based on the average file size and processing time of events, the HLT output rate limit varies from about 400 Hz to 20 kHz. 
In the same way as the L1 system, the output rate can be adjusted by changing thresholds of the triggers or by prescaling the events. 
The prescaling is done differently at the HLT than at L1. 
Instead of skipping events after the trigger decision, events are skipped before running the HLT algorithm, to reduce the average processing time of events.

Among the CMS collaboration, each physics analysis group design their own L1 triggers and HLT to select events of their specific physics interests. 
The following sections describe the trigger used in the analysis presented in this thesis.

\section{The Minimum Bias trigger}
\label{sec:MBtrigger}
Almost all trigger selections introduce a bias as they select only certain sub-set of all collision events and reject the others. 
MinimumBias (MB) events refers to events that are selected with a loose trigger which accepts a large fraction of the overall inelastic cross section of particle collisions. 
Such triggers are referred as MinimumBias triggers, which trigger on minimum detector activity to ensure the bias is very small. 
During the many years of LHC operation, the beam conditions kept changing and the CMS detector was upgraded several times. 
Therefore, different MB trigger algorithms were used to take MB events for different LHC run periods, those relevant to the analysis in this thesis are as follow:
\begin{itemize}
\item 2010 pp data taking: Events were selected by a trigger signal in each side of the BSC scintillators~\cite{Bell:2008oja}, coincident with a signal from either of the two detectors indicating the presence of at least one proton bunch crossing the interaction point at CMS. The trigger was named HLT\_L1\_BscMinBiasOR\_BptxPlusORMinus and had efficiency around 97\% for hadronic inelastic collisions.
\item 2011 PbPb data taking: The MB events were collected using coincidences between the trigger signals from both sides of either the BSC or the HF detector. The trigger was named HLT\_HIMinBiasHfOrBSC and had efficiency above 97\% for hadronic inelastic collisions.
\item 2013 pPb data taking: The relatively low pPb collision frequency (up to 0.2 MHz) provided by the LHC in the nominal run allowed the use of a track-based MB trigger, HLT\_PAZeroBiasPixel\_SingleTrack. Here, ZeroBias refers to the crossing of two beams (bunch crossing) at CMS. For every few thousand pPb bunch crossings, the detector was read out from the L1 trigger and events were accepted at the HLT if at least one track (reconstructed with only the pixel tracker information) with $\pt\ > 0.4$ GeV/c was found. The trigger had a efficiency of 99\% for hadronic inelastic collisions.
\item 2015-2016 pp data taking: A L1 fine-grain bit based HF trigger was used to select MB events. The fine-grain bit was set for each side of HF if one or more of the 6 readout towers has transverse energy ($E_T$) above a analog-to-digital converter (ADC) threshold of 7. Around 0.01\% of all events with one side of the HF fine-grain bit being set was accepted at L1, and all of them were accepted by a HLT pass-through, HLT\_L1MinimumBiasHF\_OR. The trigger efficiency was around 96\% for hadronic inelastic collisions.
\end{itemize}

\section{The high multiplicity trigger}
\label{sec:theHMtrigger}
With the goal of studying the properties of high multiplicity pPb and pp collisions, a dedicated trigger was designed and implemented since October, 2009. 
Such a trigger aimed at capturing significant samples of data covering a wide range of multiplicities, especially at the high multiplicity region. 

The high multiplicity triggers mainly involved two levels:
\begin{itemize}
\item L1: A trigger filtering on scalar sum of total transverse momentum at L1 (L1\_ETT) over the CMS calorimetry, including ECAL and HCAL, is used to select events with high multiplicity. During 2009-2010 pp data taking, the HF energy is also included in the calculation of $E_{TT}$.
\item HLT: As track reconstruction becomes available at HLT level, number of reconstructed pixel tracks is used to filter out high multiplicity events. However, a simple counting of all reconstructed pixel tracks would lead to significant contributions from pileup events, instead of high track multiplicity produced from a single collision. To reduce the number of pileup events selected, the trigger proceeds with the following sequences: the reconstructed pixel tracks with $\pt\ > 0.4$GeV, which originating within a cylindrical region of 15 cm half length and 0.2 cm in transverse radius with respect to the beamspot, are used to reconstruct vertices. The trigger then counts the number of pixel tracks with kinematic cuts of $|\eta| < 2.4$ and $\pt\ > 0.4$ GeV/c, within a distance of 0.12 cm in z-direction to the vertex associated with highest number of tracks. The position of vertices along the nominal interaction point along the beam axis is required to be within $\pm 15$cm range.
\end{itemize}

\begin{figure}
	\centering
	\includegraphics[width=0.46\textwidth]{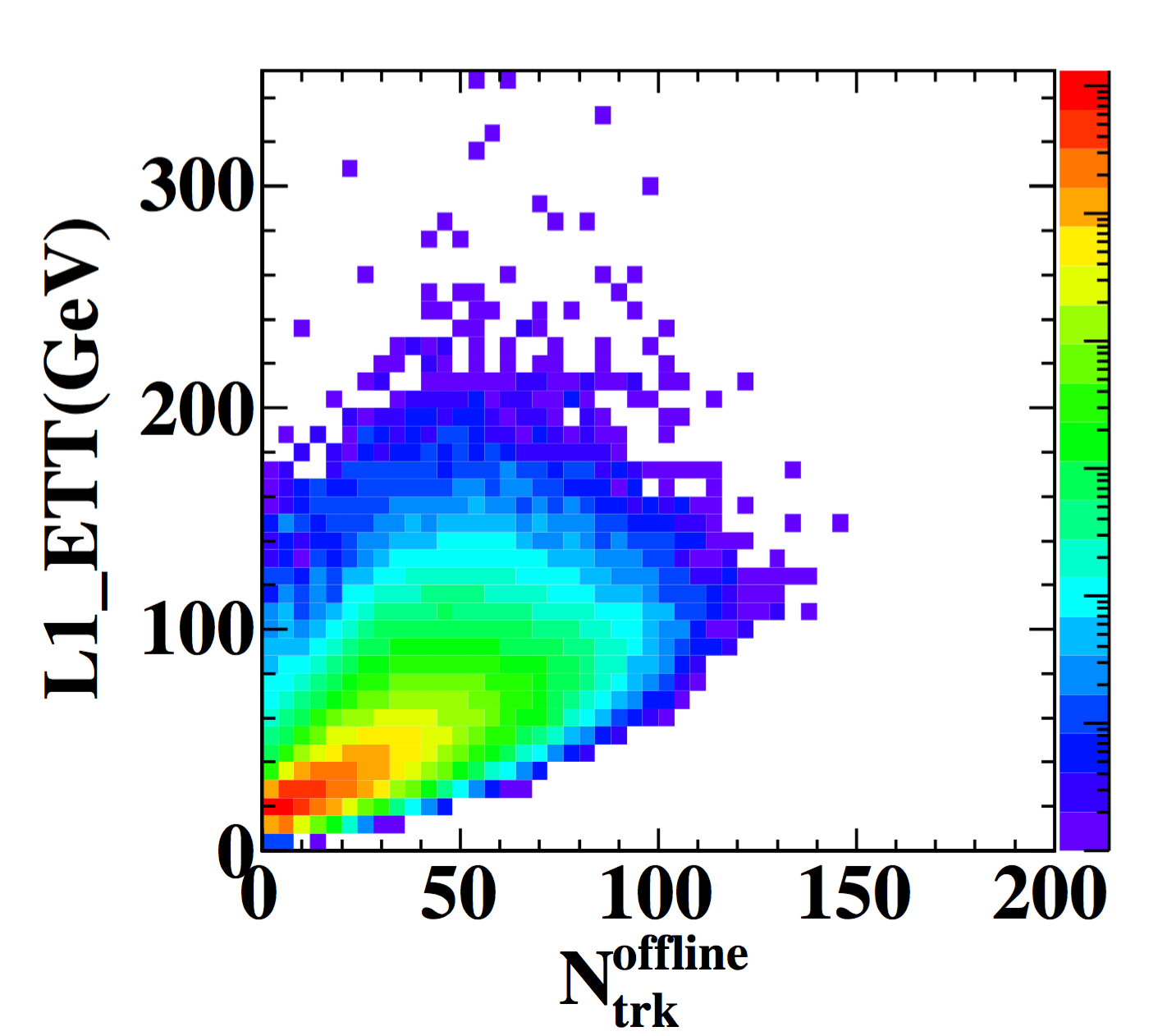}
	\includegraphics[width=0.44\textwidth]{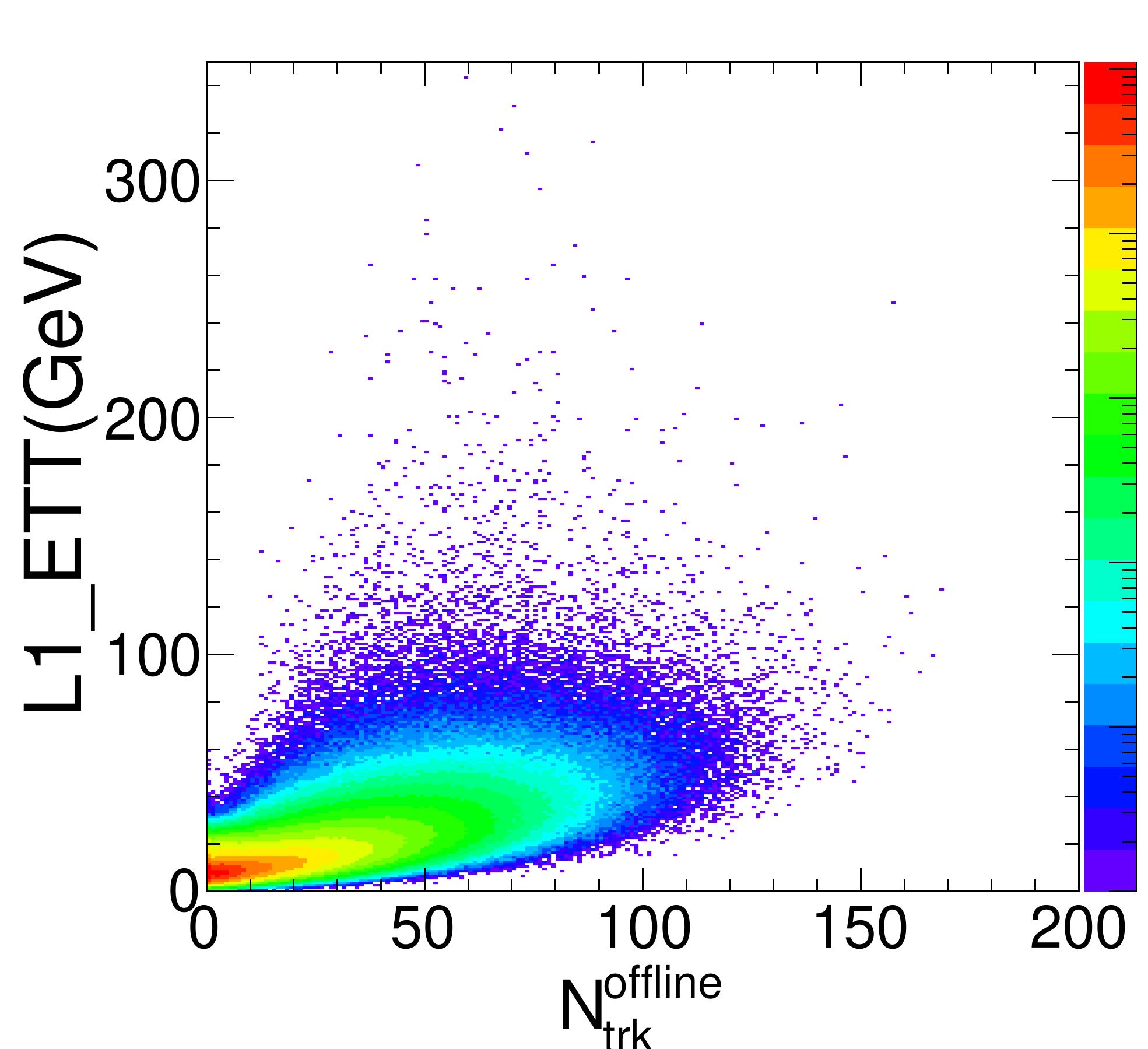}
	\caption{L1\_ETT vs. \noff\ for 7TeV pp collisions (left) and 13 TeV pp collisions during 2015 EndOfFill run (right).}
        \label{fig:ETTvsNtrk}
\end{figure}

Figure~\ref{fig:ETTvsNtrk} illustrates the correlation between L1\_ETT and \noff\ for MB events taken in 2009-2010 for 7 TeV pp collisions and in 2015 for 13 TeV pp collisions during the EndOfFill run in July. 
Here, the multiplicity of offline reconstructed tracks (described in Sec.~\ref{sec:tracking}), \noff, is counted within the kinematic cuts of $|\eta| < 2.4$ and $\pt > 0.4$ GeV. 
Due to the inclusion of HF energy in the $E_{TT}$ calculation during 2009-10 data taking, $E_{TT}$ is much larger for 7 TeV pp collisions compared to those for 13 TeV at the same \noff\ values. 
For a give region of \noff, one can always find a threshold of $E_{TT}$ such that almost all events are kept above the threshold. 
For example, for 7 TeV pp collisions, a $E_{TT}$ threshold of 60 captures almost all events with $\noff \geq 85$. 

\begin{figure}
	\centering
	\includegraphics[width=0.42\textwidth]{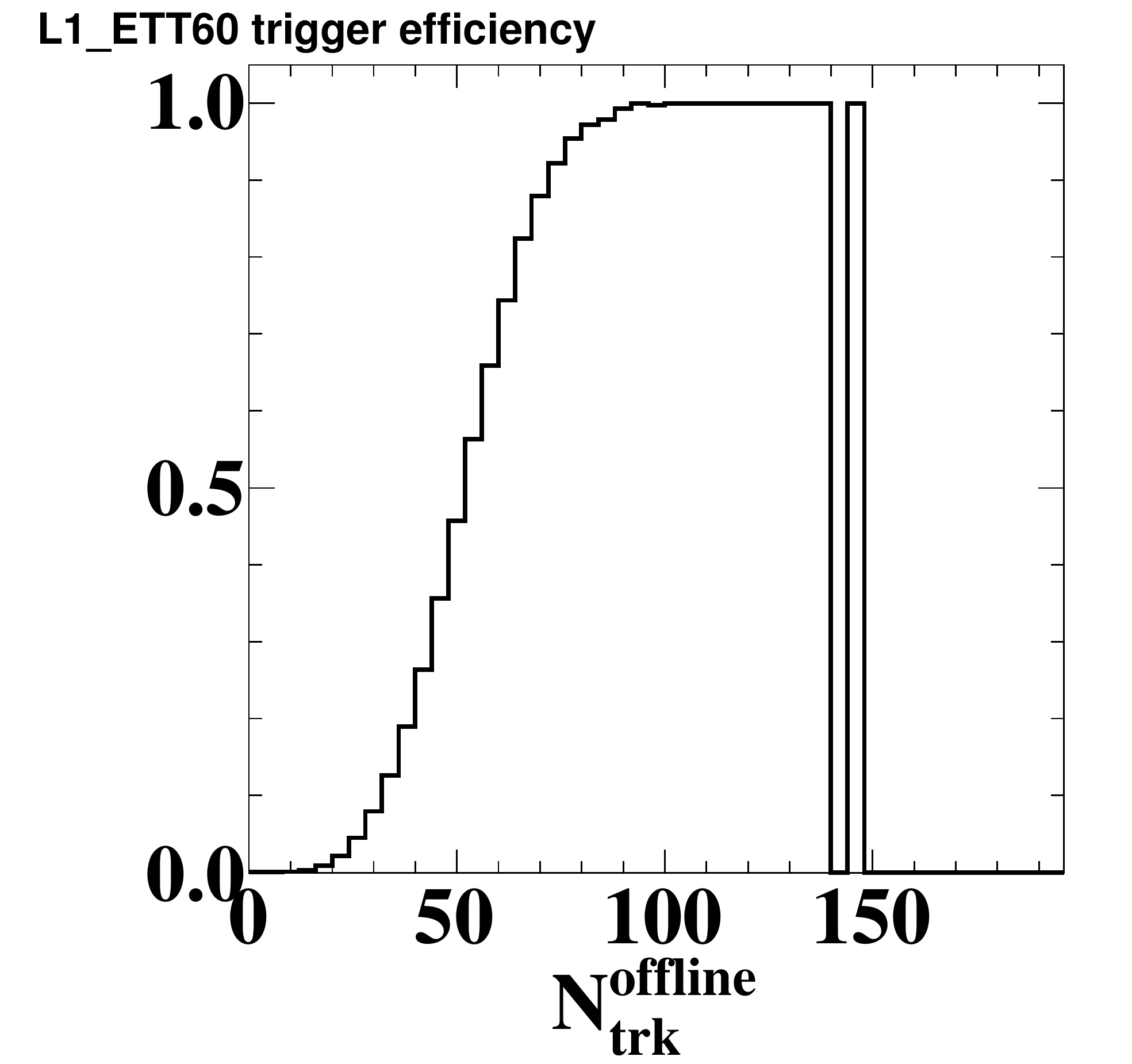}
	\includegraphics[width=0.48\textwidth]{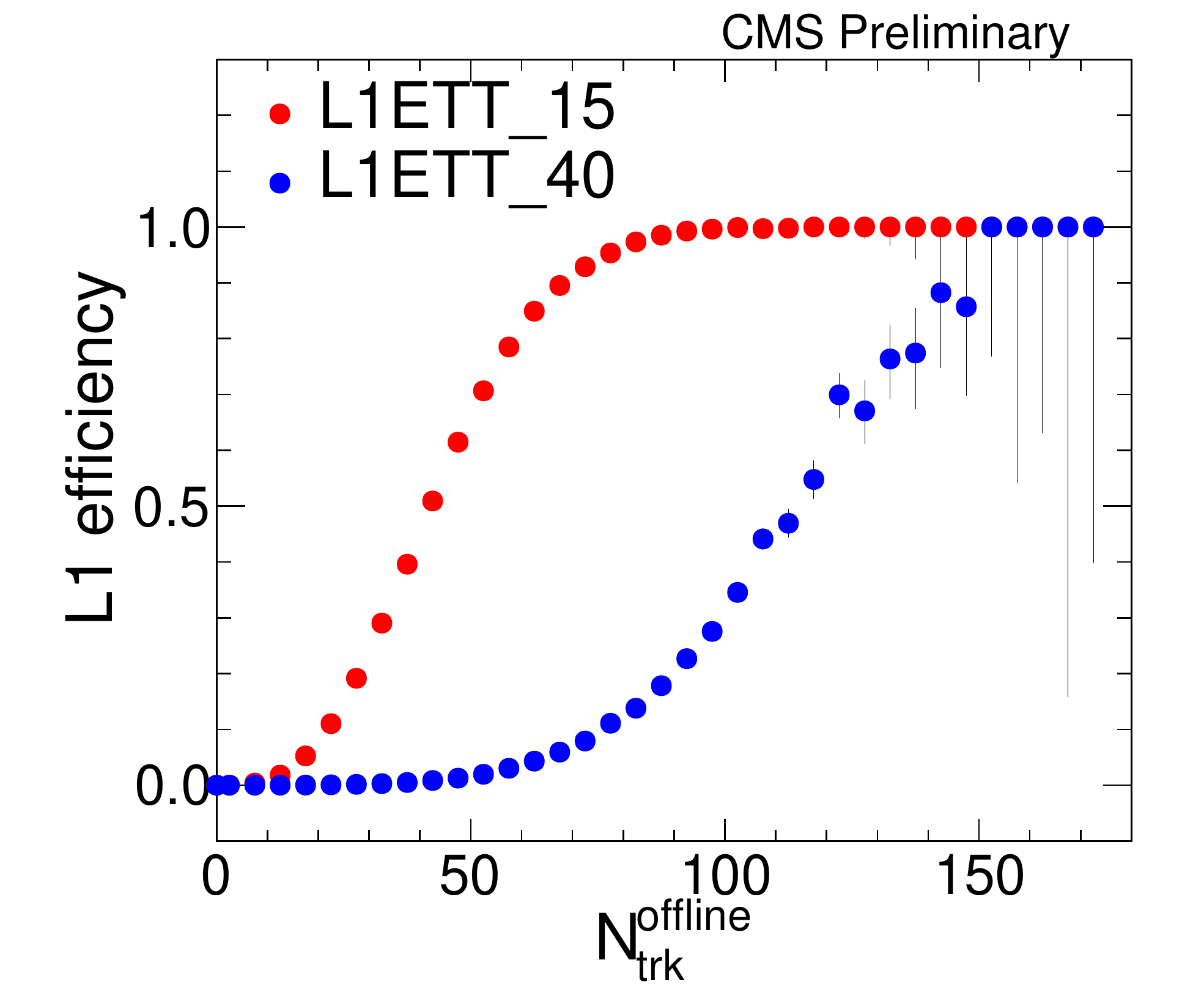}
	\caption{L1 triggering efficiency for 7TeV pp collisions (left) and 13 TeV pp collisions during 2015 EndOfFill run (right).}
        \label{fig:ETTeff}
\end{figure}

In order to reach the calorimeter, a track has to have at least $\pt\ > 0.8$GeV. 
Events that produce more high \pt\ tracks have a better chance of being accepted by the trigger. 
Therefore, a bias can be introduced in this way if L1\_ETT trigger efficiency is not 100\% at a fixed \noff\ range. 
To largely avoid such bias, the trigger setup follows a simple rule of having a L1\_ETT efficiency close to 90\% at the desired \noff\ range. 
For 7 TeV pp collisions, L1\_ETT60 is chosen for $\noff \geq 90$. 
For 13 TeV pp collisions during 2015 EndOfFill run, L1\_ETT15 is chosen for $\noff \geq 85$ and L1\_ETT40 is chosen for $\noff \geq 135$.
L1 triggering efficiencies derived from the correlation between L1\_ETT and \noff\ are shown in Fig.~\ref{fig:ETTeff} for the two runs described above. 

As the data used in this thesis are taken over a wide range of time, the detector conditions and calibrations kept changing. 
Particularly, the changes in calibrations of ECAL and HCAL affect the overall scale of $E_{TT}$. 
To keep the triggers aiming at same multiplicity range, the L1\_ETT thresholds had to be tuned from time to time. 
Table~\ref{tab:triggerseed1} summarizes the trigger setup for all the data samples used. 

\begin{landscape}
\begin{table}
\centering
\caption{L1 seeds of different HLT trigger paths for different 13 TeV pp runs.}
\begin{tabular}{ c | c | c | c | c  }
\hline
Collision & Energy & Year, run & HLT & L1 \\
\hline
\multirow{13}{*}{pp} & 5 TeV & 2015 & HLT\_PixelTracks\_Multiplicity60 & L1\_ETT40 \\
\cline{2-5}
 & \multirow{3}{*}{7 TeV} & \multirow{3}{*}{2010} & HLT\_PixelTracks\_Multiplicity70 & L1\_ETT60 \\
 & & & HLT\_PixelTracks\_Multiplicity85 & L1\_ETT60 \\
 & & & HLT\_PixelTracks\_Multiplicity100 & L1\_ETT70 \\
 \cline{2-5}
 & \multirow{9}{*}{13 TeV} & \multirow{3}{*}{2015, EndOfFill} & HLT\_PixelTracks\_Multiplicity60 & L1\_ETT15 \\
 & & & HLT\_PixelTracks\_Multiplicity85 & L1\_ETT15 \\
 & & & HLT\_PixelTracks\_Multiplicity110 & L1\_ETT40 \\
\cline{3-5}
 & & \multirow{3}{*}{2015, VdM scan} & HLT\_PixelTracks\_Multiplicity60 & L1\_ETT15 \\
 & & & HLT\_PixelTracks\_Multiplicity85 & L1\_ETT15 \\
 & & & HLT\_PixelTracks\_Multiplicity110 & L1\_ETT15 \\
\cline{3-5}
 & & \multirow{3}{*}{2015, TOTEM} & HLT\_PixelTracks\_Multiplicity60 & L1\_ETT40 \\
 & & & HLT\_PixelTracks\_Multiplicity85 & L1\_ETT45 \\
 & & & HLT\_PixelTracks\_Multiplicity110 & L1\_ETT55 \\
\hline
\multirow{4}{*}{pPb} & \multirow{4}{*}{5.02 TeV} & \multirow{4}{*}{2013} & HLT\_PixelTracks\_Multiplicity100 & L1\_ETT20 \\
 & & & HLT\_PixelTracks\_Multiplicity130 & L1\_ETT20 \\
 & & & HLT\_PixelTracks\_Multiplicity160 & L1\_ETT40 \\
 & & & HLT\_PixelTracks\_Multiplicity190 & L1\_ETT40 \\
 \hline
\end{tabular}
\label{tab:triggerseed1}
\end{table}
\end{landscape}

\begin{figure}
	\centering
	\includegraphics[width=0.6\textwidth]{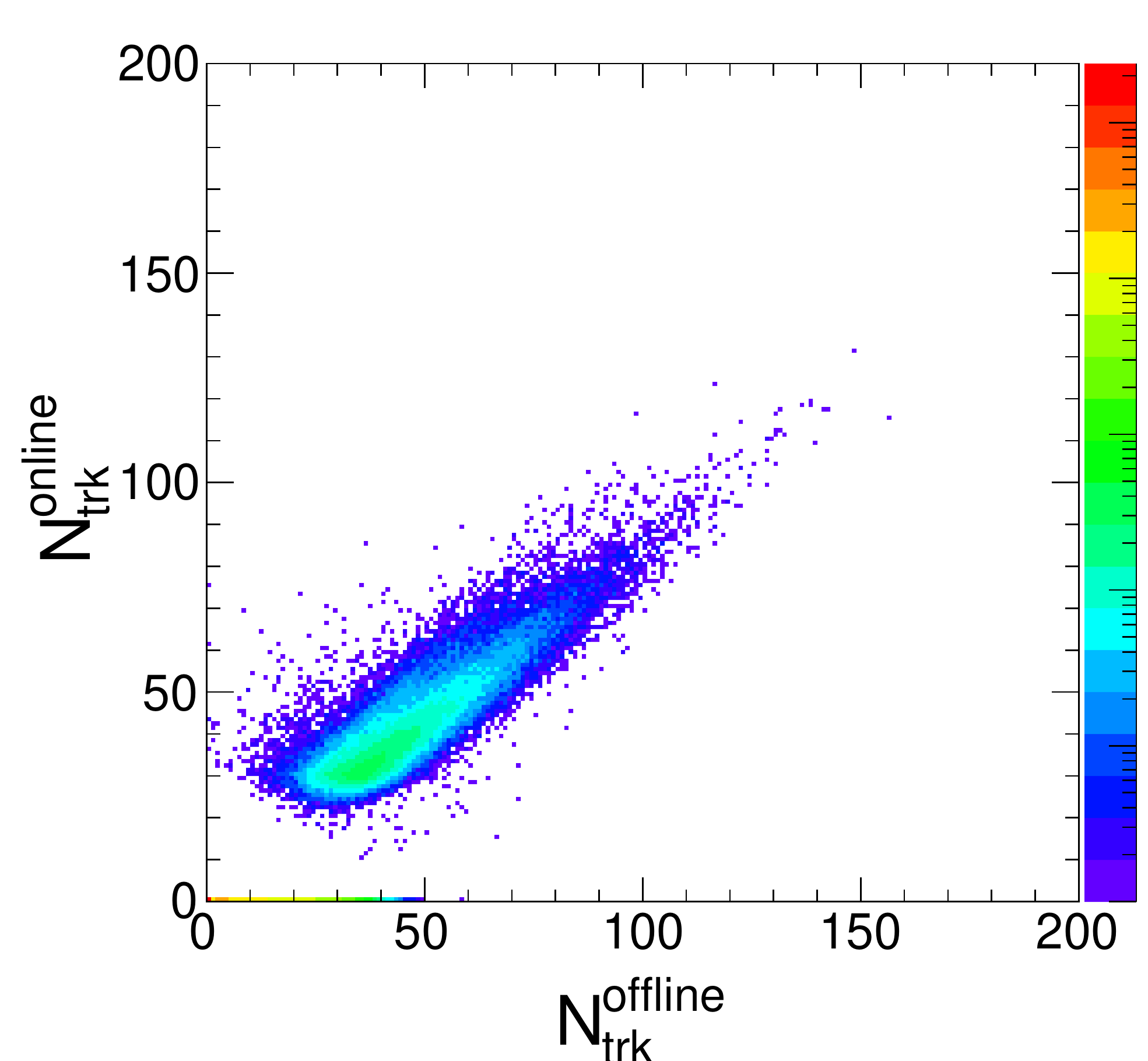}
	\includegraphics[width=0.65\textwidth]{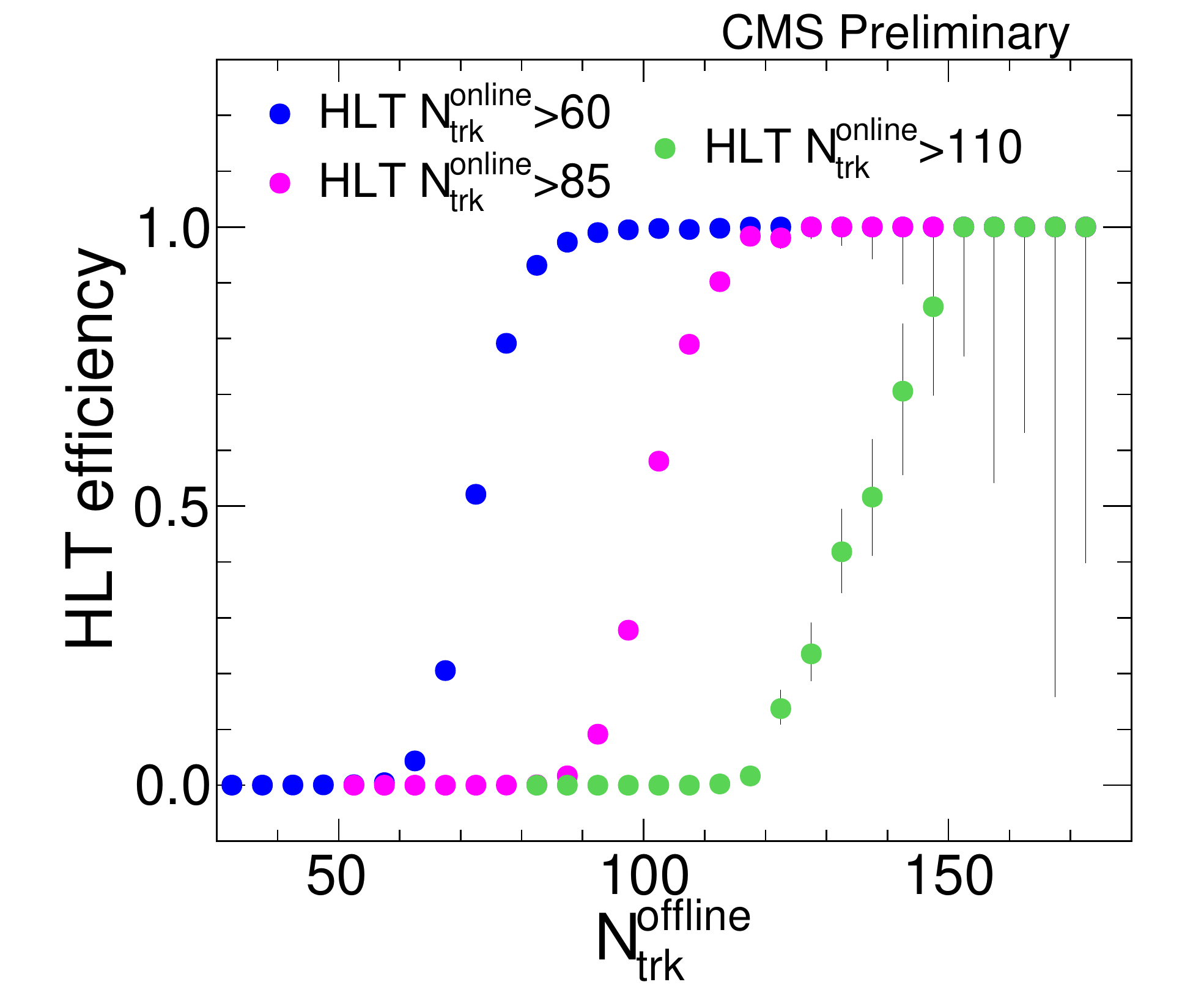}
	\caption{Correlation between \nonline\ and \noff\ (top) and HLT efficiency (bottom) for 13 TeV pp collisions during 2015 EndOfFill run.}
        \label{fig:NtrkOnlineOffline}
\end{figure}

The efficiency of HLT depends on how well the number of reconstructed pixel tracks (\nonline) is correlated with \noff. 
Fig.~\ref{fig:NtrkOnlineOffline} shows the strong correlation between \nonline\ and \noff, and the HLT efficiency for 13 TeV pp collisions during 2015 EndOfFill run. 
The de-correlation between \nonline\ and \noff\ at low \noff\ region is due to the requirement at HLT that vertex is only reconstructed when there is at least 30 tracks associated to it. 
Such a requirement is implemented to reduce the processing time of the trigger, and is not causing any efficiency loss at high \noff\ region.
Loss of efficiency at HLT is mainly due to the smearing between online and offline track reconstructions, which does not introduce any bias on the events selected. 
Therefore, to maximize the statistics of high multiplicity events, events with more than 50-60\% HLT efficiency are accepted for use in the analysis. 
Table~\ref{tab:triggerN} summarizes the corresponding \noff\ regions used for analysis for different run periods. 

\begin{landscape}
\begin{table}
\centering
\caption{HLT trigger used for each \noff\ range, for different pp and pPb runs. The numbers in curly brackets means all of those triggers are used for the corresponding \noff\ range.}
\begin{tabular}{ c | c | c | c | c  }
\hline
Collision & Energy & Year & \noff\ range & HLT \\
\hline
\multirow{10}{*}{pp} & \multirow{2}{*}{5 TeV} & \multirow{2}{*}{2015} & [0,90) & HLT\_L1MinimumBiasHF1OR \\
 & & & [90,$\infty$) & HLT\_PixelTracks\_Multiplicity60 \\
\cline{2-5}
 & \multirow{4}{*}{7 TeV} & \multirow{4}{*}{2010} & [0,90) & HLT\_L1\_BscMinBiasOR\_BptxPlusORMinus \\
 & & & [90,110) & HLT\_PixelTracks\_Multiplicity70 \\
 & & & [110,130) & HLT\_PixelTracks\_Multiplicity\{70,85\} \\
 & & & [130,$\infty$) & HLT\_PixelTracks\_Multiplicity\{70,85,100\} \\
\cline{2-5}
& \multirow{4}{*}{13 TeV} & \multirow{4}{*}{2015} & [0,85) & HLT\_L1MinimumBiasHF\_OR \\
 & & & [85,105) & HLT\_PixelTracks\_Multiplicity60 \\
 & & & [105,135) & HLT\_PixelTracks\_Multiplicity\{60,85\} \\
 & & & [135,$\infty$) & HLT\_PixelTracks\_Multiplicity\{60,85,110\} \\
 \hline
\multirow{5}{*}{pPb} & \multirow{5}{*}{13 TeV} & \multirow{5}{*}{2013} & [0,120) & HLT\_PAZeroBiasPixel\_SingleTrack \\
 & & & [120,150) & HLT\_PixelTracks\_Multiplicity100 \\
 & & & [150,185) & HLT\_PixelTracks\_Multiplicity\{100,130\} \\
 & & & [185,220) & HLT\_PixelTracks\_Multiplicity\{100,130,160\} \\
 & & & [220,$\infty$) & HLT\_PixelTracks\_Multiplicity\{100,130,160,190\} \\
 \hline
\end{tabular}
\label{tab:triggerN}
\end{table}
\end{landscape}

\begin{figure}[htb]
	\centering
	\includegraphics[width=0.7\textwidth]{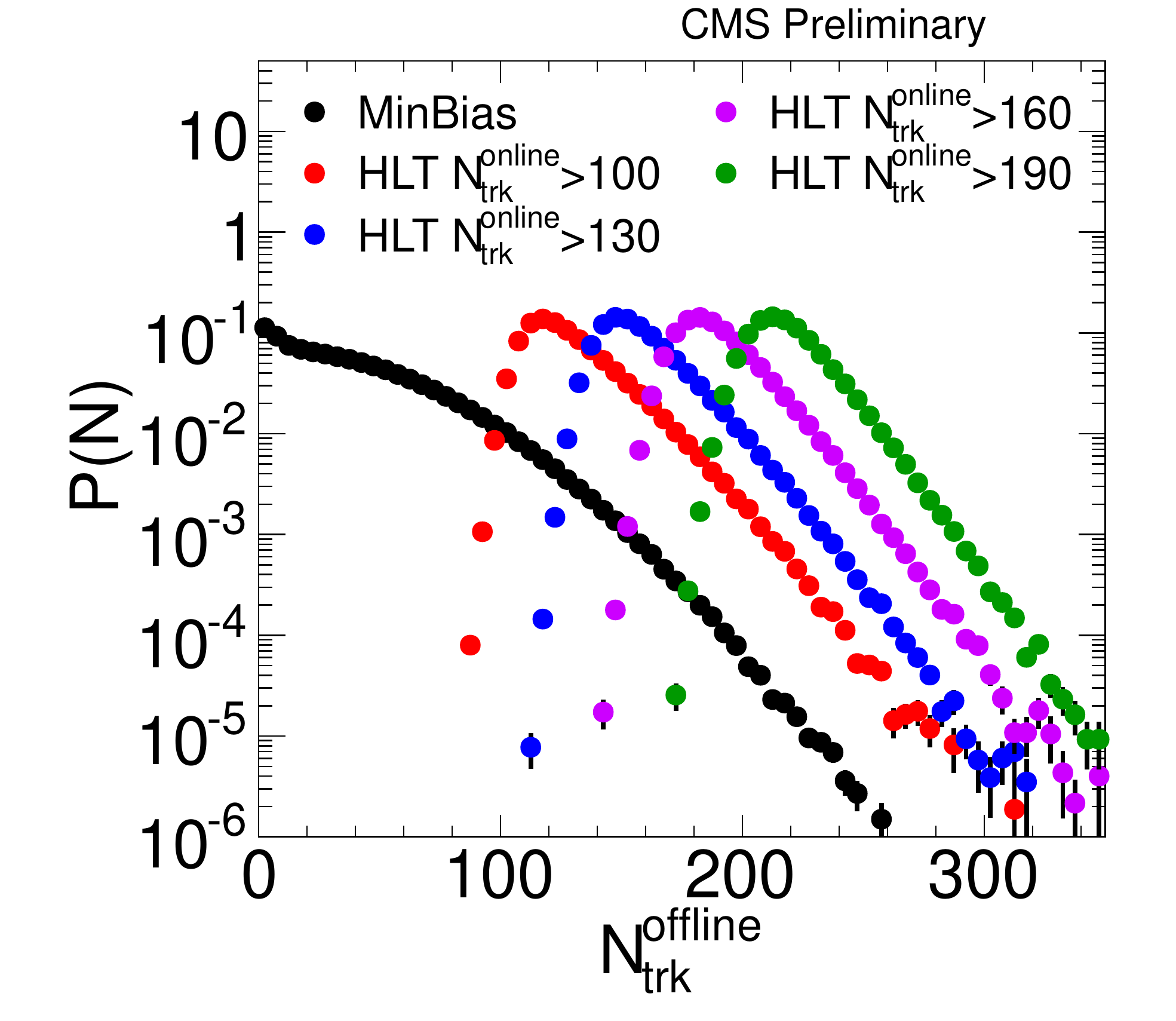}
	\caption{\noff\ distributions, normalized to unit integral, for MB and high multiplicity triggered events for 5.02 TeV pPb collision.}
        \label{fig:HMenhance}
\end{figure}

The implementation of the high multiplicity trigger largely enhances the statistics of the high multiplicity events, allowing the analysis to reach much further into the high multiplicity tail of the multiplicity distribution of the MB collisions. 
Fig.~\ref{fig:HMenhance} shows the offline track multiplicity distribution, normalized to unit integral, for MB and high multiplicity triggered events for 5.02 TeV pPb collision. 
A factor of at least $10^3$ enhancement at $\noff > 200$ region can be obtained with the high multiplicity triggers, and such enhancement is even larger at higher \noff\ regions.

Due to the limitation on the output rate of L1 and HLT, prescales have to be applied to the high multiplicity triggers. 
The prescale setup for different run periods is based on two goals:
\begin{itemize}
\item The highest multiplicity events from the collisions are always the top focus of the analysis, since they might reveal novel features. Therefore, the trigger setup is designed to always keep an un-prescaled trigger with the lowest possible multiplicity threshold.
\item  Besides the un-prescaled trigger, several lower threshold triggers are implemented in a way that all the triggers run at almost identical HLT output rate to ensure there is no intermediate multiplicity region with low statistics. 
\end{itemize}
\noindent During the pp runs, typical bandwidth assigned to high multiplicity trigger package was around 60 kHz at L1 and around 100-300 Hz at HLT. 
While those numbers were largely reduced for 2013 pPb run to around 10 kHz at L1 and 100Hz at HLT.

\clearpage

\subsection{High multiplicity trigger upgrades for 2016-2017 runs}
\label{subsec:HMtriggerUp}

To improve the performance of the high multiplicity trigger, several upgrades were done for the 2016 data taking for pp and pPb collisions. 

\begin{figure}[bh]
	\centering
	\includegraphics[width=0.45\textwidth]{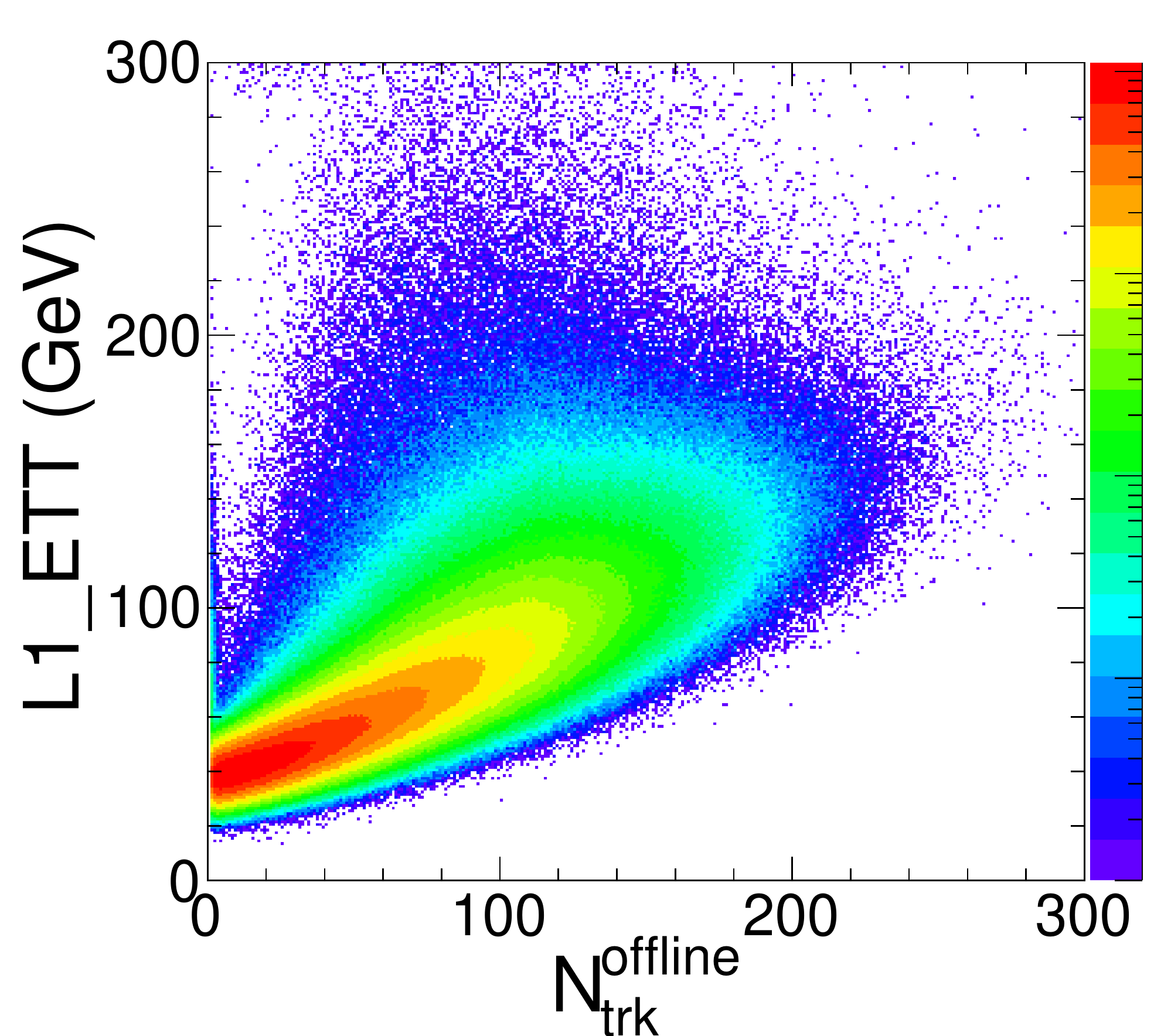}
	\includegraphics[width=0.45\textwidth]{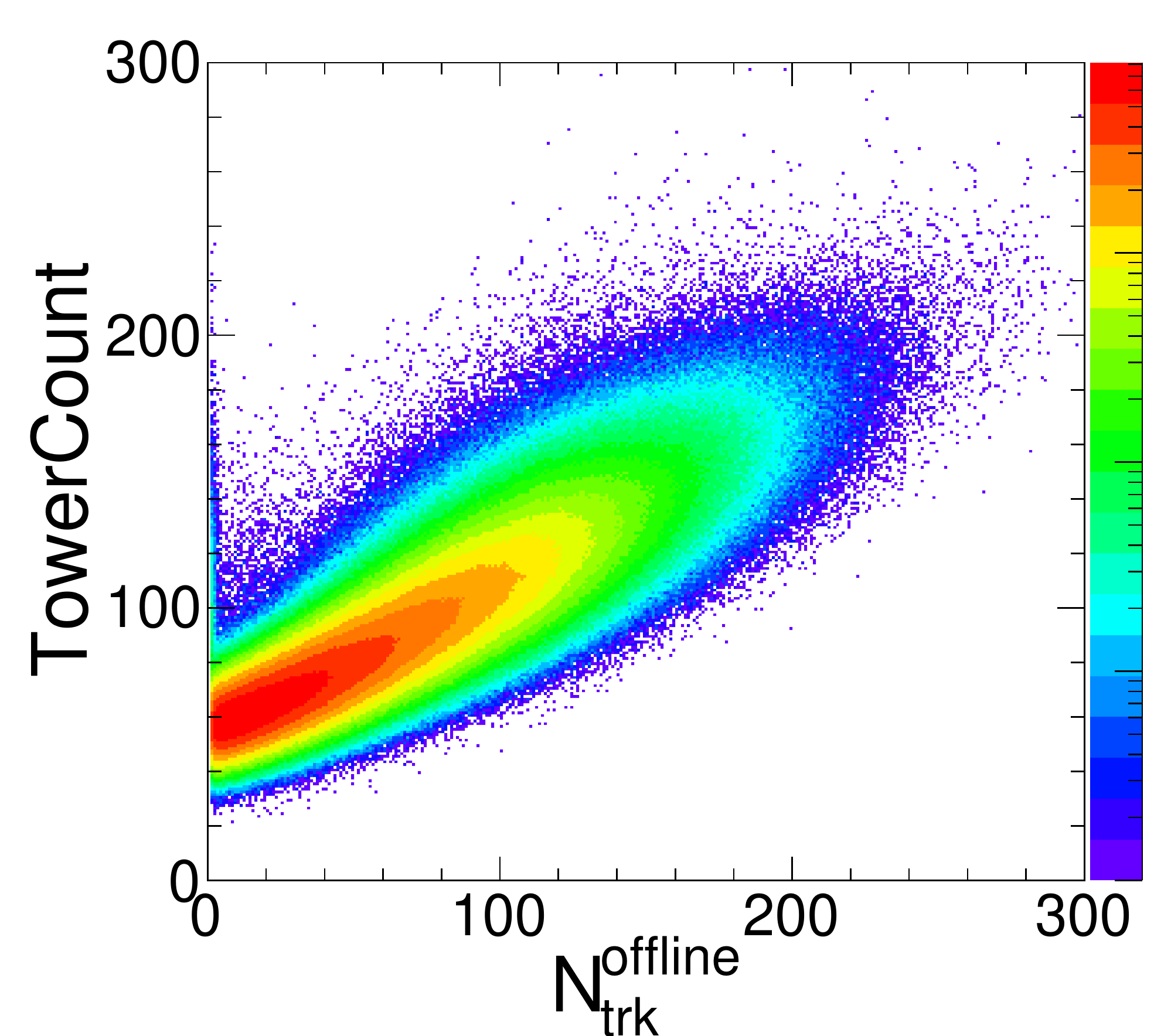}
	\caption{Correlation between $E_{TT}$ and \noff\ (left), and TC and \noff\ (right) for 8 TeV pPb collisions.}
        \label{fig:TCvsETT}
\end{figure}

At L1, a brand new algorithm, named tower count (TC), was introduced to count the number of active towers in barrel ECAL and HCAL detectors. 
An active tower is defined as a trigger tower (ECAL + HCAL) with a transverse energy greater than 0.5 GeV. 
As mentioned in Sec.~\ref{sec:theHMtrigger}, a bias could be introduced by the $E_{TT}$ trigger in a way that events produce more high \pt\ tracks have a better chance of being accepted. 
Those events end up having large values of $E_{TT}$ but low numbers of \noff. 
Such a bias is reduced in the TC trigger as higher \pt\ particles are treated equally as lower \pt\ particles in an event as long as they deposit more than 0.5 GeV energy in the trigger tower.
Fig.~\ref{fig:TCvsETT} shows the correlation between $E_{TT}$ and \noff\ and correlation between TC and \noff\ for 2016 pPb collisions. 
A better correlation with \noff\ is established by TC in a way that there are fewer events with high TC values but low \noff. 

\begin{figure}
	\centering
	\includegraphics[width=0.8\textwidth]{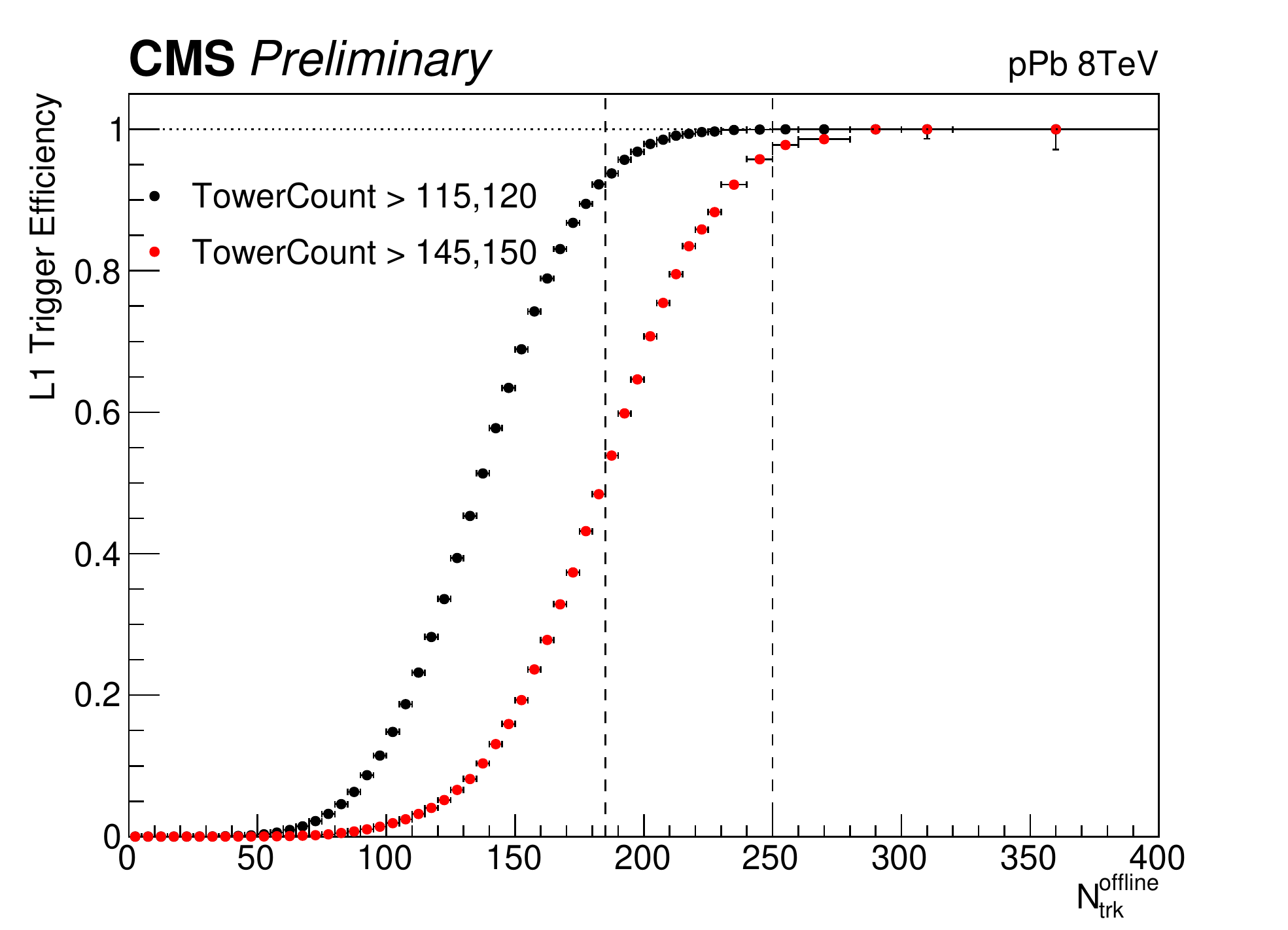}
	\caption{Efficiency of L1 tower count triggers for 8 TeV pPb collisions. Vertical dash lines indicate the region of events used for analysis.}
        \label{fig:TCeff}
\end{figure}

The TC trigger was used only during the 2016 8 TeV pPb data taking so far. 
Thresholds of 115 or 120 were used for event multiplicity between 185 and 250, and thresholds of 145 or 150 for event multiplicity above 250. 
The reason for the usage of two different TC thresholds for the same multiplicity range is related with the observation of a considerable change in the noise level of HCAL during data taking due to beam quality, which shifted the entire TC distribution by a constant of 5 GeV. 
Fig.~\ref{fig:TCeff} shows the L1 efficiency for TC triggers for 2016 pPb collisions. 
To avoid any potential bias, events with an efficiency above 95\% are considered good for analysis. 

\begin{figure}
	\centering
	\includegraphics[width=0.5\textwidth]{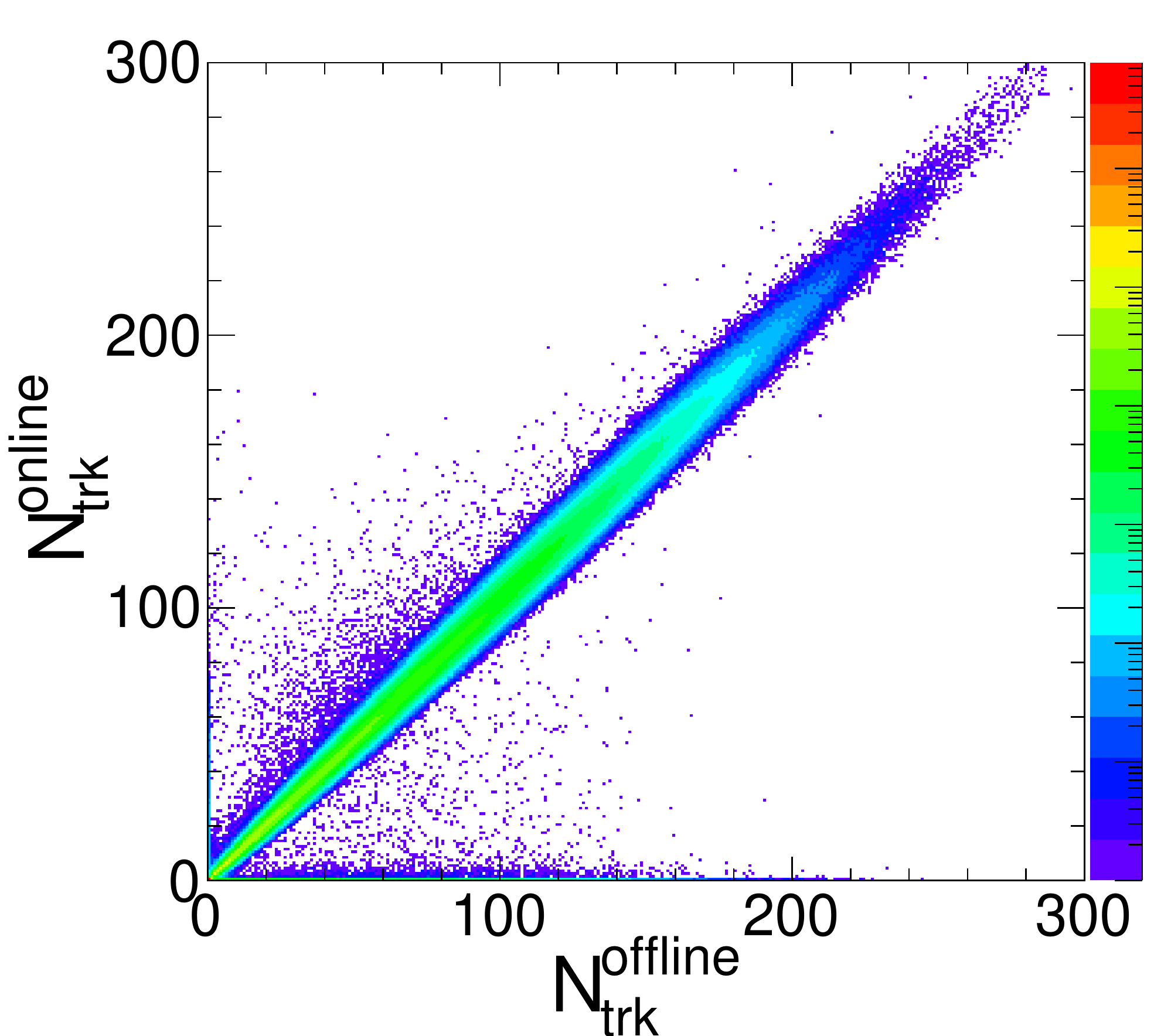}
	\includegraphics[width=0.9\textwidth]{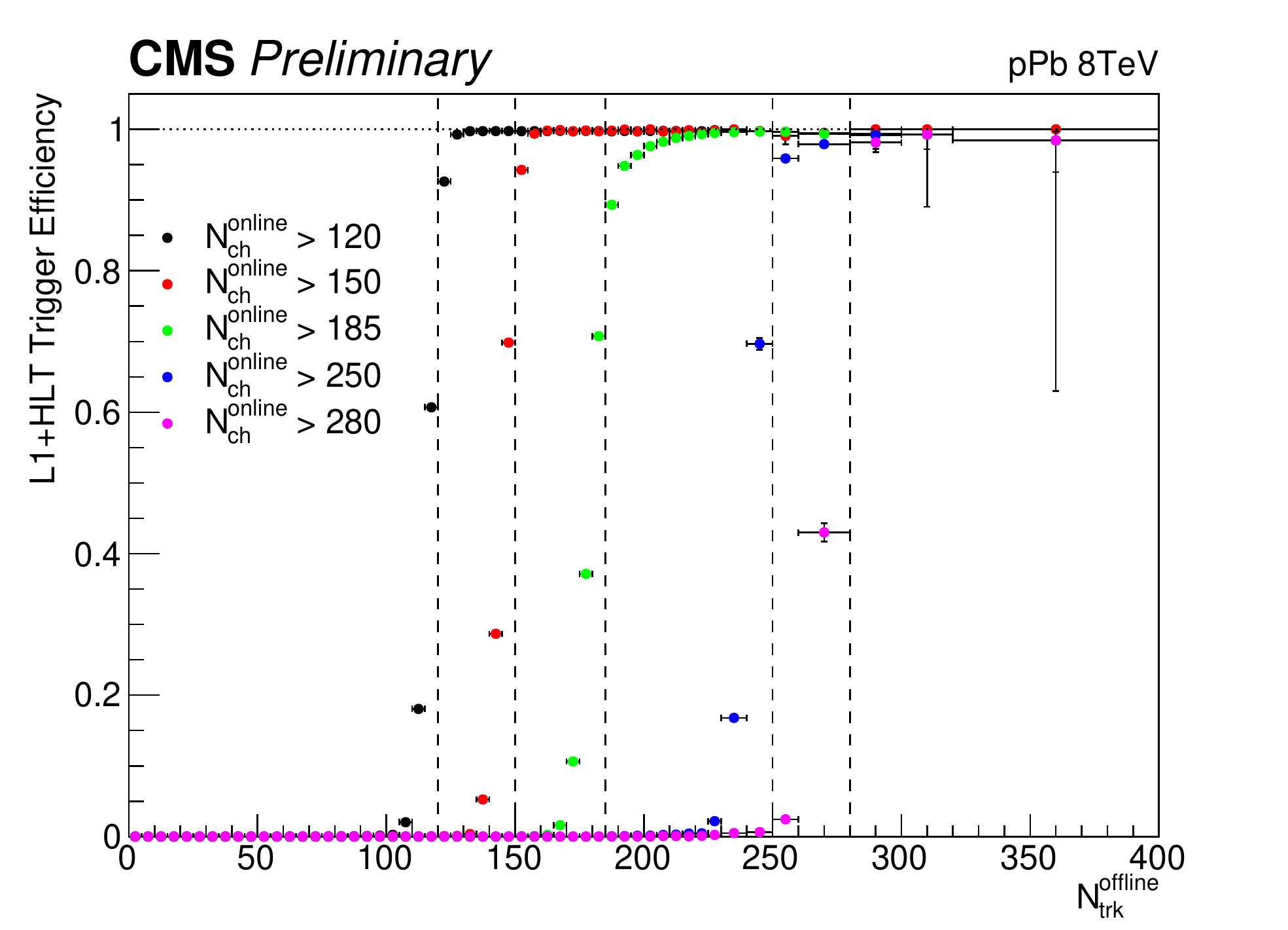}
	\caption{Correlation between \nonline\ and \noff\ (left) and HLT efficiency (right) for 8 TeV pPb collisions.}
        \label{fig:newOnlinevsOffline}
\end{figure}

At HLT, new tracking algorithm was implemented using information from the full tracking system instead of only the pixel detector. 
The track reconstruction at HLT was upgraded to be identical to the offline iterative tracking described in Sec.~\ref{sec:tracking}. 
Fig.~\ref{fig:newOnlinevsOffline} shows the correlation between \nonline\ and \noff\ with the new tracking algorithm, which is much better than what has been shown in Fig.~\ref{fig:NtrkOnlineOffline} with pixel track reconstruction. 
The HLT efficiency is also shown in the same plot.

However, the iterative tracking consumes much more processing time and memory compared to the pixel only track reconstruction. 
Since the computing resource available during data taking is limited, it is not wise to have it run on every event which passes the L1 trigger. 
Therefore, the pixel track reconstruction is still kept as a pre-filter of the full iterative tracking. 
A looser multiplicity cut is applied on the number of pixel tracks reconstructed, before full iterative tracking is executed with a tighter cut on multiplicity. 
Addition of the pre-filter reduces the average processing time by a factor of 2. 

\cleardoublepage
\chapter{Data and Monte Carlo samples}
\label{ch:datamc}

In this chapter, the data samples used for the analysis presented are introduced, together with all the Monte Carlo samples. 

\section{Data samples}

The analysis of the two-particle correlations in high multiplicity pp and pPb collisions is performed using the data recoded by CMS, 
which were certified by the CMS data certification team. 
Data are defined as good for physics analysis if all sub-detectors, trigger and physics objects (tracking, electron, muon, photon and jet) show the expected performance.
Table~\ref{tab:datasamples}-~\ref{tab:datasamplespPb} summarise the detailed information of the data samples used in this work. 
The data sample names can be found in Appendix.~\ref{apx:dataname}.

\begin{landscape}
\begin{table}
\centering
\caption{Detail information of the pp data sample used in this work, including pileup, integrated luminosity and number of events collected by the triggers.}
\begin{tabular}{ c | c | c | c | c | c | c }
\hline
Collision & Energy & Year & Pileup & Int. lumi & Trigger & Triggered events (million)  \\
\hline
\multirow{10}{*}{pp} & \multirow{2}{*}{5 TeV} & \multirow{2}{*}{2015} & \multirow{2}{*}{1.3} & \multirow{2}{*}{1.0 pb$^{-1}$} & HLT\_L1MinimumBiasHF1OR & 2500 \\
 & & & & & HLT\_PixelTracks\_Multiplicity60 & 3.7 \\
\cline{2-7}
 & \multirow{4}{*}{7 TeV} & \multirow{4}{*}{2010} & \multirow{4}{*}{0.01-0.8} & \multirow{4}{*}{6.2 pb$^{-1}$} & HLT\_L1\_BscMinBiasOR\_BptxPlusORMinus  & 41 \\
 & & & & & HLT\_PixelTracks\_Multiplicity70 & 1.5 \\
 & & & & & HLT\_PixelTracks\_Multiplicity85 & 2.1 \\
 & & & & & HLT\_PixelTracks\_Multiplicity100 & 0.6 \\
\cline{2-7}
 & \multirow{4}{*}{13 TeV} & \multirow{4}{*}{2015} & \multirow{4}{*}{0.1-1.3} & \multirow{4}{*}{0.7 pb$^{-1}$} & HLT\_L1MinimumBiasHF\_OR & 180 \\
 & & & & & HLT\_PixelTracks\_Multiplicity60 & 10.1 \\
 & & & & & HLT\_PixelTracks\_Multiplicity85 & 7.7 \\
 & & & & & HLT\_PixelTracks\_Multiplicity110 & 0.3 \\
 \hline
\end{tabular}
\label{tab:datasamples}
\end{table}
\end{landscape}

\begin{landscape}
\begin{table}
\centering
\caption{Detail information of the pPb and PbPb data sample used in this work, including pileup, integrated luminosity and number of events collected by the triggers.}
\begin{tabular}{ c | c | c | c | c | c | c }
\hline
Collision & Energy & Year & Pileup & Int. lumi & Trigger & Triggered events  \\
\hline
\multirow{5}{*}{pPb} & \multirow{5}{*}{5.02 TeV} & \multirow{5}{*}{2013} & \multirow{5}{*}{0.06} & \multirow{5}{*}{35 nb$^{-1}$} & HLT\_PAZeroBiasPixel\_SingleTrack & 31.4 \\
 & & & & & HLT\_PixelTracks\_Multiplicity100 & 19.2 \\
 & & & & & HLT\_PixelTracks\_Multiplicity130 & 18.9 \\
 & & & & & HLT\_PixelTracks\_Multiplicity160 & 17 \\
 & & & & & HLT\_PixelTracks\_Multiplicity190 & 8 \\
 \hline
 PbPb & 2.76 TeV & 2011 & 0.001 & 150 $\mu$b$^{-1}$ & HLT\_HIMinBiasHfOrBSC & 24.3 \\
\hline
\end{tabular}
\label{tab:datasamplespPb}
\end{table}
\end{landscape}

\clearpage

\section{Monte Carlo generators and samples}

The reconstruction performance of various physics objects can be tested using Monte Carlo (MC) generators. 
In order to study the reconstruction algorithm under realistic conditions, MC generators need to resemble data with similar particle production. 
In this thesis, three different MC generators are used to determine the tracking performance (including efficiency and mis-reconstruction rate), 
event selection efficiency, pileup rejection and $V^0$ reconstruction efficiency. 
\begin{itemize}
\item PYTHIA: For understanding the tracking performance and $V^0$ reconstruction efficiency in pp collisions, the dedicated high-energy particle collision generator PYTHIA (version 6.4~\cite{Sjostrand:2006za} and version 8.2~\cite{Sjostrand:2014zea}) is used. It contains theory and models for a number of physics aspects, including hard and soft interactions, parton distributions, initial- and final-state parton showers, multiparton interactions, fragmentation and decay. However, physics aspects cannot always be derived from first principles, particularly the areas of hadronization and multi-parton interactions which involve non-perturbative QCD. In order to better model the collision event, Tunes are introduced into PYTHIA generator, where each of the Tune is a set of generator parameters tuned derived from a certain kind of experimental data. For the analysis presented in this thesis,  PYTHIA6 Tune Z2~\cite{PYTHIA6Z2}, PYTHIA8 Tune 4C~\cite{PYTHIA84C} and PYTHIA8 Tune CUETP8M1~\cite{PYTHIA8CMS} are used.  
\item HIJING: The Heavy Ion Jet INteraction Generator (HIJING)~\cite{Gyulassy:1994ew} is used for understanding tracking performance and $V^0$ reconstruction efficiency in pPb collisions. HIJING 1.0 is used to reproduce the particle production with multiple nucleon-nucleon collisions.
\item EPOS: The EPOS LHC Generator~\cite{Pierog:2013ria} is used as cross-check for $V^0$ reconstruction efficiency in pPb collisions. Besides the description of particle production with multiple nucleon-nucleon collisions, it also has an implementation of collective flow.
\end{itemize}

In addition to description of particle production, it is also critical to have a good simulation of the detector. 
The detailed MC simulation of the CMS detector response is based on GEANT4~\cite{Agostinelli:2002hh}. 
Particles from generators are propagated through detector and the simulated detector signals are processed as if they are real data.

\cleardoublepage
\chapter{Reconstruction of physics objects and performance}
\label{ch:Reco}

\section{Track reconstruction}
\label{sec:tracking}

The reconstruction of tracks in the inner tracking system of CMS is one of the most important components for physics objects reconstruction. 
Track reconstruction employs a pattern recognition algorithm that converts hits in the silicon tracker into trajectories that resemble charged particles propagating 
in the magnetic field of CMS detector. 
The tracking algorithm used is known as the Combinatorial Track Finder (CTF)~\cite{Chatrchyan:2014fea}, which is an extension of the Kalman Filter~\cite{Fruhwirth:1987fm}. 

\subsection{Iterative tracking}
In each collision, there are large number of hits produced in the tracker. 
Tremendous amount of time is needed to consider all possible combinations for track reconstruction. 
To solve the combinatorial problem in a smart way, the track reconstruction procedure consists in multiple iterations of the CTF algorithm, known as iterative tracking. 
Each iteration performs track finding with a subset of hits. After finding the tracks in each iteration, the hits associated to them are removed. 
The remaining hits are considered for the next iteration of search for tracks. 
In practical, in the first iterations, tight criteria are used to identify the cleanest tracks near the beamspot position. 
Looser requirements are applied in later iterations in order to reconstruct more complex trajectories associated to low-\pt\ particles and displaced tracks. 
Each tracking iteration can be separated into four steps:
\begin{itemize}
\item Seed generation: Based on a limited number of hits in the tracker, an initial estimate (i.e. seed) of the track trajectory is determined. Track seeds are estimated with 2 or 3 hits in consecutive tracker layers, where at least one of those hits has to be from the pixel tracker. One exception is the last iteration, where information from only the strip tracker is used.
\item Pattern recognition: Seed trajectories are extrapolated to all layers of the tracker to find hits compatible with the original track. The most compatible hits are added to the hit collection associated to a given seed trajectory to form a track candidate.
\item Trajectory fitting: The final collection of hits associated to the track candidates from previous step are fitted using the CTF algorithm. The best estimation of track parameters (e.g. \pt, $\eta$) are determined from the fitting. Spurious hits with limited compatibility with the fitted track trajectory are removed from the track candidate hit collection.
\item Track quality check: A set of track-quality requirements are applied to track candidates from the previous step. Tracks are classified into different quality, such as \textit{loose}, \textit{tight} and \textit{highPurity}~\cite{Chatrchyan:2014fea}.
\end{itemize}

\subsection{Track selection}
\label{subsec:trackselect}

In the analysis presented in this thesis, the official CMS \textit{highPurity}~\cite{Chatrchyan:2014fea} tracks are used. 
For further selections, 
a reconstructed track was considered as a primary-track candidate if the impact parameter 
significance $d_{\rm xy}/\sigma(d_{\rm xy})$ and significance of z separation between the 
track and the best reconstructed primary vertex (the vertex associated with the largest 
number of tracks, or best $\chi^{2}$ probability if the same number of tracks is found) 
$d_z/\sigma(d_z)$ are both less than 3. In order to remove tracks with poor momentum estimates, 
the relative uncertainty of the momentum measurement $\sigma(\pt)/\pt$ was required
to be less than 10\%. Primary tracks that fall in the kinematic range of $|\eta|<2.4$ 
and $\pt > 0.3$ GeV were selected in the analysis to ensure a reasonable 
tracking efficiency and low fake rate. 

\section{Track reconstruction performance}
\label{sec:trackeff}
\begin{figure}[p]
	\centering
	\includegraphics[width=0.9\textwidth]{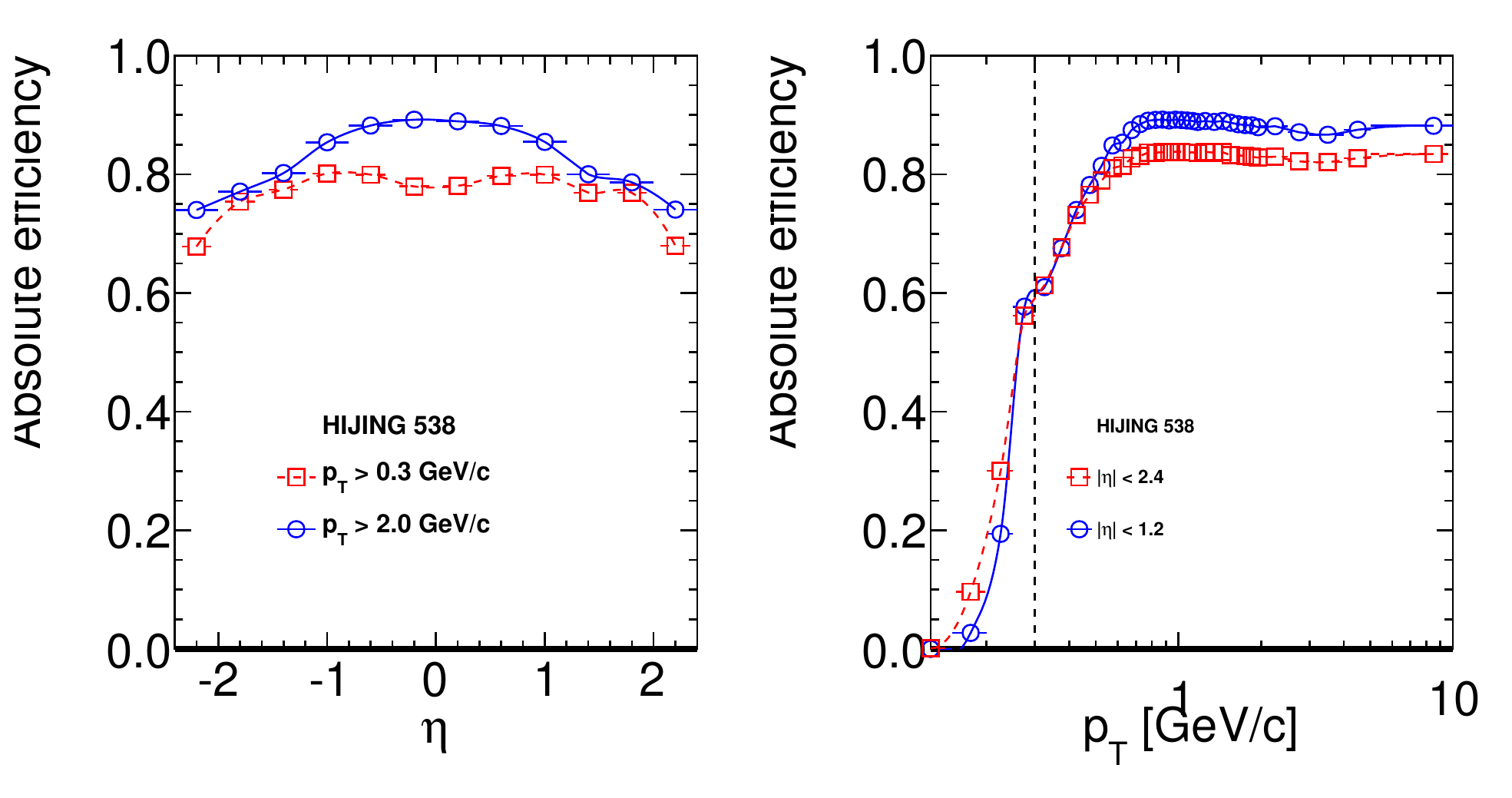}
	\caption{Projections of the tracking efficiency as a function of $\eta$ (left) and $p_{T}$ (right).  The dashed line shows the lower \pt\ limit (0.3 GeV/c) used in the analysis.}
	\label{fig:trkEff}
\end{figure}

\begin{figure}[p]
	\centering
	\includegraphics[width=0.9\textwidth]{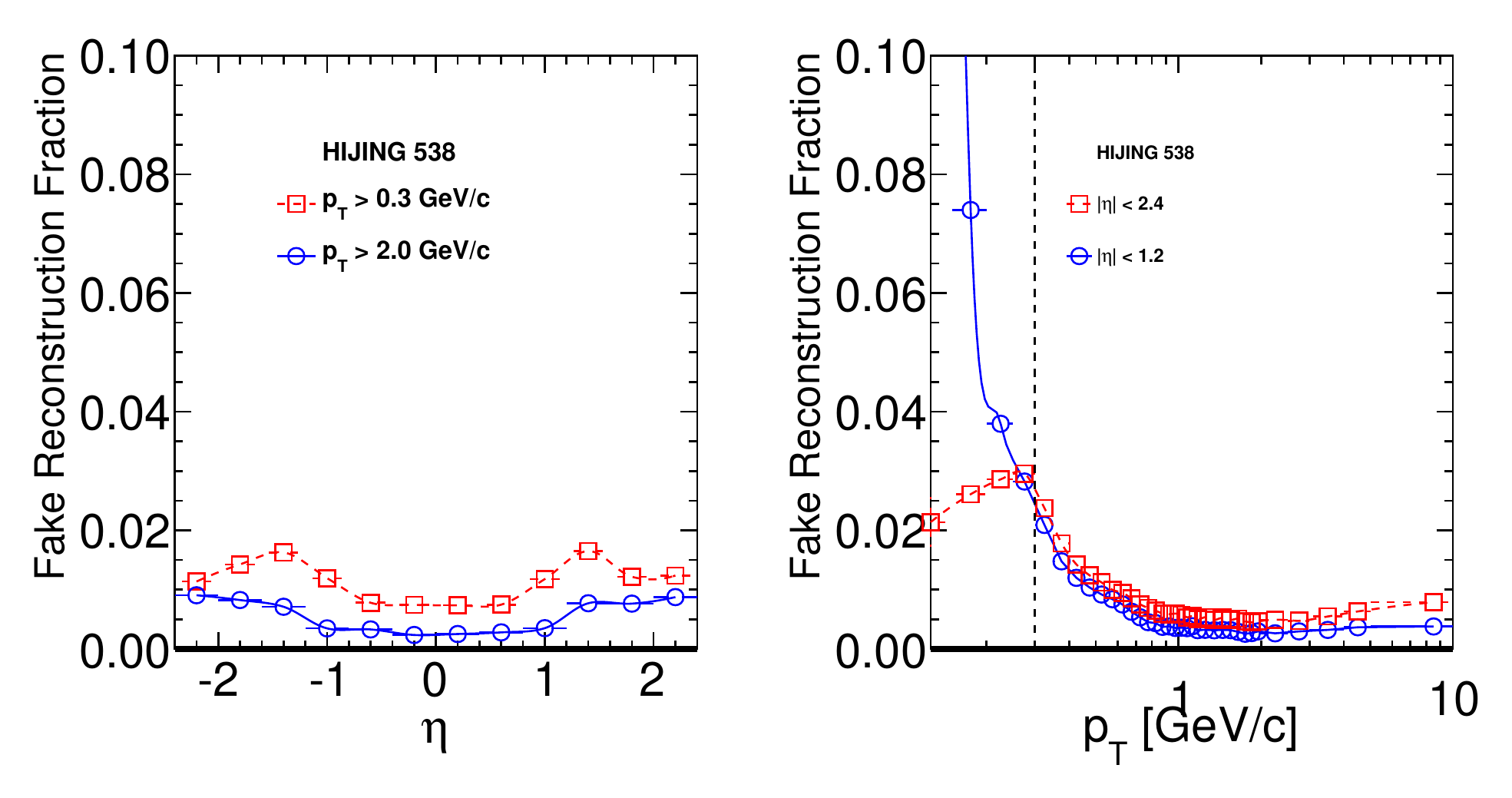}
	\caption{Projections of the fake track fraction as a function of $\eta$ (left) and $p_{T}$ (right).  The dashed line shows the lower \pt\ limit (0.3 GeV/c) used in the analysis.}
	\label{fig:trkFake}
\end{figure}

\begin{figure}[p]
	\centering
	\includegraphics[width=0.9\textwidth]{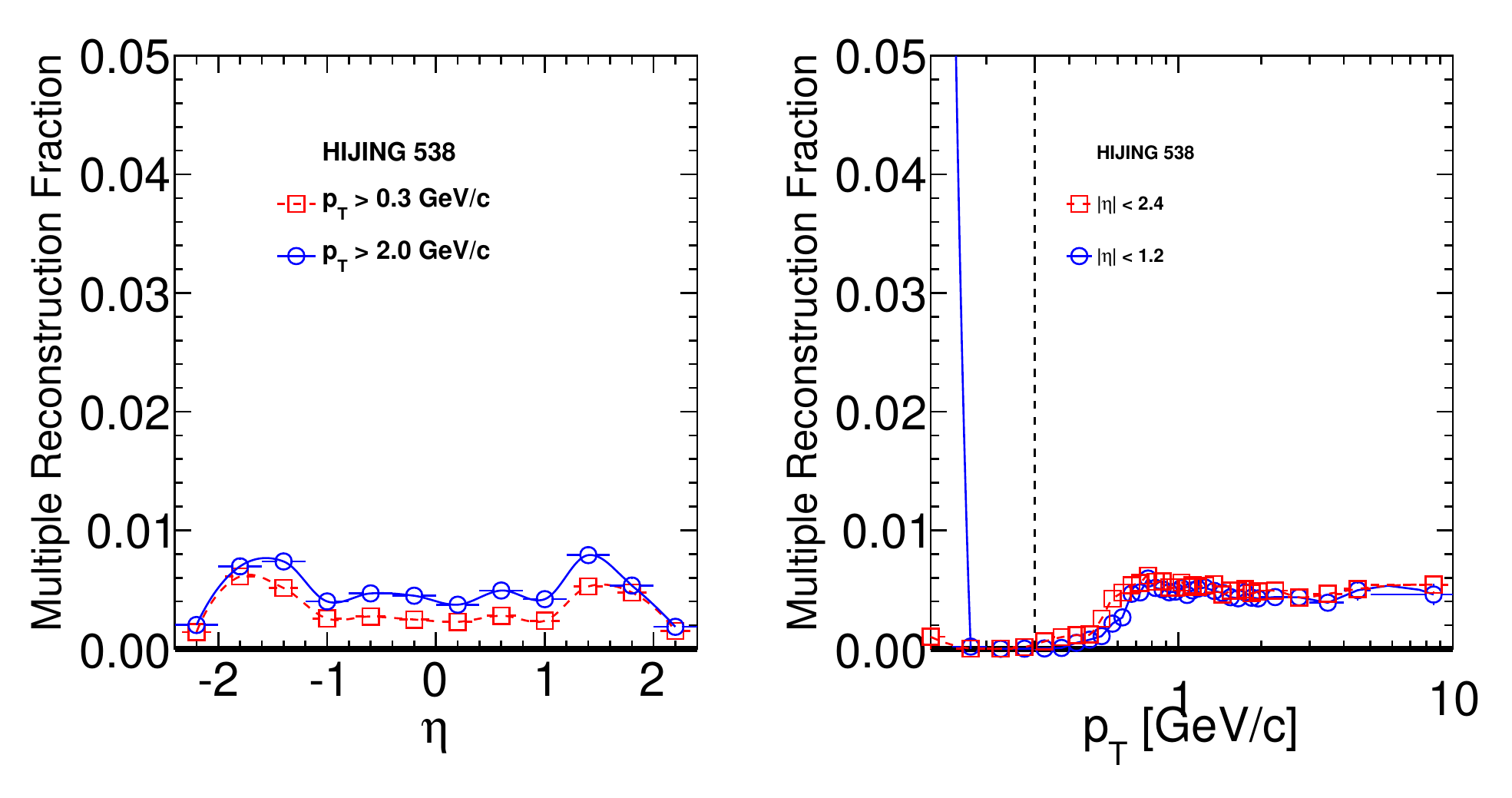}
	\caption{Projections of the multiple reconstruction fraction as a function of $\eta$ (left) and $p_{T}$ (right).  The dashed line shows the lower \pt\ limit (0.3 GeV/c) used in the analysis.}
	\label{fig:trkMul}
\end{figure}

\begin{figure}[p]
	\centering
	\includegraphics[width=0.9\textwidth]{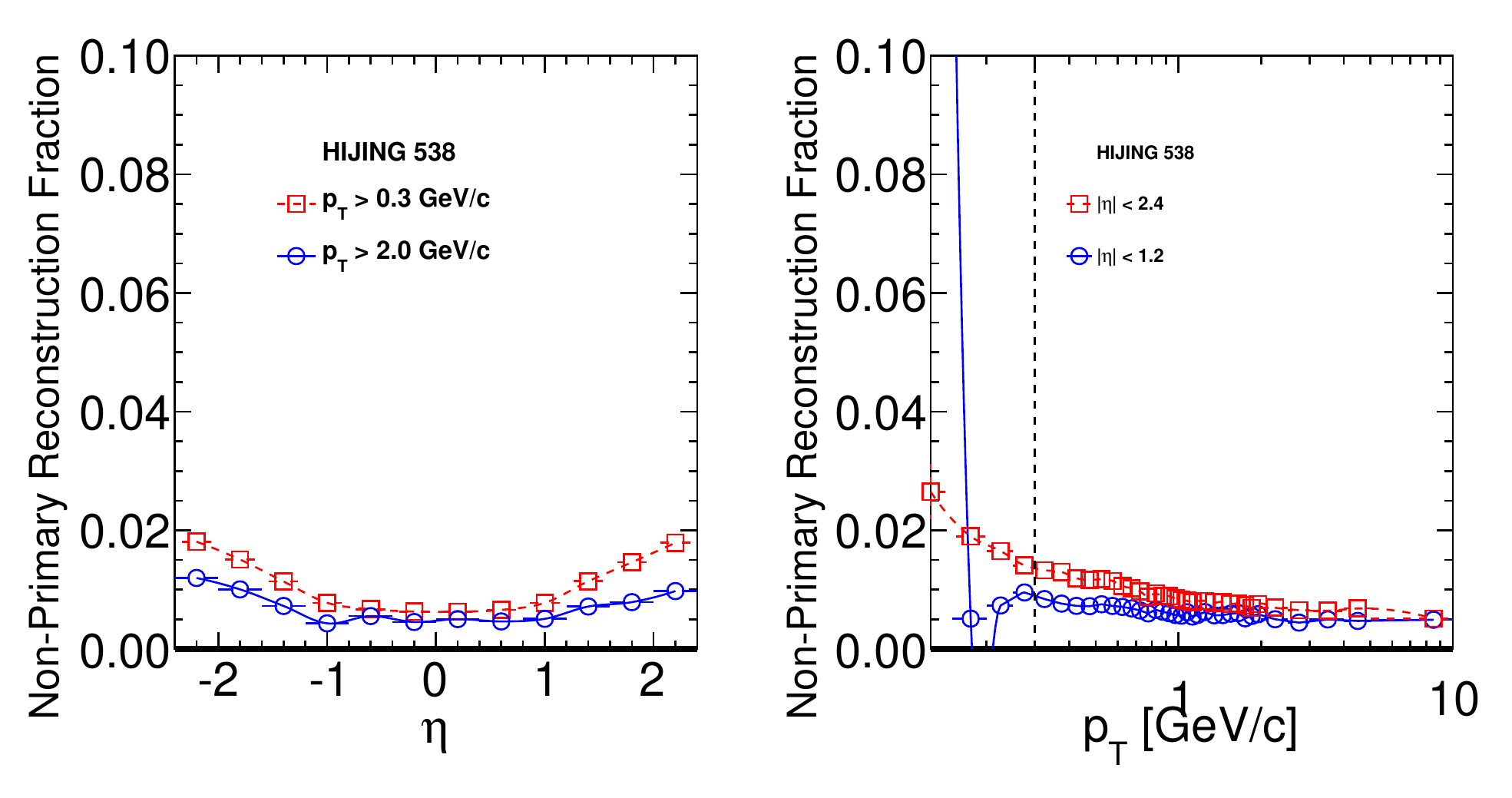}
	\caption{Projections of the non-primary reconstruction fraction as a function of $\eta$ (left) and $p_{T}$ (right).  The dashed line shows the lower \pt\ limit (0.3 GeV/c) used in the analysis.}
	\label{fig:trkNonpri}
\end{figure}

The performance of the track reconstruction is evaluated based on the matching of selected reconstructed tracks and generator level particles. 
In CMS criteria, a track is matched to a generator level charged particle if 75\% of reconstructed hits associated to the track are compatible with hits created in the simulation of a particle going through the detector. 
In order to quantify the performance of track reconstruction, several quantities are defined:
\begin{itemize}
\item Efficiency: The fraction of primary particles from generator which are matched to at least one reconstructed track. Here, primary particle is defined to be charged particles produced in the collision or are decay products of particles with a mean proper lifetime of less than 1 cm/s.
\item Fake rate: The fraction of reconstructed tracks that do not match any primary particles at generator level.
\item Multiple reconstruction rate: The fraction of generator level primary particles which match to more than one reconstructed tracks.
\item Non-primary reconstruction fraction: The fraction of reconstructed tracks matched to a non-primary particle at generator level, which is created by interactions of the primary particles with the detector.
\end{itemize}
The track reconstruction performance is more reliable when efficiency is closer to 1 and fake rate, multiple reconstruction and non-primary reconstruction rate are closer to 0. 
Figs.~\ref{fig:trkEff}-~\ref{fig:trkNonpri} shows track reconstruction performance in pseudorapidity ($\eta$) and transverse momentum ($\pt$) 
based on MC samples from HIJING pPb simulations. 
The performance is similar in pp collisions since identical reconstruction algorithm is used. 
Inelastic nuclear interactions are the main source of tracking inefficiency. 
The formation of a track can be interrupted if a hadron undergoes a large-angle elastic nuclear scattering. 
Hence the hadron can be reconstructed as a single track with fewer hits, or as two separate tracks, or even not be found at all. 
Such efficiency loss is higher at large $\eta$ regions with large material content.

\clearpage

\section{Vertex reconstruction}
Reconstructed tracks are used to determine the primary vertices associated to particle collisions. 
Positions of vertices are determined by using the extrapolated position of the track trajectories to the interaction point. 
Vertex reconstruction is performed in two steps:
\begin{itemize}
\item Vertex clustering: Based on a deterministic annealing algorithm~\cite{IEEE_DetAnnealing}, tracks are grouped into clusters, each associated to a separate collision. The algorithm is capable of resolving vertices with a longitudinal separation of approximately 1 mm.
\item Property determination: An adaptive vertex fitting technique~\cite{Speer:927395} is employed to determine the vertex properties, in particular its spatial coordinates. Based on the kinematics of the associated tracks, the algorithm fits the vertex position and reject outlier tracks. Each of the remaining tracks is assigned a weight according to the compatibility between the track kinematics and the vertex position.
\end{itemize}
The spatial resolution of the vertex position, for those reconstructed with at least 50 tracks, is between 10 $\mu$m and 12 $\mu$m for the three spatial dimensions~\cite{Chatrchyan:2014fea}.

\section{Reconstruction of \PKzS\ and \PgL/\PagL\ particles}
\label{sec:V0}
All demonstration in this section are using 5.02 TeV pPb data. The same reconstruction has also been done for pp and PbPb collisions at various collision energies.

The \PKzS\ and \PgL/\PagL\ candidates (generally referred as $V^{0}$) are reconstructed
via their decay topology by combining pairs of oppositely charged tracks that are 
detached from the primary vertex and form a good secondary vertex with an appropriate 
invariant mass. The two tracks are assumed to be pions in \PKzS\ reconstruction, and 
are assumed to be one pion and one proton in \PgL/\PagL\ reconstruction. For \PgL/\PagL\ , the 
lowest momentum track is assumed to be the pion. 

\begin{figure}[htb]
	\centering
	\includegraphics[width=0.9\textwidth]{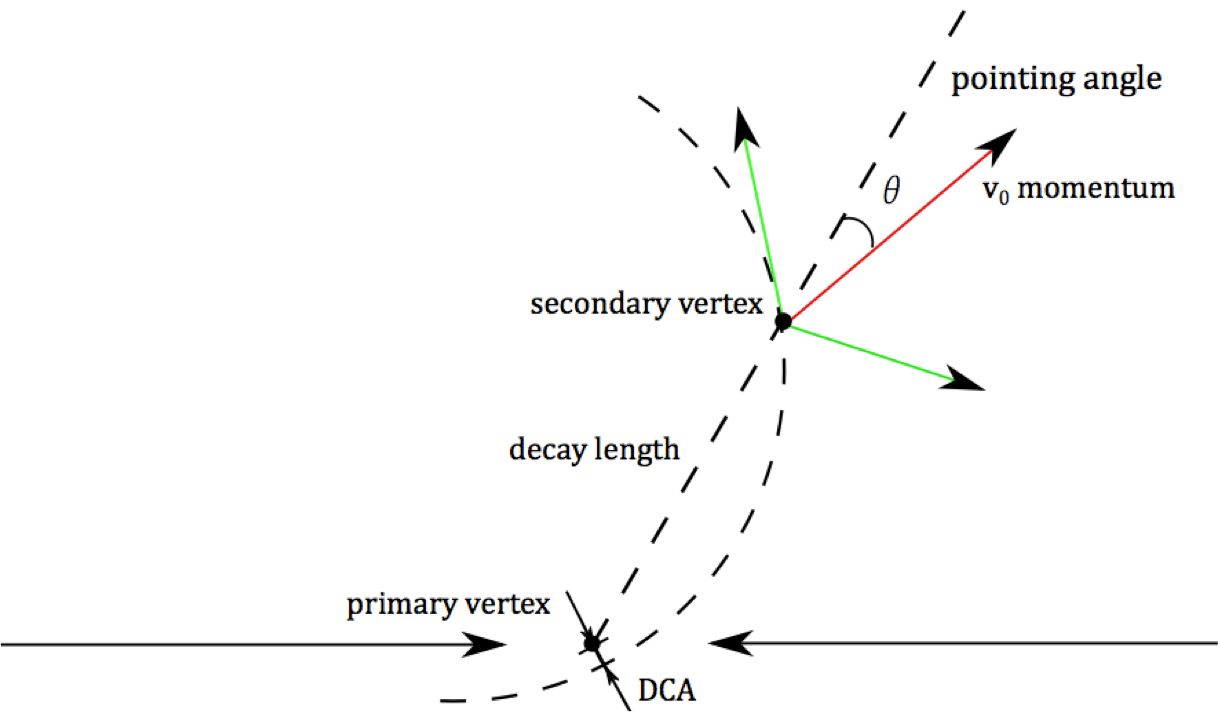}
	\caption{Demonstration of $V^{0}$ particle decay and variables used in the reconstruction.}
	\label{fig:decaydemo}
\end{figure}

To increase the efficiency for tracks with low momentum and large impact parameters,
both characteristics of the \PKzS\ and \PgL/\PagL\ decay products, the standard \textit{loose}
selection of tracks (as defined in Ref.~\cite{Chatrchyan:2014fea}) is used in reconstructing the \PKzS\ and \PgL/\PagL\ candidates. 
Fig.~\ref{fig:decaydemo} demonstrates the decay of $V^{0}$ particles and definition of various quantities used in the reconstruction. 
Main steps of reconstruction are summarized below:
\begin{itemize}
\item Oppositely charged tracks with transverse and longitudinal impact
      parameter significances (impact parameter divided by its uncertainty) greater than 1 are selected to form pairs. 
      Where impact parameter is defined as distance of closest approach of a given track to the primary vertex.
\item Distance at their closest approach (DCA) for each pair of tracks is required 
      to be less than 1~cm. Each track must consist of at least 3 valid hits.
\item The standard "KalmanVertexFitter'' is used for fitting the vertex 
      of two tracks. A normalized $\chi^2$ value less than 5 is required 
      to select good vertex candidates.
\item To suppress background and exclude the \PgL/\PagL\ contribution 
      from weak decay of $\Xi$ and $\Omega^{-}$, the $V^{0}$ momentum vector is required to point 
      back to the primary vertex. A cut on $\cos\theta^{\rm point}>$0.999 is applied,
      where pointing angle $\theta^{\rm point}$ is the angle between the $V^{0}$ momentum vector
      and vector connecting primary and $V^{0}$ vertex. This requirement also reduces the effect of nuclear interactions and random combinations of tracks.
\item Due to the long lifetime of \PKzS\ and \PgL/\PagL\ particles, the three dimensional separations between primary and $V^{0}$ vertex (decay length) are required to be
      greater than 5$\sigma$ to further suppress the background.
\end{itemize}

\begin{figure*}[thb]
\centering
\includegraphics[width=\linewidth]{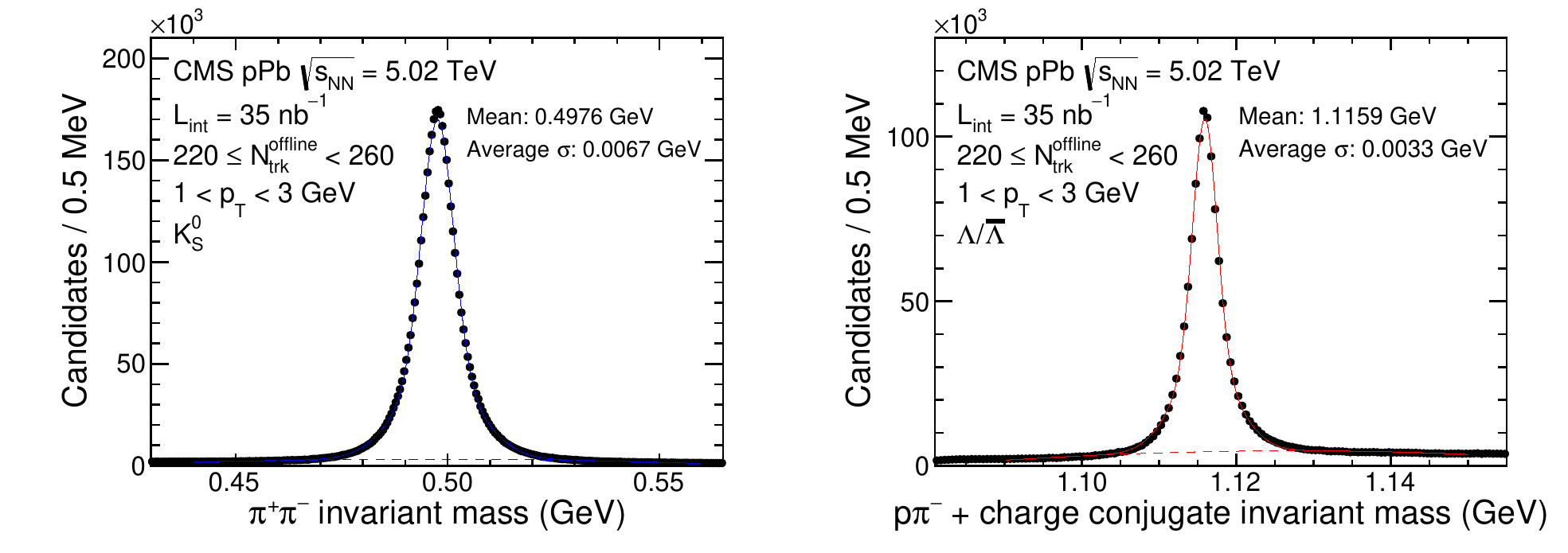}
  \caption{ \label{fig:v0mass} Invariant mass distribution of \PKzS\ (left) and \PgL/\PagL\ (right)
  candidates in the \pt\ range of 1--3 GeV/c for $220 \leq \noff < 260$ in
  pPb collisions at \rootsNN\ = 5.02 TeV. The solid line shows the fit function of
  a double Gaussian plus a 4th-order polynomial (dashed line).
   }
\end{figure*}

The resulting invariant mass distributions of reconstructed \PKzS\ and \PgL/\PagL\ 
candidates are shown in Fig.~\ref{fig:v0mass} from the 5.02 TeV pPb data, 
for $V^0$ with $1 < \pt < 3$ GeV/c for $220 \leq \noff < 260$. The $V^0$
peaks can be clearly seen with little background. 
The signal is described by a double Gaussian with a common mean, while the background is modelled by a 4th
order polynomial function. The mass peak mean value are close to PDG particle mass, and the average $\sigma$s of double Gaussian functions are calculated by:

\vspace{-0.2cm}
\begin{equation}
\label{2pcorr_incl}
\sigma_{ave} = \sqrt{ \frac{Y_{1}}{Y_{1} + Y_{2}}\sigma_{1}^{2} + \frac{Y_{2}}{Y_{1} + Y_{2}}\sigma_{2}^{2}},
\end{equation}
\vspace{-0.2cm}

\noindent where $\sigma_{1}$($\sigma_{2}$) and $Y_{1}$($Y_{2}$) are $\sigma$ and yield of first(second) Gaussian.

\subsection{Removal of mis-identified candidates}
As the identity of each track cannot be determined, the mass of each track has to be assumed
depending on the identity of $V^0$ candidate. It is possible that \PKzS\ (\PgL/\PagL) candidates 
are mis-identified as \PgL/\PagL\ (\PKzS) candidates. Especially, there is high probability a 
track assumed to be a proton in a \PgL/\PagL\ candidate is actually a pion. To select clean samples
of \PKzS\ and \PgL/\PagL\, the so-called Armenteros-Podolanski (A-P) plot is investigated.

\begin{figure}[thb]
\centering
\mbox{\includegraphics[width=\linewidth]{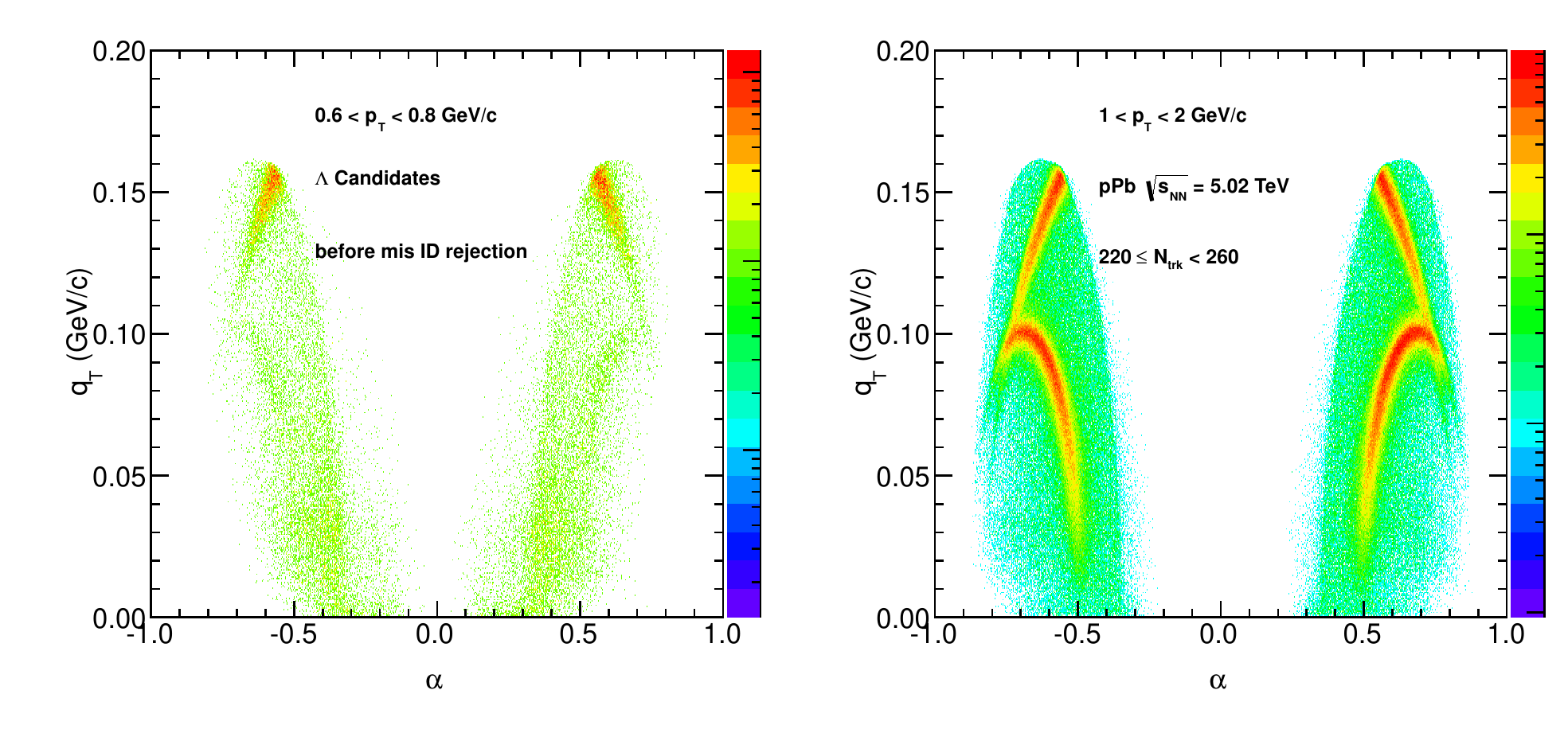}}
  \caption{ \label{fig:AP_Lambda_nocut} 
  Armenteros-Podolanski (A-P) plots for \PgL/\PagL\ candidates with $0.6 < \pt\ < 0.8$ GeV 
  (left) and $1 < \pt\ < 2$ GeV (right) before apply the mis-identified candidate mass cuts.
   }
\end{figure}

Armenteros-Podolanski (A-P) plot is a two dimensional plot, of the transverse component of the momentum 
of the positive charged decay product ($q_T$) with respect to the $V^0$ candidate versus the 
longitudinal momentum asymmetry $\alpha=(p_{L}^{+} - p_{L}^{-})/(p_{L}^{+} + p_{L}^{-})$. 
An example of A-P plot can be seen in Fig.~\ref{fig:AP_Lambda_nocut}. The obtained distribution 
can be explained by the fact that pair of pions from \PKzS\ decay have the same mass and 
therefore their momenta are distributed symmetrically on average (top band), while the proton 
(anti-proton) in \PgL/\PagL\ decay takes on average a larger part of momentum and results 
in a asymmetric distribution (two lower bands). 

Fig.~\ref{fig:AP_Lambda_nocut} shows the AP plot for \PgL/\PagL\ candidates with $0.6 < \pt\ < 0.8$ GeV 
and $1 < \pt\ < 2$ GeV. As one can see, mis-identified candidates can be clearly 
observed. The $V^0$ candidates above $q_T \approx 0.11$ are mis-identified \PKzS\ which need to be removed.

\begin{figure}[thb]
\centering
\mbox{\includegraphics[width=\linewidth]{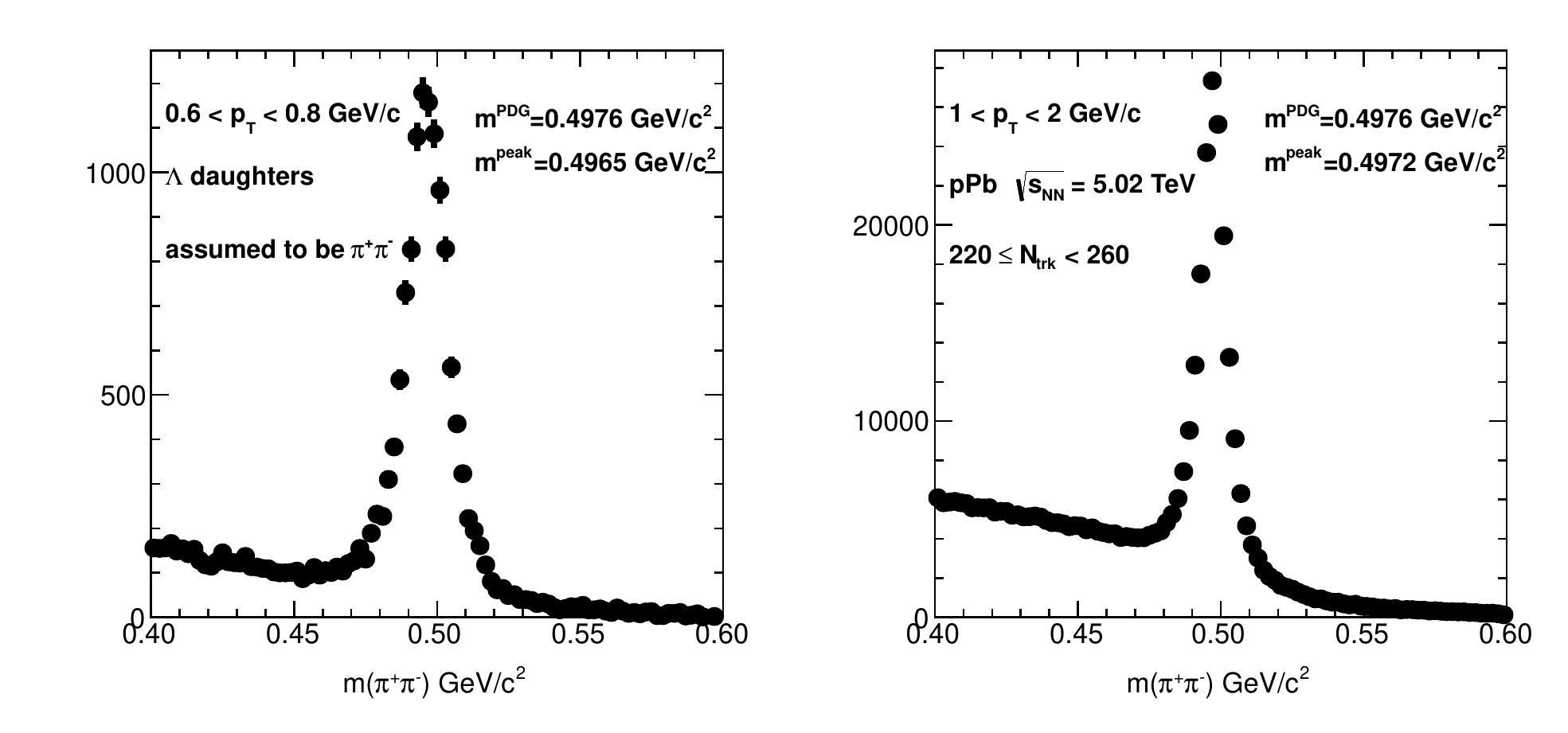}}
  \caption{ \label{fig:mass_pipi_Lambda} 
  $\pi$ -$\pi$ hypothesis re-calculated mass distributions for \PgL/\PagL\ candidates 
  with $0.6 < \pt\ < 0.8$ GeV (left) and $1 < \pt\ < 2$ GeV (right).
   }
\end{figure}

To remove the mis-identified \PKzS\ , we apply the $\pi$ -$\pi$ hypothesis to \PgL/\PagL\ 
candidates. The hypothesis assumes both daughter tracks from decay of \PgL/\PagL\ candidate 
are pions and re-calculate the invariant mass of the decayed mother particle. The re-calculated
mass distributions for \PgL/\PagL\ candidates with $0.6 < \pt\ < 0.8$ GeV and 
$1 < \pt\ < 2$ GeV are shown in Fig.~\ref{fig:mass_pipi_Lambda}. Clear peaks at 
standard \PKzS\ invariant mass, $0.497614$ GeV, are observed, which indicate some of the 
candidates are mis-identified \PKzS\ . A veto of range $0.497614 \pm 0.020$ GeV is applied 
to the re-calculated mass distribution to remove the mis-identified \PKzS. 

\begin{figure}[thb]
\centering
\mbox{\includegraphics[width=\linewidth]{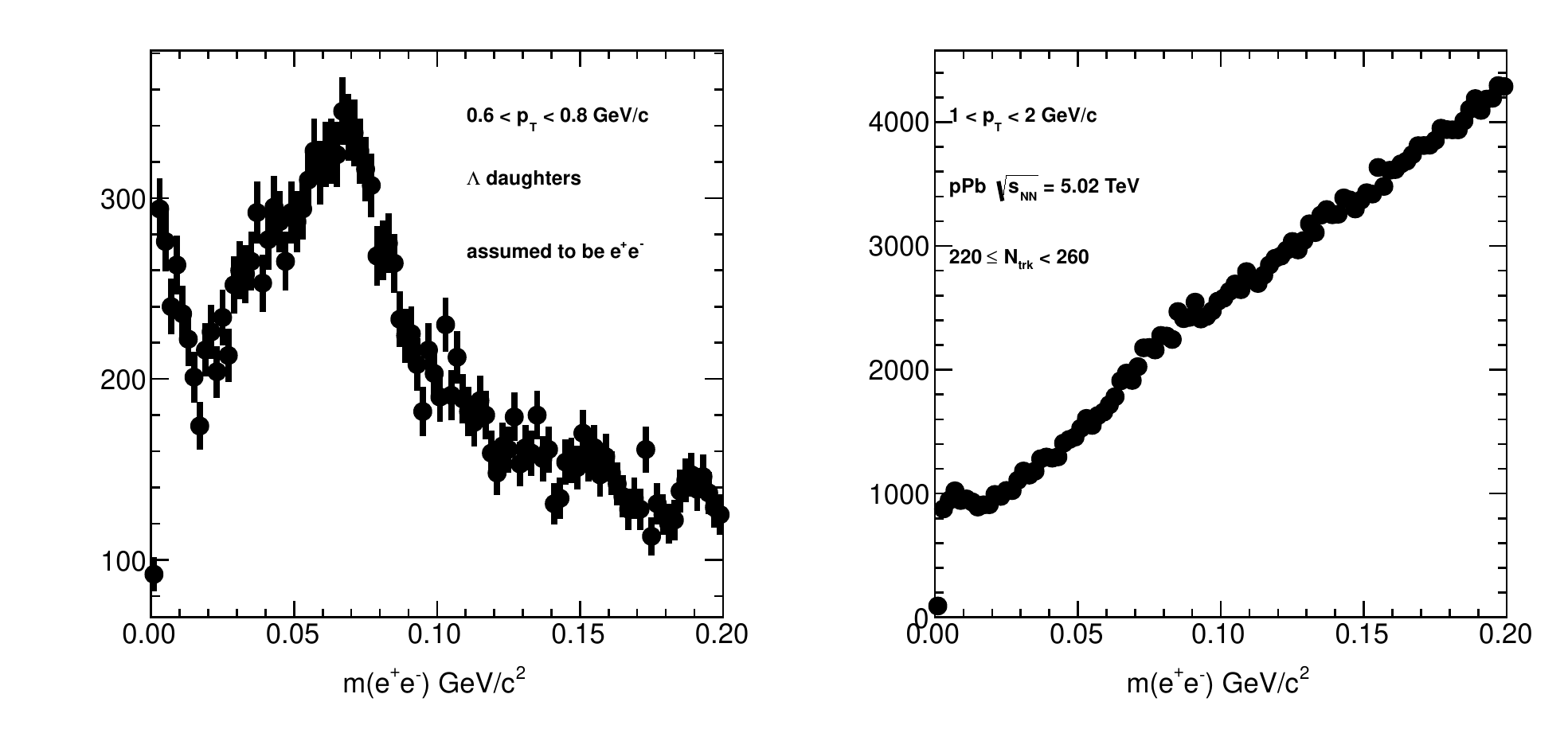}}
  \caption{ \label{fig:mass_ee_Lambda} 
  e-e hypothesis re-calculated mass distributions for \PgL/\PagL\ candidates 
  with $0.6 < \pt\ < 0.8$ GeV (left) and $1 < \pt\ < 2$ GeV (right).
   }
\end{figure}

\begin{figure}[thb]
\centering
\mbox{\includegraphics[width=\linewidth]{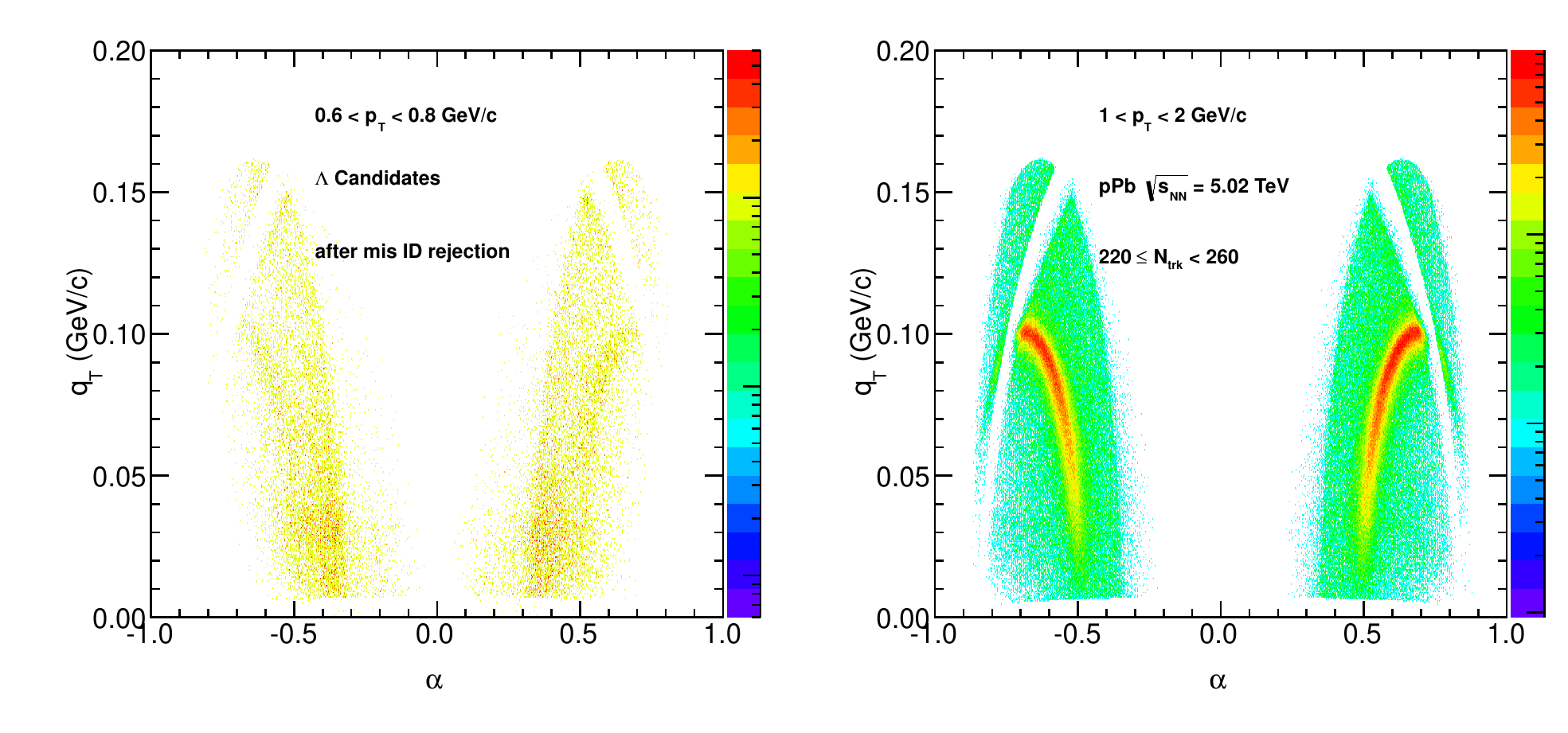}}
  \caption{ \label{fig:AP_Lambda_cut} 
  Armenteros-Podolanski (A-P) plots for \PgL/\PagL\ candidates with $0.6 < \pt\ < 0.8$ GeV 
  (left) and $1 < \pt\ < 2$ GeV (right) after apply the mis-identified candidate mass cuts.
   }
\end{figure}

There is also a chance that both of the daughter tracks are in fact electrons from photon conversion. Peaks can be seen in the e-e hypothesis re-calculated mass distributions 
in Fig.~\ref{fig:mass_ee_Lambda}. Therefore, a veto of invariant mass less than $0.015$ GeV 
is also applied to remove mis-identified photons. The AP plots after removal of mis-identified 
candidates for the same \pt\ range \PgL/\PagL\ candidates are shown in Fig.~\ref{fig:AP_Lambda_cut}. 
Although small part \PgL/\PagL\ candidates is removed, the \PKzS\ band is completely removed 
by the cuts. And there are some candidates with very low $q_T$ removed as mis-identified 
conversion photons, which has very little effect to the \PgL/\PagL\ candidates.

\clearpage

\section{Reconstruction efficiency of $V^{0}$ candidates}
The performance of reconstructing the $V^{0}$ candidates are evaluated based on 
MC simulations. Two different approaches are deployed to study the $V^{0}$ efficiency:
(1) direct fitting and counting of the number of reconstructed $V^{0}$ candidates
under the mass peak, as is done for the real data; (2) matching $V^{0}$ daughter
tracks to simulated tracks in GEANT4. The standard track matching criteria are used,
which require a reconstructed track to share at least 75\% of its hits with a simulated
primary particle. 

\begin{figure}[thb]
\centering
\mbox{\includegraphics[width=\linewidth]{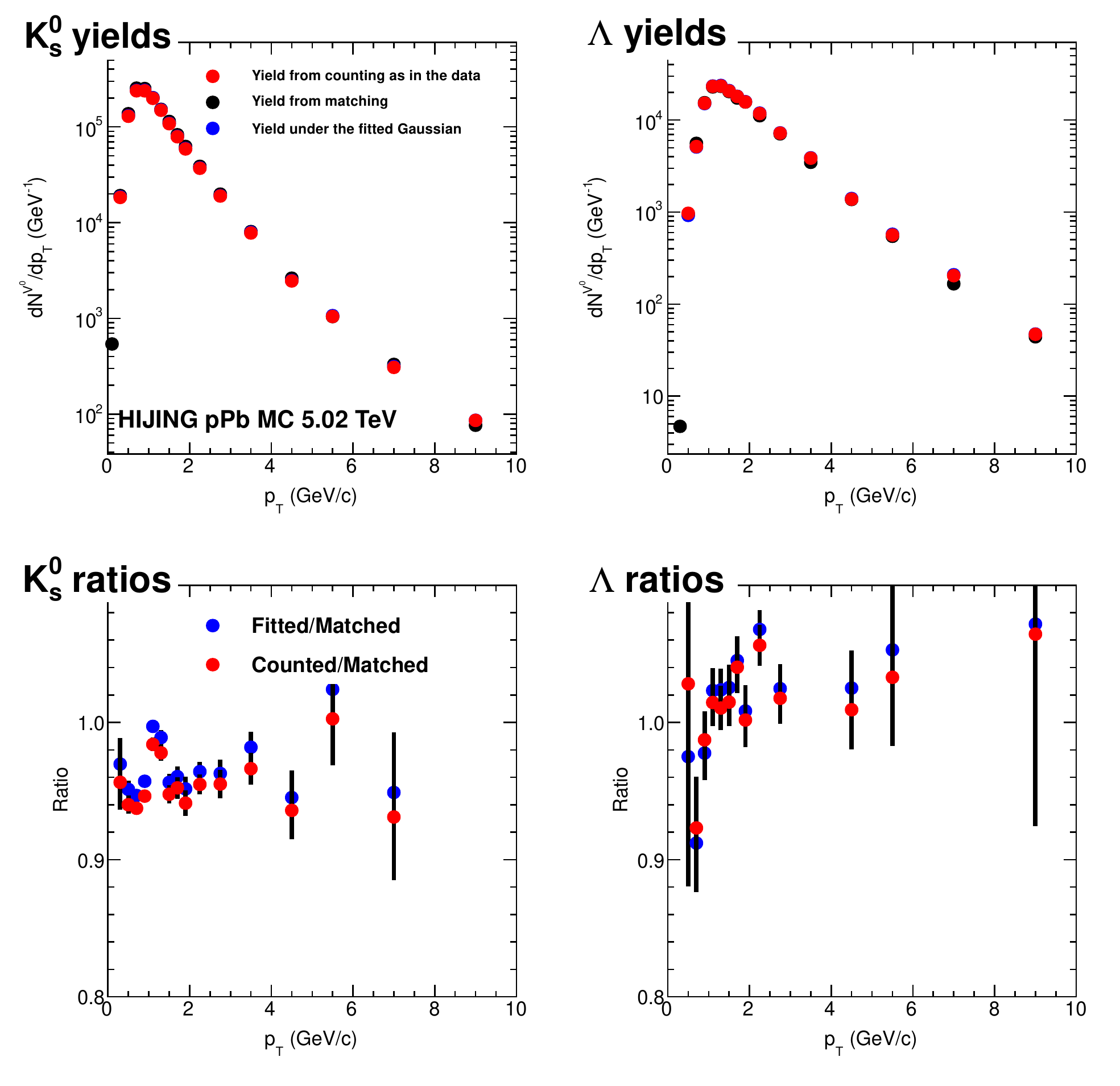}}
  \caption{ \label{fig:YieldAndRatioComp_Hijing}
    Yields of \PKzS\ (left) and \PgL/\PagL\ (right) particles in minimum bias HIJING 
    pPb events using three different methods: direct counting of the number of $V^{0}$ candidates
     under the mass peak within $\pm 2\sigma$ window, as is done for the real data (red), 
     full integral of the fitted double Gaussian function (blue), and matching of $V^{0}$ 
     candidates' daughter tracks with simulated MC tracks (black). The relative ratios of
     each method are shown in the bottom.
   }
\end{figure}

Figure~\ref{fig:YieldAndRatioComp_Hijing} shows the extracted yields of \PKzS\ and \PgL/\PagL\
particles in HIJING pPb events using three different methods, calculated 
within $\pm 2\sigma$ mass window: 
directly counting the number of reconstructed $V^{0}$ candidates 
after subtracting the background (red), an integral of the fitted double 
Gaussian function (blue), and matching procedure of $V^{0}$ candidates' daughter tracks 
with simulated MC tracks (black). The first method corresponds to the analysis on the 
actual data, and thus is used for calculating efficiency as will be shown below. 
The ratios of the first two methods to the matching method are shown in the bottom 
of Fig.~\ref{fig:YieldAndRatioComp_Hijing}. Three methods show consistent results 
within about 5\%. First two methods give almost identical results. 
The small discrepancy to the matching method is expected as it depends also on 
the matching criteria. 

\begin{figure}[thb]
\centering
\mbox{\includegraphics[width=\linewidth]{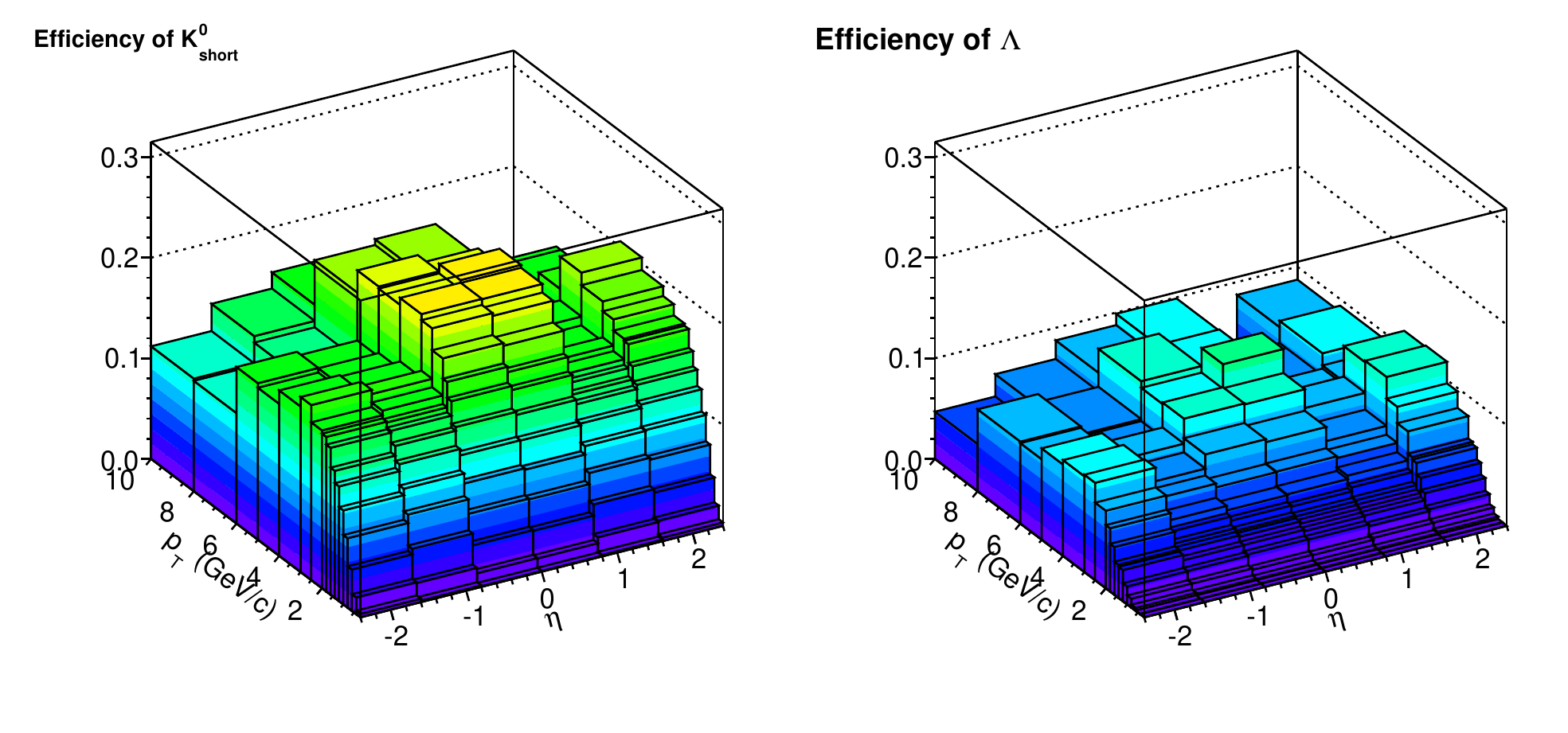}}
  \caption{ \label{fig:EfficiencyVsEtaPt_V0} Efficiency of \PKzS\ (left) and \PgL/\PagL\ (right)
  reconstruction as a function of \pt\ and $\eta$, derived from HIJING pPb MC events.
   }
\end{figure}

\begin{figure}[thb]
\centering
\mbox{\includegraphics[width=\linewidth]{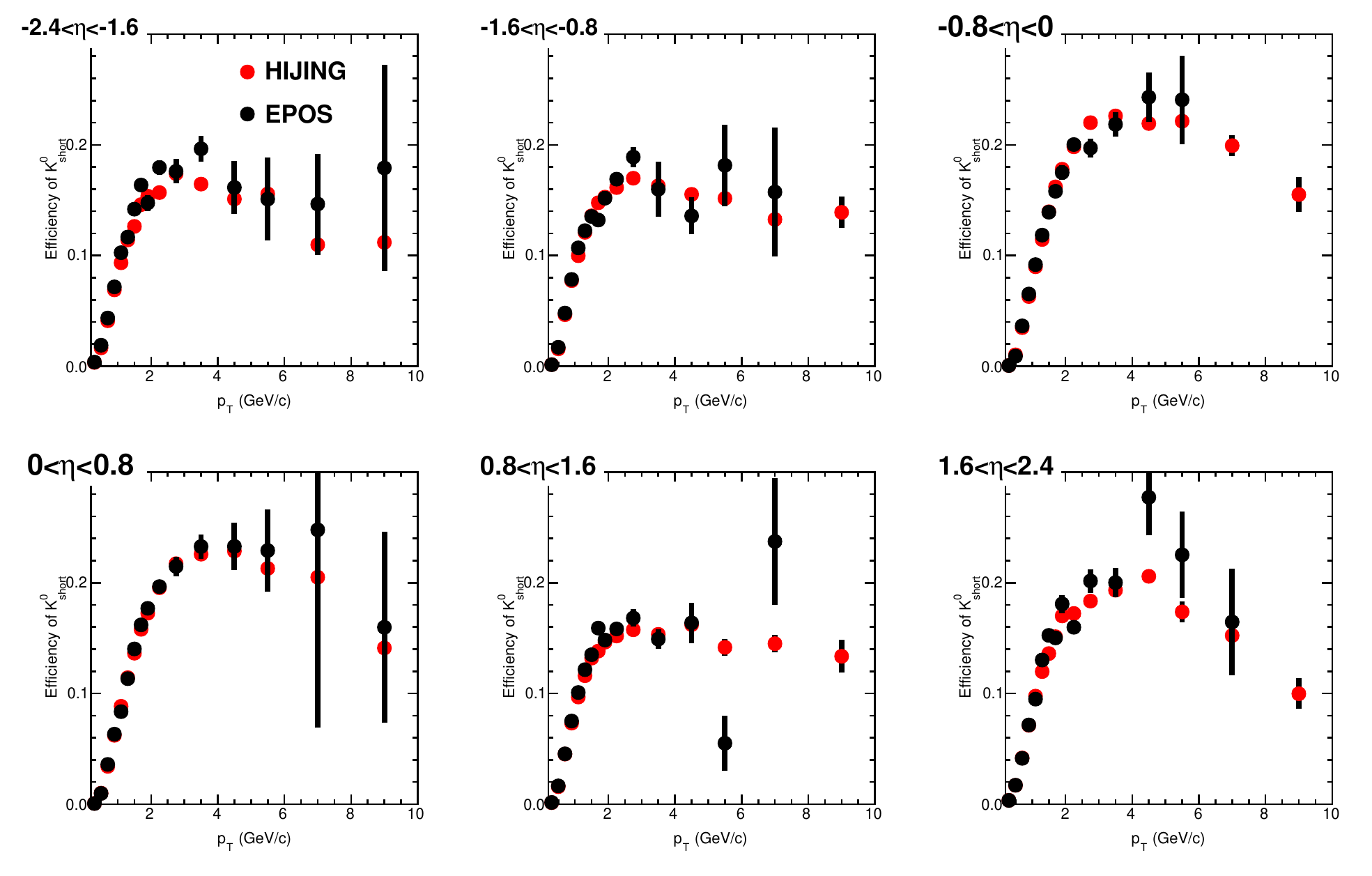}}
  \caption{ \label{fig:EfficiencyVsPt_Ks} Efficiency of \PKzS\ reconstruction as a function 
  of \pt\ in six bins of $\eta$, derived from HIJING and EPOS pPb MC events.
   }
\end{figure}

\begin{figure}[thb]
\centering
\mbox{\includegraphics[width=\linewidth]{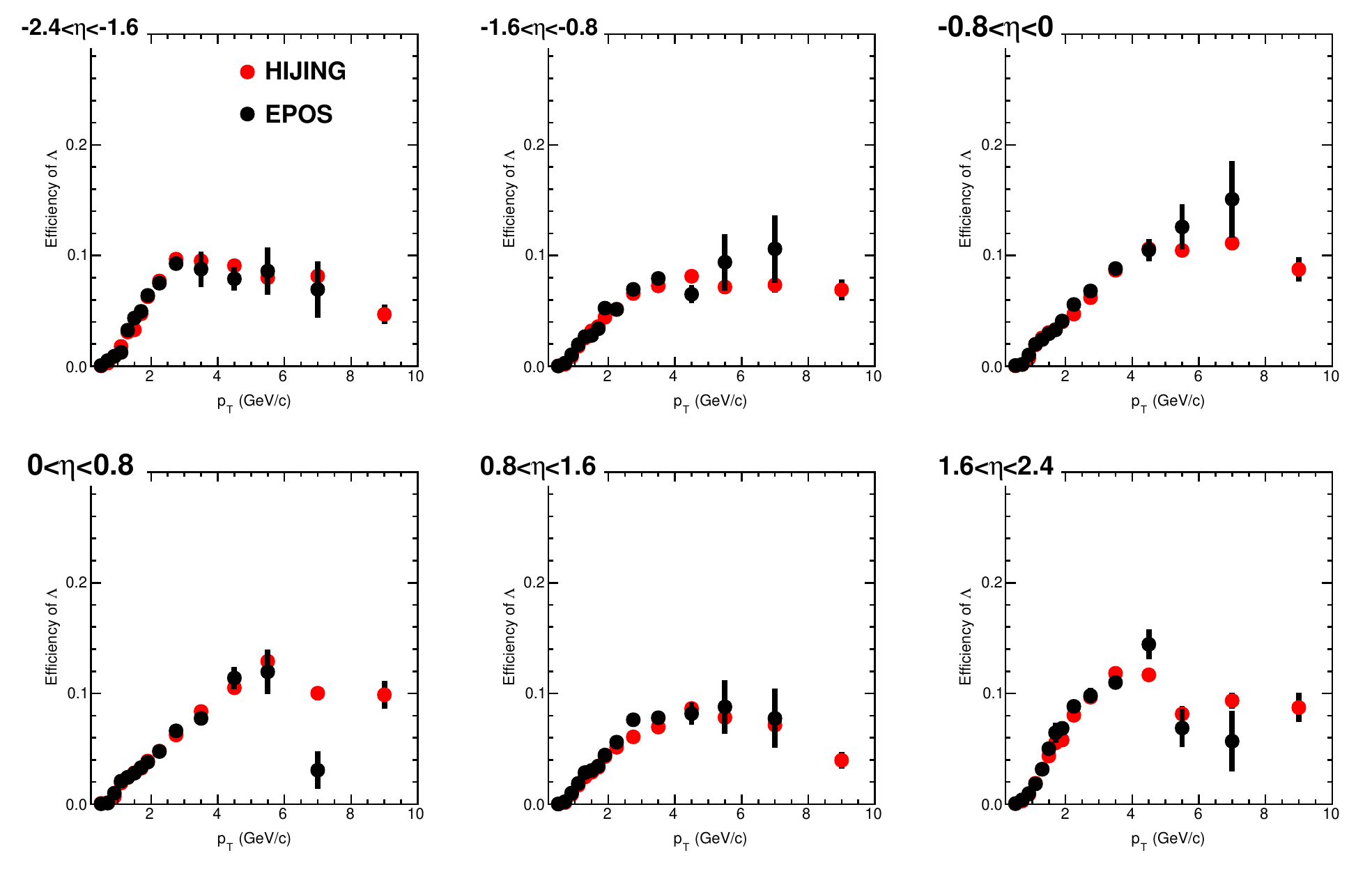}}
  \caption{ \label{fig:EfficiencyVsPt_Lambda} Efficiency of \PgL/\PagL\ reconstruction 
  as a function of \pt\ in six bins of $\eta$, derived from HIJING and EPOS pPb MC events.
   }
\end{figure}

Following the method of directly counting the number of $V^{0}$ candidates 
within the mass peak, the reconstruction efficiency of \PKzS\ and \PgL/\PagL\ 
is shown in Fig.~\ref{fig:EfficiencyVsEtaPt_V0} as a function of \pt\ and $\eta$
using HIJING event generator.
The projected efficiencies as a function of \pt\ in each $\eta$ bin are also 
shown in Fig.~\ref{fig:EfficiencyVsPt_Ks} and Fig.~\ref{fig:EfficiencyVsPt_Lambda},
compared between HIJING and EPOS generators.
The estimated efficiency is applied as an inverse weight correction factor to 
the calculation of two-particle correlation functions. Efficiency derived from HIJING is used for correction
as the EPOS sample has limited statistics. 

\cleardoublepage
\chapter{Event selection and classification}
\label{ch:evtsel}

\section{Offline selection of collision events}
\label{sec:evtselMB}

Events selected by the triggers described in Sec.~\ref{sec:trigger} include those which are not of physics interest of the analysis, such as diffractive events and beam-induced background events. 
To reject those events, a series of offline selections are applied.

To preferentially select non-single-diffractive events, a coincidence of at least one calorimeter tower with more than 3 GeV total energy on each of the positive and negative $\eta$ sides of the HF is required. 
Beam induced background events producing an anomalously large number of pixel hits are excluded by rejecting events with a requirement of high purity track fraction greater than 0.25 for events more then 10 tracks. 
Finally, events were required to contain at least one reconstructed primary vertex that falls within $\pm 15$ cm window along the beam axis and a radius of $\rho < 0.15$ cm in the transverse plane relative to the average vertex position over all events, with at least two fully reconstructed tracks associated to it. 

\begin{figure}
	\centering
	\includegraphics[width=0.7\textwidth]{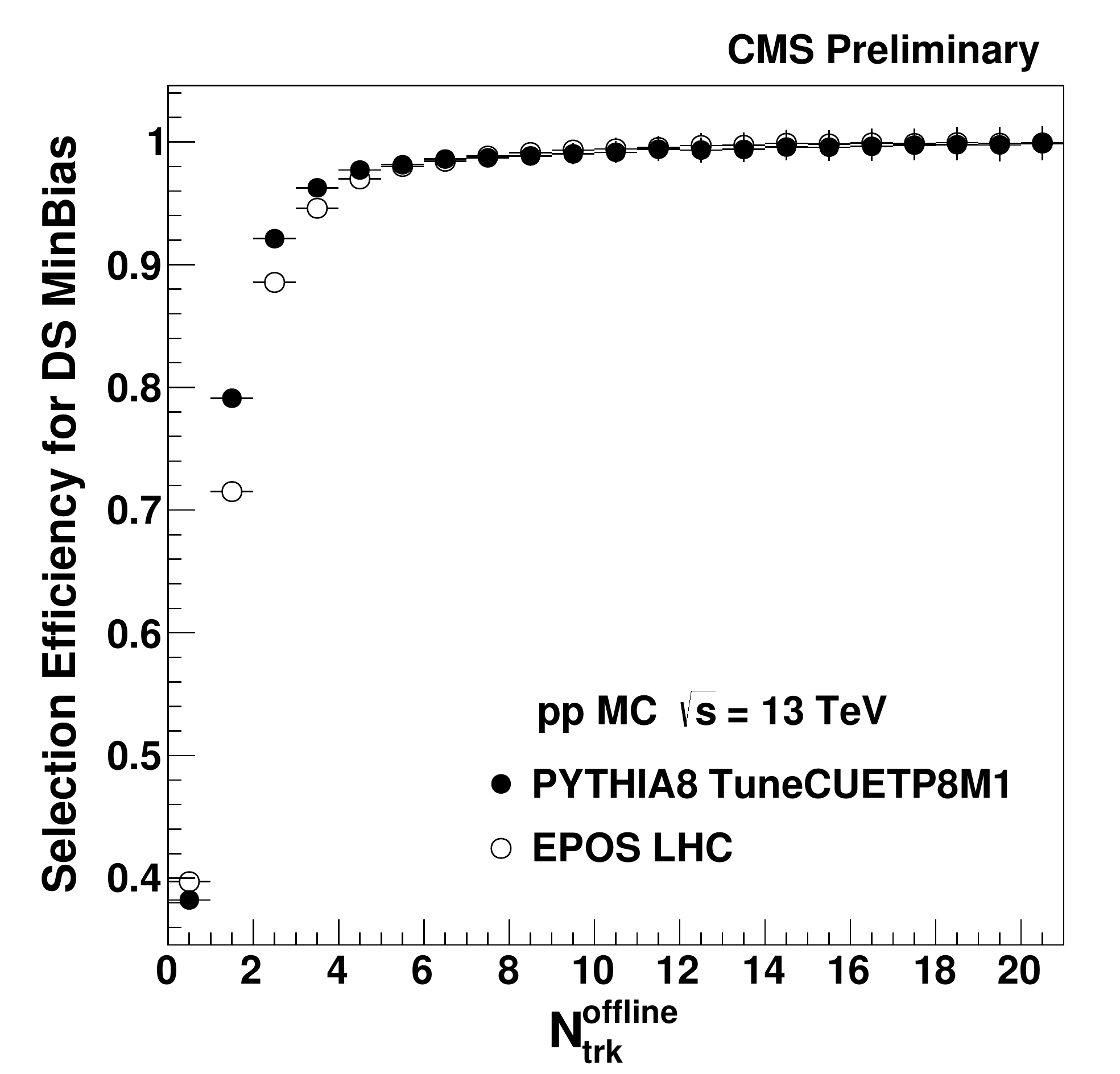}
	\caption{Event selection efficiency as a function of \noff\ derived from PYTHIA and EPOS for 13 TeV pp collisions.}
        \label{fig:offlineMBeff}
\end{figure}

The efficiency for selecting double-sided (DS) events derived from MC generators is illustrated in Fig.~\ref{fig:offlineMBeff} for 13 TeV pp collisions. 
Here, double-sided events are defined as those pp interactions which have at least one primary particle with total energy greater than 3 GeV in both $\eta$ range of $-5 < \eta < -3$ and $3 < \eta < 5$ (compatible with HF $\eta$ range). 
The efficiency reaches almost 100\% for multiplicity larger than 10 and the overall efficiency is around 96\%. 
Identical event selections are applied to 5 and 7 TeV pp collisions and 5 TeV pPb collisions, where the overall efficiency is found to be 96-97\%.

\clearpage

\section{Pileup rejection}
\label{sec:pileup}

The LHC circulates particles not in a continuous stream but in several closely packed bunches. 
Every time these bunches cross one another, more than one collisions can take place, which is known as pileup.
Pileup events present serious challenges to physics analyses which need to distinguish single high-multiplicity collisions in those events. 
Therefore, events must be rejected when it is not possible to distinguish multiple collisions in them. 
A dedicated pileup rejection algorithm is developed for pp and pPb collisions with a relative small pileup around 1-3. 
In this section, the pileup rejection mechanism is described for 13 TeV pp collisions, where it applies to pp and pPb collisions in general. 

\begin{figure}[thb]
  \begin{center}
    \includegraphics[width=0.7\linewidth]{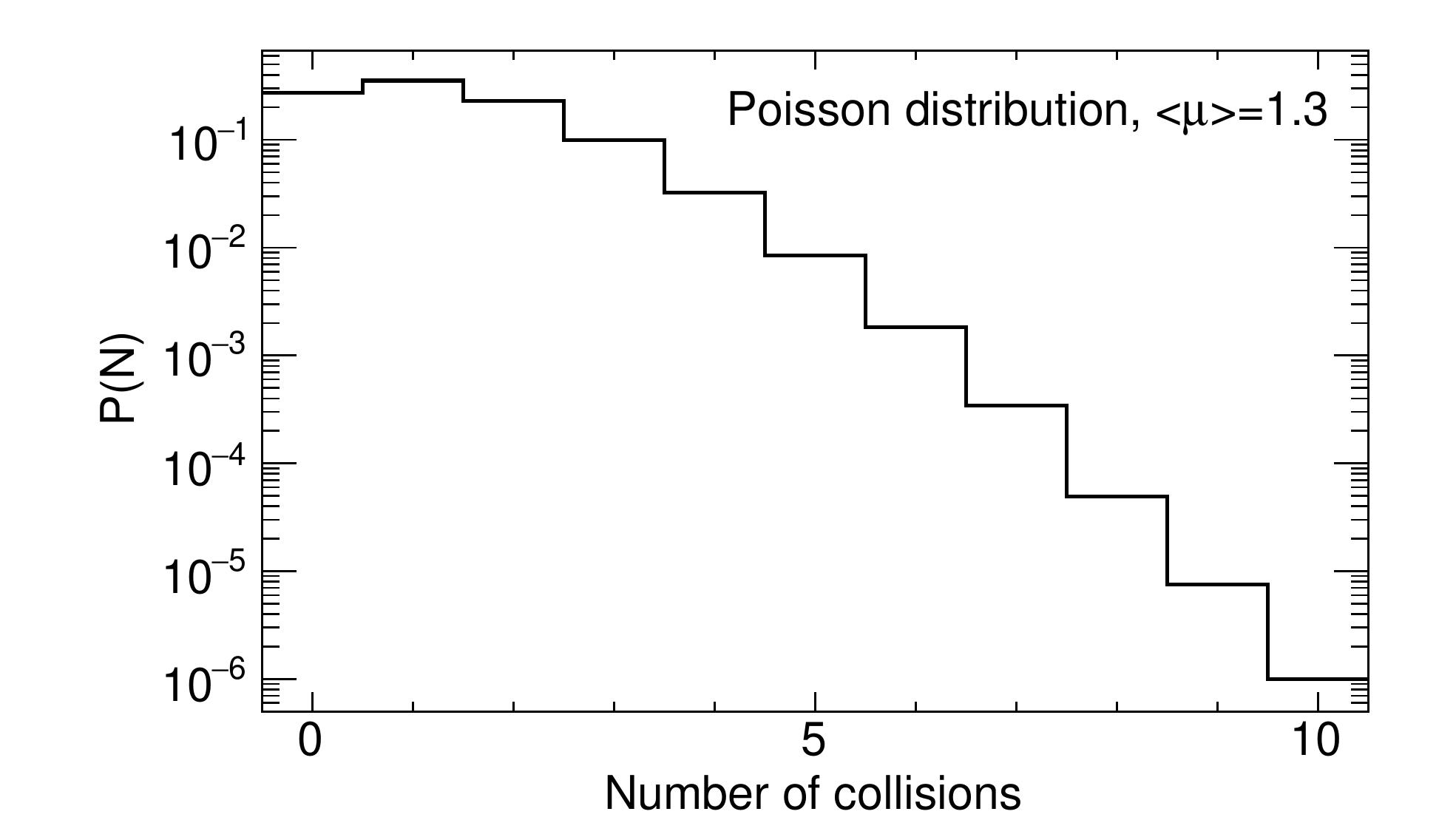}
    \caption{ Poisson distribution with mean of 1.3, which corresponds to the 
    probability distribution of number of concurrent interactions in the same bunch crossing. }
    \label{fig:poisson}
  \end{center}
\end{figure}

During the 13 TeV pp data taking in 2015, the average number of collisions per bunch crossing
is about 1.3, 0.4 and 0.1 for EndOfFill, VdM scan and TOTEM runs . The probability distribution for having various  
interactions, or pileups, in the same bunch crossing is shown in 
Fig.~\ref{fig:poisson} for EndOfFill run, where Poisson distributions have been assumed.
Therefore, the probability of having
two or more collisions is 37.3\%. Such level of pileup is not negligible, 
especially for very high multiplicity triggered events,
which deals with a large number of reconstructed tracks close to a primary vertex, two pileup collisions
that are very close to each other could contaminate the physics results.

In order to study clean high multiplicity pp collisions, a procedure of
rejecting pileup events is developed. The general goals we aim 
at for rejecting pileup events include the following:
\begin{itemize}
\item Reject multiple collision events as much as possible, while keeping events with several vertices that are far apart 
from each other in the z vertex. As in the analysis we require tracks directly from the primary vertex, having another vertex far apart will not affect the results.
\item Avoid removing events with split vertices. The reconstruction algorithm allows
for obtaining secondary vertices even if there is only one pp interaction. Therefore, 
if only one reconstructed primary vertex is required to be present, a significant 
fraction of good single-interaction events will be lost, especially at high multiplicity.
\item Accept some contamination of pileup interactions with small multiplicity. Those
interactions will not having a significant impact on the multiplicity of the primary vertex.
\item Evaluate the systematic uncertainties associated with vertex merging, although it is not 
possible to directly identify them.
\end{itemize}

\begin{figure}[thb]
  \begin{center}
    \includegraphics[width=0.85\linewidth]{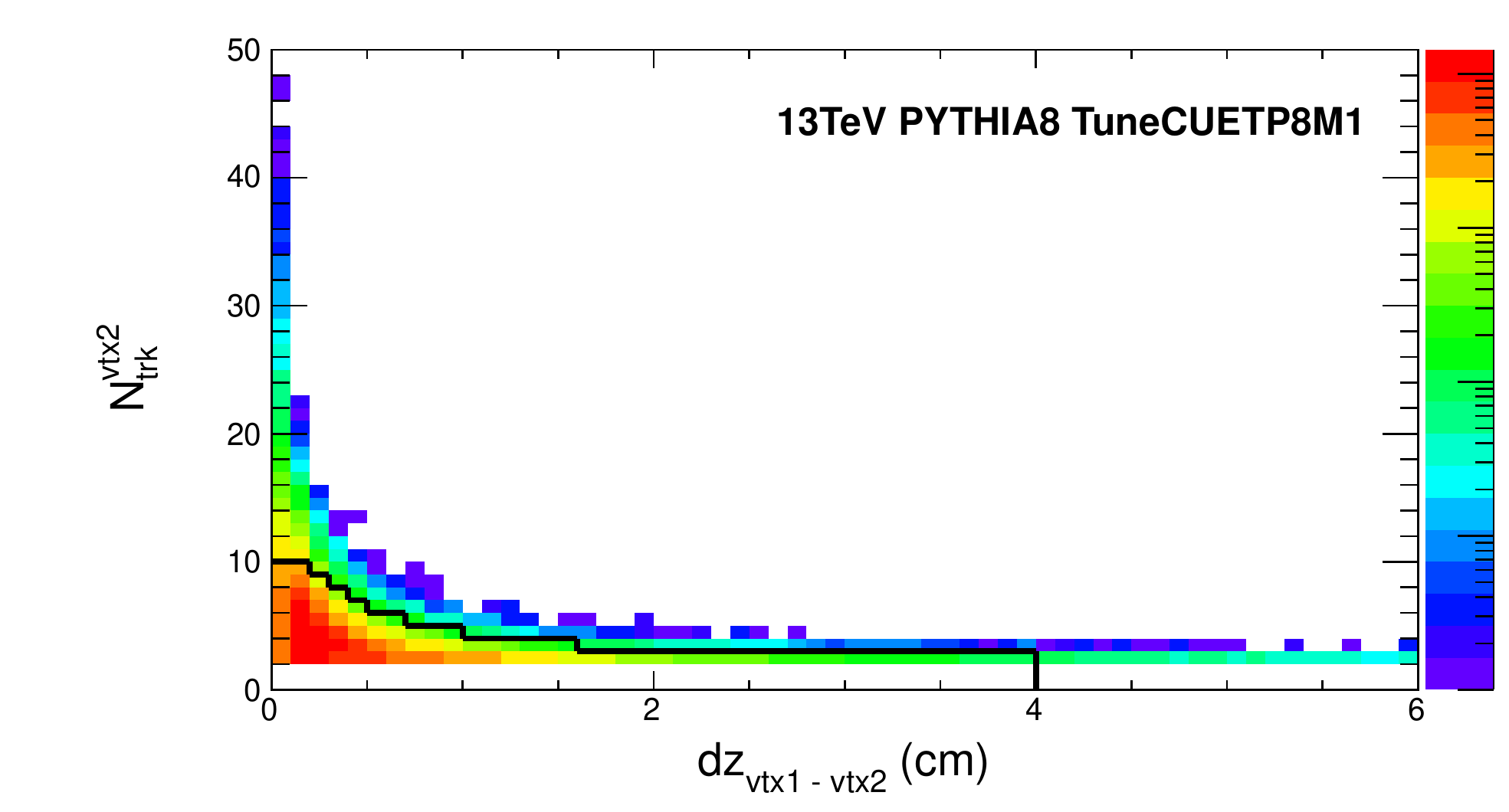}
    \caption{ the longitudinal displacement ($dz$) between additional primary vertices and the
lead primary vertex versus the number of tracks ($N_{trk}^{vtx2}$) associated
with the additional primary vertex for the pythia8 MinBias sample. }
    \label{fig:pileup_dzntrk_mc}
  \end{center}
\end{figure}

\begin{figure}[thb]
  \begin{center}
    \includegraphics[width=0.85\linewidth]{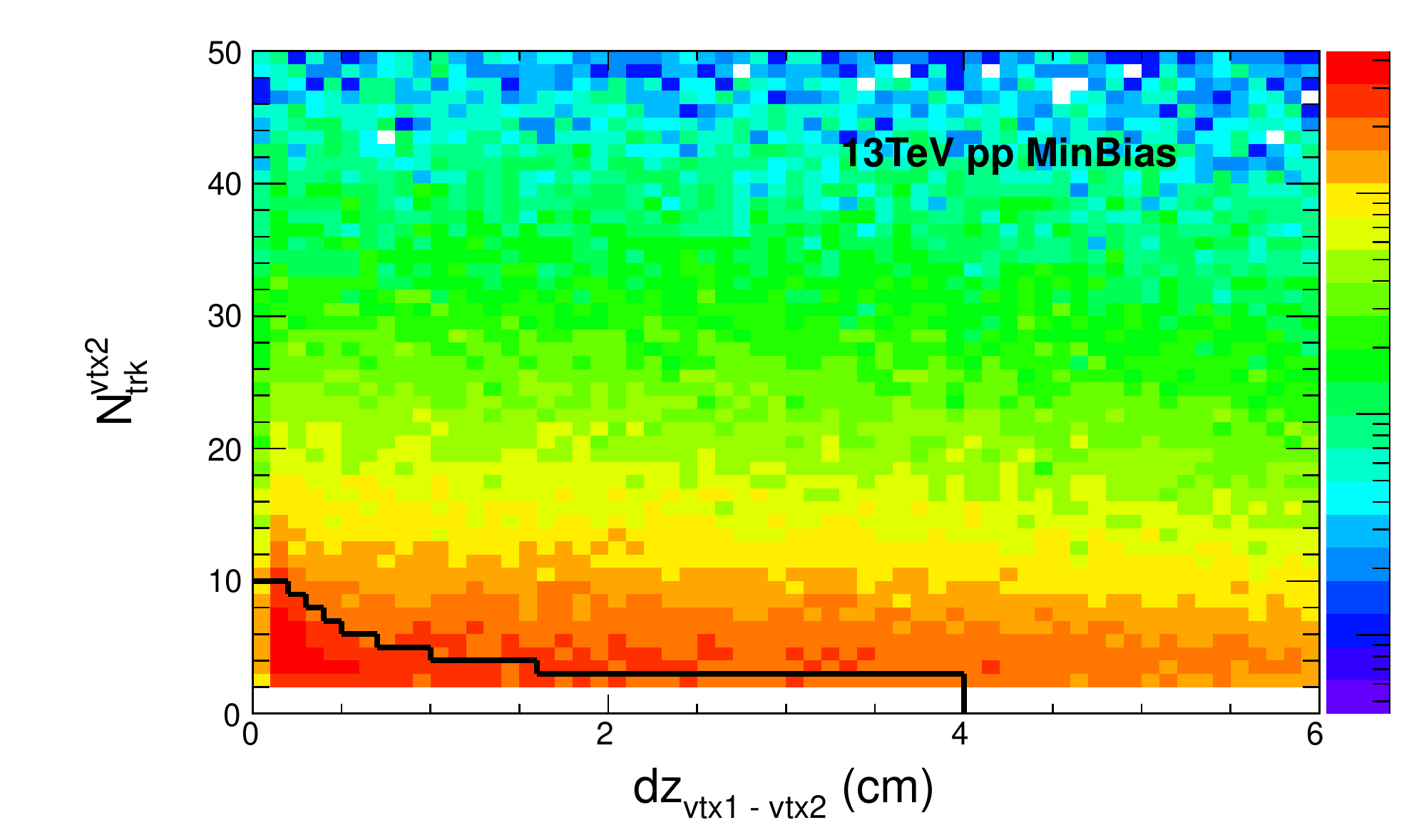}
    \caption{ the longitudinal displacement ($dz$) between additional primary vertices and the
lead primary vertex versus the number of tracks ($N_{trk}^{vtx2}$) associated
with the additional primary vertex for 13 TeV pp data with average pile up 1.3. }
    \label{fig:pileup_dzntrk_data}
  \end{center}
\end{figure}

Vertices from different collisions can be distinguished from split or
secondary decay vertices from a single collision by looking at the number of
tracks associated with each vertex. The lead primary vertex is defined as the
vertex with the highest $\sum {p_{T}^{trk}}^{2}$. In Fig.\ref{fig:pileup_dzntrk_mc},
the longitudinal displacement ($dz$) between additional primary vertices and the
lead primary vertex is plotted against the number of tracks ($N_{trk}^{vtx2}$) associated
with the additional primary vertex for the PYTHIA8 MinBias sample. As there is no pile up 
events in this sample, all additional vertices are from split or secondary decay. 
In Fig.\ref{fig:pileup_dzntrk_data}, the longitudinal displacement ($dz$) between additional primary vertices and the
lead primary vertex is plotted against the number of tracks ($N_{trk}^{vtx2}$) associated
with the additional primary vertex for the 13 TeV pp data where the average pile-up is 1.3. From these figures,
one can see that for additional primary vertices from splitting of a single collision
or secondary particle decays, there is a strong inverse relationship between
$dz$ and $N_{trk}^{vtx2}$. The primary vertices resulting from additional collisions in
an event are more randomly distributed. In order to exclude the vertices that
may have arisen from a single collision, events with additional vertices
are only cut if the additional vertex has a minimum $dz$ value. The specific value of $dz$ is dependent on
$N_{trk}^{vtx2}$ as shown in Table~\ref{tab:gpluscut}, and is shown as the black lines
in Fig.~\ref{fig:pileup_dzntrk_mc} and in Fig.~\ref{fig:pileup_dzntrk_data}.

\begin{table}[ht]
\centering
\caption{ Minimum longitudinal displacement ($dz$) as a function of the number of
associated tracks ($nTrk$) of an additional primary vertex required to remove an event}
\begin{tabular}{ l | l | l | l }
\hline
$nTrk$  & $dz$ (cm) & $N_{trk}$ & $dz$ (cm) \\
\hline
0-1 & N/A & 6 & 0.5  \\
\hline
2 & 4.0 & 7 & 0.4 \\
\hline
3 & 1.6 & 8 & 0.3 \\
\hline
4 & 1.0 & 9 & 0.2 \\
\hline
5 & 0.7 & 10+ & 0.0 \\
\hline
\end{tabular}
\label{tab:gpluscut}
\end{table}

\begin{figure}[thb]
  \begin{center}
    \includegraphics[width=\linewidth]{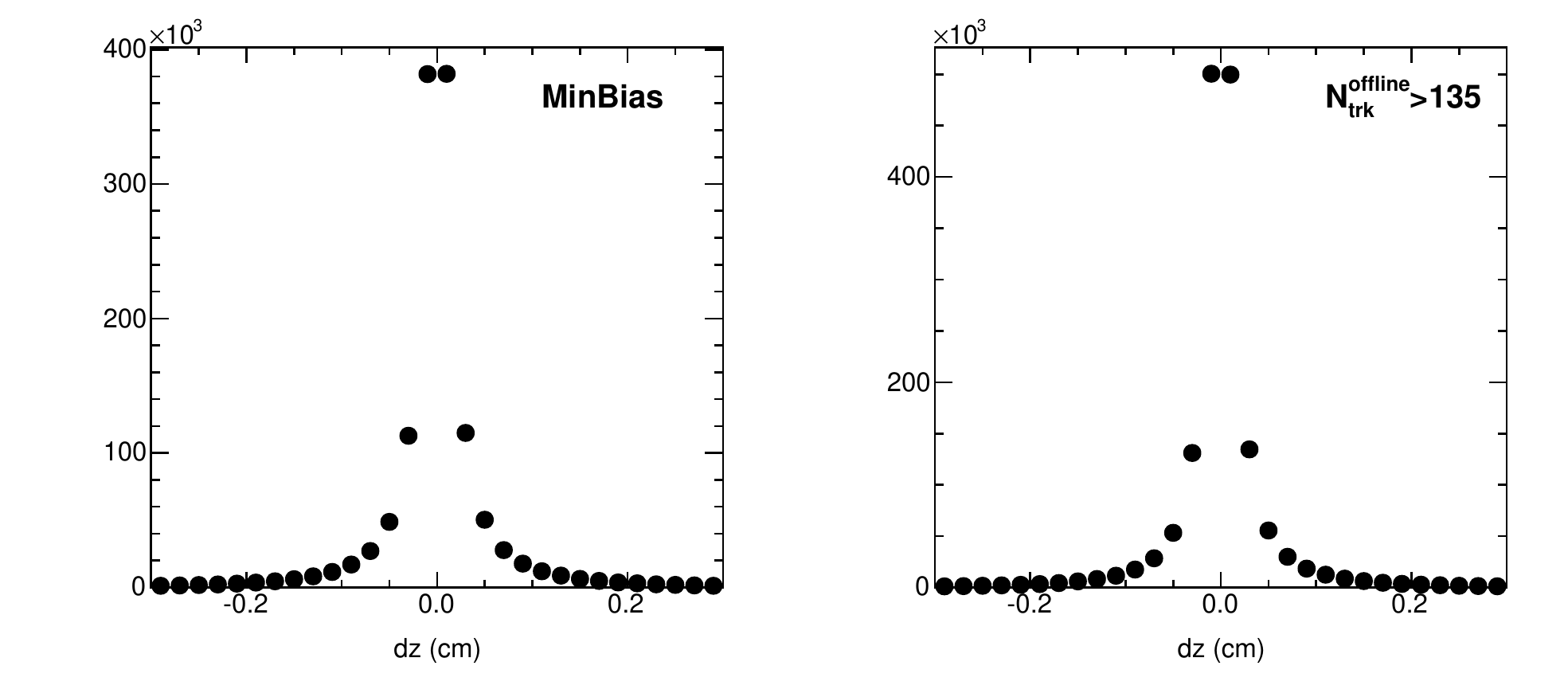}
    \caption{ Longitudinal displacement ($dz$) between the primary vertex and tracks selected by standard track selection described in Sec.~\ref{subsec:trackselect} for 13 TeV pp data for MinBias events (left) and events with $\noff > 135$ triggered by HLT\_PixelTracks\_Multiplicity110 (right). }
    \label{fig:pileup_vtxtrackz}
  \end{center}
\end{figure}

Furthermore, as in the analysis we require tracks directly from the primary vertex, having another vertex far apart will not affect the results. Fig.~\ref{fig:pileup_vtxtrackz} shows the the longitudinal displacement ($dz$) between the primary vertex and tracks selected by standard track selection described in Sec.~\ref{sec:tracking} for 13 TeV pp data for MinBias events (left) and events with $\noff > 135$ triggered by HLT\_PixelTracks\_Multiplicity110 (right). Over 97\% of tracks have $|dz| < 0.2 cm$. Therefore, events are also accepted if there is no multiple vertices within 1 cm from each other in the z vertex. In such a way, tracks from additional vertices are not used neither for the definition of \noff\ nor for the particle correlations.

\begin{figure}[thb]
  \begin{center}
    \includegraphics[width=\linewidth]{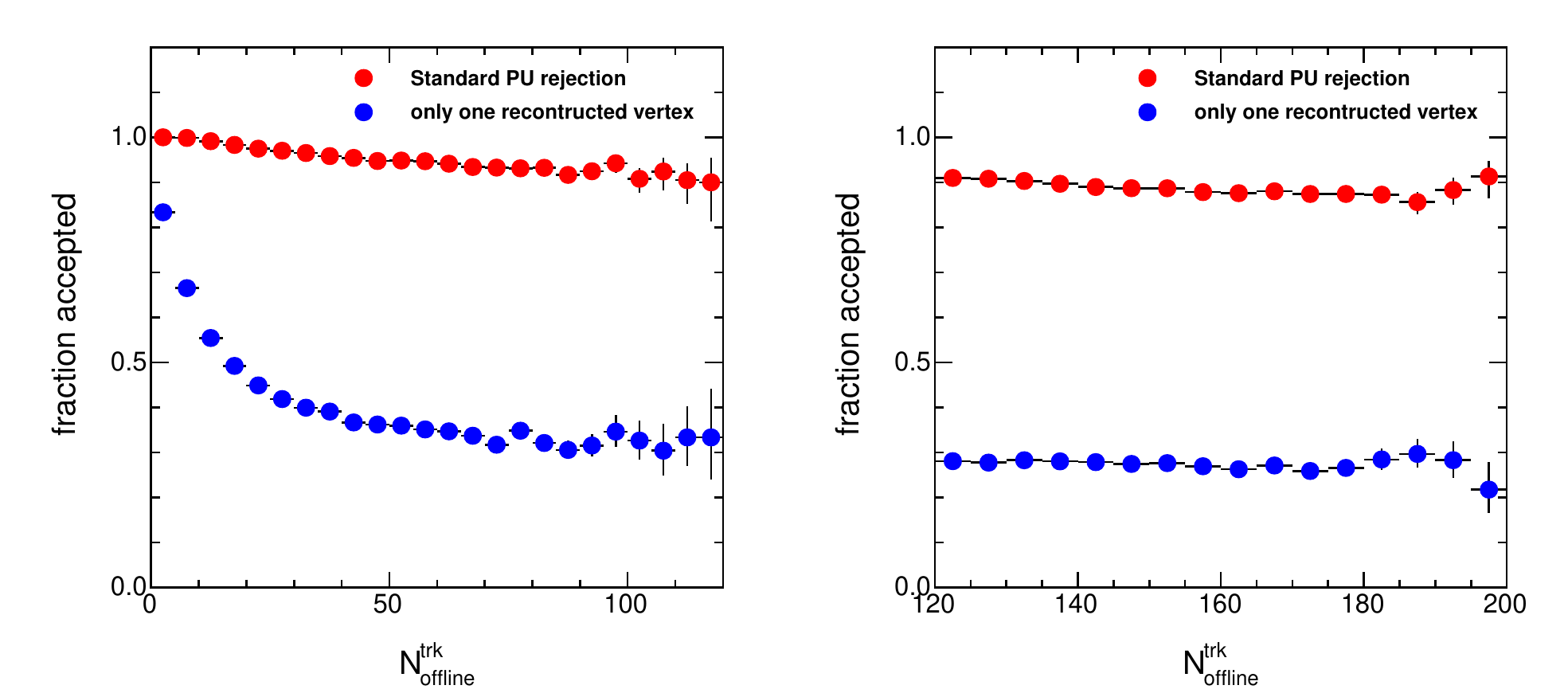}
    \caption{ Fraction of events accepted by different pileup rejection algorithm for MinBias events (left) and events triggered by HLT\_PixelTracks\_Multiplicity110 (right). Standard pileup rejection refers to the pileup filter used in this analysis.}
    \label{fig:pileup_fraction}
  \end{center}
\end{figure}

Figure~\ref{fig:pileup_fraction} shows the fraction of events accepted by different pileup rejection algorithm for MinBias events (left) and events triggered by HLT\_PixelTracks\_Multiplicity110 (right). An algorithm which rejects all events with more than one reconstructed vertices is chosen for comparison, this algorithm rejects all pileup events but also rejects events with split or
secondary decay vertices form a single collision. By accepting events where there is no multiple vertices within 1 cm from each other in the z vertex, a much larger fraction of events are accepted comparing to requiring only one reconstructed vertex. 

\clearpage

\section{Multiplicity classification}
\label{sec:mult}

In order to compare different collision systems on the same foot, events are classified by track multiplicity, hence that results can be compared at fixed multiplicity. 
The multiplicity of offline reconstructed tracks, \noff, is counted within the kinematic cuts of $|\eta| < 2.4$ and $\pt > 0.4$ GeV. 
The full track multiplicity range is divided into different multiplicity bins. 
The fractions of total number of events for each multiplicity bin, as well as the average track multiplicity before (\noff) and after (\ncorr) tracking efficiency corrections, 
are summarized in Tab.~\ref{tab:newmultbinningpPb} and Tab.~\ref{tab:newmultbinningpp} for pPb and pp data. 
The uncertainties on \ncorr\ come from the tracking efficiency correction procedure which introduce a total systematic uncertainty of 3.8\%.

\begin{table}[h]
\centering
\caption{\label{tab:newmultbinningpPb} Fraction of the full event sample
in each multiplicity bin and the average multiplicity per bin for 5 TeV pPb data.}
\begin{tabular}{ l | l | l | l}
\hline
Multiplicity bin (\noff) & Fraction & $\left<\noff \right>$ & $\left<N_\mathrm{trk}^\mathrm{corrected}\right>$\\
\hline
MB           &   1.00  &  40   & 50$\pm$2  \\
$[0, 20)$    &   0.31  &  10   & 12$\pm$1  \\
$[20, 30)$   &   0.14  &  25   & 30$\pm$1  \\
$[30, 40)$   &   0.12  &  35   & 42$\pm$2  \\
$[40, 50)$   &   0.10  &  45   & 54$\pm$2  \\
$[50, 60)$   &   0.09  &  54   & 66$\pm$3  \\
$[60, 80)$   &   0.12  &  69   & 84$\pm$4  \\
$[80, 100)$  &   0.07  &  89   & 108$\pm$5 \\
$[100, 120)$ &   0.03  &  109  & 132$\pm$6 \\
$[120, 150)$ &   0.02  &  132  & 159$\pm$7 \\
$[150, 185)$ &   $4 \times 10^{-3}$   &  162  & 195$\pm$9 \\
$[185, 220)$ &   $5 \times 10^{-4}$   &  196  &  236$\pm$10  \\
$[220, 260)$ &   $6 \times 10^{-5}$   &  232  &  280$\pm$12  \\
$[260, 300)$ &   $3 \times 10^{-6}$   &  271  &  328$\pm$14  \\
$[300, 350)$ &   $1 \times 10^{-7}$   &  311  &  374$\pm$16  \\
\hline
\end{tabular}
\end{table}

\begin{table}[h]\renewcommand{\arraystretch}{1.2}\addtolength{\tabcolsep}{-1pt}
\centering
\caption{ \label{tab:newmultbinningpp} Fraction of MB triggered events after event selections
in each multiplicity bin, and the average multiplicity of reconstructed tracks per bin
with $|\eta|<2.4$ and $\pt >0.4$ GeV, before (\noff) and after 
($N_\mathrm{trk}^\mathrm{corrected}$) efficiency correction, for pp data at 
\roots\ = 5.02, 7 and 13 TeV.}
\begin{tabular}{ l | l | l | l | l | l | l | l | l | l}
\hline
\multirow{2}{*}{\noff} & \multicolumn{3}{c|}{Fraction} & \multicolumn{3}{c|}{$\left<N_\mathrm{trk}^\mathrm{offline}\right>$} & \multicolumn{3}{c}{$\left<N_\mathrm{trk}^\mathrm{corrected}\right>$} \\\cline{2-10}
 & 5 TeV  & 7 TeV & 13 TeV & 5 TeV & 7 TeV & 13 TeV & 5 TeV & 7 TeV & 13 TeV  \\
\hline
MB & 1.0 & 1.0 & 1.0 & 13 & 15 & 16 & 15$\pm$1 & 17$\pm$1 & 19$\pm$1  \\
$[0, 10)$  & 0.48 & 0.44 & 0.43 & 4.8 & 4.8 & 4.8 & 5.8$\pm$0.3 & 5.5$\pm$0.2 & 5.9$\pm$0.3 \\
$[10, 20)$ & 0.29 & 0.28 & 0.26 & 14 & 14 & 14 & 17$\pm$1 & 16$\pm$1 & 17$\pm$1 \\
$[20, 30)$ & 0.14 & 0.15 & 0.15 & 24 & 24 & 24 & 28$\pm$1 & 28$\pm$1 & 30$\pm$1 \\
$[30, 40)$ & 6e-2 & 8e-02 & 8e-02 & 34 & 34 & 34 & 41$\pm$2 & 40$\pm$2 & 42$\pm$2 \\
$[40, 60)$ & 3e-2 & 5e-02 & 7e-02 & 47 & 47 & 47 & 56$\pm$2 & 54$\pm$2 & 58$\pm$2 \\
$[60, 85)$ & 3e-3 & 7e-03 & 2e-02 & 66 & 67 & 68 & 80$\pm$3 & 78$\pm$3 & 83$\pm$3 \\
$[85, 95)$ & 9e-5 & 3e-04 & 1e-03 & 88 & 89 & 89 & 106$\pm$4 & 103$\pm$4 & 109$\pm$4 \\
$[95, 105)$ & 2e-5 & 9e-05 & 5e-04 & 98 & 99 & 99 & 118$\pm$5 & 114$\pm$4 & 121$\pm$5 \\
$[105, 115)$ & 5e-6 & 2e-05 & 2e-04 & 108 & 109 & 109 & 130$\pm$5 & 126$\pm$5 & 133$\pm$5 \\
$[115, 125)$ & 1e-6 & 8e-06 & 6e-05 & 118 & 118 & 119 & 142$\pm$6  & 137$\pm$5 &145$\pm$6 \\
$[125, 135)$ & 2e-7 & 2e-06 & 2e-05 & 126 & 128 & 129 & 153$\pm$6 & 149$\pm$6 & 157$\pm$6  \\
$[135, 150)$ & 5e-8 & 4e-07 & 8e-06 & 139 & 140 & 140 & 167$\pm$7 & 162$\pm$6 & 171$\pm$7 \\
$[150, \infty)$ & 5e-9 & 8e-08 & 2e-06 & 155 & 156 & 158 & 186$\pm$8 & 181$\pm$7 & 193$\pm$8 \\
\hline
\end{tabular}
\end{table}

\cleardoublepage
\chapter{Two-particle correlations and anisotropy Fourier harmonics}
\label{ch:technique}

As discussed in Sec.~\ref{subsec:vnflow}, azimuthal anisotropic flow $v_n$ can be extracted from Fourier expansion of the particle distributions, 
\begin{equation}
\frac{dN}{d\phi} = \frac{N}{2\pi} \left(1+2\sum_{n=1}^{\infty} v_{n}\cos[n(\phi-\Psi_{RP})]\right),
\end{equation}
\noindent where $\Psi_{RP}$ is the reaction plane angle. 
The straight forward approach for $v_n$ measurement is to determine the reaction plane.
However, the reaction plane is not directly measurable in heavy ion collision. 
Instead, the experimentally reconstructed event plane is used. 
The $n^{th}$ harmonic event plane $\Psi_n$ can be obtained from the emitted particles as 
\begin{equation}
\Psi_n = \frac{1}{n}tan^{-1}\frac{\sum_{i}w_{i}\sin{n\phi_{i}}}{\sum_{i}w_{i}\cos{n\phi_{i}}},
\end{equation}
\noindent where $\phi$ is the azimuthal angle of a particle and $w_{i}$ are the weights to optimize the event plane resolution~\cite{Poskanzer:1998yz}. 
The sum runs over the particles used in the event plane determination. 
A perfect event plane determination, i.e. $\Psi_n = \Psi_{RP}$, requires infinite number of emitted particles. 
In reality, the finite number of detected particles produces a limited resolution in the measurement of the reaction plane, especially in pPb and pp collisions where the number of final state particle is relatively small compared to that in AA collisions. 

The $v_n$ coefficients can be measured using azimuthal correlations between observed particles to avoid the determination of event plane. 
The method used in this thesis is the two-particle azimuthal correlation, which can be written as 
\begin{equation}
\label{eq:fact}
\begin{split}
\left\langle\left\langle e^{in(\phi_a-\phi_b)}\right\rangle\right\rangle& = \left\langle\left\langle e^{\phi_a-\Psi_{n}-(\phi_b-\Psi_{n})}\right\rangle\right\rangle \\
												& = \left\langle\left\langle e^{\phi_a-\Psi_{n}}\right\rangle\left\langle e^{-(\phi_b-\Psi_{n})}\right\rangle\right\rangle \\
												& = \left\langle\left\langle\cos{n(\phi_a-\Psi_{n})}\right\rangle\left\langle\cos{n(\phi_b-\Psi_{n})}\right\rangle\right\rangle \\
												& = \left\langle v_{n,a}v_{n,b}\right\rangle,
\end{split}
\end{equation}
\noindent where the double brackets denote an average over all particles within an event, followed by an average over all events. 
The average product $<v_{n,a}v_{n,b}>$ can be extracted as the Fourier coefficients from Fourier expansion of the two-particle azimuthal difference distribution, 
\begin{equation}
\frac{dN^{pair}}{d\Delta\phi} = \frac{N^{pair}}{2\pi} \left(1+2\sum_{n=1}^{\inf} v_{n,a}v_{n,b}cos[n(\phi_a-\phi_b)]\right).
\end{equation}

A key assumption for the factorization of single particle $v_n$ in Eq.~\ref{eq:fact} is that the event plane angle is a global phase angle for all particles of the entire event. 
A significant breakdown of the factorization assumption up to 20\%, was recently observed for pairs of particles, separated by more than 2 units of $\eta$, from different \pt\ ranges in ultra-central (0-0.2\% centrality) PbPb collisions~\cite{CMS:2013bza}. 
Such an effect is referred to as factorization breakdown and is found to increase with the difference in \pt\ and $\eta$ between the two particles~\cite{Khachatryan:2015oea}.
The $v_n$ measurements presented in this thesis do not correct for factorization breakdown effect, any precise theory comparison should take this fact into account.

The following sections detail the procedure of two-particle correlation construction and $v_n$ extraction used in the analyses. 

\section{Two-particle $\Delta\eta$-$\Delta\phi$ correlation functions}
\label{sec:twopar}

Two-particle $\Delta\eta$-$\Delta\phi$ correlations measure the angular distribution of the associated particles relative to the trigger particle. 
For each track multiplicity bin, trigger particles are defined as identified $V^{0}$ or charged particles originating
from the primary vertex, with $|\eta| < 2.4$ and in a specified \pttrg\ range.
The number of trigger particles in the event is denoted by $N_{\rm trig}$, and there 
 may be more than one trigger particle per event.
Particle pairs are formed by associating with every trigger particle 
the remaining charged particles with $|\eta| < 2.4$ and in a specified \ptass\ 
range. The per-trigger-particle associated yield distribution is then defined by:

\vspace{-0.2cm}
\begin{equation}
\label{2pcorr_incl}
\frac{1}{N_{\rm trig}}\frac{d^{2}N^{\rm pair}}{d\Delta\eta d\Delta\phi}
= B(0,0)\times\frac{S(\Delta\eta,\Delta\phi)}{B(\Delta\eta,\Delta\phi)},
\end{equation}
\vspace{-0.2cm}

\noindent where $\Delta\eta$ and $\Delta\phi$ are the differences in $\eta$ 
and $\phi$ of the pair, respectively. The signal distribution, $S(\Delta\eta,\Delta\phi)$, is 
the measured per-trigger-particle distribution of same-event pairs, i.e.,

\vspace{-0.2cm}
\begin{equation}
\label{eq:signal}
S(\Delta\eta,\Delta\phi) = \frac{1}{N_{\rm trig}}\frac{d^{2}N^{\rm same}}{d\Delta\eta d\Delta\phi}.
\end{equation}
\vspace{-0.2cm}

\noindent
A signal pair is rejected if the associated track is found to be a daughter of 
the $V^0$ candidate, which is relevant only for small \deta\ and \dphi\ region.
The mixed-event background distribution, 

\vspace{-0.2cm}
\begin{equation}
\label{eq:background}
B(\Delta\eta,\Delta\phi) = \frac{1}{N_{\rm trig}}\frac{d^{2}N^{\rm mix}}{d\Delta\eta d\Delta\phi},
\end{equation}
\vspace{-0.2cm}

\noindent is constructed by pairing the trigger particles in each event with the 
associated particles from 10 different random events, excluding 
the original event. The symbol $N^{\rm mix}$ denotes the number of 
pairs resulting from the event mixing. The normalization of both signal and background distributions 
by dividing by $N_{\rm trig}$ is done event by event.

The background distribution is used to account
for random combinatorial background and pair-acceptance effects.
The normalization factor $B(0,0)$ is 
the value of $B(\Delta\eta,\Delta\phi)$ at $\Delta\eta=0$ and $\Delta\phi=0$ 
(with a bin width of 0.3 in $\Delta\eta$ and $\pi/16$ in $\Delta\phi$), representing
the mixed-event associated yield for both particles of the pair 
going in approximately the same direction, 
thus having full pair acceptance. 
Therefore, the ratio $B(0,0)/B(\Delta\eta,\Delta\phi)$
is the pair-acceptance correction factor used to derive the corrected
per-trigger-particle associated yield distribution. Equation~(\ref{2pcorr_incl}) is calculated
in 2~cm wide bins of the vertex position ($z_{\rm vtx}$) 
along the beam direction and averaged over the range 
$|z_{\rm vtx}| < 15$~cm. 

Each reconstructed track or $V^{0}$ particle is weighted by the inverse of the 
efficiency factor, $\varepsilon_{\rm trk}(\eta,\pt)$,
as a function of the track's pseudorapidity and transverse momentum.
The efficiency weighting factor accounts for the detector 
acceptance $A(\eta,\pt)$, the reconstruction efficiency $E(\eta,\pt)$, 
and the fraction of misidentified tracks, $F(\eta,\pt)$,

\vspace{-0.2cm}
\begin{equation}
\label{eq:eff_correction}
\varepsilon_\text{trk}(\eta,\pt) = \frac{A(\eta,\pt) E(\eta,\pt)}{1-F(\eta,\pt)},
\end{equation}
\vspace{-0.2cm}

\noindent These factors are derived from MC simulations.

\begin{figure}[hbtp]
  \begin{center}
    \hspace{1cm}\includegraphics[width=\textwidth]{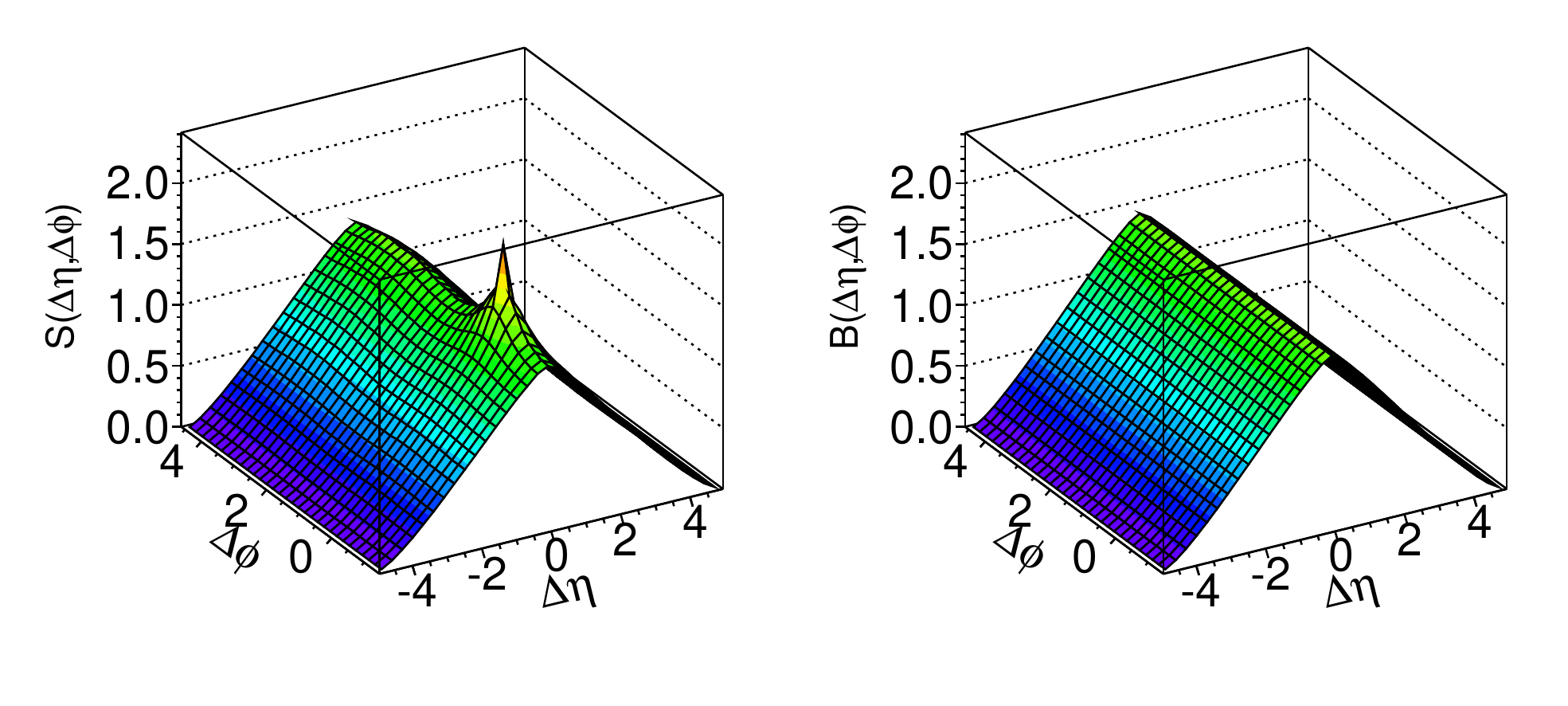}
    \caption{ Signal and mixed-event background 2-D distributions versus $\Delta\eta$ and $\Delta\phi$ for 
    $1<\pttrg<2$ GeV/c, $1<\ptass<2$ GeV/c 
    in 5.02~TeV pPb data with \noff\ $\geq$ 110.}
    \label{fig:signalbackground}
  \end{center}
\end{figure}

An example of signal and background pair two-dimensional (2-D) distributions for hadron-hadron
correlations in $\Delta\eta$ and $\Delta\phi$ is shown in Fig.~\ref{fig:signalbackground} 
for $1 < \pt < 2$ GeV/c in 5 TeV pPb data for $\noff\ \geq 110$. The triangular shape 
in $\Delta\eta$ is due to the limited acceptance in $\eta$ such that the phase space for obtaining 
a pair at very large $\Delta\eta$ drops almost linearly toward the edge of 
the acceptance. 

\begin{figure}[hbtp]
  \begin{center}
    \hspace{1cm}\includegraphics[width=0.8\textwidth]{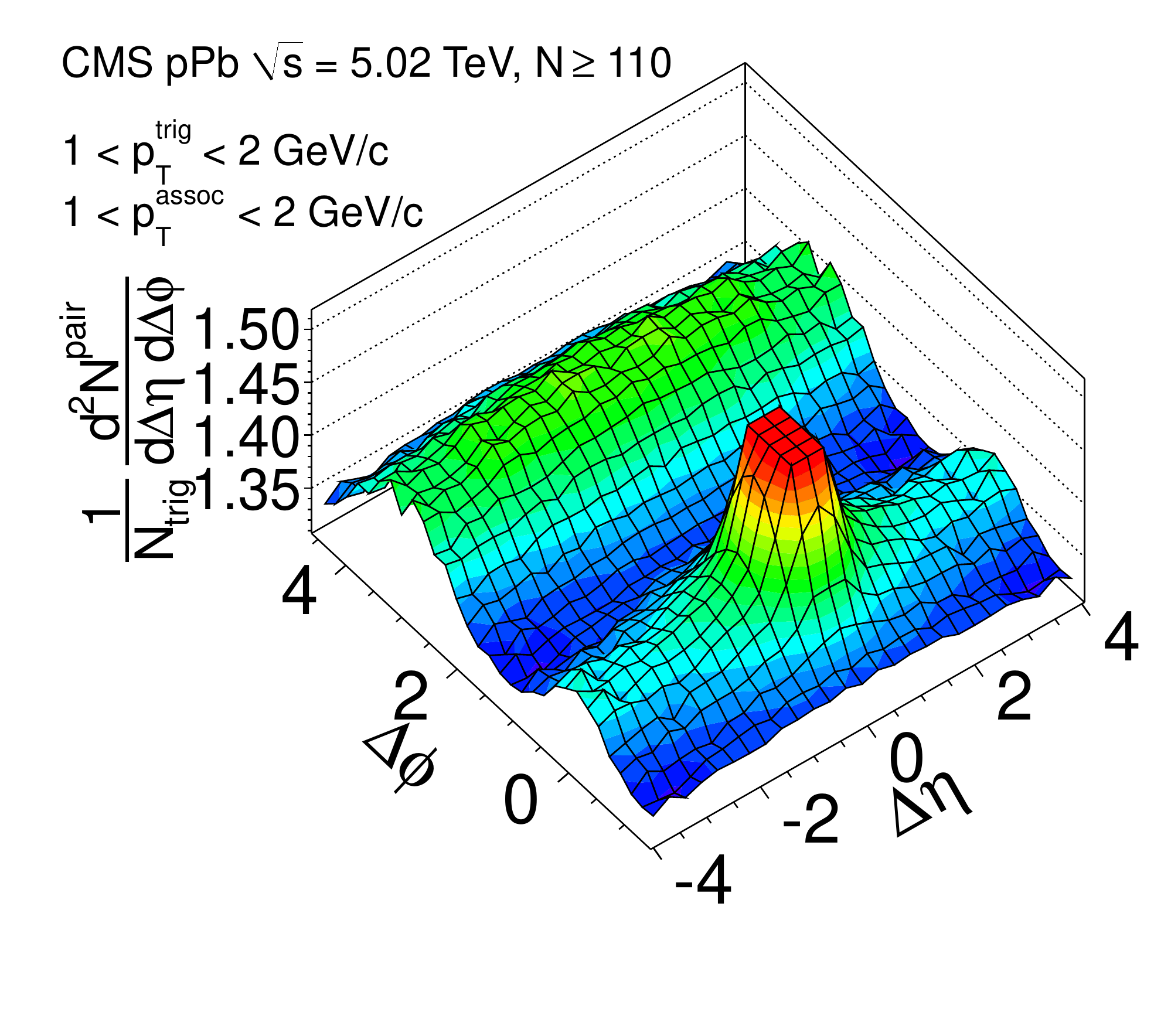}
    \caption{ 2-D two-particle correlation function for 
    $1<\pttrg<2$ GeV/c, $1<\ptass<2$ GeV/c 
    in 5.02~TeV pPb data with \noff\ $\geq$ 110.}
    \label{fig:corr2D}
  \end{center}
\end{figure}

The corresponding per-trigger associated yield distribution is 
shown in Fig.~\ref{fig:corr2D} as a function of $\Delta\eta$ and $\Delta\phi$. 
A large near side peak at $\deta \approx 0$ and $\dphi \approx 0$ is observed in the distribution. 
Such a peak mainly reflects the short range correlation from jet fragmentation, but also contains contributions from high-\pt\ resonance decay and Bose-Einstein correlations, where particles are expected to be produced spatially close to each other. 
Besides the near side peak, an elongated double-ridge structure is also presented at $\dphi \approx 0$ and $\dphi \approx \pi$, extending over a range of at least 4 units in $|\deta|$. 
As discussed in Sec.~\ref{sec:qgpsmall}, the existence of the ridge structure is an indication of collective motion of the system. 

\clearpage

\section{Azimuthal anisotropy harmonics from two-particle correlations}
\label{sec:vn}

\begin{figure}[hbtp]
  \begin{center}
    \hspace{1cm}\includegraphics[width=0.6\textwidth]{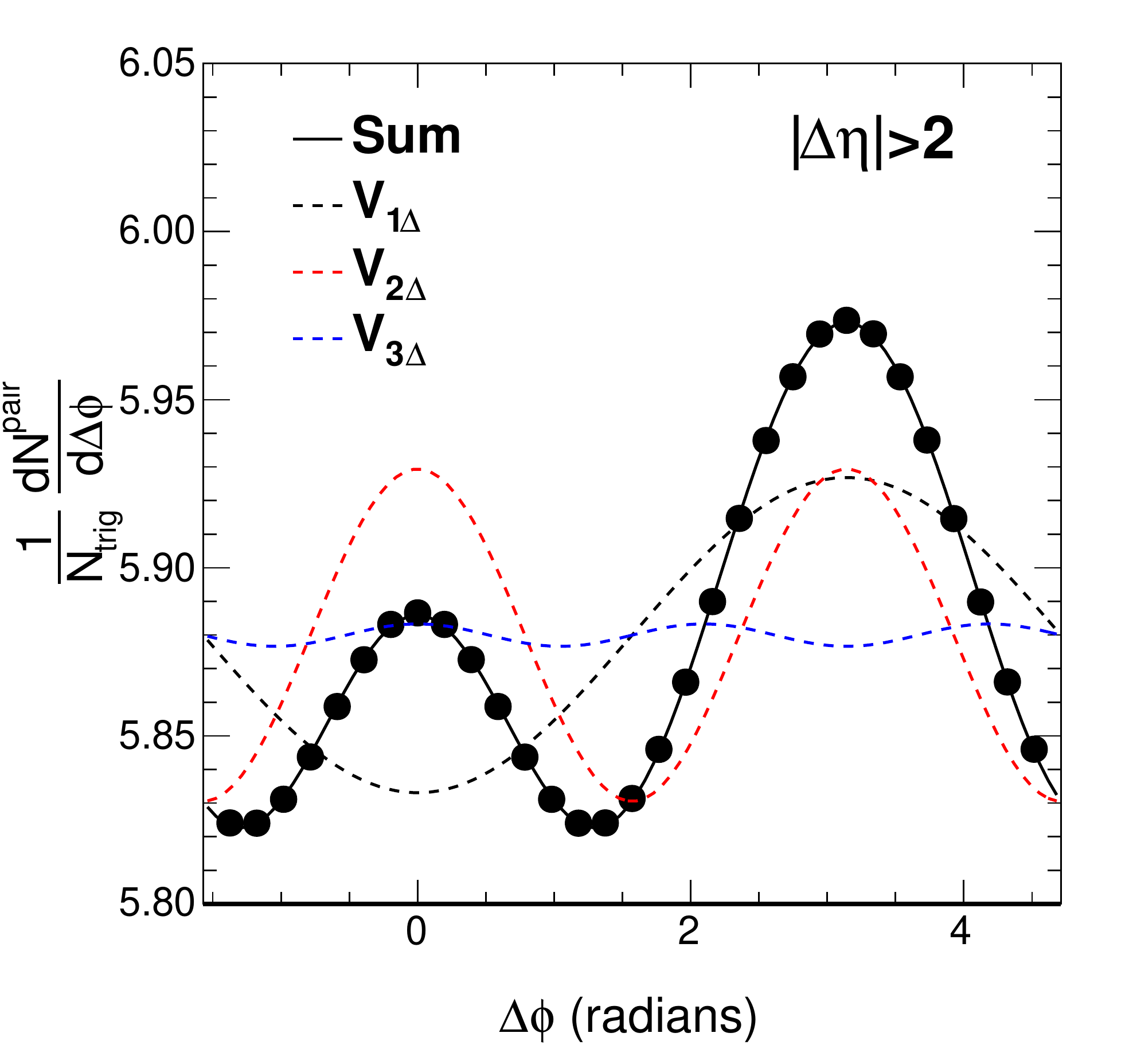}
    \caption{The 1D $\Delta\phi$ two-particle correlation function for charged particles with $1<\pt<3$ GeV/c in pPb collisions with $120<\noff<150$ at 5.02 TeV. The dashed curves show various orders of $V_{n\Delta}$, while the solid curve is the sum of all $V_{n\Delta}$ components.}
    \label{fig:fourierfit}
  \end{center}
\end{figure}

To further quantify the correlation structure, the 2-D distributions are reduced to
one-dimensional (1-D) distributions in $\Delta\phi$ by averaging over the $\Delta\eta$ 
range~\cite{CMS:2012qk,Khachatryan:2010gv,Chatrchyan:2011eka,Chatrchyan:2012wg}.
The azimuthal anisotropy harmonics are determined from a Fourier decomposition 
of long-range ($|\deta|>2$ to remove most of short-range correlations)
two-particle \dphi\ correlation functions, 

\begin{equation}
\label{eq:Vn}
Y(\Delta\phi)=\frac{1}{N_\text{trig}}\frac{d N^\text{pair}}{d\Delta\phi}=\frac{N_{\rm assoc}}{2\pi} \left\{1+\sum\limits_{n} 2V_{n\Delta} \cos (n\Delta\phi)\right\},
\end{equation}

\noindent where $V_{n\Delta}$ 
are the Fourier coefficients and $N_{\rm assoc}$ represents the total number of pairs per trigger 
particle for a given $(\pttrg, \ptass)$ bin. The first three Fourier terms are included
in the fits. Including additional terms have negligible effects on the fit results. 
Fig.~\ref{fig:fourierfit} shows an example Fourier fit with the first three Fourier terms plotted separately. 

As discussed above, $V_{n\Delta}$ can
be factorized into a product of single-particle Fourier harmonics, $v_{n}(\pttrg)$,
for trigger particles and $v_{n}(\ptass)$, for associated particles:
\begin{equation}
\label{eq:factorization}
V_{n\Delta}=v_{n}(\pttrg) \times v_{n}(\ptass).
\end{equation}

\noindent In this way, the elliptic and triangular 
anisotropy harmonics, $v_{2}\{2,|\deta|>2\}$ and $v_{3}\{2,|\deta|>2\}$, 
from two-particle correlation method can be extracted from the fitted Fourier coefficients 
as a function of \pt\ by assuming the factorization relation:

\begin{equation}
\label{eq:Vnpt}
v_{n}\{2,|\deta|>2\}(\pt) = \frac{V_{n\Delta}(\pt,\ptref)}{\sqrt{V_{n\Delta}(\ptref,\ptref)}},   n=2, 3
\end{equation}

\noindent Here, a reference particle \ptref\ range is chosen to be $0.3 < \pt < 3.0$ GeV/c. 
The $V_{n\Delta}(\ptref,\ptref)$ is derived by correlating unidentified charged hadrons
both from $0.3 < \pt < 3.0$ GeV/c range (reference particles), while $V_{n\Delta}(\pt,\ptref)$
represents Fourier coefficients by correlating a trigger particle with dedicated \pt\ range with a reference particle.
This can be understood if, in Eq.~\ref{eq:factorization}, one first choose \pttrg = \ptass = \ptref,
\begin{equation}
\label{eq:fact_ref}
V_{n\Delta}(\ptref,\ptref)=v_{n}(\ptref) \times v_{n}(\ptref),
\end{equation}

\noindent and then, choose  \ptass = \ptref\ and \pttrg = \pt, 

\begin{equation}
\label{eq:fact_trg}
V_{n\Delta}(\pt,\ptref)=v_{n}(\pt) \times v_{n}(\ptref).
\end{equation}

\noindent After plugging Eq.~\ref{eq:fact_ref} into Eq.~\ref{eq:fact_trg}, one arrives at Eq.~\ref{eq:Vn}.

\subsection{Extraction of \vnsig\ for \PKzS\ and \PgL/\PagL}
\label{subsec:V0vn}

\begin{figure}[hbt]
  \begin{center}
    \hspace{1cm}\includegraphics[width=0.7\textwidth]{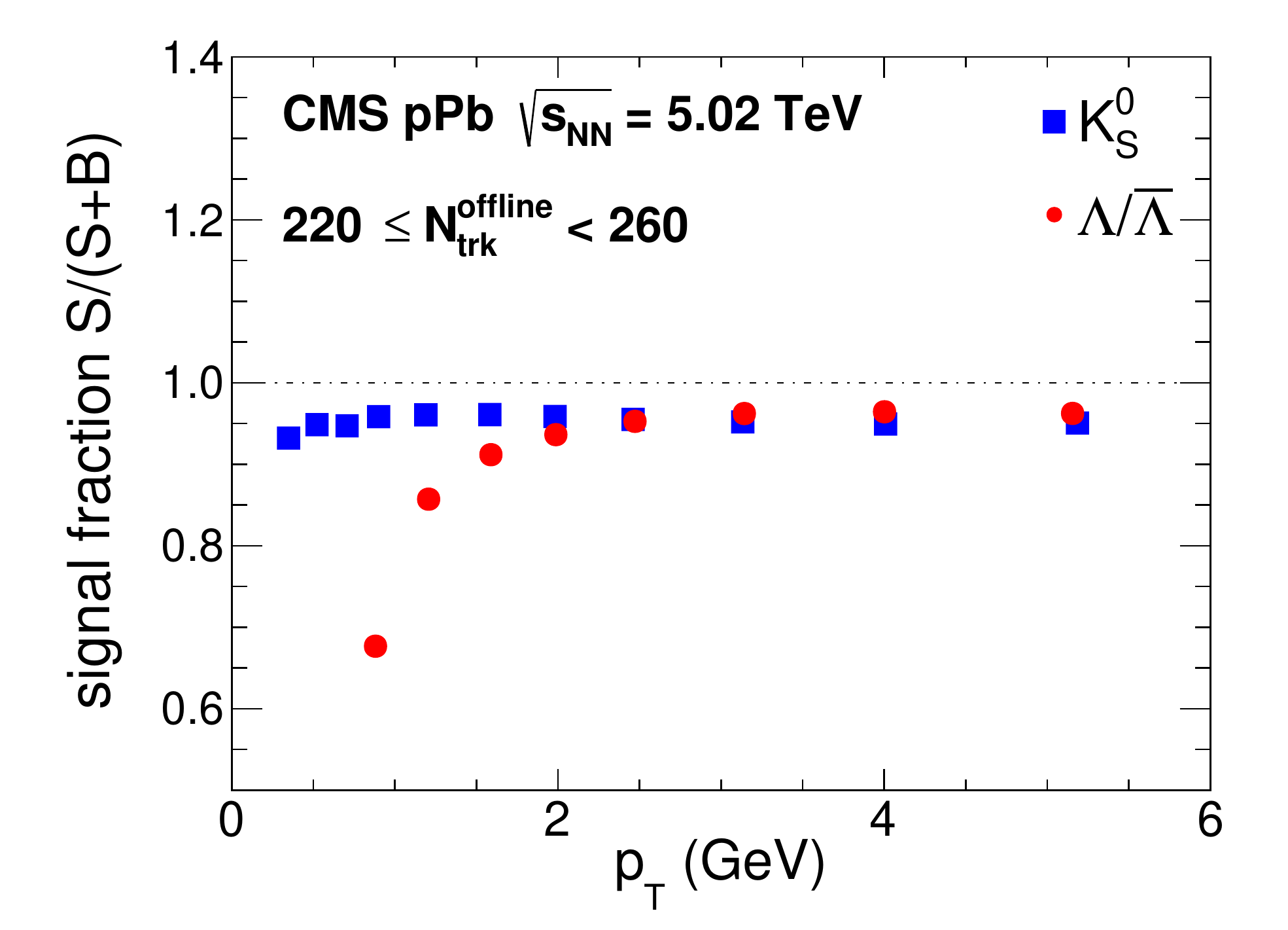}
    \caption{ $f_{\text{sig}}$ values for \PKzS\ and \PgL/\PagL\ as a function 
    of \pt\ in 5 TeV pPb collisions for $220 \leq \noff < 260$.}
    \label{fig:sig_frac}
  \end{center}
\end{figure}

\begin{figure}[hbt]
  \begin{center}
    \hspace{1cm}\includegraphics[width=0.7\textwidth]{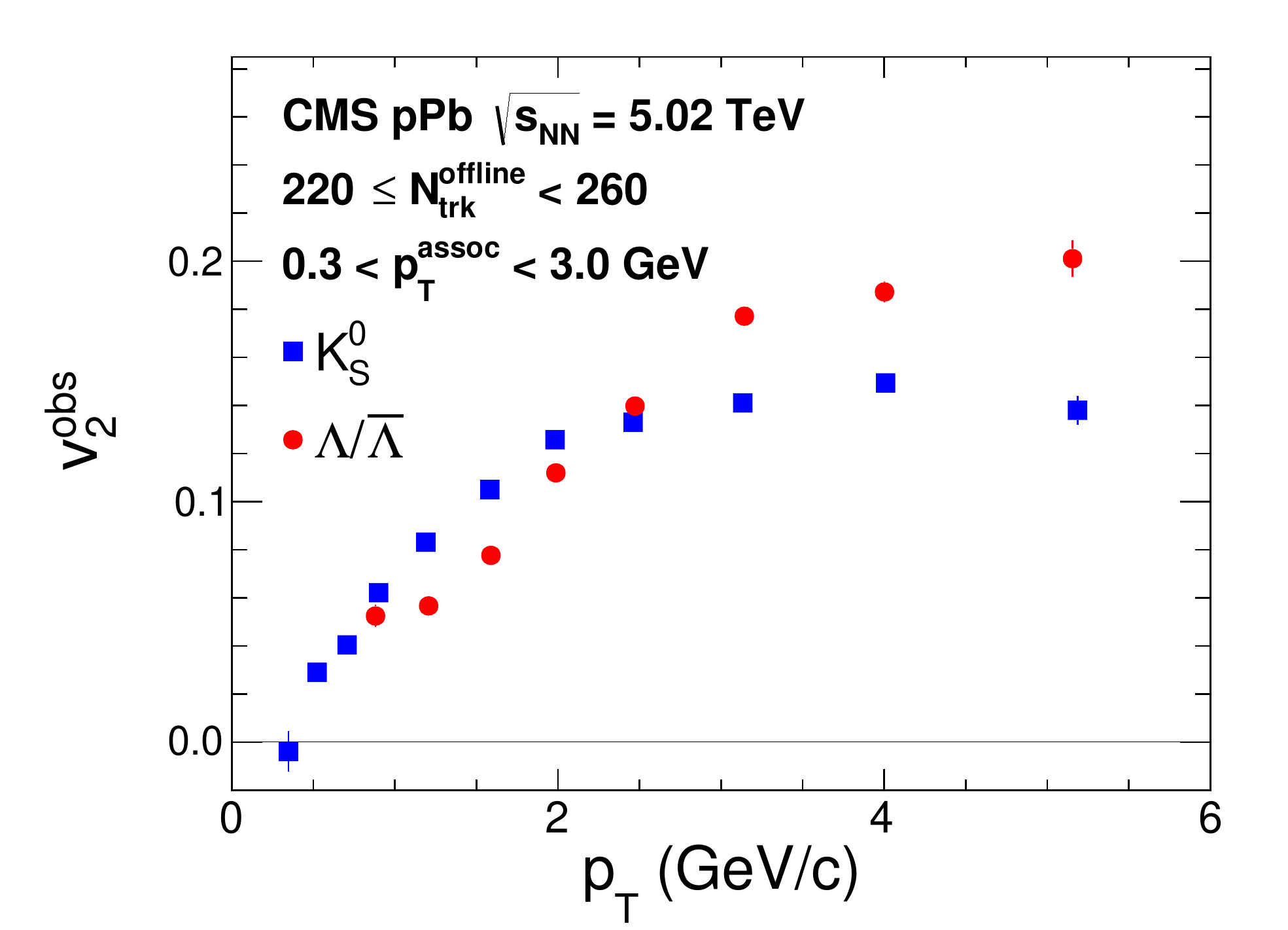}
    \caption{ $v_2^{obs}$ values for \PKzS\ and \PgL/\PagL\ as a function 
    of \pt\ in 5 TeV pPb collisions for $220 \leq \noff < 260$.}
    \label{fig:v2obs}
  \end{center}
\end{figure}

\begin{figure}[hbt]
  \begin{center}
    \hspace{1cm}\includegraphics[width=0.7\textwidth]{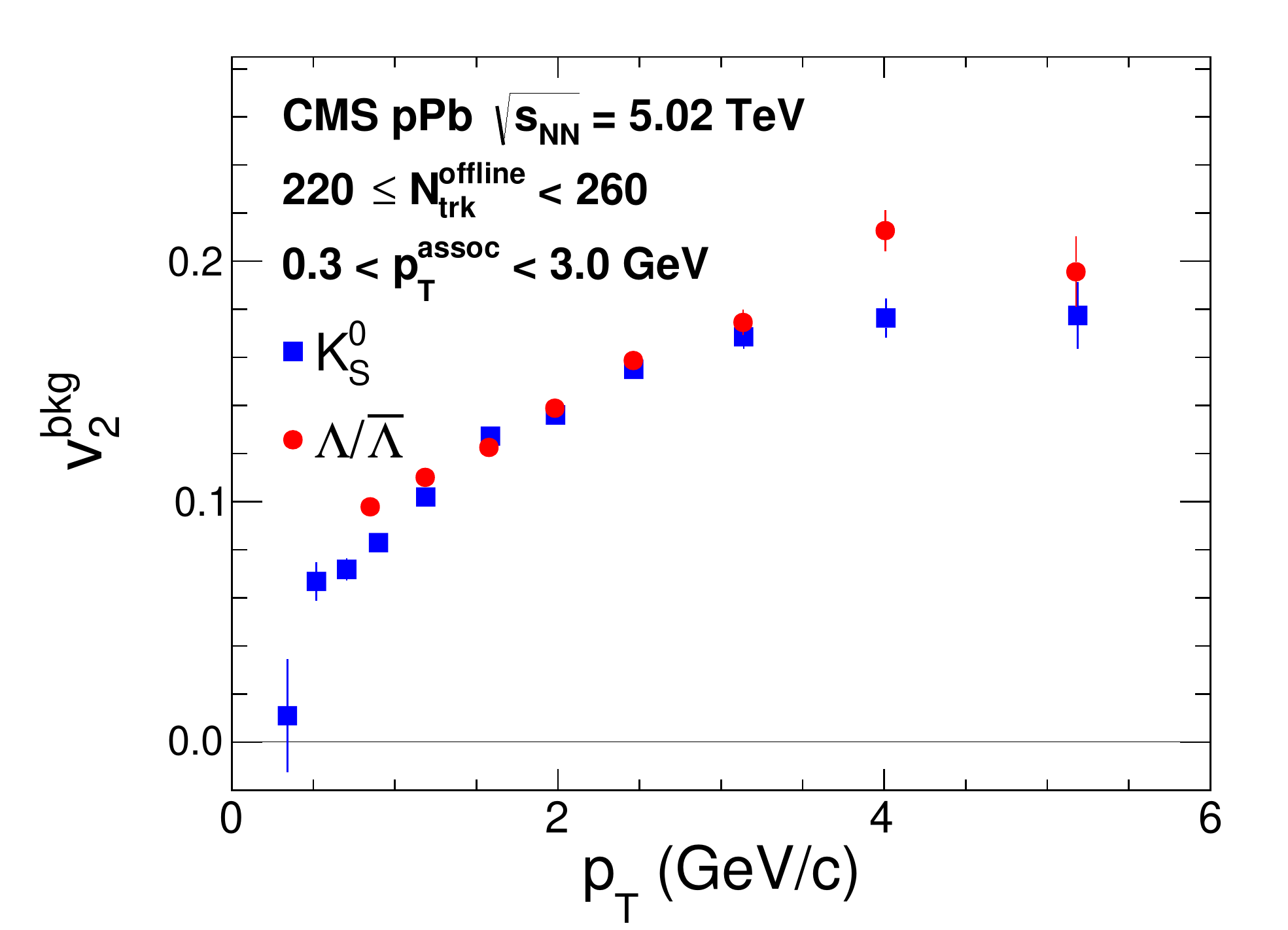}
    \caption{ $v_2^{bkg}$ values for \PKzS\ and \PgL/\PagL\ as a function 
    of \pt\ in 5 TeV pPb collisions for $220 \leq \noff < 260$.}
    \label{fig:v2bkg}
  \end{center}
\end{figure}

To extract \vnsig\ (true $v_{n}$) for \PKzS\ and \PgL/\PagL, the effect from background 
candidates in the reconstructed $V^0$s must be removed. 
As shown in Fig.~\ref{fig:v0mass}, 
the region between the $\pm 2\sigma$ mass cut containing both signal and background 
candidates is defined as peak region, and the region $> 3\sigma$ away from the peak region containing 
only background candidates is defined as background region. Two-particle correlation functions 
are constructed and azimuthal anisotropy 
harmonics are extracted for both the peak region, denoted as \vnobs, and 
the background region, \vnbkg. The \vnobs\ signal contains contributions
from real $V^0$ candidates, denoted as \vnsig, and from background candidates, 
\vnbkg, via the following equation:

\begin{equation}
\vnobs = f_{\text{sig}} \vnsig + (1-f_{\text{sig}}) \vnbkg
\end{equation}

\noindent where $f_{\text{sig}}$ is the fraction of signal yield extracted in the peak region
from fit to the mass distribution. Typical values of signal fraction as a function of \pt\ are 
shown in Fig.~\ref{fig:sig_frac} in 5 TeV pPb collisions for $220 \leq \noff < 260$. The extracted
$v_2^{obs}$ values are shown in Fig.~\ref{fig:v2obs} and $v_2^{bkg}$ values are shown 
in Fig.~\ref{fig:v2bkg} for \PKzS\ and \PgL/\PagL\ as a function of \pt, 
also in 5 TeV pPb collisions for $220 \leq \noff < 260$.

With $f_{\text{sig}}$, \vnobs\ and \vnbkg, the \vnsig\ value of $V^0$ candidates can 
be derived as:

\begin{equation}
\label{eq:vnsignal}
\vnsig = \frac{\vnobs - (1-f_{\text{sig}}) \vnbkg}{f_{\text{sig}}},
\end{equation}

\noindent results of \vnsig\ for \PKzS\ and \PgL/\PagL\ as a function of \pt\ for pp, pPb and PbPb collisions are shown in Chs.~\ref{ch:resultpPb} and~\ref{ch:resultpp}.

\clearpage

\paragraph{Propagation of uncertainties of \vnsig.} The Fourier coefficients $V_{n\Delta}$ and their uncertainties $\Delta V_{n\Delta}$ 
are extracted from the Fourier decomposition fits of two-particle \dphi\ correlation 
functions. Some notations are defined as follow:

\begin{itemize}
\item $V_{n}^{\text{ref}}$, $\Delta V_{n}^{\text{ref}}$: $V_{n}$ coefficients and uncertainties
of hadron-hadron correlations.
\item $V_{n}^{\text{obs}}$, $\Delta V_{n}^{\text{obs}}$: $V_{n}$ coefficients and uncertainties
of $V^0$-hadron correlations from peak region.
\item $V_{n}^{\text{bkg}}$, $\Delta V_{n}^{\text{bkg}}$: $V_{n}$ coefficients and uncertainties
of $V^0$-hadron correlations from background region.
\end{itemize}

The $v_{n}^{\text{ref}}$ for reference particles as well as it uncertainty is derived as 
 
\begin{equation}
v_{n}^{\text{ref}} = \sqrt{V_{n}^{\text{ref}}},
\end{equation}

\noindent and

\begin{equation}
\Delta v_{n}^{\text{ref}} = \frac{1}{2}\sqrt{V_{n}^{\text{ref}}}\frac{\Delta V_{n}^{\text{ref}}}{V_{n}^{\text{ref}}}.
\end{equation}

The \vnobs\ and \vnbkg\ and their uncertainties are calculated as

\begin{equation}
v_{n}^{\text{obs,bkg}} = \frac{V_{n}^{\text{obs,bkg}}}{v_{n}^{\text{ref}}},
\end{equation}

\noindent and

\begin{equation}
\Delta v_{n}^{\text{obs,bkg}} = v_{n}^{\text{obs,bkg}}\sqrt{\left( {\frac{\Delta V_{n}^{\text{obs,bkg}}}{V_{n}^{\text{obs,bkg}}}} \right)^{2} + \left( {\frac{\Delta v_{n}^{\text{ref}}}{v_{n}^{\text{ref}}}} \right)^{2}},
\end{equation}

\noindent which are shown as error bars in Fig.~\ref{fig:v2obs} and Fig.~\ref{fig:v2bkg}.

Finally, uncertainties of single-particle \vnsig\ for \PgL/\PagL\ and \PKzS\ can be 
derived from Eq.~\ref{eq:vnsignal} with values of $\Delta \vnobs$ and $\Delta \vnbkg$,

\begin{equation}
\Delta \vnsig = \frac{\sqrt{\left( \Delta \vnobs \right)^{2} + \left( \Delta \vnbkg\ (1-f_{\text{signal}}) \right)^{2}}}{f_{\text{signal}}}.
\end{equation}


\section{Systematic uncertainties}
\label{sec:syst}

This section summarize various systematic effect studies for results shown in Chs.~\ref{ch:resultpPb} and~\ref{ch:resultpp}. 
The experimental systematic effects are evaluated by varying conditions in extracting $v_n$ values. 

For inclusive charged particle results, the systematic uncertainties are found to have
no significant dependence on \pt\ and collision energy so they are quoted to be constant
percentages over the entire \pt\ range for all collision energies.
Each of the systematic uncertainty study is described as follow:
\begin{itemize}
\item Track selection. Experimental systematic uncertainties due to track quality requirements are
examined by varying the track selection thresholds for $d_z/\sigma(d_z)$ and $d_{xy}/\sigma(d_{xy})$
from 2 to 5.
\item Vertex position. The sensitivity of the results to the primary vertex position ($z_\mathrm{vtx}$) is quantified by comparing results at
different $z_\mathrm{vtx}$ locations over a 30 cm wide range. 
\item Trigger efficiency. Results extracted from data taken with different triggers are compared to evaluate the systematic uncertainties from variation of trigger efficiency. For each of the multiplicity bin where the trigger efficiency is not 100\%, data from a lower threshold trigger is used to study the effect of trigger in-efficiency.
\item Pileup. To investigate potential residual pileup effect after pileup rejection described in Sec.~\ref{sec:pileup}, the analysis is repeated by requiring only one reconstructed vertex present in the event. This is an extreme way of removing pileup events and is used to evaluate the systematic uncertainty by comparing to the results with standard pileup rejection. 
\end{itemize}

The dominant sources of systematic uncertainties for the \vsecsig\ and \vtrdsig\ measurements for $\PKzS$ and $\PgL$/$\PagL$ particles are related to their reconstruction. 
The systematic effect are found to have no dependence on \pt\ or multiplicity. 
Therefore, they are assumed to be constant percentage over the entire \pt\ and multiplicity range. 
The systematic uncertainty study are described as follow: 
\begin{itemize}
\item $V^{0}$ mass distribution range used in fit. The range of the $V^{0}$ mass distribution in fitting the signal plus background is varied by 10\%, which could affect the value of $f_{\text{signal}}$, to study the systematic effect.
\item Size of $V^{0}$ mass region for signal. As discussed in Sec.~\ref{subsec:V0vn}, \vnobs\ and $f_{\text{signal}}$ have dependence on the range of peak region, which results in an uncertainty in \vnsig. The range of peak region from $\pm 1\sigma$ to $\pm 3\sigma$ to evaluate the effects on $\vnsig$ results.
\item Size and location of $V^{0}$ mass sideband region. Systematic uncertainties due to the selection of different sideband mass regions, which could change \vnbkg\ are estimated by vary the range of background region from $> 3\sigma$ to $> 5\sigma$. 
\item Misidentified $V^{0}$ mass region. In the mis-identified candidate removal procedure described in Sec.~\ref{sec:V0}, different cuts applied to $\pi$ -$\pi$ and $\pi$ -p hypothesis remove different amount of $V^0$ candidates. The invariant mass range used to reject mis-identified $V^{0}$ candidates is varied by 25\% to evaluate the systematic effect. 
\item $V^{0}$ selection criteria. Systematic effects related to selection of the $V^{0}$ candidates are evaluated by varying the requirements on the decay length significance and $\cos\theta^{\text{point}}$. 
\item Tracker misalignment. As misalignment of the tracker detector elements can affect the $V^0$ reconstruction performance, an alternative detector geometry is studied. Compared to the standard configuration, this alternative has the two halves of the barrel pixel detector shifted in opposite directions along the beam by a distance on the order of 100 $\mu$m. 
\item MC closure test. To test the procedure of extracting the $V^{0}$ signal $v_2$ from Eq.~(\ref{eq:vnsignal}), a study using EPOS LHC pPb MC events is performed to compare the extracted \vsecsig\ results with the generator-level \PKzS\ and \PgL/\PagL\ values.
\end{itemize}

Systematic uncertainties originating from different independent sources are added in quadrature
to obtain the overall systematic uncertainty shown as boxes in the figures in Ch.~\ref{ch:resultpPb} and~\ref{ch:resultpp}.
Because of insufficient statistical precision, the uncertainties in $v_3$ are assumed to be the same as those in $v_2$,
as was done in Refs.~\cite{Chatrchyan:2013nka,Khachatryan:2014jra}.
For the same reason, the systematic uncertainties
on the $v_{2}$ results for $V^{0}$ particles in pp collision in Ch.~\ref{ch:resultpp} are obtained from studies performed for pPb collisions in Ch.~\ref{ch:resultpPb}, while those resulting from systemtic bias of the HLT trigger and jet subtraction method are taken to be the same as for the inclusive charged particles.
The relative systematic uncertainties for the
two-particle $V_{n\Delta}$ coefficients as a function of \noff\ in Fig.~\ref{fig:C2C3} (described in Section~\ref{sec:vncharge})
are exactly twice those for the corresponding $v_{n}$ harmonics, since $V_{n\Delta} = v_{n}^{2}$
when trigger and associated particles are selected from the same \pt\ range.

Table~\ref{tab:systpPb} summarizes systematic uncertainties in \vnsig\ for $V^{0}$ particles from the above sources for pp, pPb and PbPb data. 
Table~\ref{tab:syst-table3} summarizes systematic uncertainties for multiplicity-dependent inclusive charged particle results in pp collision. 
The same sources apply to \pt\ differential results, leading to total experimental systematic uncertainty of 5\%.

For pp results in Ch.~\ref{ch:resultpp}, there are additional systematic effects from the jet subtraction procedure which will be described in detail in Sec.~\ref{subsec:jetsubsyst}.

\begin{table*}[hbt]
\caption{\label{tab:systpPb} Summary of systematic uncertainties in \vnsig\ for pPb and PbPb data.}

\begin{center}
\begin{tabular}{lcc}
 Source			&	 pPb (\%)		&	PbPb (\%) \\
\hline
 $V^0$ mass distribution range used in fit 	&   1	&	1	\\
 Size of $V^0$ mass region for signal	  &  2   &	2	\\
 Size and location of $V^0$ mass sideband region 	&  2.2	&	2.2	\\
 Misidentified $V^0$ mass region	& 2	&	2	\\
 $V^0$ selection criteria		&	3	&	3	\\
 Tracker misalignment		&	2	&	2	\\
 MC closure	test			&	4	&	4	\\
 Trigger efficiency			&	2	&	---	\\
 Pile-up			&	1	&	---	\\
\hline
 Total		&	6.9	&	6.6	\\
\end{tabular}
\end{center}
\end{table*}

\begin{table*}[ht]
\caption{\label{tab:syst-table3} Summary of systematic uncertainties
for multiplicity-dependent $v_{n}$ from two-particle correlations (after correcting for jet
correlations) in pp collisions.
Different multiplicity ranges are represented as $[m,n)$.}
\centering
\begin{tabular}{l|c|c|c|c|c|c}
\hline
\multirow{2}*{Source} & \multicolumn{3}{c|}{$v_{2}$ (\%)} & \multicolumn{3}{c}{$v_{3}$ (\%)} \\
\cline{2-7}
          & [0,40) & [40,85) & [85,$\infty$) & [0,40) & [40,85) & [85,$\infty$) \\
\hline
 HLT trigger bias		&  --  & --  & 2 &  --  & --  & 2 \\
 Track quality cuts		&  1 & 1 & 1 &  1 & 1 & 1  \\
 Pileup effects		    &  1.5 & 1.5 & 1.5 &  1.5 & 1.5 & 1.5  \\
 Vertex dependence		&  1.5 & 1.5 & 1.5 &  1.5 & 1.5 & 1.5  \\
 Jet subtraction		&  18 & 9.5 & 6.5 &  26.8 & 17 & 8.5 \\
\hline
 Total		&	18.2 & 9.8 & 7.2	&	27 & 17.3 & 8.8	 \\
\hline
\end{tabular}
\end{table*}

\cleardoublepage
\chapter{Long-range two-particle correlations in pPb and PbPb collisions}
\label{ch:resultpPb}

This chapter presents measurements of two-particle angular correlations between an identified strange hadron ($\PKzS$ or $\PgL$/$\PagL$)
and a charged particle in pPb collisions at \ensuremath{\sqrt{s_{_{NN}}}} = 5 TeV. The results
are compared to semi-peripheral PbPb collision data at \ensuremath{\sqrt{s_{_{NN}}}} = 2.76 TeV,
covering similar charged-particle multiplicities in the events. The observed azimuthal
correlations at large relative pseudorapidity are used to extract the second-order ($v_2$) and third-order ($v_3$) anisotropy
harmonics of $\PKzS$ and $\PgL$/$\PagL$ particles following the procedure described in Chapter~\ref{ch:technique}.
These quantities are studied as a function of the charged-particle multiplicity in the event and
the transverse momentum of the particles. 

The majority of work presented in this chapter is published in Ref.~\cite{Khachatryan:2014jra}. 

\section{Two-particle correlation function}
\label{sec:resultpPbcorrfcn}

\begin{figure*}[hb]
\centering
\includegraphics[width=0.8\linewidth]{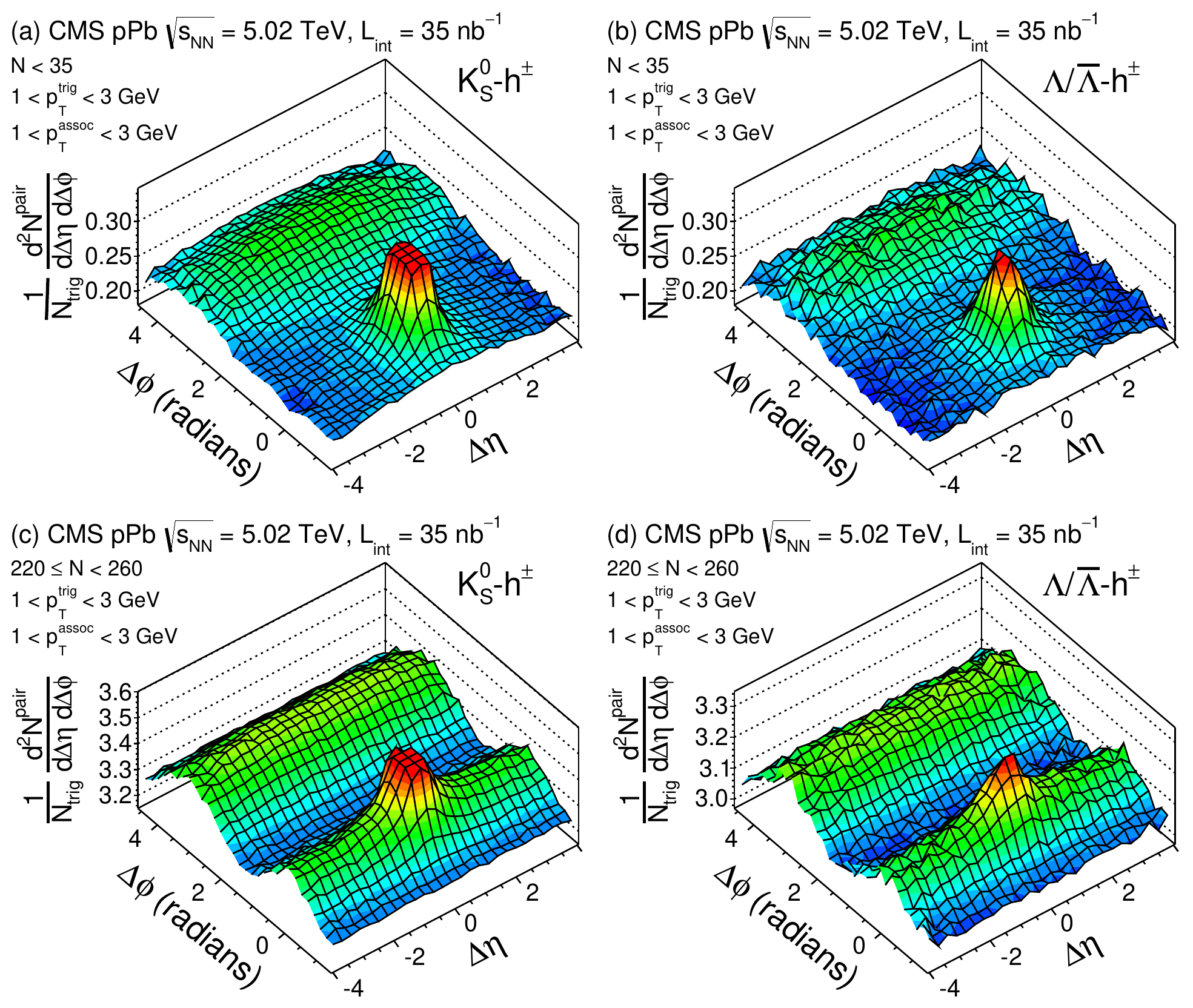}
  \caption{ \label{fig:Corr_Fcn_Fig1} The 2D two-particle correlation functions
  in pPb collisions at \rootsNN\ = 5 TeV for pairs of a \PKzS\ (a,c) or \PgL/\PagL\ (b,d)
  trigger particle and a charged associated particle ($h^{\pm}$), with $1<\pttrg<3$ GeV
  and $1<\ptass<3$ GeV, in the multiplicity ranges $\noff< 35$ (a, b) and
  $220 \leq \noff< 260$ (c, d). The sharp near-side peak from jet correlations
  is truncated to emphasize the structure outside that region.
   }
\end{figure*}

The 2D two-particle correlation functions measured in pPb collisions for
pairs of a \PKzS\ (left) and \PgL/\PagL\ (right) trigger particles
and a charged associated particle ($h^{\pm}$) are shown in Fig.~\ref{fig:Corr_Fcn_Fig1}
in the \pt\ range of 1--3 GeV. Following the same approach of correcting $v_n$ in Eq.~(\ref{eq:vnsignal}), the 2D correlation functions are corrected
for the background $V^{0}$ candidates.
The correction is negligible in this \pt\ range because
of the high signal fraction of $V^{0}$ candidates.
For low-multiplicity events ($\noff < 35$, Figs.~\ref{fig:Corr_Fcn_Fig1} (a) and (b)),
a sharp peak near $(\deta, \dphi) = (0, 0)$ due to jet fragmentation (truncated
for better illustration of the full correlation structure) can be clearly observed
for both $\PKzS$--$h^{\pm}$ and $\PgL/\PagL$--$h^{\pm}$ correlations. Moving to
high-multiplicity events ($220 \leq \noff< 260$, Figs.~\ref{fig:Corr_Fcn_Fig1}
(c) and (d)), in addition to the peak from jet fragmentation,
a pronounced long-range structure is seen at $\dphi \approx 0$, extending at least 4 units
in $|\deta|$. This structure was previously observed in high-multiplicity ($\noff \sim 110$)
pp collisions at \roots\ = 7 TeV~\cite{Khachatryan:2010gv} and pPb collisions
at \rootsNN\ = 5.02 TeV~\cite{CMS:2012qk,alice:2012qe,atlas:2012fa,Chatrchyan:2013nka}
for inclusive charged particles, and also for identified charged pions, kaons, and
protons in pPb collisions at \rootsNN\ = 5.02 TeV~\cite{ABELEV:2013wsa}. A similar long-range
correlation structure has also been extensively studied in AA collisions over a
wide range of energies~\cite{Adams:2005ph,Abelev:2009af,Alver:2008gk,Alver:2009id,Abelev:2009jv,Chatrchyan:2011eka,Chatrchyan:2012wg,Aamodt:2011by,ATLAS:2012at},
where it is believed to arise primarily from collective flow of a strongly interacting
medium~\cite{Voloshin:1994mz}.

\begin{figure*}[thb]
\centering
\includegraphics[width=0.8\linewidth]{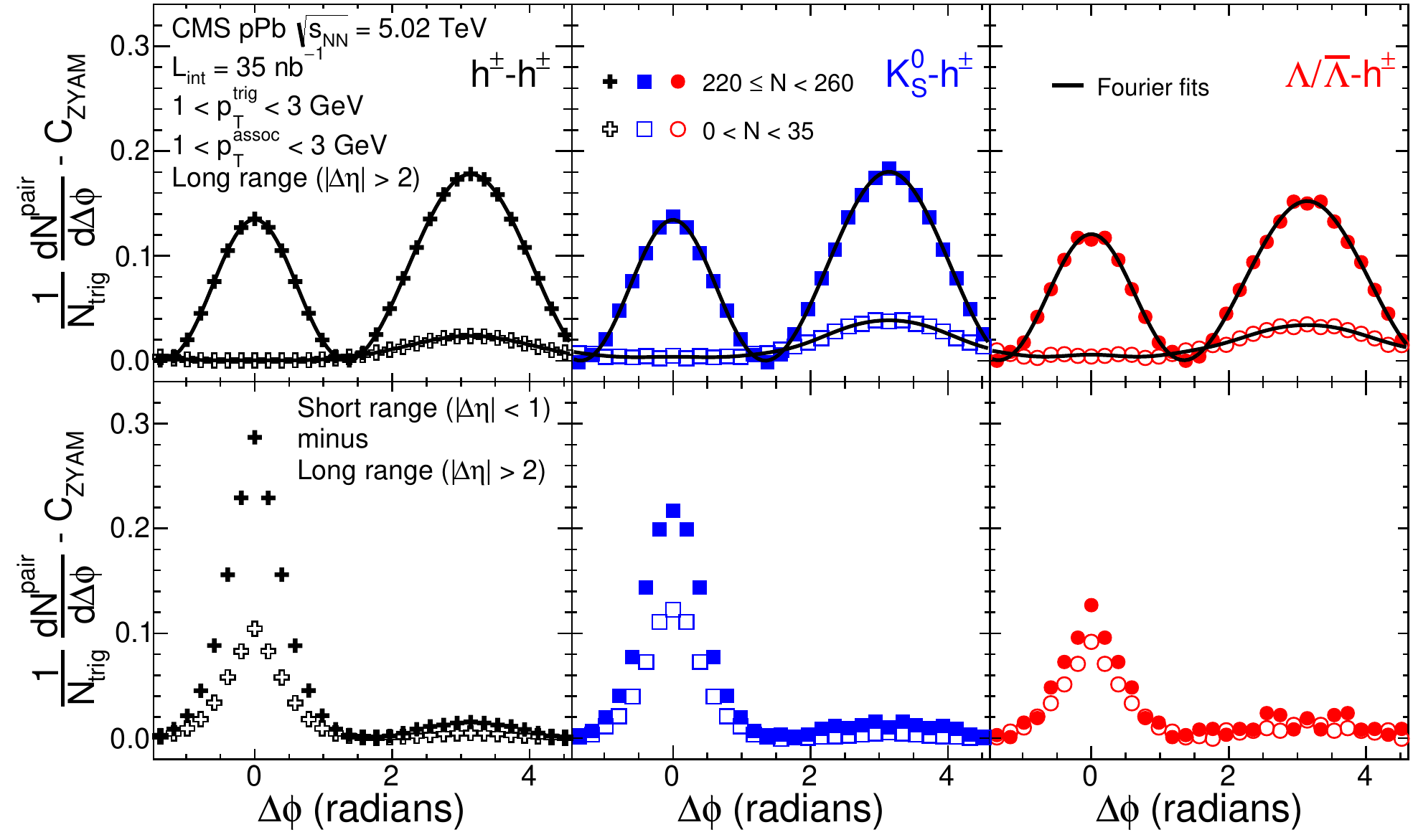}
  \caption{ \label{fig:Corr_Proj_Fig2}
     The 1D \dphi\ correlation functions from pPb data after applying the ZYAM procedure,
     in the multiplicity range $\noff\ < 35$ (open) and $220 \leq \noff\ < 260$ (filled),
     for trigger particles composed of inclusive charged particles (left), \PKzS\ particles
     (middle), and \PgL/\PagL\ particles (right).
     Selection of a fixed \pttrg\ and \ptass\ range of both 1--3 GeV is shown for the long-range region ($|\deta|>2$) on top and the short-range ($|\deta|<1$) minus long-range region on the bottom.
     The curves on the top panels correspond to the Fourier fits including the first three terms. Statistical uncertainties
     are smaller than the size of the markers.
   }
\end{figure*}

To investigate the correlation structure for different species of particles in
detail, one-dimensional (1D) distributions in $\Delta\phi$ are found
by averaging the signal and mixed-event 2D distributions over $|\deta| < 1$ (defined
as the "short-range region") and $|\deta| > 2$ (defined as the "long-range region"),
as done in Refs.~\cite{Khachatryan:2010gv,Chatrchyan:2011eka,Chatrchyan:2012wg,CMS:2012qk,Chatrchyan:2013nka}.
Fig.~\ref{fig:Corr_Proj_Fig2} shows the
1D $\Delta\phi$ correlation functions from pPb data for trigger
particles composed of inclusive charged particles (left)~\cite{Chatrchyan:2013nka}, \PKzS\
particles (middle), and \PgL/\PagL\ particles (right), in
the multiplicity range $\noff\ < 35$ (open) and $220 \leq \noff\ < 260$
(filled). The curves show the Fourier fits from Eq.~(\ref{eq:Vn}) to the long-range region,
which will be discussed in detail later. Following the standard zero-yield-at-minimum (ZYAM) procedure~\cite{Chatrchyan:2013nka}, each distribution is shifted to have zero
associated yield at its minimum to represent the correlated portion of the associated yield. Selection of fixed \pttrg\ and \ptass\ ranges of 1--3 GeV is shown for the long-range region (top) and for the difference of the short- and long-range regions (bottom) in Fig.~\ref{fig:Corr_Proj_Fig2}.
As illustrated in Fig.~\ref{fig:Corr_Fcn_Fig1}, the near-side long-range
signal remains nearly constant in
$\Delta\eta$. Therefore, by taking a difference of 1D \dphi\ projections between
the short- and long-range regions, the near-side jet correlations can be extracted.
As shown in the bottom panels of Fig.~\ref{fig:Corr_Proj_Fig2},
due to biases in multiplicity selection toward higher \pt\ jets, a larger jet peak yield is observed
for events selected with higher multiplicities. Because charged particles are directly used in determining the multiplicity in the event,
this selection bias is much stronger for charged particles than \PKzS\ and \PgL/\PagL\ particles.
For $\noff\ < 35$, no near-side correlations are observed in the long-range region for
any particle species. The PbPb data show qualitatively the same behavior as the pPb data, and thus are not presented here.
\clearpage

\section{Mass ordering of $v_2$}
\label{sec:massv2}

\begin{figure*}[thb]
\centering
\includegraphics[width=0.8\linewidth]{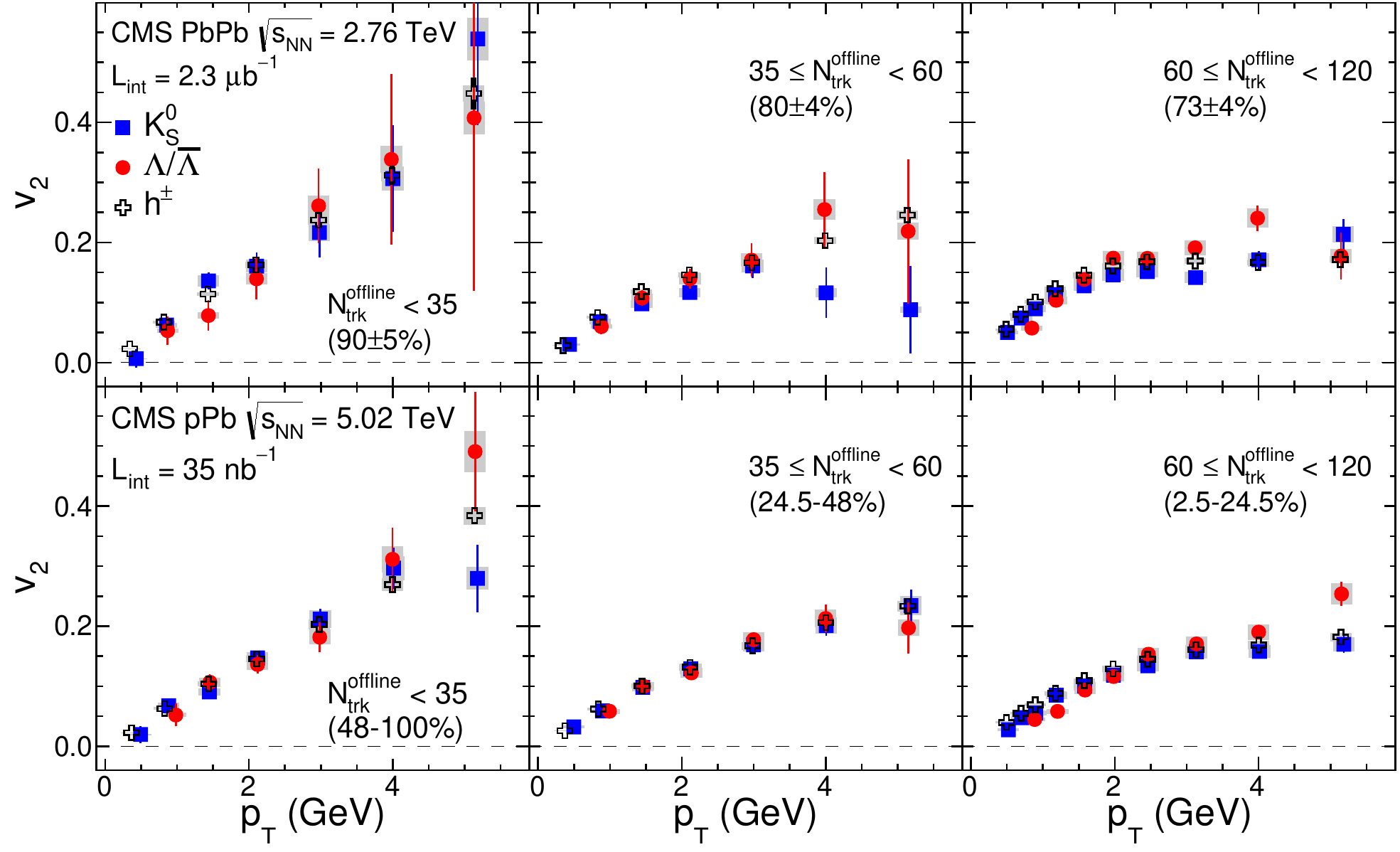}
  \caption{ \label{fig:v2_PID_lowN}
    The $v_2$ results for \PKzS\ (filled squares) and \PgL/\PagL\ (filled circles) particles
    as a function of \pt\
    for three multiplicity ranges obtained from minimum bias triggered PbPb
    sample at \rootsNN\ = 2.76 TeV (top row) and pPb sample at \rootsNN\ = 5.02 TeV
    (bottom row). The error bars correspond to statistical uncertainties,
    while the shaded areas denote the systematic uncertainties.
    The values in parentheses give the mean and standard deviation of the
    HF fractional cross section for PbPb and the range of the fraction of the full multiplicity distribution included for pPb.
   }
\end{figure*}

Recently, the $v_2$ anisotropy harmonics for charged pions, kaons, and protons
have been studied using two-particle correlations in pPb collisions~\cite{ABELEV:2013wsa},
and are found to be qualitatively consistent with hydrodynamic models~\cite{Werner:2013ipa,Bozek:2013ska}.
In this paper, the elliptic ($v_2$) and triangular ($v_3$) flow harmonics
of \PKzS\ and \PgL/\PagL\ particles are extracted from the Fourier decomposition of 1D
$\Delta\phi$ correlation functions for the long-range region ($|\Delta\eta| > 2$)
in a significantly larger sample of pPb collisions such that the particle
species dependence of $v_n$ can be investigated in detail. In Fig.~\ref{fig:v2_PID_lowN},
the $v_2^{\text{sig}}$ of \PKzS\ and \PgL/\PagL\ particles
are plotted as a function of \pt\ for the three lowest multiplicity ranges in PbPb
and pPb collisions. These data were recorded
using a minimum bias trigger. The range of the fraction of the
full multiplicity distribution that each multiplicity selection corresponds to, as determined in
Ref.~\cite{Chatrchyan:2013nka}, is also specified in the figure. In contrast to most
other PbPb analyses, the present work uses multiplicity to classify events, instead of
the total energy deposited in HF (the standard procedure of
centrality determination in PbPb)~\cite{Chatrchyan:2012xq,Chatrchyan:2013nka}.
By examining the HF energy distribution for PbPb events in each of the multiplicity ranges, the corresponding
average HF fractional cross section (and its standard deviation) can be determined, which are
presented for PbPb data in the figure.

In the low multiplicity region (Fig.~\ref{fig:v2_PID_lowN}),
the $v_2$ values of \PKzS\ and \PgL/\PagL\ particles are compatible within
statistical uncertainties. As there is no evident long-range near-side correlation
seen in these low-multiplicity events, the extracted $v_2$ most likely reflects
back-to-back jet correlations on the away side. Away-side jet correlations
typically appear as a peak structure around $\dphi \approx \pi$,
which contributes to various orders of Fourier terms.
%

\begin{figure*}[thb]
\centering
\includegraphics[width=\linewidth]{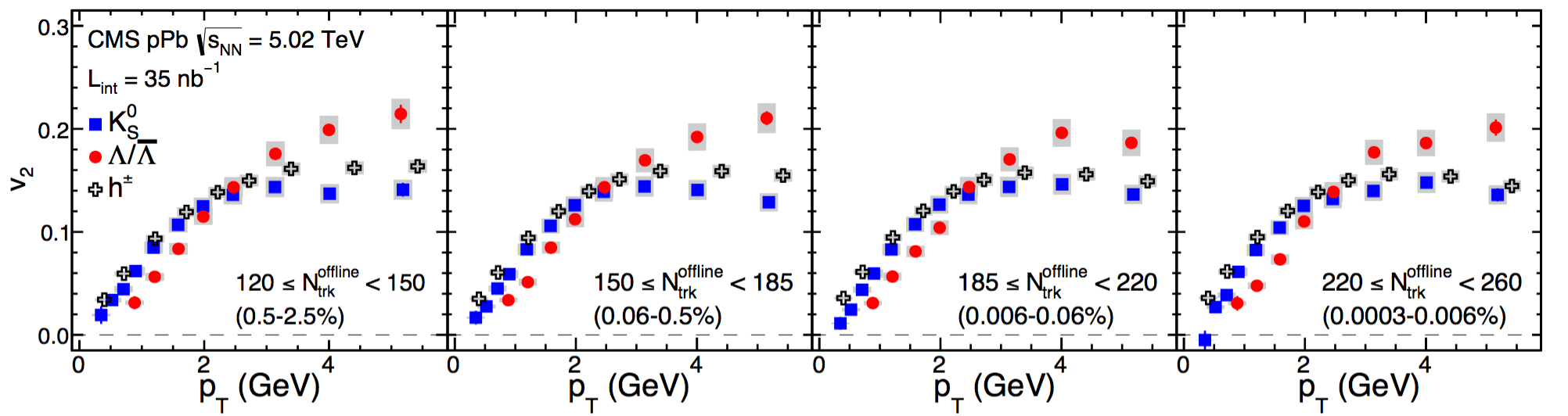}
  \caption{ \label{fig:v2_PID_highN_pPb}
    The $v_2$ results for \PKzS\ (filled squares), \PgL/\PagL\ (filled circles),
    and inclusive charged particles (open crosses) as a function of \pt\
    for four multiplicity ranges obtained from high-multiplicity triggered pPb sample at \rootsNN\ = 5.02 TeV.
    The error bars correspond to statistical uncertainties, while the shaded areas denote
    the systematic uncertainties. The values in parentheses give the range of the
    fraction of the full multiplicity distribution included for pPb.
   }
\end{figure*}

Fig.~\ref{fig:v2_PID_highN_pPb} shows the measured $v_2$ values for \PKzS\ and \PgL/\PagL\ particles as a function of \pt\ from the high multiplicity pPb data, along with the previously published results for inclusive charged particles~\cite{Chatrchyan:2013nka}.
In the $\pt\lesssim 2$ GeV region
for all high-multiplicity ranges, the $v_2$ values of \PKzS\ particles are larger than those for \PgL/\PagL\ particles at
each \pt value. Both of them are consistently below the $v_2$ values of inclusive
charged particles. As most charged particles are pions in this \pt\ region,
the data indicate that lighter particle species exhibit a stronger azimuthal anisotropy signal. 
A similar trend was first observed in AA collisions at RHIC~\cite{Adler:2003kt,Adler:2001nb}.
This mass ordering behavior is consistent with expectations in hydrodynamic models~\cite{Werner:2013ipa,Bozek:2013ska} and the observation
in 0--20\% centrality pPb collisions~\cite{ABELEV:2013wsa}. 
The same effect is also qualitatively reproducible by non-hydrodynamic models, such as AMPT through parton escape mechanism~\cite{Li:2016ubw}, UrQMD through hadronic interaction~\cite{Zhou:2015iba} and an alternative initial state interpretation with CGC~\cite{Schenke:2016lrs}.
At higher \pt, the $v_2$ values of \PgL/\PagL\ particles are larger than those of \PKzS.
The inclusive charged particle $v_2$ values fall between the values of the
two identified strange hadron species but are much closer to the $v_2$ values for \PKzS\ particles.
Note that the ratio of baryon to meson yield in pPb collisions is enhanced at higher \pt,
an effect that becomes stronger as multiplicity increases~\cite{Chatrchyan:2013eya,Abelev:2013haa}.
This should also be taken into account when comparing $v_n$ values between inclusive and identified particles. Comparing the results in Fig.~\ref{fig:v2_PID_lowN} and Fig.~\ref{fig:v2_PID_highN_pPb}, the dependence of $v_2$ on the particle species may already be emerging in the multiplicity range of $60 \leq \noff < 120$.

\begin{figure*}[thb]
\centering
\includegraphics[width=\linewidth]{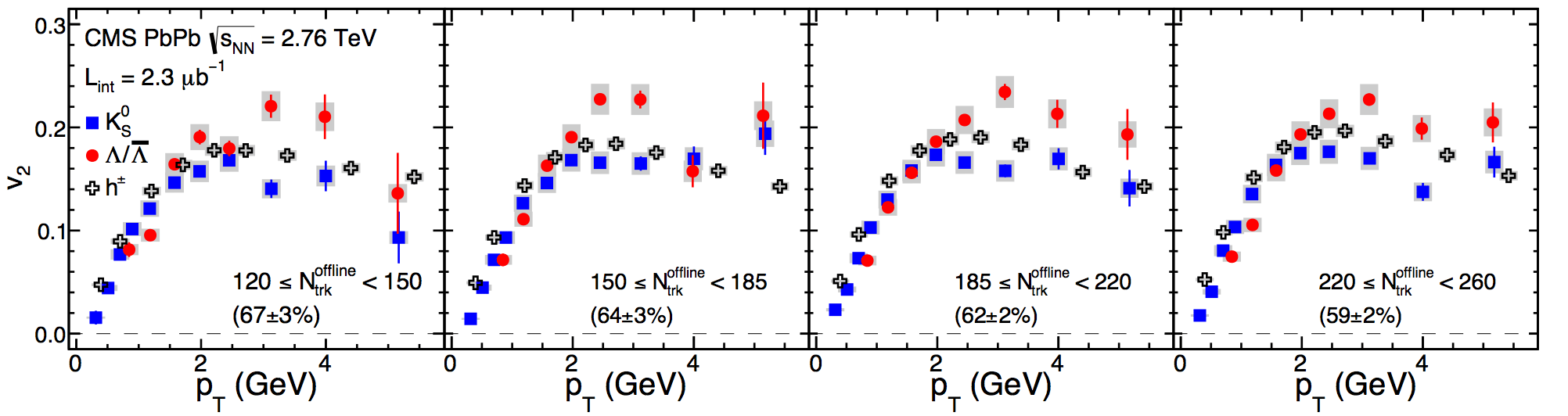}
  \caption{ \label{fig:v2_PID_highN_PbPb}
    The $v_2$ results for \PKzS\ (filled squares), \PgL/\PagL\ (filled circles),
    and inclusive charged particles (open crosses) as a function of \pt\
    for four multiplicity ranges obtained from minimum bias triggered PbPb sample at \rootsNN\ = 2.76 TeV.
    The error bars correspond to statistical uncertainties, while the shaded areas denote
    the systematic uncertainties. The values in parentheses give the mean and standard deviation of the HF fractional cross section
    for PbPb.
   }
\end{figure*}

The particle species dependence of $v_2$ is also studied
in PbPb data over the same multiplicity ranges as for
the pPb data, as shown in Fig.~\ref{fig:v2_PID_highN_PbPb}. The mean and standard
deviation of the HF fractional cross section of the PbPb data are indicated on the plots, 
which are mostly in the peripheral range of 50--100\% centrality.
Qualitatively, a similar particle-species dependence of $v_2$
is observed. However, the mass ordering effect is found to be less
evident in PbPb data than in pPb data for all multiplicity ranges. In hydrodynamic models such as those presented in Refs.~\cite{Bozek:2012qs,Karpenko:2012yf}, this
behavior, together with results on particle production~\cite{Khachatryan:2016yru}, 
can be interpreted as a stronger radial flow is developed in the pPb system as its energy density is higher
than that of a PbPb system due to having a smaller size system at the same multiplicity.
\clearpage

\section{Number of constituent quark scaling}
\label{sec:ncqpPb}

\begin{figure*}[thb]
\centering
\includegraphics[width=\linewidth]{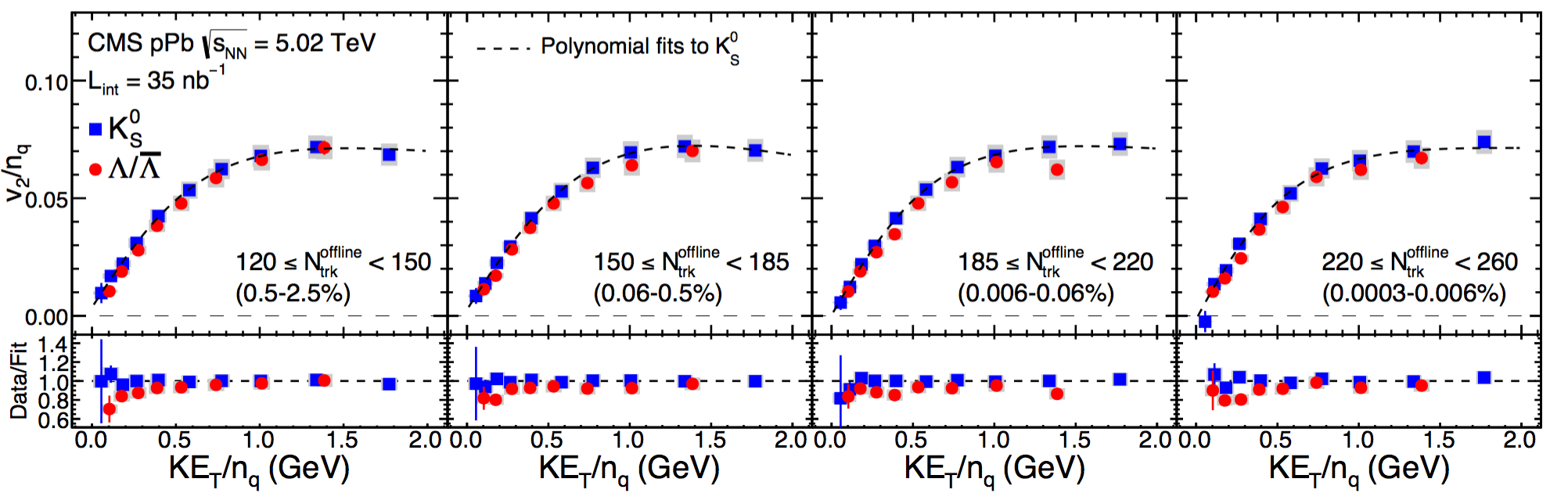}
  \caption{ \label{fig:v2_PID_highN_pPb_ncq}
    Top row: the $v_2/n_q$ ratios for \PKzS\ (filled squares) and \PgL/\PagL\ (filled circles) particles
    as a function of $\ket/n_q$, along with a fit to the \PKzS\ results using a polynomial function.
    Bottom row: ratios of $v_2/n_q$ for \PKzS\ and \PgL/\PagL\ particles to
    the fitted polynomial function as a function of $\ket/n_q$.
    The error bars correspond to statistical uncertainties, while the shaded areas denote
    the systematic uncertainties. The values in parentheses give the range of the
    fraction of the full multiplicity distribution included for pPb.
   }
\end{figure*}

The scaling behavior of $v_2$ divided by the number of constituent quarks as a function of
transverse kinetic energy per quark, $\ket/n_{q}$, is investigated for
high-multiplicity pPb events in the top row of Fig.~\ref{fig:v2_PID_highN_pPb_ncq}.
After scaling by the number of constituent quarks, the $v_2$ distributions for
\PKzS\ and \PgL/\PagL\ particles are found to be in agreement. The
top row of Fig.~\ref{fig:v2_PID_highN_pPb_ncq} also shows the result of fitting a polynomial
function to the \PKzS\ data. The bottom row of Fig.~\ref{fig:v2_PID_highN_pPb_ncq} shows the
$n_q$-scaled $v_2$ results for \PKzS\ and \PgL/\PagL\ particles divided by
this polynomial function fit, indicating that the scaling is valid over most of the $\ket/n_{q}$ range,
except for $\ket/n_{q}<0.2$ GeV.
In AA collisions, this approximate scaling behavior is conjectured to be related to quark
recombination~\cite{Molnar:2003ff,Greco:2003xt,Fries:2003vb}, which postulates
that collective flow is developed among constituent quarks before they combine
into final-state hadrons. Note that the scaling of $v_2$ with the number of constituent quarks was originally observed as a function
of \pt, instead of \ket, for the intermediate \pt\ range of a few GeV~\cite{Adams:2003am},
and interpreted in a simple picture of quark coalescence~\cite{Molnar:2003ff}.
However, it was later discovered that when plotted as a function of \ket\ in order to
remove the mass difference of identified hadrons, the scaling appears to hold over the entire
kinematic range~\cite{Abelev:2007qg,Adare:2006ti}. However, this scaling
behavior is not expected to be exact at low \pt\ in hydrodynamic models because of the impact of radial flow.
As the $v_n$ data tend to approach a constant value as a function of \pt\ or \ket\ for \pt\ $\gtrsim$ 2 GeV, the scaling behavior
in terms of \pt\ and \ket\ cannot be differentiated in that regime. Therefore, the $n_{q}$-scaled $v_n$ results in this thesis
are presented as a function of $\ket/n_{q}$ in order to explore the scaling behavior over a wider kinematic range.

\begin{figure*}[thb]
\centering
\includegraphics[width=\linewidth]{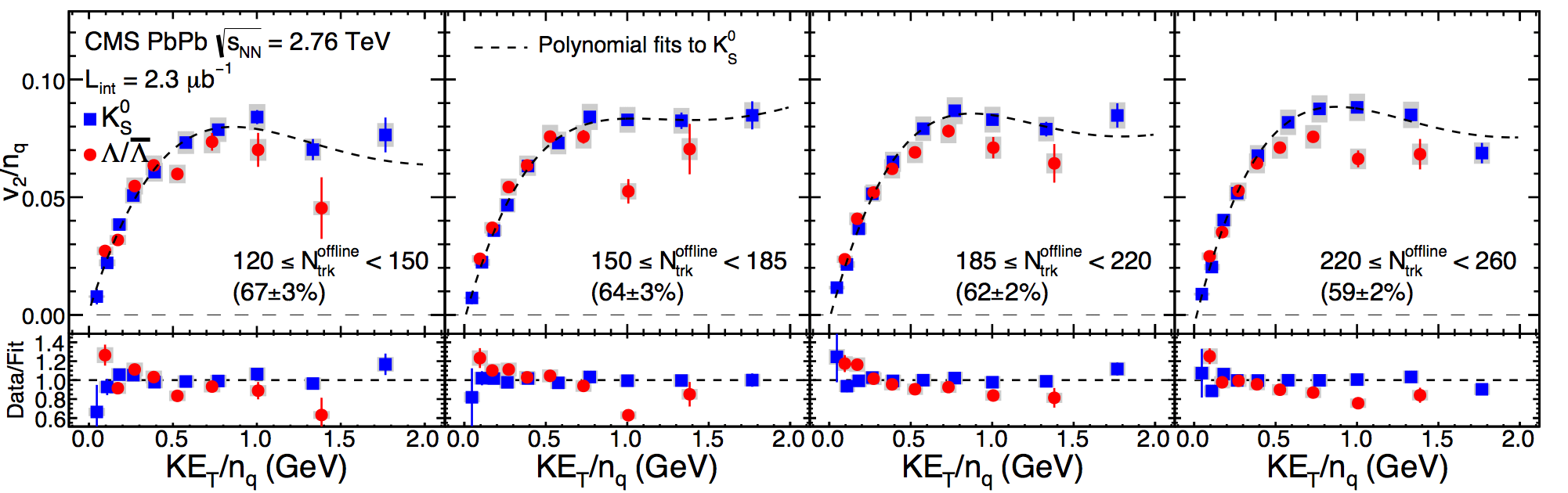}
  \caption{ \label{fig:v2_PID_highN_PbPb_ncq}
    Top row: the $v_2/n_q$ ratios for \PKzS\ (filled squares) and \PgL/\PagL\ (filled circles) particles
    as a function of $\ket/n_q$, along with a fit to the \PKzS\ results using a polynomial function.
    Bottom row: ratios of $v_2/n_q$ for \PKzS\ and \PgL/\PagL\ particles to
    the fitted polynomial function as a function of $\ket/n_q$.
    The error bars correspond to statistical uncertainties, while the shaded areas denote
    the systematic uncertainties. The values in parentheses give the mean and standard deviation of the HF fractional cross section
    for PbPb.
   }
\end{figure*}

The scaling behavior is also studied
in PbPb data over the same multiplicity ranges as for
the pPb data, as shown in Fig.~\ref{fig:v2_PID_highN_PbPb_ncq}.
The $n_{q}$-scaled $v_2$ data in PbPb at similar multiplicities suggest
a stronger violation of constituent quark number scaling than is observed in pPb,
especially for higher $\ket/n_{q}$ values. This is also observed in peripheral AuAu collisions
at RHIC, while the scaling applies more closely for central AuAu collisions~\cite{Adare:2012vq}.
\clearpage

\section{Triangular flow $v_3$}

\begin{figure*}[thb]\centering
\centering
\includegraphics[width=0.45\linewidth]{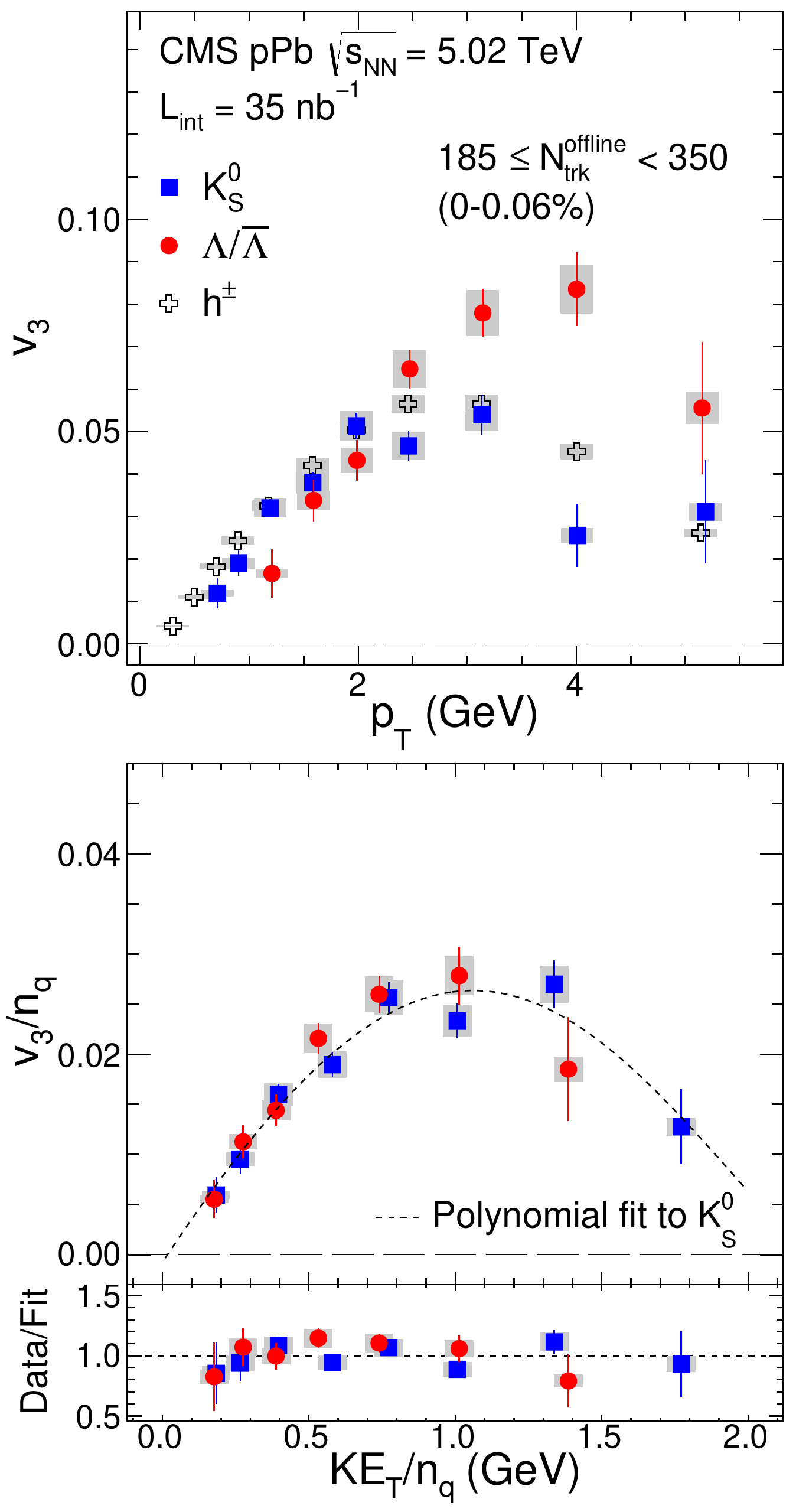}
\includegraphics[width=0.45\linewidth]{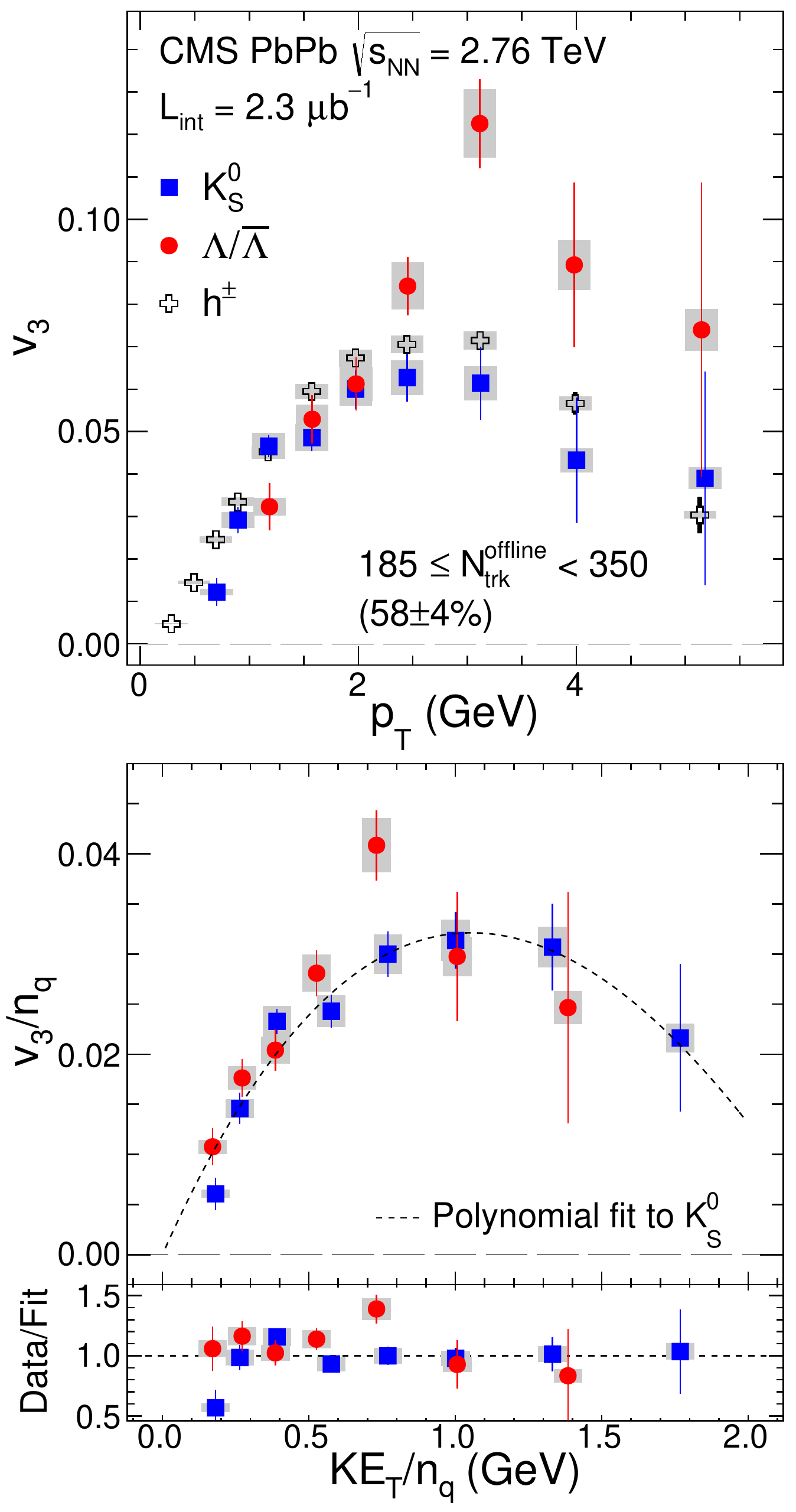}
  \caption{ \label{fig:v3_PID_highN}
    Top: the $v_3$ results for \PKzS\ (filled squares), \PgL/\PagL\ (filled circles),
    and inclusive charged particles (open crosses) as a function of \pt\ for
    the multiplicity range $185 \leq \noff < 350$ in pPb collisions at
    \rootsNN\ = 5.02 TeV (left) and in PbPb collisions at \rootsNN\ = 2.76 TeV (right).
    Bottom: the $n_{q}$-scaled $v_3$ values of \PKzS\ (filled squares)
    and \PgL/\PagL\ (filled circles) particles as a function of $\ket/n_q$ for the same two systems. Ratios of $v_n/n_q$ to
    a smooth fit function of $v_n/n_q$ for \PKzS\ particles as a function of $\ket/n_q$ are also shown.
    The error bars correspond to statistical uncertainties, while the shaded areas denote
    the systematic uncertainties. The values in parentheses give the mean and standard deviation of the HF fractional cross section
    for PbPb and the range of the fraction of the full multiplicity distribution included for pPb.
   }
\end{figure*}

The triangular flow harmonic, $v_3$, of \PKzS\ and \PgL/\PagL\ particles is also extracted in pPb
and PbPb collisions, as shown in Fig.~\ref{fig:v3_PID_highN}. Due to limited statistical
precision, only the result in the multiplicity range $185 \leq \noff < 350$ is presented.
A similar species dependence of $v_3$ to that of $v_2$ is observed and,
within the statistical uncertainties, the $v_3$ values scaled by the constituent quark number
for \PKzS\ and \PgL/\PagL\ particles match over the full $\ket/n_{q}$ range.
To date, no calculations of the quark number scaling of triangular flow, $v_3$, have been performed in the
parton recombination model.

\clearpage

\section{Summary}
\label{sec:resultpPbSum}

Measurements of two-particle
correlations with an identified \PKzS\ or \PgL/\PagL\ trigger particle have been presented over a broad
transverse momentum and
pseudorapidity range in pPb collisions at \rootsNN\ = 5.02 TeV and PbPb
collisions at \rootsNN\ = 2.76 TeV. The identified particle correlation data
in pPb collisions are explored over a broad particle multiplicity range,
comparable to that covered by 50--100\% centrality PbPb collisions.
The long-range ($|\deta|>2$) correlations are quantified in terms of azimuthal
anisotropy Fourier harmonics ($v_n$) motivated by hydrodynamic models.
In low-multiplicity pPb and PbPb events, similar $v_2$ values of \PKzS\ and \PgL/\PagL\ particles
are observed, which likely originate from back-to-back jet correlations.
For higher event multiplicities, a particle species dependence of $v_2(\pt)$ and $v_3(\pt)$
is observed. For $\pt \lesssim 2$ GeV,
the values of $v_n$ for \PKzS\ particles are found to be larger than those of \PgL/\PagL\ particles, while this order
is reversed at higher \pt. This behavior is consistent with RHIC and LHC results in
AA collisions and for identified charged hadrons in pPb and dAu collisions.
For similar event multiplicities, the particle species dependence of $v_2$ and $v_3$ at low \pt\
is observed to be more pronounced in pPb than in PbPb collisions. In the context of hydrodynamic models,
this may indicate that a stronger radial flow boost is developed in pPb collisions.
Furthermore, constituent quark number scaling of $v_2$ and $v_3$
between \PKzS\ and \PgL/\PagL\ particles is found to apply for PbPb and high-multiplicity
pPb events. The constituent quark number scaling is found to hold better in pPb collisions than PbPb collisions, for similar event multiplicities. 
The results presented provide important input to the further exploration of the possible collective flow origin
of long-range correlations in pPb collisions, and can be used to evaluate models of quark recombination in a deconfined medium
of quarks and gluons.

\cleardoublepage
\chapter{Long-range two-particle correlations in pp collisions}
\label{ch:resultpp}

Including the results presented in the previous chapter, significant progress has been made in unrevealing the nature of the ridge correlations in high multiplicity pPb collisions. 
However, in high multiplicity pp collisions, the nature of the observed long range correlation still remains poorly understood. 
Long range correlations in pp collision were first observed in 2010~\cite{Khachatryan:2010gv}, since then no further study has been made on such correlations for years.
The analysis presented in this chapter was one of the first to do detail measurements of anisotropy Fourier harmonics in pp collisions. 

This chapter presents measurements of two-particle angular correlations in pp collisions
at $\sqrt{s} = 5$, 7, and 13 TeV. 
The second-order ($v_2$) and third-order ($v_3$) azimuthal anisotropy harmonics of
unidentified charged particles, as well as $v_2$ of $\PKzS$ and $\PgL/\PagL$ particles,
are extracted from long-range two-particle correlations as functions of particle multiplicity
and transverse momentum. 
A jet subtraction method, known as low multiplicity subtraction, is implemented to account for the contribution from jet correlation. 

The majority of work presented in this chapter is published in Ref.~\cite{Khachatryan:2016txc}, except Sec.~\ref{sec:ATLAScomp}.

\section{Two-particle correlation function}
\label{sec:resultppcorrfcn}

\begin{figure*}[thbp]
  \begin{center}
    \includegraphics[width=0.39\textwidth]{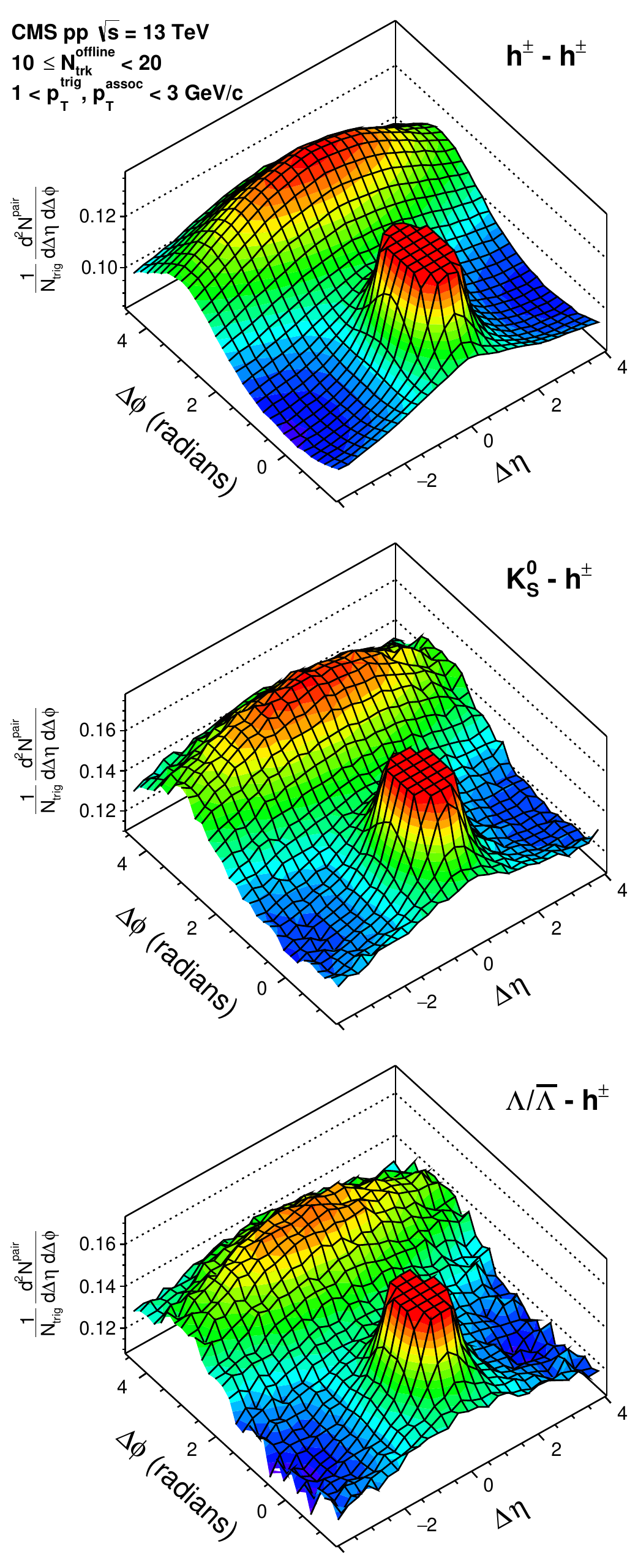}
    \includegraphics[width=0.39\textwidth]{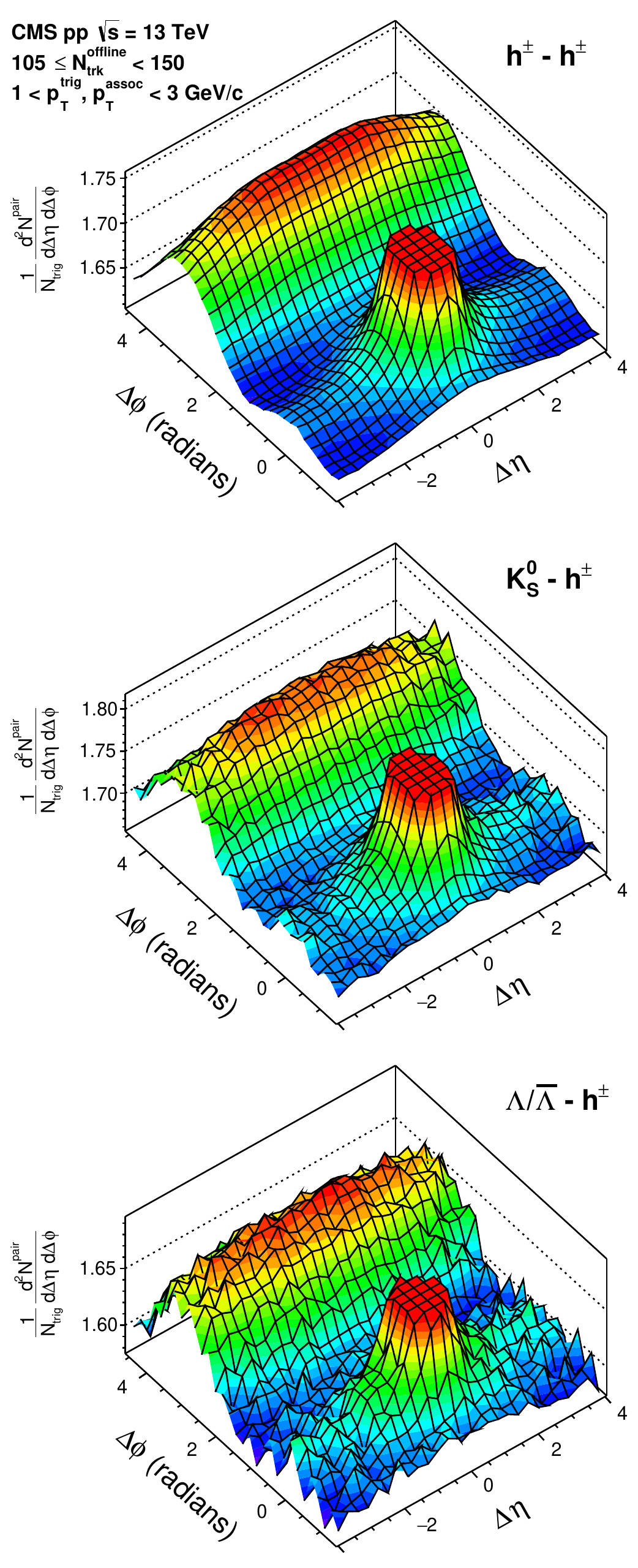}
    \caption{ The 2D two-particle correlation functions for inclusive charged
    particles (top), $\PKzS$ particles (middle), and \PgL/\PagL\ particles (bottom),
    with $1<\pttrg<3$ GeV/c and associated charged particles with $1<\ptass<3$ GeV/c,
    in low-multiplicity ($10 \leq \noff < 20$, left) and high-multiplicity
    ($105 \leq \noff < 150$, right) pp collisions at $\roots = 13$ TeV.
    }
    \label{fig:corr2D_pp}
  \end{center}
\end{figure*}

Figure~\ref{fig:corr2D_pp} shows the 2D \deta--\dphi correlation functions,
for pairs of a charged (top), a $\PKzS$ (middle), or a \PgL/\PagL\ (bottom)
trigger particle with a charged associated particle, in low-multiplicity ($10 \leq \noff < 20$, left)
and high-multiplicity ($105 \leq \noff < 150$, right) pp collisions at $\roots = 13$ TeV.
Both trigger and associated particles are selected from the \pt range of 1--3 GeV/c.
For all three types of particles at high multiplicity, in addition to the correlation
peak near $(\Delta\eta, \Delta\phi) = (0, 0)$ that results from jet fragmentation,
a long-range ridge structure is seen at $\dphi \approx 0$ extending at least 4 units
in $|\deta|$, while such a structure is not observed in low multiplicity events.

\begin{figure*}[t!hb]
\centering
\includegraphics[width=\linewidth]{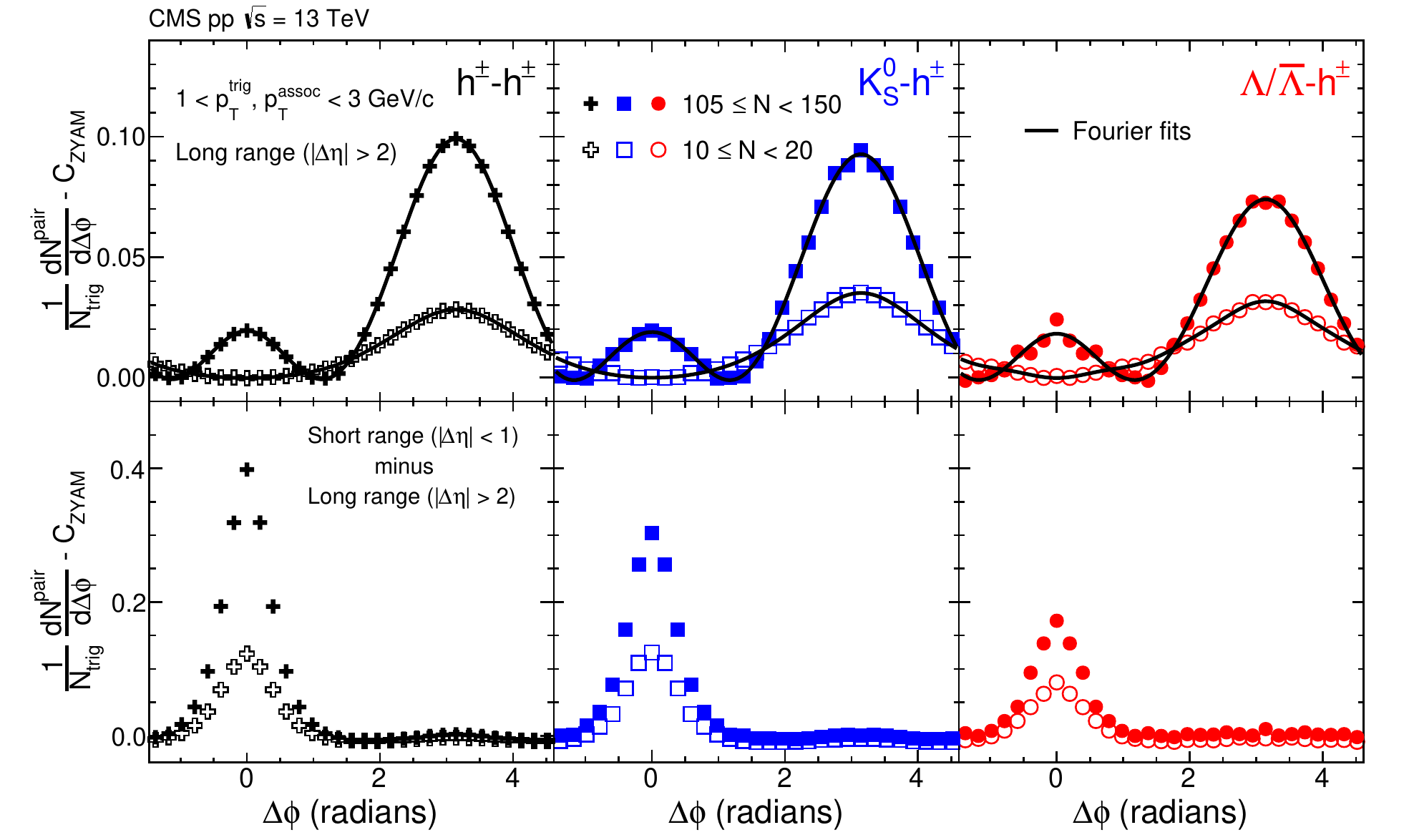}
  \caption{ \label{fig:Corr_Proj_pp}
     The 1D \dphi correlation functions for the long-range (top) and short- minus long-range (bottom) regions
     after applying the ZYAM procedure
     in the multiplicity range $10 \leq \noff < 20$ (open symbols) and
     $105 \leq \noff < 150$ (filled symbols) of pp collisions at $\roots = 13$ TeV,
     for trigger particles composed of inclusive charged particles (left, crosses), $\PKzS$ particles
     (middle, squares), and \PgL/\PagL\ particles (right, circles).
	A selection of 1--3 GeV/c for both \pttrg\ and \ptass\ is used in all cases.
     }
\end{figure*}

To investigate the observed correlations in finer detail, the 2D distributions
shown in Fig.~\ref{fig:corr2D_pp} are reduced to
one-dimensional (1D) distributions in $\Delta\phi$ by averaging over $|\Delta\eta| < 1$
(defined as the "short-range region") and $|\Delta\eta| > 2$ (defined as the "long-range region"),
respectively, as done in Refs.~\cite{CMS:2012qk,Khachatryan:2010gv,Chatrchyan:2011eka,Chatrchyan:2012wg}.
Figure~\ref{fig:Corr_Proj_pp} shows examples of 1D \dphi correlation
functions for trigger particles composed of inclusive charged particles (left), $\PKzS$ particles (middle), and \PgL/\PagL\ particles (right), in the multiplicity range
$10 \leq \noff < 20$ (open symbols) and $105 \leq \noff < 150$ (filled symbols).
The curves show the Fourier fits from Eq.~(\ref{eq:Vn}) to the long-range region,
which will be discussed in detail in Section~\ref{sec:vncharge}.
To represent the correlated portion of the associated yield, each distribution is shifted to have zero
associated yield at its minimum following the standard zero-yield-at-minimum (ZYAM) procedure~\cite{Chatrchyan:2013nka}.
An enhanced correlation at
$\Delta\phi \approx 0$ in the long-range region is observed for $105 \leq \noff < 150$,
while such a structure is not presented for $10 \leq \noff < 20$. As illustrated in
Fig.~\ref{fig:corr2D_pp} (right), the near side long-range ridge structure remains nearly
constant in $\Delta\eta$. Therefore, as shown in the bottom panels in Fig.~\ref{fig:Corr_Proj_pp}, the near side jet correlation can be extracted by
taking a difference of 1D $\Delta\phi$ projections between the short- and long-range regions, which is useful in the jet subtraction procedure discussed in the following section.

\clearpage

\section{Jet contribution subtraction}
\label{sec:lowsubpp}

On the away side ($\Delta\phi \approx \pi$) of the correlation functions shown in Fig.~\ref{fig:corr2D_pp}, a long-range
structure is also seen and found to exhibit a much larger magnitude compared to that on the
near side for this \pt\ range. This away side correlation structure contains
contributions from back-to-back jets, which need to be accounted for before extracting any
other source of correlations.

By assuming that the shape of the jet-induced correlations is invariant with event
multiplicity, a procedure of removing jet-like correlations in pPb collisions
was proposed in Refs.~\cite{alice:2012qe,Aad:2012gla}.
The method consists of subtracting the results for low-multiplicity events, where the
ridge signal is not present, from those for high-multiplicity events.
For this analysis, a very similar low-multiplicity subtraction method developed for pPb collisions~\cite{Chatrchyan:2013nka} is employed.
The Fourier
coefficients, $V_{n\Delta}$, extracted from Eq.~(\ref{eq:Vn}) for $10 \leq \noff < 20$
are subtracted from the $V_{n\Delta}$ coefficients extracted in the higher-multiplicity region, with

\begin{equation}
\label{eq:vnsubperiph}
V^\text{sub}_{n\Delta}=V_{n\Delta}-V_{n\Delta}(10\leq\noff<20)\times\frac{N_\text{assoc}(10\leq\noff<20)}{N_\text{assoc}}\times\frac{Y_\text{jet}}{Y_\text{jet}(10\leq\noff<20)}.
\end{equation}

Here, $Y_\text{jet}$ represents the near side jet yield obtained by integrating
the difference of the short- and long-range event-normalized associated yields for each multiplicity class
as shown for $105 \leq \noff < 150$ in
Fig.~\ref{fig:Corr_Proj_pp} over $|\Delta\phi| < 1.2$.
The ratio, $Y_\text{jet}/Y_\text{jet}(10\leq\noff<20)$,
is introduced to account for the enhanced jet correlations resulting from the
selection of higher-multiplicity events. This jet subtraction procedure is verified using PYTHIA6 (Z2)
and PYTHIA8 tune CUETP8M1 pp simulations, where no jet modification from initial-
or final-state effects is present. The residual $V_{n\Delta}$ after subtraction is
found to be consistent with zero.
The azimuthal anisotropy harmonics $v_n$ after correcting for back-to-back jet correlations estimated from
low-multiplicity data (denoted as $v_{n}^\text{sub}$)
can be extracted from $V^\text{sub}_{n\Delta}$ using Eq.~(\ref{eq:Vnpt}) and~(\ref{eq:vnsignal}).
In this thesis, both the $v_{n}$ and $v_{n}^\text{sub}$ results are presented.

\begin{figure*}[thb!p]
\centering
\includegraphics[width=\linewidth]{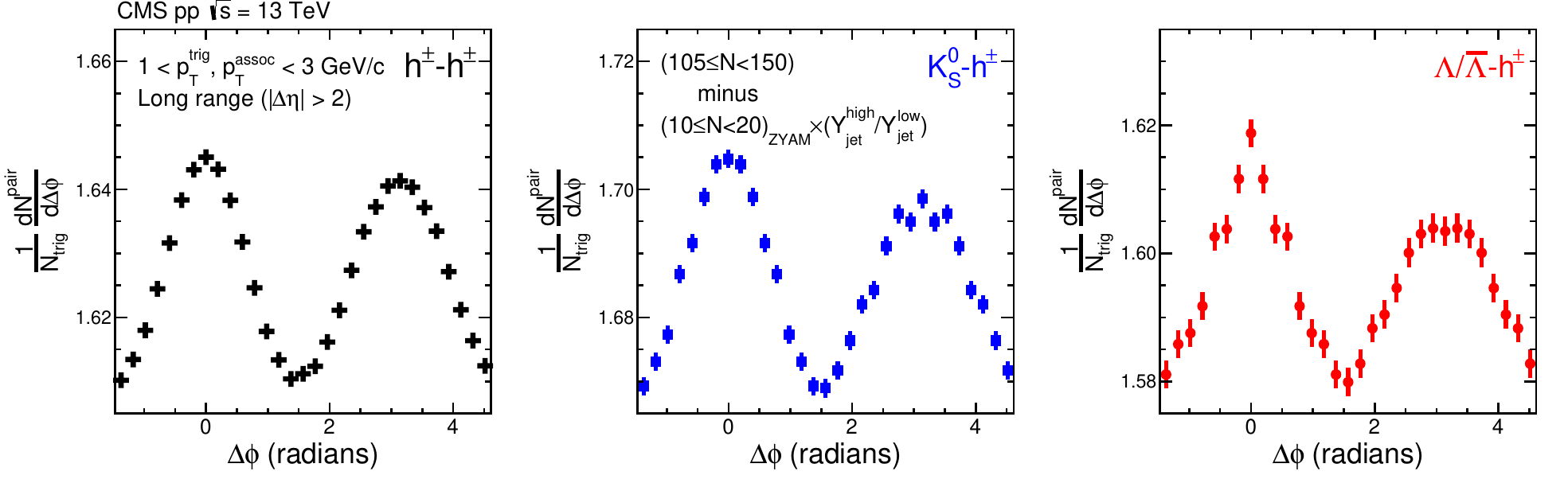}
  \caption{ \label{fig:Corr_Proj_sub_pp}
     The 1D \dphi correlation functions for the long-range regions
     in the multiplicity range $105 \leq \noff < 150$
     of pp collisions at $\roots = 13$ TeV, after subtracting scaled results from $10 \leq \noff <20$ with the ZYAM procedure applied.
     A selection of 1--3 GeV/c for both \pttrg\ and \ptass\ is used in all cases.}
\end{figure*}

After subtracting the results, with the ZYAM procedure applied, for low-multiplicity $10 \leq \noff <20$
scaled by $Y_\text{jet}/Y_\text{jet}(10\leq\noff<20)$ as in Eq.~(\ref{eq:vnsubperiph}),
the long-range 1D \dphi correlation functions in the high-multiplicity range
$105 \leq \noff < 150$ for pp collisions at $\roots = 13$ TeV are shown in
Fig.~\ref{fig:Corr_Proj_sub_pp}, for trigger particles composed of inclusive
charged particles (left), $\PKzS$ (middle), and \PgL/\PagL\ (right) particles.
A ``double-ridge" structure on the near and away side is observed after subtraction
of jet correlations. The shape of this structure, which is dominated by a second-order Fourier component,
is similar to what has been observed in pPb~\cite{CMS:2012qk,alice:2012qe,Aad:2012gla,Aaij:2015qcq}
and PbPb~\cite{Chatrchyan:2011eka,Chatrchyan:2012wg,Aamodt:2011by,ATLAS:2012at,Chatrchyan:2012ta} collisions.

\subsection{Jet subtraction systematic uncertainties}
\label{subsec:jetsubsyst}

In the jet subtraction procedure for $v_{n}\{2\}$ measurements, while
the factor $Y_\text{jet}/Y_\text{jet}(10\leq\noff<20)$ accounts for any bias in the magnitude
of jet-like associated yield due to multiplicity selection, a change in the \dphi\ width of
away side yields could lead to residual jet effects in $v_{n}\{2\}$ results. This systematic
uncertainty is evaluated by integrating the associated yields in the $|\Delta\eta| > 2$ region over
fixed \dphi windows of $|\dphi| < \pi/3$ and $|\dphi-\pi| < \pi/3$ on the near and away sides, respectively.
When extracting $v_{n}^\text{sub}$ results, 
the $Y_\text{jet}$ parameter in Eq.~(\ref{eq:vnsubperiph}) is then replaced by this difference of the near and away side yields.
By taking the difference of the yields in two \dphi\ windows symmetric around $\dphi=\pi/2$,
contributions from the second and fourth Fourier components are cancelled. By choosing the
\dphi\ window size to be $2\pi/3$, any contribution from the third Fourier component to the
near and away side associated yields is also cancelled. Any dependence of this yield difference on the event
multiplicity (beyond that induced by the $Y_\text{jet}/Y_\text{jet}(10\leq\noff<20)$ factor)
would indicate a modification of jet correlation width in \dphi.
The systematic uncertainty of $v_n$
due to this effect is estimated to be 16\%, 9\%, and 6\% for $\noff<40$, $40\leq\noff<85$, and $\noff>85$, respectively.
In the same sense, any multiplicity dependence of the \deta\ distribution of the away side would indicate a modification of the jet correlation.
The \deta\ distribution is investigated in a fixed window $|\dphi-\pi| <\pi/16$ for different \noff ranges, resulting in systematic uncertainties of
8\%, 3\%, and 2.5\% for $\noff<40$, $40\leq\noff<85$, and $\noff>85$, respectively.
The same studies apply to \pt-differential results,
leading to total uncertainties of 9\%, 13\%, 23\%, and 37\%
for $\pttrg<2.2$ GeV/c, $2.2 \leq \pttrg<3.6$ GeV/c, $3.6 \leq \pttrg<4.6$ GeV/c, and $\pttrg \geq 4.6$ GeV/c, respectively. 
For \PKzS\ results, the above systematic effects lead to total uncertainties of 6.4\%, 10.8\%, 15.3\% and 43.1\% for $\pttrg<1.8$ GeV/c, $1.8 \leq \pttrg<2.8$ GeV/c, $2.8 \leq \pttrg<4.0$ GeV/c, and $\pttrg \geq 4.0$ GeV/c, respectively.
For \PgL/\PagL\ results, the total uncertainties are 6.4\% and 21.2\% for $\pttrg<4$ GeV/c and $\pttrg \geq 4.0$ GeV/c.

\begin{figure*}[thb!p]
\centering
\includegraphics[width=\linewidth]{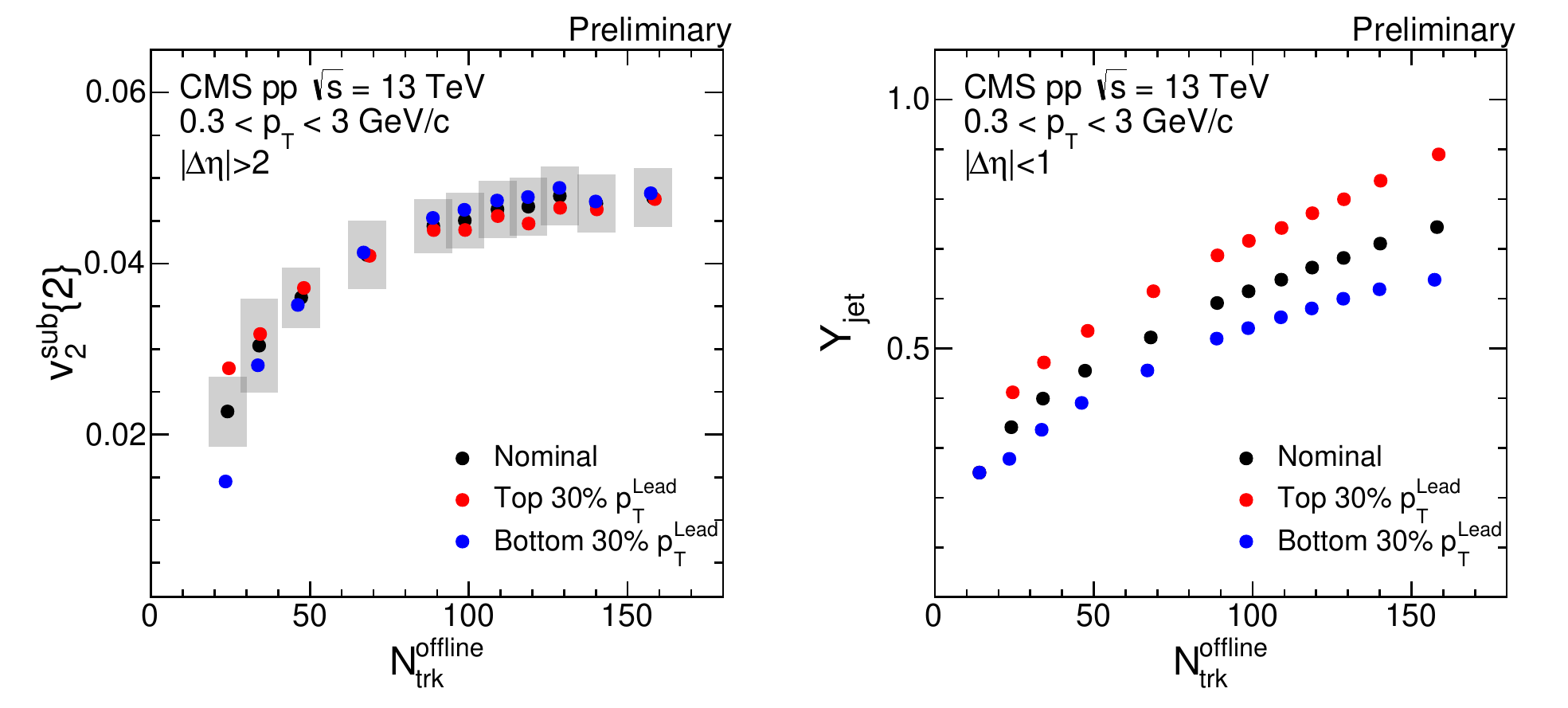}
  \caption{ \label{fig:jetvary} Left: $v_{2}^\text{sub}$ results after low multiplicity subtraction as function of multiplicity for all events and events with top (bottom) 30\% leading particle \pt\ for pp collision at 13 TeV. Systematic uncertainties on the nominal results (described in Sec.~\ref{sec:vncharge}) are shown as shaded areas. 
	Right: Near side jet yield as function of multiplicity for all events and events with top (bottom) 30\% leading particle \pt\ for pp collision at 13 TeV.
     }
\end{figure*}

In addition, by separating events in a given multiplicity range into two groups corresponding to the top and bottom
30\% in the leading particle \pt\ distribution, jet correlations are either strongly
enhanced or suppressed in a controlled manner. 
Fig.~\ref{fig:jetvary} (right) shows the near side jet yield for the two groups of events, a variation of almost factor of 2 in the jet correlation is achieved.
After applying the subtraction procedure,
$v_{2}^\text{sub}$ results for the two event groups are consistent within 5\%, as shown in Fig.~\ref{fig:jetvary} (left). 
This observation confirms that low multiplicity subtraction is robust against any multiplicity dependent bias on jet mechanism.
\clearpage

\section{$v_2$, $v_3$ as function of multiplicity and \pt}
\label{sec:vncharge}

\begin{figure*}[thbp]
\centering
\includegraphics[width=0.98\textwidth]{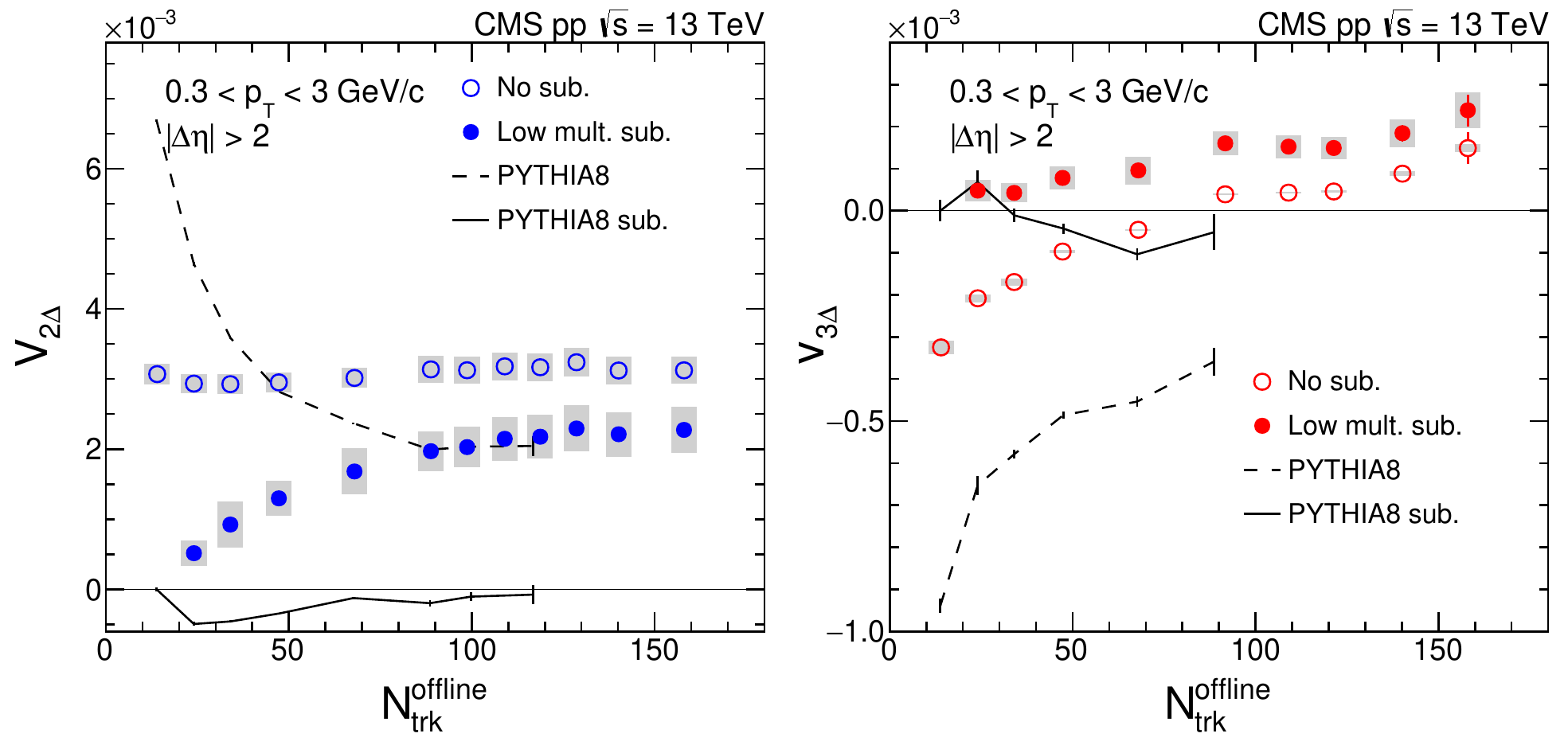}
  \caption{ \label{fig:C2C3}
     The second-order (left) and third-order (right) Fourier coefficients,
     $V_{2\Delta}$ and $V_{3\Delta}$, of
     long-range ($|\deta|>2$) two-particle \dphi\ correlations
     as a function of \noff\ for charged particles, averaged over $0.3 < \pt < 3.0$ GeV/c,
     in pp collisions at $\roots = 13$ TeV, before (open) and after (filled)
     correcting for back-to-back jet correlations, estimated from the $10 \leq \noff < 20$ range.
     Results from PYTHIA8 tune CUETP8M1 simulation are shown as curves.
     The error bars correspond to statistical uncertainties, while the shaded areas denote the systematic uncertainties.
   }
\end{figure*}

Fourier coefficients, $V_{n\Delta}$, extracted from 1D $\Delta\phi$ two-particle correlation functions
for the long-range \deta\ region using Eq.~(\ref{eq:Vn}), are first studied for inclusive
charged hadrons. Figure~\ref{fig:C2C3} shows the $V_{2\Delta}$ and $V_{3\Delta}$ values for
pairs of inclusive charged particles averaged over $0.3  < \pt < 3.0$ GeV/c as a function of multiplicity
in pp collisions at $\roots = 13$ TeV, before and after correcting for back-to-back jet
correlations estimated from low-multiplicity data ($10 \leq \noff < 20$).

Before low-multiplicity subtraction, the $V_{2\Delta}$ coefficients are found to remain relatively
constant as a function of multiplicity. This behavior is very different from the PYTHIA8 tune CUETP8M1
MC simulation, where the only source of long-range correlations is back-to-back jets and
the $V_{2\Delta}$ coefficients decrease with \noff. The $V_{3\Delta}$ coefficients found
using the PYTHIA8 simulation are always negative because of dominant contributions at $\dphi \approx \pi$
from back-to-back jets~\cite{Aamodt:2011by}, with their magnitudes decreasing as a function of \noff.
A similar trend is seen in the data for the
low multiplicity region, $\noff<90$. However, for $\noff\geq90$, the $V_{3\Delta}$ coefficients
in pp data change to positive values. This transition directly indicates
a new phenomena that is not present in the PYTHIA8 simulation.
After applying the low-multiplicity subtraction detailed in Section~\ref{sec:lowsubpp},
$V_{2\Delta}$ exhibits an increase with multiplicity for $\noff \lesssim 100$, and
reaches a relatively constant value for the higher \noff\ region. The $V_{3\Delta}$ values
after subtraction of jet correlations become positive over the entire multiplicity range
and increase with multiplicity.

\begin{figure}[thb]
\centering
\includegraphics[width=0.8\linewidth]{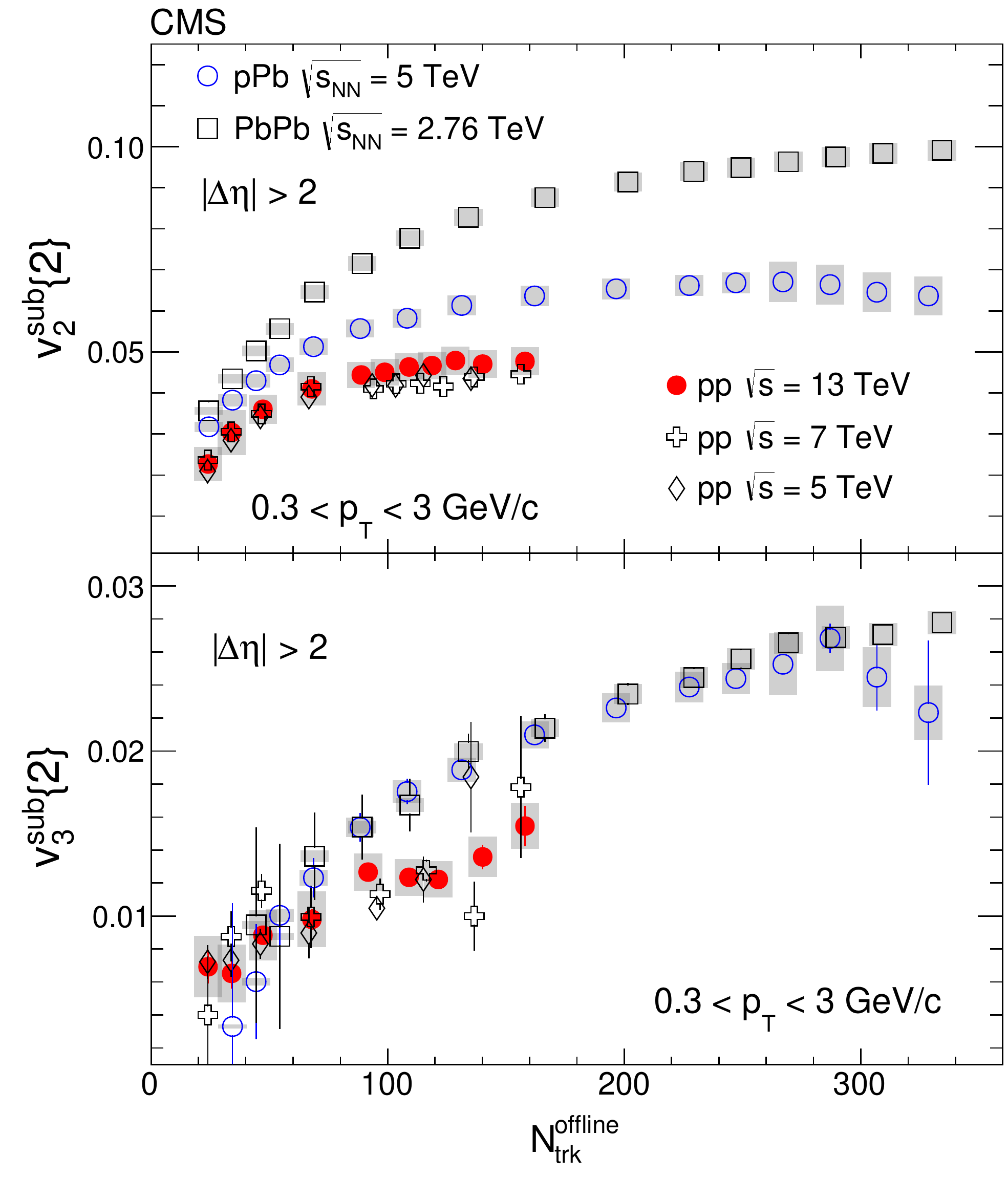}
  \caption{\label{fig:v2v3vsN}
     The $v_{2}^\text{sub}$ (top) and $v_{3}^\text{sub}$ (bottom) results of
     charged particles as a function of \noff, averaged over $0.3<\pt<3.0$ GeV/c,
     in pp collisions at $\roots = 5$, 7, and 13 TeV, pPb collisions
     at $\rootsNN = 5$ TeV, and PbPb collisions $\rootsNN = 2.76$ TeV,
     after correcting for back-to-back jet correlations estimated from low-multiplicity data.
     The error bars correspond to the statistical uncertainties, while the shaded areas denote the systematic uncertainties.
     Systematic uncertainties are found to have no dependence on \roots\ for pp results and therefore are only shown for 13 TeV.
   }
\end{figure}

The elliptic ($v_2$) and triangular ($v_3$) flow harmonics for charged particles
with $0.3 < \pt < 3.0$ GeV/c, after applying the jet correction procedure, are then
extracted from the two-particle Fourier coefficients obtained using Eq.~(\ref{eq:Vnpt}),
and are shown in Fig.~\ref{fig:v2v3vsN} for pp collisions at $\roots = 5$, 7, and 13 TeV.
The previously published pPb data at $\rootsNN = 5$ TeV and PbPb data at
$\rootsNN = 2.76$ TeV~\cite{Chatrchyan:2013nka} are also shown for comparison among
different collision systems.

Within experimental uncertainties, for pp collisions at all three
energies, there is no or only a very weak energy dependence for the $v_{2}^\text{sub}$ values.
The $v_{2}^\text{sub}$ results for pp collisions
show a similar pattern as the pPb results, becoming relatively constant as \noff\ increases,
while the PbPb results show a moderate increase over
the entire \noff\ range shown in Fig.~\ref{fig:v2v3vsN}. Overall, the pp data show a smaller $v_{2}^\text{sub}$ signal
than pPb data over a wide multiplicity range, and both systems show smaller
$v_{2}^\text{sub}$ values than for the PbPb system.

The $v_{3}^\text{sub}$ values of
the pp data are comparable to those observed in pPb and PbPb collisions
in the very low multiplicity region $\noff < 60$, although
systematic uncertainties are large for all the three systems. At higher \noff,
$v_{3}^\text{sub}$ in pp collisions
increases with multiplicity, although at a slower rate than observed in pPb and PbPb collisions.

\begin{figure}[thbp]
\centering
\includegraphics[width=0.49\textwidth]{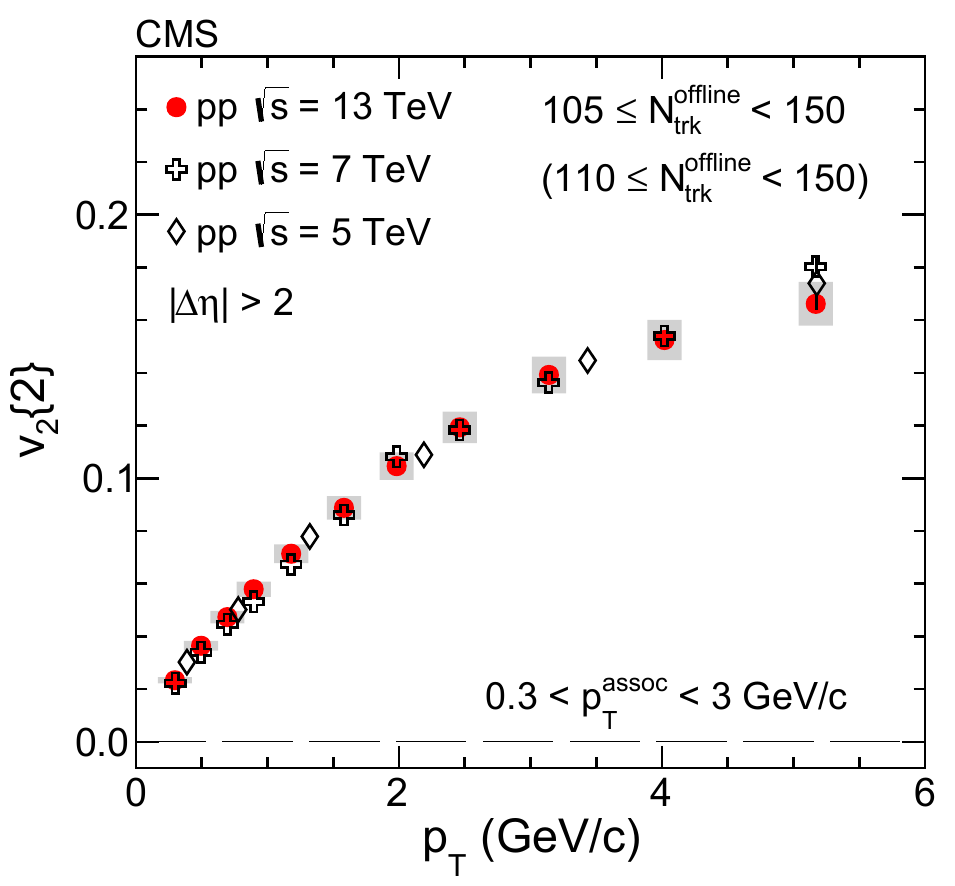}
\includegraphics[width=0.49\textwidth]{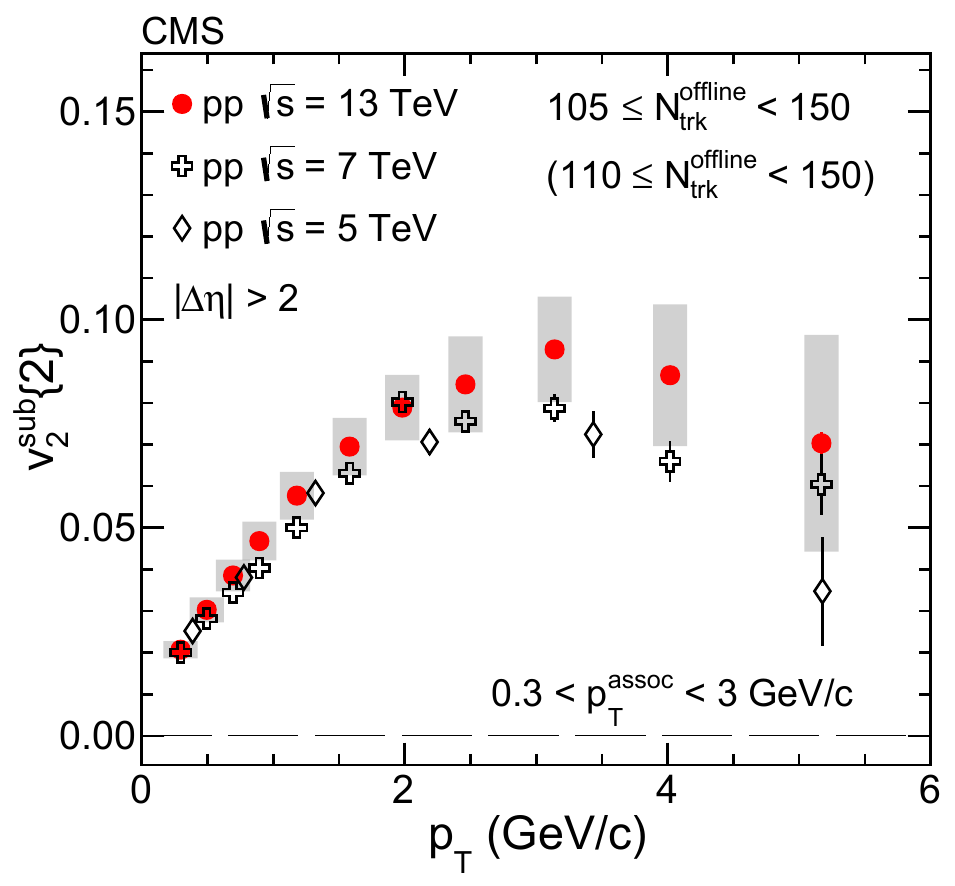}
  \caption{\label{fig:v2vspT}
  The $v_2$ results of inclusive charged particles, before (left) and after (right)
  subtracting correlations from low-multiplicity events, as a function of \pt\  in pp collisions at $\roots = 13$ TeV for $105 \leq \noff < 150$ and at $\roots = 5$, 7 TeV for $110 \leq \noff < 150$.
 The error bars correspond to the statistical uncertainties, while the shaded areas denote the systematic uncertainties.
 Systematic uncertainties are found to have no dependence on \roots\ for pp results and therefore are only shown for 13 TeV.
   }
\end{figure}

The $v_2$ results as a function of \pt\ for high-multiplicity
pp events at $\roots = 5$, 7, and 13 TeV are shown in Fig.~\ref{fig:v2vspT}
before (left) and after (right)
correcting for jet correlations.
To compare results with similar average \noff, $105 \leq \noff < 150$ is chosen for 13 TeV while $110 \leq \noff < 150$ is chosen for 5 and 7 TeV.
Little energy dependence is observed for the $\pt$-differential $v_{2}$ results,
especially before correcting for jet correlations, as shown in Fig.~\ref{fig:v2vspT} (left).
This conclusion also holds after jet correction procedure for $v_{2}^\text{sub}$ results (Fig.~\ref{fig:v2vspT}, right)
within systematic uncertainties, although systematic uncertainties for $v_{2}^\text{sub}$ are
significantly higher at high \pt\ because of the large magnitude of the subtracted term.
This observation is consistent with the energy independence of associated long-range yields
on the near side reported in Ref.~\cite{Khachatryan:2015lva}.
The observed \pt\ dependence of $v_{2}^\text{sub}$, in high-multiplicity pp events with peak values at 2--3 GeV/c
at various energies, is similar to that in pPb~\cite{Chatrchyan:2013nka,Aad:2014lta,ABELEV:2013wsa} and PbPb~\cite{Chatrchyan:2012wg,ATLAS:2011ah,Adam:2016izf} collisions.

\clearpage

\section{Mass ordering of $v_2$}

\begin{figure}[thbp]
\centering
\includegraphics[width=0.49\textwidth]{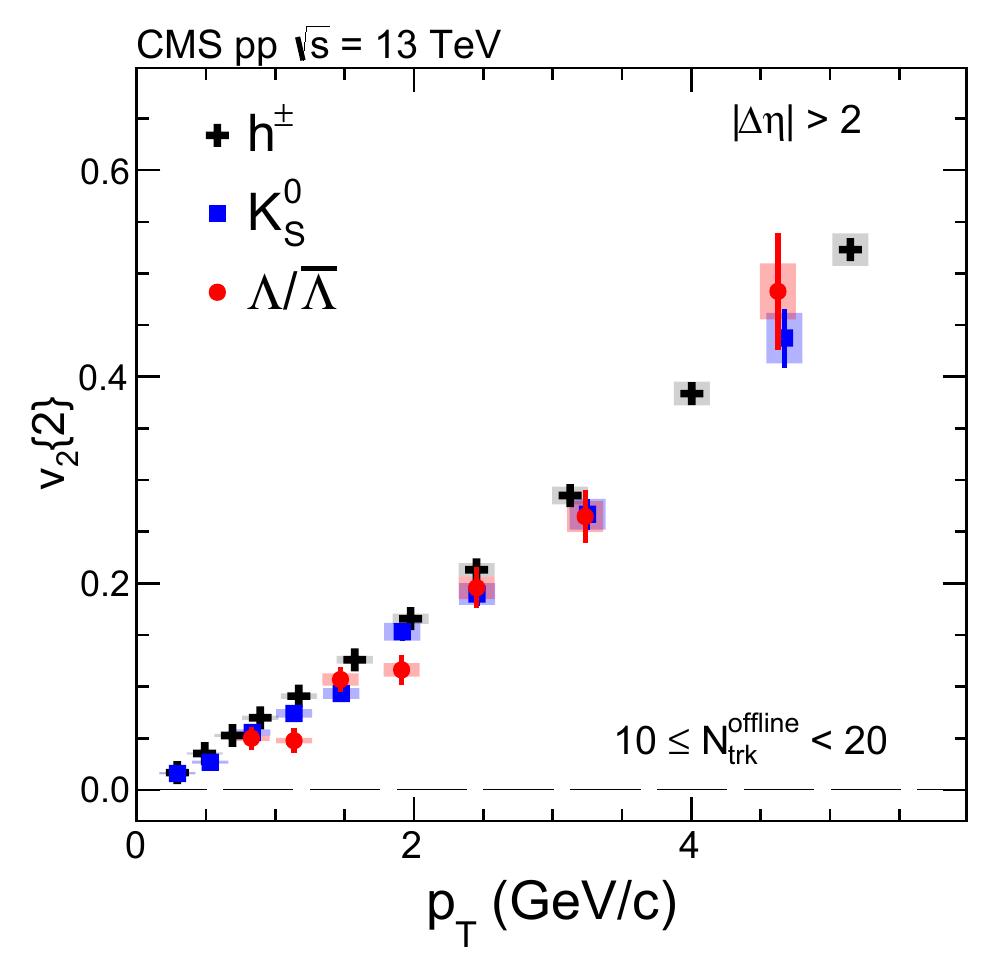}
\includegraphics[width=0.49\textwidth]{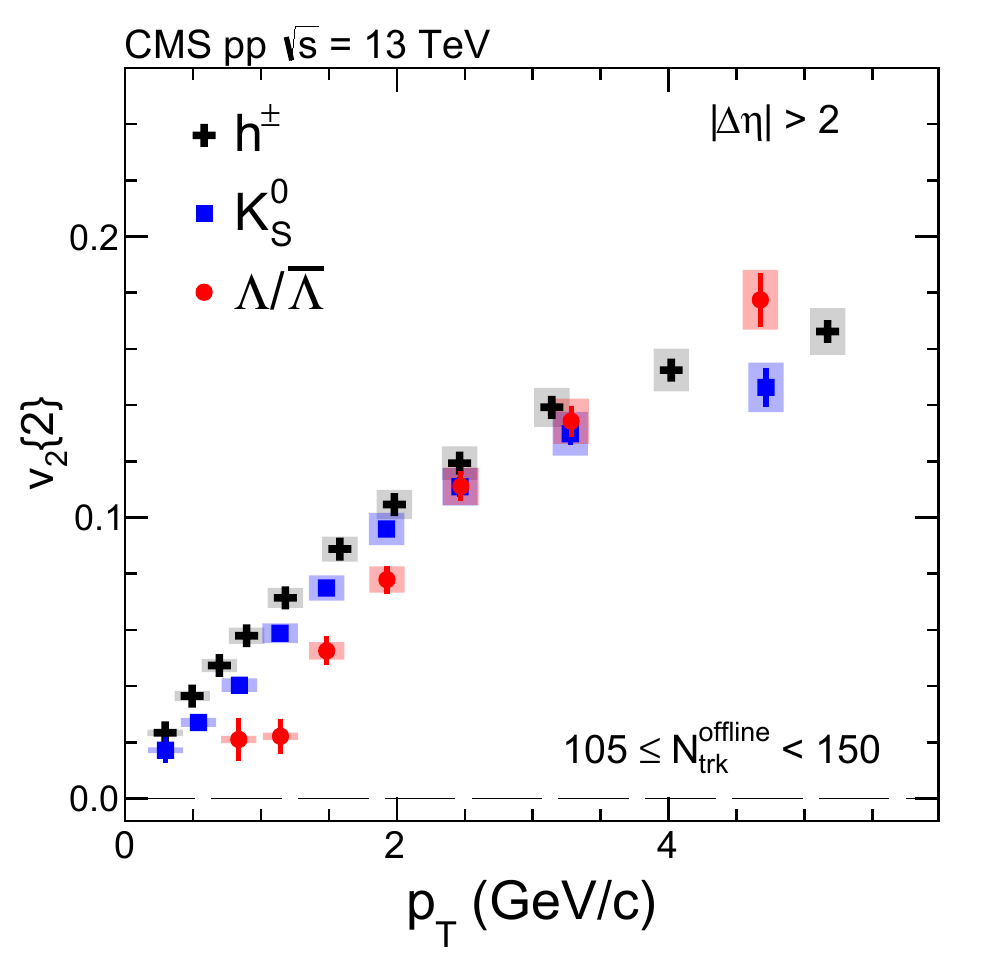}
  \caption{\label{fig:v2vspTPID}
     The $v_2$ results for inclusive charged particles, $\PKzS$ and \PgL/\PagL\ particles
     as a function of \pt\ in pp collisions at $\roots = 13$ TeV,
     for $10 \leq \noff < 20$ (left) and $105 \leq \noff < 150$ (right).
     The error bars correspond to the statistical uncertainties, while the shaded areas denote the systematic uncertainties.
   }
\end{figure}

\begin{figure}[thb]
\centering
\includegraphics[width=0.6\textwidth]{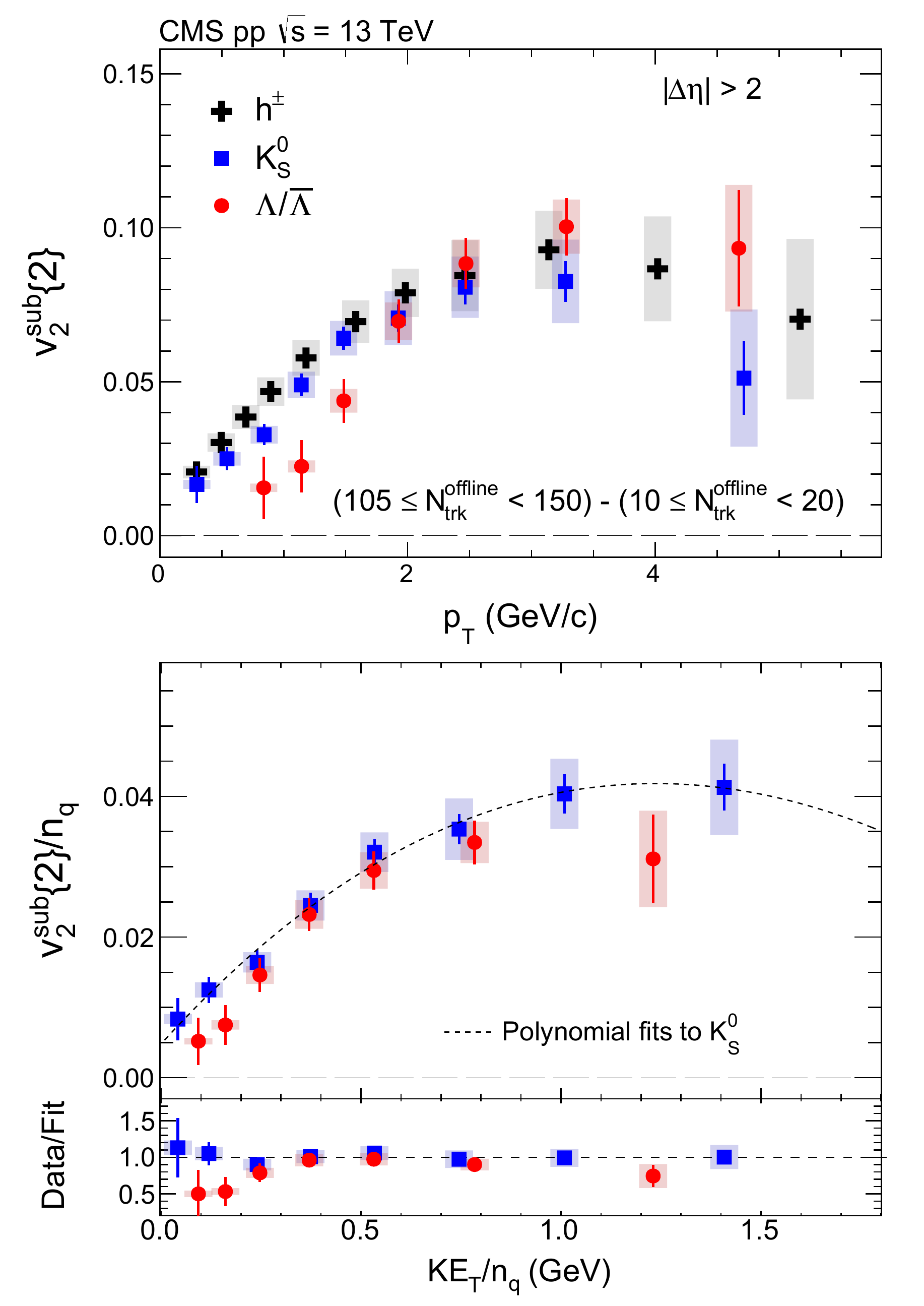}
  \caption{\label{fig:NCQsub}
     Top: the $v_{2}^\text{sub}$ results of inclusive charged particles, $\PKzS$ and
     \PgL/\PagL\ particles as a function of \pt\ for $105 \leq \noff < 150$,
     after correcting for back-to-back jet correlations
     estimated from low-multiplicity data. Bottom: the $n_\mathrm{q}$-scaled $v_{2}^\text{sub}$ results
     for $\PKzS$ and \PgL/\PagL\ particles as a function of $\ket/n_\mathrm{q}$.
     Ratios of $v_{2}^\text{sub}/n_\mathrm{q}$ for $\PKzS$ and \PgL/\PagL\ particles to
     a smooth fit function of data for $\PKzS$ particles are also shown.
     The error bars correspond to the statistical uncertainties, while the shaded areas denote the systematic uncertainties.
   }
\end{figure}

The dependence of the elliptic flow harmonic on particle species can shed further light on the nature of the correlations.
The $v_2$ data as a function of \pt\ for identified $\PKzS$ and \PgL/\PagL\
particles are extracted for pp collisions at $\roots = 13$ TeV.
Figure~\ref{fig:v2vspTPID} shows the results for a low ($10 \leq \noff < 20$)
and a high ($105 \leq \noff < 150$) multiplicity range before applying the
jet correction procedure.

At low multiplicity (Fig.~\ref{fig:v2vspTPID} left), the $v_{2}$ values
are found to be similar for charged particles, $\PKzS$ and \PgL/\PagL\ hadrons
across most of the \pt\ range within statistical uncertainties, similar to the
observation in pPb collisions at $\rootsNN = 5$ TeV~\cite{Khachatryan:2014jra}, described in Sec.~\ref{sec:massv2}.
This would be consistent with the expectation that back-to-back jets are the
dominant source of long-range correlations on the away side in low-multiplicity
pp events. Moving to high-multiplicity pp events
($105 \leq \noff < 150$, Fig.~\ref{fig:v2vspTPID} right),
a clear deviation of $v_{2}$ among various particle species
is observed. In the lower \pt\ region of $\lesssim 2.5$ GeV/c,
the $v_{2}$ value of $\PKzS$ is greater than that of \PgL/\PagL\ at a given \pt\ value.
Both are consistently below the inclusive charged particle $v_{2}$ values.
Since most charged particles are pions in this \pt\ range, this indicates that
lighter particle species exhibit a stronger azimuthal anisotropy signal.
A similar trend was first observed in AA collisions at RHIC~\cite{Abelev:2007qg,Adare:2012vq},
and later also seen in pPb collisions at the LHC~\cite{ABELEV:2013wsa,Khachatryan:2014jra}.
This behavior is found to be qualitatively consistent with both hydrodynamic
models~\cite{Werner:2013ipa,Bozek:2013ska} and non-hydrodynamic models, such as AMPT through parton escape mechanism~\cite{Li:2016ubw}, UrQMD through hadronic interaction~\cite{Zhou:2015iba} and an alternative initial state interpretation with CGC~\cite{Schenke:2016lrs}.
At $\pt>2.5$ GeV/c, the $v_{2}$
values of \PgL/\PagL\ particles tend to become greater than those of $\PKzS$ particles.
This reversed ordering of $\PKzS$ and \PgL/\PagL\ at high \pt\ is similar
to what was previously observed in pPb and PbPb collisions~\cite{Khachatryan:2014jra}.

After applying the correction for jet correlations, the $v_{2}^\text{sub}$ results as a
function of \pt\ for $105 \leq \noff < 150$ are shown in Fig.~\ref{fig:NCQsub} (top) for the identified particles and charged hadrons.
The $v_{2}^\text{sub}$ values for all three types of particles are found to increase
with \pt, reaching 0.08--0.10 at $2<\pt<3$ GeV/c, and then show a trend of decreasing $v_{2}^\text{sub}$ values for
higher \pt\ values. The particle mass ordering of $v_{2}$ values in the lower \pt region is also observed after applying jet correction procedure, while at higher \pt the ordering tends to reverse.
\clearpage

\section{Number of constituent quark scaling}

As done in Sec.~\ref{sec:ncqpPb} (Ref.~\cite{Khachatryan:2014jra}), the scaling
behavior of $v_{2}^\text{sub}$ divided by the number of constituent quarks, $n_\mathrm{q}$, as a function
of transverse kinetic energy per quark, $\ket/n_\mathrm{q}$, is investigated for high-multiplicity
pp events in Fig.~\ref{fig:NCQsub} (bottom). The dashed curve corresponds to a polynomial
fit to the $\PKzS$ data. The ratio of $n_\mathrm{q}$-scaled $v_{2}^\text{sub}$ results for $\PKzS$ and \PgL/\PagL\ particles divided by this polynomial function fit is also shown
in Fig.~\ref{fig:NCQsub} (bottom). An approximate scaling is seen for $\ket/n_\mathrm{q} \gtrsim 0.2$ GeV/c.

\clearpage

\section{Comparison to multi-particle correlation results across different collision systems}

\begin{figure*}[thb]
\centering
\includegraphics[width=\linewidth]{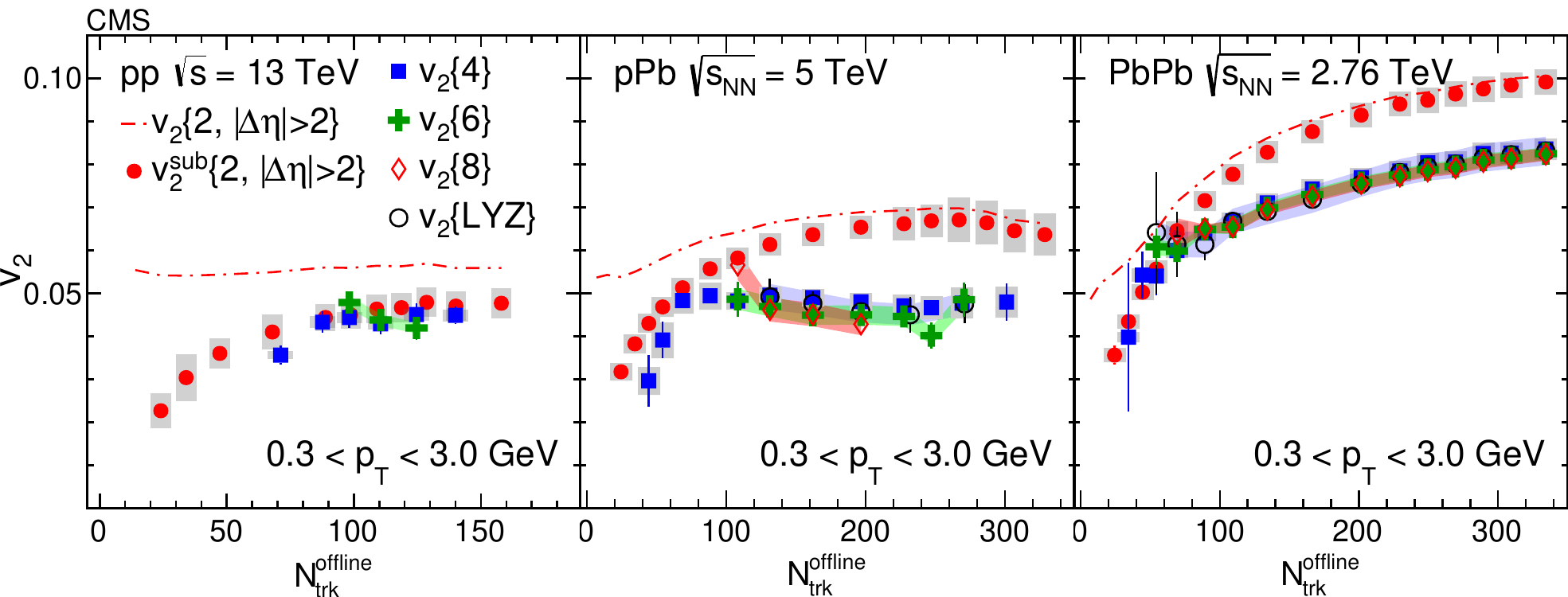}
  \caption{\label{fig:v24}
     Left: The $v_{2}^\text{sub}\{2, |\Delta\eta|>2\}$, $v_{2}\{4\}$ and $v_{2}\{6\}$ values as a function
     of \noff for charged particles, averaged over $0.3 < \pt < 3.0$ GeV/c
     and $|\eta|<2.4$, in pp collisions at $\roots = 13$ TeV.
     Middle: The $v_{2}^\text{sub}\{2, |\Delta\eta|>2\}$, $v_{2}\{4\}$, $v_{2}\{6\}$, $v_{2}\{8\}$,
     and $v_{2}\{\mathrm{LYZ}\}$ values in pPb collisions at $\rootsNN = 5$ TeV~\cite{Chatrchyan:2013nka}.
     Right: The $v_{2}^\text{sub}\{2, |\Delta\eta|>2\}$, $v_{2}\{4\}$, $v_{2}\{6\}$, $v_{2}\{8\}$,
     and $v_{2}\{\mathrm{LYZ}\}$ values in PbPb collisions at $\rootsNN = 2.76$ TeV~\cite{Chatrchyan:2013nka}.
     The error bars correspond to the statistical uncertainties, while the shaded areas denote the systematic uncertainties.
   }
\end{figure*}

The $v_{2}\{4\}$ and $v_{2}\{6\}$ results, extracted from multi-particle cumulant method, for pp collisions at $\roots = 13$ TeV are measured in Ref.~\cite{Khachatryan:2016txc}.
The left panel of Fig.~\ref{fig:v24} shows the comparison between $v_{2}^\text{sub}\{2, |\Delta\eta|>2\}$, $v_{2}\{4\}$ and $v_{2}\{6\}$ 
as a function of event multiplicity. Within experimental uncertainties, the multi-particle cumulant
$v_{2}\{4\}$ and $v_{2}\{6\}$ values in high-multiplicity pp collisions are
consistent with each other, similar to what was observed previously in
pPb and PbPb collisions (shown in Fig.~\ref{fig:v24} middle and right panels~\cite{Khachatryan:2015waa}). This provides strong
evidence for the collective nature of the long-range correlations observed in pp collisions.
However, unlike for pPb and PbPb collisions where
$v_{2}^\text{sub}\{2, |\Delta\eta|>2\}$ values show a larger magnitude than multi-particle cumulant $v_2$ results,
the $v_2$ values obtained from two-, four-, and six-particle correlations are
comparable in pp collisions at $\roots = 13$ TeV within uncertainties. 

In the context of hydrodynamic models, i.e. $v_n$ is proportional to the initial eccentricity $\epsilon_n$ of the medium, 
the relative difference of $v_2$ among two- and various
orders of multi-particle correlations provide insights to the details of initial-state
geometry fluctuations in pp and pPb systems. 
In AA collision, the event-by-event distribution of $\epsilon$ (drop n for simplicity) is shown to be well modelled by a ``Bessel-Gaussian" function~\cite{Voloshin:2007pc}, 
\begin{equation}
P(\epsilon) = \frac{2\epsilon}{\sigma^2}I_{0}\left(\frac{2\epsilon\bar{\epsilon}}{\sigma^2}\right)exp\left(-\frac{\epsilon^2+\bar{\epsilon}^2}{\sigma^2}\right),
\end{equation}
\noindent where $I_0$ is the modified Bessel function, $\sigma$ is the variance of the distribution and $\bar{\epsilon}$ is the average eccentricity in the reaction plane. 
From this distribution, the calculation of $v_{2}\{2\}$ and $v_{2}\{2n, n>1\}$ reveals that 
\begin{equation}
\begin{split}
v_{2}\{2\} \propto \sqrt{\bar{\epsilon}^2+\sigma^2},\\
v_{2}\{2n, n>1\} \propto \bar{\epsilon}.
\end{split}
\end{equation}
Therefore, if the same interpretation is carried from AA collision to pp collision, the results in Fig.~\ref{fig:v24} would indicate initial-state fluctuation in pp collision is close to 0. 
Such an interpretation is apparently in contradiction to the expectation that the large $v_2$ signal observed in pp collision is a result of initial-state fluctuation. 
However, as shown in Ref.~\cite{Yan:2013laa}, event-by-event distribution of $\epsilon$ in small collision system is better modelled by a power law distribution, 
\begin{equation}
P(\epsilon) = 2\alpha\epsilon(1-\epsilon^2)^{\alpha-1},
\end{equation}
\noindent where $\alpha = (N_{\mathrm{source}}-1)/2$ and $N_{\mathrm{source}}$ is the total number of point-like fluctuating sources in the system. 
This distribution provides relation between $v_{2}\{2n\}$ and $\alpha$ as 
\begin{equation}
\begin{gathered}
v_{2}\{2\} \propto \frac{1}{\sqrt{1+\alpha}},\\
v_{2}\{2n, n>1\} \propto \left[\frac{2}{(1+\alpha)^2(2+\alpha)}\right]^{1/4}.
\end{gathered}
\end{equation}
\noindent Therefore, the ratio of $v_{2}\{4\}$ to $v_{2}^\text{sub}\{2, |\Delta\eta|>2\}$ is related to the total number of
fluctuating sources, $N_{\mathrm{source}}$, in the initial stage of a collision,
\begin{equation}
\frac{v_{2}\{4\}}{v_{2}\{2\}} = \left(\frac{2}{1+N_{\mathrm{source}}/2}\right)^{1/4}.
\end{equation}
\noindent The comparable magnitudes of $v_{2}^\text{sub}\{2, |\Delta\eta|>2\}$ and $v_{2}\{4\}$ signals
observed in pp collisions, compared to pPb collisions at similar multiplicities,
may indicate a smaller number of initial fluctuating sources that drive
the long-range correlations seen in the final state. 
Even with $v_2$ results before the jet subtraction, shown in Fig.~\ref{fig:v24} as dash lines, 
$v_{2}\{2\}$ to $v_{2}\{4\}$ ratio is smaller in pp collisions compared to pPb collisions.

Meanwhile, it remains to be seen whether other
proposed mechanisms~\cite{Dusling:2015gta,Dusling:2012wy,Dusling:2012cg}
in interpreting the long-range correlations in pPb and PbPb collisions can also describe
the features of multi-particle correlations seen in pp collisions.

\clearpage

\section{Comparison to ATLAS $v_2$ result}
\label{sec:ATLAScomp}

\begin{figure}[thb]
\centering
\includegraphics[width=0.8\textwidth]{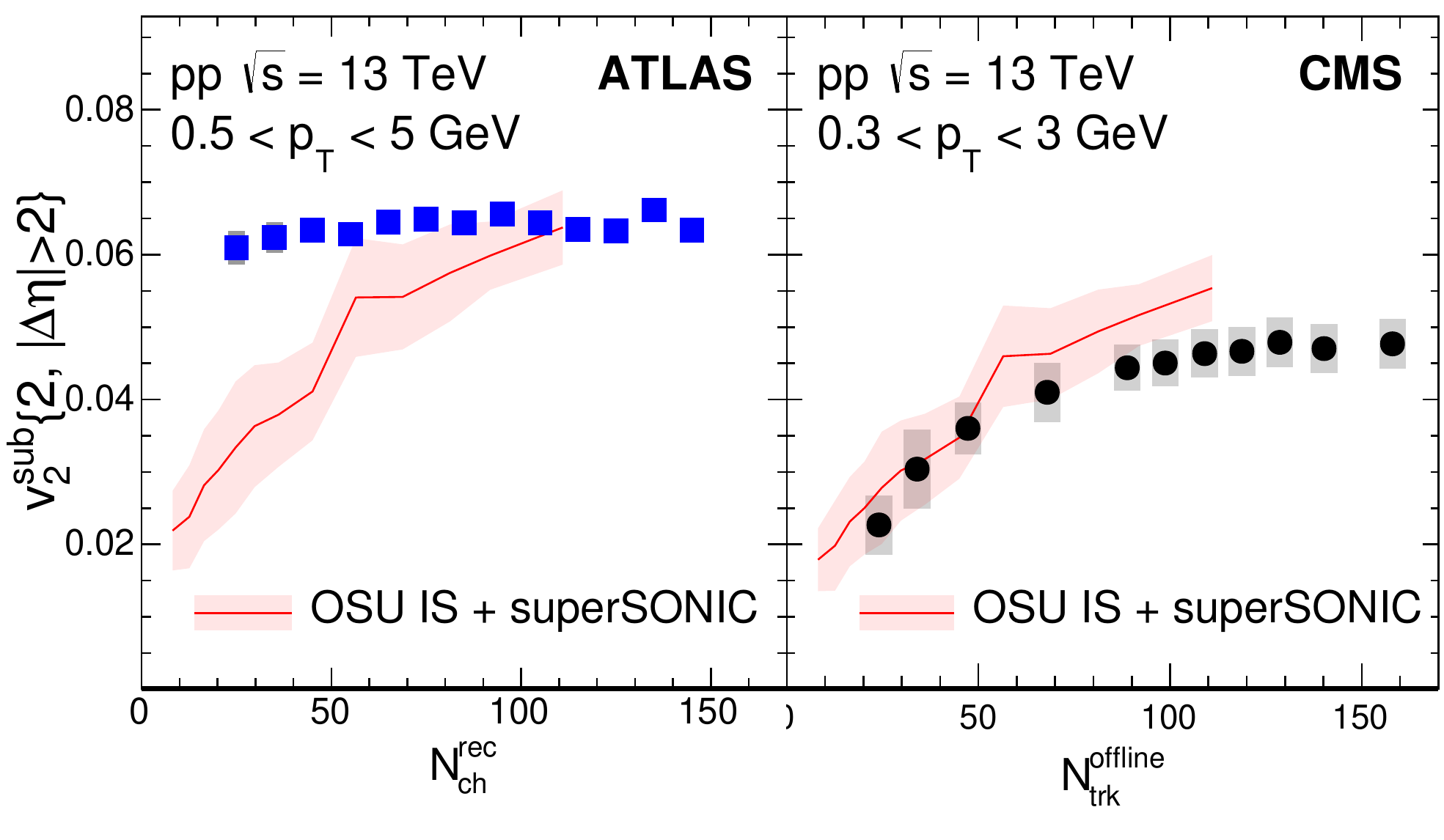}
  \caption{ \label{fig:CMSATLAS}
     The $v_{2}$ results as function of multiplicity from ATLAS~\cite{Aaboud:2016yar} and CMS.}
\end{figure}

Recently, ATLAS collaboration reported two-particle correlation $v_{2}$ results in pp collisions at 13 TeV~\cite{Aaboud:2016yar} 
with a new template-fitting method to account for jet correlation contributions. 
Although the $v_2$ results as function of \pt\ are consistent with what has been presented in Sec.~\ref{sec:vncharge}, 
the results as function of multiplicity show large difference at low multiplicity region. 
Instead of a decreasing trend towards low multiplicity ATLAS reported a constant $v_2$ value over the entire multiplicity range, as shown in Fig.~\ref{fig:CMSATLAS}. 
Such an inconsistency, especially the $v_2$ values at low multiplicity, makes the results hard to constrain the theory interpretations, e.g. the hydro calculations from SuperSONIC~\cite{Weller:2017tsr} shown in Fig.~\ref{fig:CMSATLAS}.

It has been understood that the difference is coming from the way of subtracting jet contributions, 
and that the results are model-dependent at low multiplicity region. 
The following sub-sections discuss the limitation of the low multiplicity subtraction method and the template-fitting method. 

\subsection{Limitation of the low multiplicity subtraction method}
\label{subsec:CMSlimit}

The detail description of the low multiplicity subtraction method is given in Sec.~\ref{sec:lowsubpp}. 
By assuming the Fourier coefficients, $V_{n\Delta}$, can be decomposed into contributions from collective correlation ($V_{n\Delta}^{col}$) and jet correlation ($V_{n\Delta}^{jet}$), Eq.~\ref{eq:vnsubperiph} can be written as 
\begin{equation}
\label{eq:lowsubexp}
\begin{split}
V^\text{sub}_{n\Delta}= &[V_{n\Delta}^{col}+V_{n\Delta}^{jet}]-[V_{n\Delta}^{col}(10\leq\noff<20)+V_{n\Delta}^{jet}(10\leq\noff<20)]\\
&\times\frac{N_\text{assoc}(10\leq\noff<20)}{N_\text{assoc}}\times\frac{Y_\text{jet}}{Y_\text{jet}(10\leq\noff<20)}\\
\\
= &V_{n\Delta}^{col}-V_{n\Delta}^{col}(10\leq\noff<20)\times\frac{N_\text{assoc}(10\leq\noff<20)}{N_\text{assoc}}\\
&\times\frac{Y_\text{jet}}{Y_\text{jet}(10\leq\noff<20)}\\
\\
&+V_{n\Delta}^{jet}-V_{n\Delta}^{jet}(10\leq\noff<20)\times\frac{N_\text{assoc}(10\leq\noff<20)}{N_\text{assoc}}\\
&\times\frac{Y_\text{jet}}{Y_\text{jet}(10\leq\noff<20)}.
\end{split}
\end{equation}
\noindent As stated in Sec.~\ref{sec:lowsubpp}, jet contribution at a multiplicity range of interest can be modelled by scaling up the contribution at $10\leq\noff<20$, all the terms involving jet contribution are cancelled. 
The leftover in Eq.~\ref{eq:lowsubexp} indicates that one measures the exact collective contribution to $V_{n\Delta}$ only when there is no such contribution at the low multiplicity region ($10\leq\noff<20$). 
Therefore, the decreasing trend of $v_2^\text{sub}$ towards low multiplicity is by construction, 
and results from the low multiplicity subtraction method is over-subtracted if there is collective behavior developing at low multiplicity region. 
Such a over-subtraction becomes less prominent towards high multiplicity where the fraction being subtracted tends to be smaller, as shown in Fig.~\ref{fig:C2C3}.

\subsection{Limitation of the template-fitting method}
\label{subsec:ATLASlimit}

To separate the ridge from angular correlations present in low-multiplicity pp collisions, 
a template fit function is used by ATLAS to fit the long-range two-particle \dphi\ correlation function, $Y(\Delta\phi)$,

\begin{equation}
\label{eq:ATLAStemplatefit}
Y(\Delta\phi) = F Y_\text{low}(\Delta\phi) + Y^\text{ridge}(\Delta\phi),
\end{equation}

\noindent where $Y_\text{low}(\Delta\phi)$ is the correlation function at low multiplicity and

\begin{equation}
\label{eq:ATLAStemplatefitRidge}
Y^\text{ridge}(\Delta\phi) = G [1 + 2V_{n\Delta}^\text{fit} \cos(n\Delta\phi)].
\end{equation}

\noindent Fourier decomposition (using Eq.~\ref{eq:Vn}) of Eq.~\ref{eq:ATLAStemplatefit} provides 
relation between Fourier coefficients before and after correcting jet contribution,

\begin{equation}
\label{eq:ATLASVn}
V_{n\Delta}^\text{fit} = \frac{N^\text{assoc}}{G}V_{n\Delta} - \frac{F N^\text{assoc}_\text{low}}{G}V_{n\Delta}^\text{low},
\end{equation}

\noindent where $G = N^\text{assoc} - F N^\text{assoc}_\text{low}$.

Decomposing $V_{n\Delta}$ into $V_{n\Delta}^{col}+V_{n\Delta}^{jet}$ and assume all the terms for jet contribution are cancelled, 
Eq.~\ref{eq:ATLASVn} can be written as 
\begin{equation}
\label{eq:ATLAScorrvn}
\begin{split}
V_{n\Delta}^\text{fit}= &\frac{N^\text{assoc}}{G}V_{n\Delta}^{col}-\frac{F N^\text{assoc}_\text{low}}{G}V_{n\Delta}^{\text{low},col}\\
= &V_{n\Delta}^{col}+\frac{F N^\text{assoc}_\text{low}}{G}(V_{n\Delta}^{col}-V_{n\Delta}^{\text{low},col}).
\end{split}
\end{equation}

\noindent The above equation indicates that one measures the exact collective contribution to $V_{n\Delta}$ only when there is equal amount of such contribution at the low multiplicity region.
Therefore, results from the template-fitting method is under-subtracted if the collective behavior developing at low multiplicity region is less prominent compared to those at the high multiplicity region. 

\subsection{Monte Carlo test of jet subtraction}
\label{subsec:MCsub}

\begin{figure}[thb]
\centering
\includegraphics[width=0.6\textwidth]{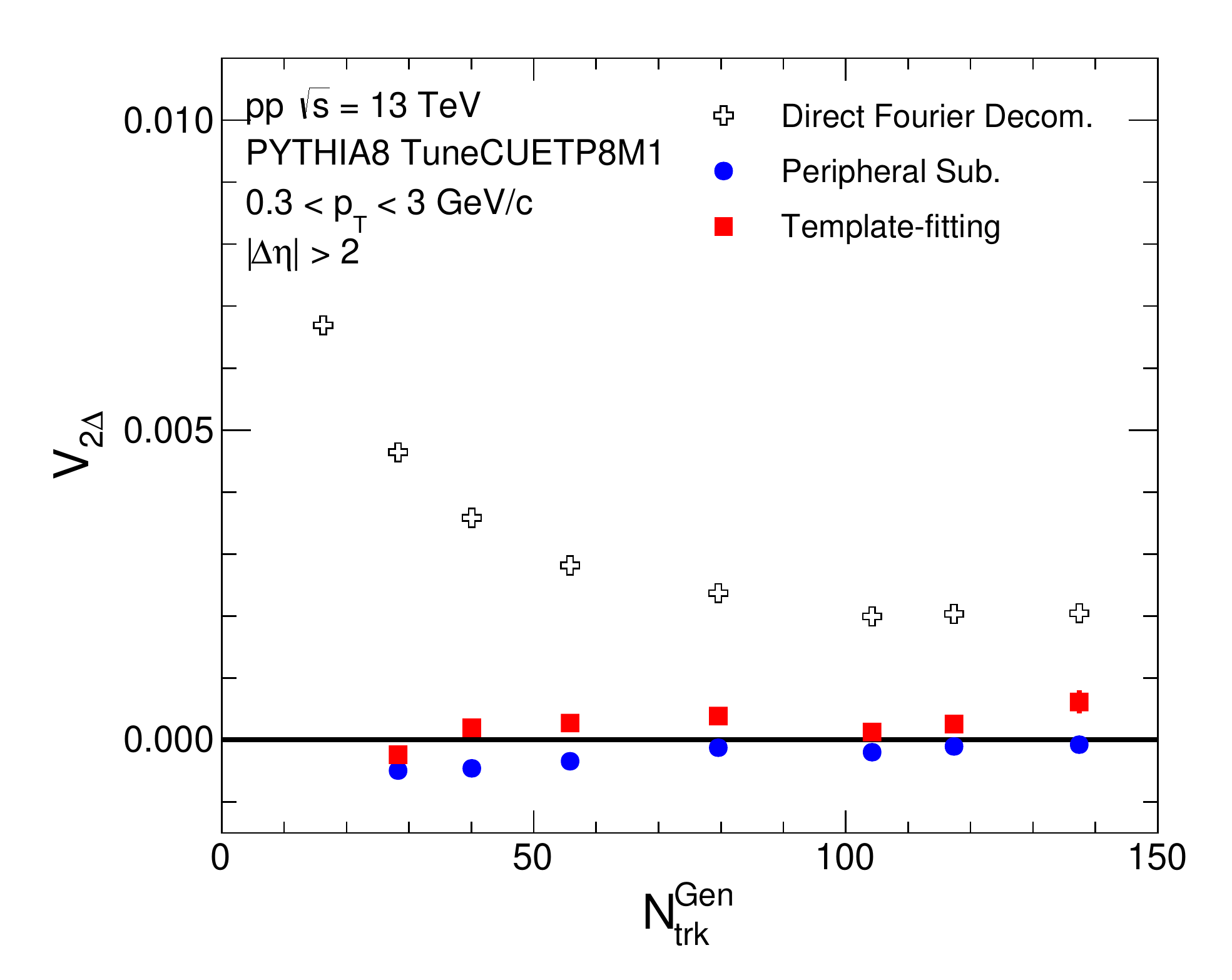}
  \caption{ \label{fig:SubCompare}
     The $V_{2\Delta}$ results as function of \ngen\ from direct Fourier decomposition, 
     peripheral subtraction and template-fitting for PYTHIA8 simulation.}
\end{figure}

The two methods of jet subtraction are tested in PYTHIA8 pp simulations.
Due to statistical limitation, tests are done for the second order harmonics $v_2$, 
while the behavior of higher harmonics are expected to be the same. 
As there is no collective correlation in PYTHIA8 events, it is expected that $v_2$ results are consistent with 0 after applying the two methods. 
Figure.~\ref{fig:SubCompare} shows $V_{2\Delta}$ results as function of \ngen\ (defined as number of generator level charged particles) 
from direct Fourier decomposition and after applying the low multiplicity subtraction and template-fitting.
The two methods work equally well while a slight over-subtraction for peripheral subtraction and a slight under-subtraction for template-fitting. 

\begin{figure}[thb]
\centering
\includegraphics[width=0.45\textwidth]{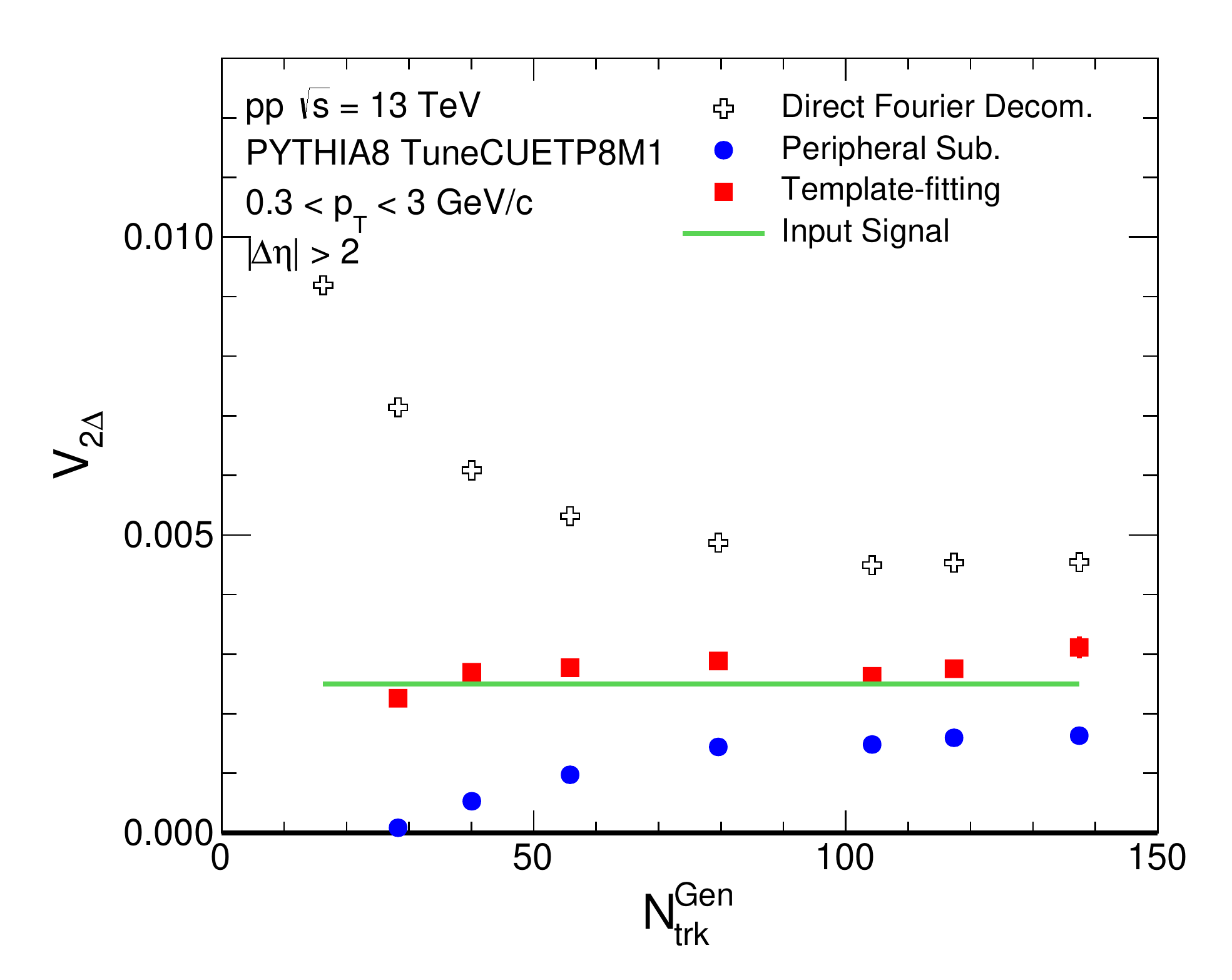}
\includegraphics[width=0.45\textwidth]{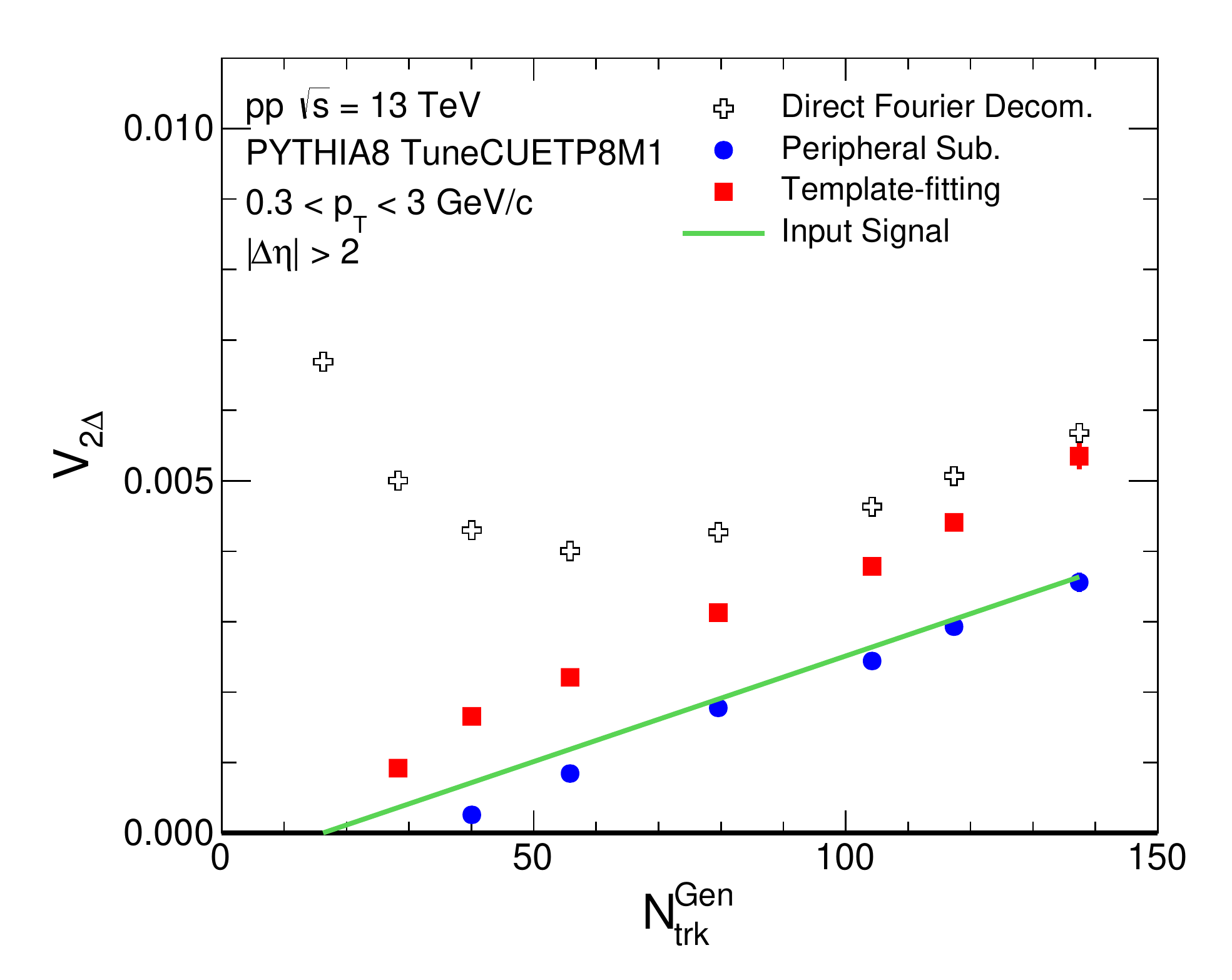}
  \caption{ \label{fig:SubCompareWithSignal}
     The $V_{2\Delta}$ results as function of \ngen\ from direct Fourier decomposition, 
     peripheral subtraction and template-fitting for PYTHIA8 simulation 
     with constant additional $V_{2\Delta}$ signal (left) and an additional $V_{2\Delta}$ 
     signal that increases with multiplicity (right).}
\end{figure}

To further investigate the behavior of the two methods, 
additional $V_{2\Delta}$ signal is put into 1D azimuthal correlation function by adding a $2N^\text{assoc}V_{2\Delta} \cos(2\Delta\phi)$ 
term to mimic the collective contribution in real data. 
Additional $V_{2\Delta}$ signal with different multiplicity dependence are added according to the $v_{2}$ results reported by ATLAS and CMS.
Results are shown in Figure.~\ref{fig:SubCompareWithSignal} for a constant additional $V_{2\Delta}$ signal (case 1) and an additional $V_{2\Delta}$ signal (case 2) that increases with multiplicity. 

In case 1, results from low multiplicity subtraction have smaller values than the input signal. 
As discussed in Sec.~\ref{subsec:CMSlimit}, there is an assumption made in the low multiplicity subtraction method that there is 
negligible collective correlation in very low multiplicity events. 
Any sizeable collective correlation in low multiplicity events results in an over subtraction of collective signal at higher multiplicity.
On the 	other hand, low multiplicity subtraction provides accurate $V_{2\Delta}$ signal results in case 2 when the assumption is fulfilled, 
while template-fitting yields results larger than the true value. 

The template-fitting method provides results consistent with the input signal in case 1, where equal amount of collective contribution is present at the low and high multiplicity region, as discussed in Sec.~\ref{subsec:ATLASlimit}. 
In case 2, the results after jet subtraction tends to be larger than the input signal, which is understood to be an under-subtraction. 

The above studies indicate that $v_n$ results reported by ATLAS and CMS after jet subtraction are model dependent. 
There are potential under-subtraction in ATLAS results and over-subtraction in CMS results. 
These effects are prominent at low multiplicity and high \pt\ region where jet contribution is large, and are reduced at high multiplicity and low \pt\ region.
Without further constrains on the magnitude of collective correlation at low multiplicity region, 
one should consider ATLAS results as the upper limit of the true collective $v_n$, and CMS results as the lower limit. 
\clearpage

\section{Summary}
\label{sec:resultppSum}
The CMS detector has been used to measure two-particle azimuthal correlations
with $\PKzS$, $\PgL/\PagL$ and inclusive charged particles
over a broad pseudorapidity and transverse momentum range
in pp collisions at $\roots = 5$, 7, and 13 TeV. The elliptic ($v_2$) and triangular ($v_3$) flow
Fourier harmonics are extracted from long-range two-particle correlations.
After subtracting contributions from back-to-back jet correlations estimated using
low-multiplicity data, the $v_2$ and $v_3$ values are found to increase with
multiplicity for $\noff \lesssim 100$, and reach a relatively constant value
at higher values of \noff. The \pt\ dependence of the $v_2$ harmonics in high-multiplicity pp events
is found to have no or very weak dependence on the
collision energy. In low-multiplicity events, similar $v_2$ values as a function of
\pt\ are observed for inclusive charged particles, $\PKzS$ and \PgL/\PagL,
possibly reflecting a common back-to-back jet origin of the correlations for all particle species.
Moving to the higher-multiplicity region, a mass ordering of $v_2$ is
observed with and without correcting for jet correlations. For $\pt \lesssim 2$ GeV/c,
the $v_2$ of $\PKzS$ is found to be larger than that of \PgL/\PagL. This behavior
is similar to what was previously observed for identified particles
produced in pPb and AA collisions at RHIC and the LHC. 
These observations provide strong evidence supporting the interpretation of a collective origin for the observed long-range correlations in high-multiplicity pp collisions.

\cleardoublepage
\chapter{ Conclusion and outlook}
\label{ch:conclusion}

In this thesis two analyses on two-particle correlation and azimuthal anisotropy $v_n$ in pPb and pp collisions were discussed. 
Comparing to the profound knowledge of the existence and dynamic properties of the hot and dense medium created in nucleus-nucleus (AA) collisions, 
the underlying mechanism for the observed correlation phenomena in smaller collision systems remains poorly understood. 
A better understanding requires detailed study of the properties of the $v_n$ in pPb and pp collisions. 
In particular, their dependence on particle species, and other aspects related to their possible collective nature, are the key to scrutinize various theoretical interpretations. 

The measurements described here are carried out with three different collision systems, pp, pPb and PbPb, using the data collected by CMS detector at the LHC. 
With the implementation of a dedicated high-multiplicity trigger, the pp and pPb data sample gives access to multiplicity comparable to those in semi-peripheral PbPb collisions. 
Detailed study of two-particle azimuthal correlations with unidentified charged particles, as well as correlations of reconstructed $\PKzS$ and \PgL/\PagL\ particles are performed in pPb collision of total integrated luminosity of 35 nb$^{-1}$ at nucleon-nucleon center of mass energy of 5.02 TeV, and in pp collisions of total integrated luminosity of 1.0 pb$^{-1}$, 6.2 pb$^{-1}$ and 0.7 pb$^{-1}$ at center of mass energy of 5 TeV, 7TeV and 13 TeV, respectively. 
The results of $v_2$ and $v_3$, extracted from two-particle correlations, are studies as function of particle \pt\ and event multiplicity. 
In pp collisions, the residual contribution to long-range correlations or back-to-back jet correlations is estimated and removed by subtracting correlations obtained from very low multiplicity pp events. 
To examine the validity of constituent quark number scaling, $v_2/n_q$ and $v_3/n_q$ are obtained as function of $KE_T/n_q$ for both $\PKzS$ and \PgL/\PagL\ particles. 

In pp collisions, the $v_2$ and $v_3$ values of inclusive charged particles are found to increase with
multiplicity for $\noff \lesssim 100$, and reach a relatively constant value
at higher values of \noff. 
Comparing to results in pPb and PbPb collisions, a strong system size dependence is observed in the $v_2$ results, where smaller system shows smaller $v_2$ at same multiplicity. 
On the other hand, $v_3$ results show no or very weak dependence on system size between pPb and PbPb collisions, while the $v_3$ is generally smaller in pp collision at high multiplicity. 
These observations provide constrains on the initial state fluctuation of the three systems. 

In high multiplicity pPb and pp collisions, a mass ordering of $v_2$ and $v_3$ is observed at low \pt\ region of $\lesssim 2.5$ GeV/c. 
This behavior is similar to what was previously observed for identified particles produced in AA collisions at RHIC and the LHC, which is understood to be developed during the hydro expansion of the perfect fluid medium.
However, in small collision systems, the observation can also be qualitatively explained by AMPT model through parton scattering, UrQMD model through hadronic interaction and Color Glass Condensate (CGC) + fragmentation model. 
Nevertheless, all the possible interpretations point to a collective system. 

Furthermore, constituent quark number (NCQ) scaling is found to hold for $\PKzS$ and \PgL/\PagL\ particles for $v_2$ in high multiplicity pp and pPb collisions, and for $v_3$ in high multiplicity pPb collision. 
The observation is reminiscent of the NCQ scaling among large amount of particle species in AA collisions, which is conjectured to be related to quark recombination, hence considered as an evidence of deconfinement and the existence of strongly interacting medium. 
At this point, although the NCQ scaling in small collision systems needs more precise measurement over more particle species to be conclusive, it can already shed light on the underlying mechanism of particle production in small collision system.

The results presented in this thesis, together with other cutting edge measurements in high multiplicity small collision systems in the heavy ion community, strongly point to a collective nature of the correlation developed in those collisions.
However, whether there is Quark Gluon Plasma formed in the small but dense collisions and how the systems evolve are still open questions to the future.

One important observation for Quark Gluon Plasma that is missing in small collision systems is the jet medium interaction. 
If the collectivity observed in small system is suggestive of a strongly interacting medium formed in small collision systems, 
one expects jet quenching like in AA collisions. 
The parton energy loss in a QGP medium is expected to depend on temperature T and path length L (or equivalently the system size), 
$\Delta E \sim T^{3}L^{2}$~\cite{Baier:1996kr,Zakharov:1997uu}. 
Comparing to AA collisions at same multiplicity, although the system size is smaller in pp and pA collisions, 
the temperature is larger since smaller system possesses a higher entropy density. 
Therefore, for high-multiplicity pp and pA collisions, the parton energy loss should be comparable to that for peripheral AA collisions. 
Recent calculations~\cite{Shen:2016egw}, which combine jet energy loss Monte Carlo with a hydrodynamic background describing the bulk of pPb collisions, 
has predicted significant parton energy loss in 0-1\% central pPb collisions. 
However, the experimental search for jet quenching in small systems is difficult due to the non-trivial correlation between underlying event multiplicity and jet production~\cite{ATLAS:2014cpa}. 
On the other hand, a measurement of $v_2$ at very high \pt\ using multi-particle cumulants might reveal possible path length dependence of parton energy loss as was done in AA collisions~\cite{Sirunyan:2017pan}.

Further interesting question to ask is whether the collectivity extends to non-hadronic collisions. 
While high-multiplicity final state seems to be necessary for collectivity, the initial colliding particles might not be restricted to hadronic but also for electromagnetic probes. 
Observation of collective flow in high-multiplicity electron-proton, electron-nucleus and even electron-positron collisions will open up unique opportunities for studying of many body QCD system.
It remains to be seen in what collision systems and at what final state multiplicities the QCD vacuum can be excited to flow collectively like a perfect fluid. 

\cleardoublepage

\appendix
\chapter{Data sample names}
\label{apx:dataname}

Tab.~\ref{tab:datasamples7TeVpp}-~\ref{tab:data5TeV} summarises the names of the data samples for 5, 7, 13 TeV pp collisions, 5 TeV pPb collisions and 2.76 TeV PbPb collisions. 

\begin{table}[h]\renewcommand{\arraystretch}{1.2}\addtolength{\tabcolsep}{-1pt}
\centering
\caption{Official data samples for 7 TeV collisions analysis.}
\begin{tabular}{ l | l }
\hline
Run Range & Dataset  \\
\hline
132440 - 135735 & /MinimumBias/Commissioning10-May19ReReco-v1/RECO  \\
\hline
135808 - 144114 & /MinimumBias/Run2010A-Apr21ReReco-v1/RECO  \\
\hline
144919 - 149711 & /MinimumBias/Run2010B-Apr21ReReco-v1/RECO \\
\hline
\end{tabular}
\label{tab:datasamples7TeVpp}
\end{table}

\begin{table}[h]\renewcommand{\arraystretch}{1.2}\addtolength{\tabcolsep}{-1pt}
\centering
\caption{Official data samples for 5 TeV pPb collisions}
\begin{tabular}{ l | l }
\hline
Run & Dataset  \\
\hline
210498--210658 & /PAHighPt/HIRun2013-28Sep2013-v1/RECO \\
210498--210658 & /PAMinBiasUPC/HIRun2013-28Sep2013-v1/RECO \\
210676--211631 & /PAHighPt/HIRun2013-PromptReco-v1/RECO \\
210676--211631 & /PAMinBiasUPC/HIRun2013-PromptReco-v1/RECO \\
\hline
\end{tabular}
\label{tab:datasamples5TeVpPb}
\end{table}

\begin{table}[ht]\renewcommand{\arraystretch}{1.2}\addtolength{\tabcolsep}{-1pt}
\centering
\caption{Official data samples for 2.76 TeV PbPb collisions}
\begin{tabular}{ l | l }
\hline
Run & Dataset  \\
\hline
181611--183013 & /HIMinBiasUPC/HIRun2011-12Jun2013-v1/RECO \\
\hline
\end{tabular}
\label{tab:datasamples_PbPb}
\end{table}

\begin{table}[h]\renewcommand{\arraystretch}{1.2}\addtolength{\tabcolsep}{-1pt}
\centering
\caption{Official data samples for 13 TeV analysis.}
\begin{tabular}{ l | l }
\hline
Run Range & Dataset  \\
\hline
251721 & /ZeroBias\{1-8\}/Run2015B-PromptReco-v1/AOD  \\
(July EndOfFill) & /HighMultiplicity/Run2015B-16Oct2015-v1/AOD  \\
 & /HighMultiplicity85/Run2015B-16Oct2015-v1/AOD  \\
\hline
254986--255031 & /L1MinimumBiasHF\{1-8\}/Run2015C-PromptReco-v1/AOD \\
(August Vdm scan) & /HighMultiplicity/Run2015C\_25ns-05Oct2015-v1/AOD \\
 & /HighMultiplicity85/Run2015C\_25ns-05Oct2015-v1/AOD \\
\hline
259152--259431 & /L1MinimumBiasHF\{1-8\}/Run2015D-PromptReco-v4/AOD \\
(October TOTEM run) & /HighMultiplicity/Run2015D-PromptReco-v4/AOD \\
\hline
\end{tabular}
\label{tab:data13TeV}
\end{table}

\begin{table}[h]\renewcommand{\arraystretch}{1.2}\addtolength{\tabcolsep}{-1pt}
\centering
\caption{Official data samples for 5 TeV analysis.}
\begin{tabular}{ l | l }
\hline
Run Range & Dataset  \\
\hline
262163--262273 & /MinimumBias\{1-20\}/Run2015E-PromptReco-v1/AOD \\
(reference run) & /HighMultiplicity/Run2015E-PromptReco-v1/AOD \\
\hline
\end{tabular}
\label{tab:data5TeV}
\end{table}

\cleardoublepage


\bibliographystyle{ieeetr}
\bibliography{Phd_bibliography}
\end{document}